\documentstyle[preprint,aps,harvard,floats,epsfig]{revtex}
\citationstyle{dcu}

\begin{document}
\tightenlines

\renewcommand{\thefootnote}{\fnsymbol{footnote}}

\begin{flushright}
hep-ph/9702381\\
revised version, as it appeared in:\\ 
Rev. Mod. Phys. {\bf 71}, 513-574 (1999)
\end{flushright}

\vspace{1.5cm}

{\centerline {\large \bf Top quark condensation}}

\vspace{1.cm}

{\centerline {G.~Cveti\v{c}}}\footnote[1]{e-mail address:
cvetic@doom.physik.uni-dortmund.de}

{\centerline {\em Institut f\"ur Physik, Universit\"at Dortmund,
44221 Dortmund, Germany}}
{\centerline {\em and Department of Physics, Universit\"at Bielefeld, 
33501 Bielefeld, Germany}}

\vspace{1.5cm}

\renewcommand{\thefootnote}{\arabic{footnote}}

\begin{abstract}

Top quark condensation, in particular the minimal framework 
where the neutral Higgs scalar is (predominantly) 
an effective ${\bar t} t$ condensate of the standard model,
is reviewed. Computational approaches
are compared and similarities, differences
and deficiencies pointed out. Extensions of the minimal
framework, including scenarios with 
two composite Higgs doublets,
additional neutrino condensates, and
${\bar t} t$ condensation arising 
from four-fermion interactions with 
enlarged symmetries, are described. 
Possible renormalizable models of underlying  
physics potentially responsible for the condensation,
including topcolor assisted technicolor frameworks, 
are discussed. Phenomenological implications of 
top condensate models are outlined.
Outstanding theoretical issues and problems for future
investigation are pointed out. 
Progress in the field after this article was accepted
has been briefly covered in a Note added at the end.

\end{abstract}

\vspace{1.cm}

\newpage

\noindent {\large \bf Explanatory list of abbreviations 
used in the paper}

\noindent BHL -- Bardeen, Hill and Lindner;
\noindent BS equation -- Bethe--Salpeter equation;
\noindent CDF -- Collider detector at Fermilab;
\noindent CKM matrix -- Cabibbo--Kobayashi--Maskawa matrix;
\noindent CGS -- Chivukula, Golden and Simmons;
\noindent DEWSB -- dynamical electroweak symmetry breaking;
\noindent d.o.f.'s -- degrees of freedom;
\noindent DS equation -- Dyson--Schwinger equation;
\noindent DSB -- dynamical symmetry breaking;
\noindent EW -- electroweak;
\noindent EWSB -- electroweak symmetry breaking;
\noindent FCNC -- flavor-changing neutral current;
\noindent FD -- flavor democracy;
\noindent IRFP -- infrared fixed point;
\noindent KM -- King and Mannan;
\noindent LPA -- local potential approximation;
\noindent LEP -- Large electron-positron collider;
\noindent LHC -- Large hadron collider;
\noindent MSM -- minimal standard model;
\noindent MSSM -- minimal supersymmetric standard model;
\noindent MTY -- Miransky, Tanabashi and Yamawaki;
\noindent NGB -- Nambu--Goldstone boson;
\noindent NJLVL -- Nambu--Jona-Lasinio--Vaks--Larkin;
\noindent NNT enhancement -- Nagoshi--Nakanishi--Tanaka enhancement;
\noindent NPRG -- nonperturbative renormalization group;
\noindent NTL -- next-to-leading (in $1/{N_{{\rm c}}}$ 
expansion);
\noindent PQ symmetry -- Peccei--Quinn symmetry;
\noindent PS relation -- Pagels--Stokar relation;
\noindent RG -- renormalization group;
\noindent RGE -- renormalization group equation;
\noindent SLC -- Stanford (SLAC) Linear Collider;
\noindent SM -- standard model;
\noindent SSB -- spontaneous symmetry breaking;
\noindent SUSY -- supersymmetry;
\noindent TC -- technicolor;
\noindent TC2 -- topcolor assisted technicolor;
\noindent TSM -- top-mode standard model;
\noindent 2HDSM -- two-Higgs-doublet standard model;
\noindent 2HDSM(II) -- two-Higgs-doublet standard model type II;
\noindent VEV -- vacuum expectation value.   

\newpage

\tableofcontents

\section{Introduction}
\label{Intr}
\setcounter{equation}{0}

The most important outstanding problem in the standard 
model (SM)
of strong and electroweak interactions 
is the source of the electroweak
symmetry breaking (EWSB), and the related problem of
fermion mass hierarchies.
The EWSB mechanism should explain the masses of 
EW gauge bosons $Z$ and $W$, 
and at the same time
the generation of masses of fermions, particularly the
heavy quarks. An understanding of the origin of
EWSB is expected to provide a window to
physics beyond the SM.

The SM, as currently defined, contains an elementary
$SU(2)_L$-doublet Higgs scalar sector which is appended to the
model in an {\em ad hoc\/} way. 
In order to accommodate the gauge boson
masses $M_Z$ and $M_W$ while keeping
the model formally renormalizable, a Higgs field mass term
and a quartic self-interaction term are
added to the Lagrangian density.
Parameters associated with those
terms are adjusted to obtain a nonzero EW vacuum 
expectation value (VEV) for the neutral CP-even component of 
the Higgs field. The VEV is then responsible
for the nonzero $Z$ and $W$ masses.
This is called the Higgs mechanism.
Further, when Higgs-fermion-antifermion
Yukawa terms are added to the model
with {\em ad hoc\/} coupling strengths, 
the nonzero VEV leads to the masses of fermions.

The outlined Higgs mechanism 
of the SM is unsatisfactory in several
ways. The origin of a Higgs field 
is not explained in a fundamental way. This is
disturbing since the residual Higgs particle 
(${\cal {H}}$) has yet to be observed. Further, emergence of the
required nonzero VEV is obtained
simply by adjusting parameters
in the Higgs potential. Another disturbing feature
is that the known values of fermion masses are obtained
in an {\em ad hoc\/} manner, by adjusting phenomenologically
introduced Yukawa parameters. Also, the Higgs sector of the
SM requires extreme fine-tuning in order to preserve the
perturbative renormalizability of the model.
Namely, loop-induced corrections
to the mass of the Higgs (and hence to its VEV) grow 
violently ($\propto$$ {\Lambda}^{\!2}$) 
when the ultraviolet cutoff 
${\Lambda}$ of the theory increases. 
To maintain the renormalizability in a formal sense
(i.e., with ${\Lambda}\!\to\! \infty$), an extreme fine-tuning
is needed to cancel the various 
${\Lambda}^{\!2}$-terms, or equivalently,
the bare mass parameter $M^2_{\Phi}(\Lambda)$ 
of the Higgs isodoublet must be fine-tuned
at each loop order in the perturbative expansion.

One solution to some of the above problems is the introduction of
supersymmetry. In that case fine-tuning of the bare
mass parameters need be performed only once, at the tree level. 
The mentioned ${\Lambda}^{\!2}$-terms are then not
present at higher (loop) levels, due to the cancellation of 
the ${\Lambda}^{\!2}$ radiative contributions of particles and their 
superpartners. This solves to a large extent the problem of 
fine-tuning. In that case, the cutoff is
of the order of the grand unified scale
$E_{{\rm GUT}}$$\sim$$10^{16}$ GeV. Thus, a
``great desert'' in an energy interval $[E_{{\rm SUSY}},
E_{{\rm GUT}}] $$\sim$$ [10^4 \mbox{ GeV}, 
10^{16} \mbox{ GeV}]$ generally emerges in these
scenarios. While eliminating several of the
free parameters of the SM, supersymmetry introduces
many new parameters and elementary
particles which haven't been observed (yet). The Higgs field is
generally elementary in most supersymmetric frameworks.

The Higgs sector of the SM appears to be just an effective,
Ginzburg-Landau-type, description of low energy (SM) physics
represented by a composite (nonelementary) isodoublet scalar field(s).
This is the basic idea of top quark condensation,
as well as technicolor (TC) models. In such
frameworks, the Higgs isodoublet is a condensate
of fermion-antifermion pair(s), 
the constituents being predominantly $t_R$, $t_L$ 
and $b_L$, or pairs of technifermions. The focus of the present 
review article is such models involving primarily ${\bar t} t$
condensation, although scenarios including TC are 
also discussed.

Models involving ${\bar t} t$ condensation generally start
with minimal assumptions about physics beyond the
SM. They assume that the underlying physics above
a compositeness scale ${\Lambda}$ leads at energies
${\mu}$$\sim$${\Lambda}$ to effective four-quark
interactions strong enough to induce quark-antiquark
condensation into composite Higgs field(s), leading
thus to an effective SM at ${\mu} $$<$$ {\Lambda}$.
The attraction of such effective dynamical frameworks lies 
in the simplicity of their assumptions and in the fact 
that these frameworks can connect the
dynamical generation of the heavy top quark mass
and all or part of the (dynamical) EWSB. In principle,
such models can even lead to dynamical generation of
the lighter fermion masses.
Unfortunately, the more we want to explain, the more
assumptions about the four-fermion interactions we have to
make.

To explain the origin of the forementioned effective
four-fermion terms, 
several models of underlying dynamics
have been constructed in the literature. In most cases,
these models are renormalizable, and some are 
quite promising.
The present article
does not provide a detailed review and discussion of
such models. Instead, it concentrates more on
effective models of relatively simple four-fermion
interactions, and on methods of calculating
the dynamical generation of fermionic masses and
dynamical EWSB. Since the physical mechanism which
such methods investigate is nonperturbative, it is understandable
that the methods
are still relatively crude, and many outstanding
questions remain regarding their applicability.
In the opinion of this author, these questions are very
pressing. To make further progress,
a reliable and systematic method to
investigate condensation mechanisms, particularly in cases when
the compositeness scale ${\Lambda}$ is low ($<$$ 10^8$ GeV), 
phenomenologically more attractive, is required.

In Section~\ref{MRGE}, we review the simplest 
(minimal) framework, in which ${\bar t} t$ condensation
alone is assumed to be responsible for the generation
of the full dynamical EWSB and top quark mass.
We also discuss studies of that phenomenon involving
perturbative renormalization group equations (RGE's). 
These methods are used 
in the literature to investigate condensation frameworks,
and were applied prominently in the original minimal framework.
The actual dynamics of condensation
in such an approach is either not investigated directly,
or examined only schematically --
in the lowest, quark loop, approximation. 
The latter approximation then provides
compositeness boundary conditions for RGE's at
an energy ${\mu}$$\sim$${\Lambda}$, due to a
strong attraction in the top quark sector.
For very large values of ${\Lambda}$ 
(${\Lambda} $$\stackrel{>}{\sim}$$ 10^8$ GeV
in the minimal framework), low energy predictions 
$m_t^{\rm phys.}$ and $m_H^{\rm phys.}$ do not depend
on details of the actual physical condensation
mechanism, due to the infrared fixed-point behavior of the 
relevant perturbative RGE's. This
feature makes the method relevant for the
${\bar t} t$ condensation program, but only
for large values of ${\Lambda}$.
The method can be used for any realization of
an effective strong attractive interaction, not just
four-fermion terms.

In Section~\ref{MDSPS}, we describe a different approach for
studying quark condensation effects. 
This method employs Dyson--Schwinger (DS) integral
equations to relate dynamically
generated quark masses with the strengths of the
four-quark terms, as well as the Pagels--Stokar (PS) equations
relating the dynamical quark masses with the 
decay constant $F_{\pi}$ of the Nambu--Goldstone bosons 
(NGB's). PS relations are
closely related to the Bethe--Salpeter (BS) equations
for the bound state of NGB.
The method is illustrated within
the minimal ${\bar t} t$ condensation framework 
(where $F_{\pi}$$=$ EW VEV),
and within its extension involving
also ${\bar b} b$ condensation.
The approach described in this Section, in
contrast to the RGE method of Section~\ref{MRGE}, addresses
the strong dynamics of the condensation mechanism more directly. 
However, the integral DS equations are complicated, and a full
DS$+$PS analysis has been performed in the literature
only in the leading-$N_{{\rm c}}$ (quark loop $+$ QCD)
approximation. Further, in any known systematic
approximation scheme (e.g., $1/N_{\rm c}$ expansion),
this method is fraught with the problem of gauge
noninvariance.

In Section~\ref{NTLEE}, we discuss
next-to-leading (NTL) effects 
in the $1/N_{{\rm c}}$ expansion, within
the minimal condensation framework.
The method of calculating and minimizing the effective
potential yields a ``hard bare mass''
(i.e., nonvariational) version of the DS equation,
usually called the gap equation. This equation,
combined with subsequent renormalization of the
bare dynamical ``hard mass'' $m_t({\Lambda})$, is
expected to be a good approximation to the corresponding
variational NTL version of the DS equation
involving a running dynamical mass ${\Sigma}_t({\bar p}^2)
$$\equiv$$ m_t(|{\bar p}|)$, as long as the compositeness scale
${\Lambda}$ is not very high. 
A full systematic NTL version of the
Bethe--Salpeter (BS) equation and its application
to the $t {\bar t}$ condensate
framework has not appeared in the literature.
Inclusion of the latter equation would be needed
for a full NTL analysis of the condensation. Besides a fully
nonperturbative renormalization group (NPRG) approach 
(proposed in the literature, cf.~Sec.~\ref{MRGE5}, 
but not yet applied to ${\bar t} t$ condensation),
the $1/N_{{\rm c}}$ expansion -- an extension of 
approaches of DS$+$PS-type beyond the 
leading-$N_{{\rm c}}$
level -- appears to be at this time the
only viable approach to calculate condensation
effects when the compositeness scale is not 
very high: ${\Lambda} $$<$$ 10^8$ GeV.

In Section~\ref{CCVAMF}, we compare and comment 
on the various approaches to calculating condensation effects.
We also review work on mass-dependent 
perturbative RGE approach,
and discuss possible contributions of
four-quark interactions with dimension higher than six.

The minimal $t {\bar t}$ condensation
framework appears to be
ruled out, since it predicts a too high mass
$m_t$ when the full dynamical ${\bar t} t$-induced EWSB is 
implemented. For that reason,
many extensions of the minimal framework
have been investigated in the literature
-- cf.~Secs. \ref{EWESG}--\ref{SPTAQ}.
Extensions usually involve, in addition to
${\bar t} t$, other condensates 
which also contribute to the EWSB (i.e., to EW
VEV $v $$\approx$$ 246$ GeV).
This feature can bring the mass $m_t^{\rm dyn.}$ 
down to acceptable values. 
Beyond understanding $m_t$ {\em and\/} EWSB dynamically,
other reasons for investigating extensions
include hopes to understand and/or predict
dynamically the mass spectra of fermions other than $t$,
CKM mixing, parity violation, larger spectra of composite
particles, and breaking of higher gauge symmetries.

In Section~\ref{EWESG}, we review 
generalizations of the minimal condensation framework
which don't involve an extension of the SM gauge group --
among them effective four-quark
scenarios with two Higgs doublets, 
colored composite scalars, and effective scenarios involving,
in addition to heavy quarks, 
leptonic condensation.

In Section~\ref{EGSG}, we discuss works on effective 
four-fermion interaction models 
with extended symmetries and the 
resulting dynamical symmetry breaking (DSB) patterns.

In Section~\ref{RMUP}, we review
those extensions of the minimal framework
which embed ${\bar t} t$ condensation
in fully renormalizable frameworks of underlying physics.
At some compositeness scale $\sim$${\Lambda}$,
these models become strongly coupled and effectively 
lead to strong four-fermion terms and thus to
condensation. We include a discussion
of topcolor assisted technicolor (TC2). These models
combine features of ${\bar t} t$ condensation
and those of the technicolor (TC) -- TC is
here predominantly responsible for the
dynamical EWSB and for giving masses $\sim$${\Lambda}$
to the topcolor gauge sector, and the latter
leads to effective four-quark terms and to
${\bar t}t$ condensation and $m_t^{{\rm dyn.}}$.

In Section~\ref{SPTAQ}, we discuss phenomenological
predictions of scenarios involving ${\bar t} t$ condensation,
including TC2 scenarios. We also describe
some outstanding theoretical questions arising in
condensation frameworks.

Section~\ref{SO} contains a short summary,
outlines prospects for further development, and
emphasizes some remaining unresolved problems.

\section{The method of renormalization group equations}
\label{MRGE}
\setcounter{equation}{0}

\subsection{The starting Lagrangian -- truncated top-mode
standard model}
\label{MRGE1}
Bardeen, Hill and Lindner (BHL) (1990),
motivated by an idea of Nambu (1989)\footnote{
A brief discussion of Nambu's ``bootstrap'' idea for 
${\bar t} t$ condensation is included in Sec.~\ref{EWESG2}.
} 
and by works of Miransky, Tanabashi and Yamawaki
(MTY) (1989a, 1989b),\footnote{
Works and methods applied by MTY (1989a, 1989b)
are discussed in Sec.~\ref{MDSPS}.}
considered top quark condensation induced by a truncated 
four-quark interaction involving only the heaviest generation
of quarks\footnote{
In fact, a prediction for the mass of $t$
as a dynamical mass arising via condensation
was first made by Terazawa (1980): 
$m_t \approx 132$ GeV, and $m_H \approx 264$ GeV.   
This prediction was based on a sum rule
within a specific model where the Higgs and all the gauge
bosons (including their transverse components) are composite
-- collective excitations of lepton--antilepton or
quark--antiquark pairs (Terazawa, Akama, and Chikashige, 1977).
This footnote is the only part of this manuscript 
that is {\em not\/} included in the published version 
[Rev. Mod. Phys. {\bf 71}, 513 (1999)].
The author apologizes to Terazawa for
having noted his paper (Terazawa, 1980) 
too late to include it in the published version.}
\begin{equation}
{\cal {L}}^{(\Lambda)}={\cal {L}}_{{\rm kin}}^0
+G \left( {\overline \Psi} _{{\rm L}}^{ia}
 t_{{\rm R}}^a\right) 
 \left( \bar t_{{\rm R}}^b
 \Psi _{{\rm L}}^{ib}\right) \ ,
\label{TSM}
\end{equation}
where $a$ and $b$ are color and $i$ isospin indices,
and $\Psi_L^T $$=$$ (t_L, b_L)$. ${\cal {L}}_{{\rm kin}}^0$
represents the usual gauge-invariant kinetic terms for fermions and
gauge bosons. 
The four-quark term is also invariant under the gauge group
$SU(3)_c\!\times\!SU(2)_L\!\times\!U(1)_Y$ of the standard model (SM);
here, the subscript $c$ denotes the color of the strong
interaction, $L$ refers to the left-handed sector, and $Y$ to
the hypercharge sector of the electroweak interactions.
As denoted
above, the model assumes a finite upper cutoff scale ${\Lambda}$
where ${\bar t} t$ condensation is supposed to occur,
and all constants and fields in (\ref{TSM}) are the 
``bare'' quantities in this theory: $G $$=$$ G({\Lambda})$,
${\Psi}_L  $$=$$ {\Psi}_L^{({\Lambda})}$, etc. The four-quark term
is assumed to have a strong coupling parameter
${\Lambda}^{\!2} G $ responsible for creation 
of a composite Higgs doublet ${\Phi}$:
\begin{equation} 
\Phi \propto  \frac{1}{2} \left(
\begin{array}{c}
\bar b^a (1+\gamma_5) t^a \\
- \bar t^a (1+ \gamma_5) t^a
\end{array} \right) 
= i {\tau}_2 \left( {\bar t}_R {\Psi}_L \right)^{\dagger T} \ ,
\quad {\tilde {\Phi}} \equiv i {\tau}_2 {\Phi}^{\dagger T}
\propto {\bar t}_R {\Psi}_L \ .
\label{HD1}
\end{equation}
Other four-quark terms omitted in (\ref{TSM}) are assumed to be
unimportant. That assumption is based on the belief that
$t$, being by far the heaviest known quark, is perhaps
the only one responsible for generating
its own mass and possibly also for 
the electroweak symmetry breaking
($\langle {\cal {H}} \rangle_0 $$\approx$$ 246$ GeV). 
The discussion in Sec.~\ref{MDSPS} will
substantiate this assumption.
Framework (\ref{TSM}), called also the (truncated) top-mode
standard model (TSM), is a specific version of the
Nambu--Jona-Lasinio--Vaks--Larkin (NJLVL) type
\cite{NJL61,VaksLarkin61},
but regarded
here not as a low energy QCD framework 
(${\Lambda}$$\sim$$10^2$ MeV), but a framework for the dynamical
electroweak symmetry breaking (DEWSB, ${\Lambda} $
$\stackrel{>}{\sim}$ $1$ TeV).\footnote{
In this article, we will refer to dimension six
four-fermion contact interactions, i.e., those
without any derivatives, as NJLVL terms.}
The underlying physics responsible
for the four-quark term (\ref{TSM})
is not specified. It could, for example, be a
theory with massive
gauge bosons ($M$$\sim$${\Lambda}$) that couple strongly to the    
third generation quarks.

The Lagrangian density (\ref{TSM}) of the truncated TSM
can be rewritten with an
additional, auxiliary, scalar $SU(2)_L$-isodoublet 
$\Phi $, by adding to (\ref{TSM}) a quadratic term
\begin{eqnarray}
{\cal {L}}^{(\Lambda)}_{{\rm new}}& = &
{\cal {L}}^{(\Lambda)}_{{\rm old}}-\left[ 
 M_0 \tilde \Phi ^{i\dagger }+\sqrt{G}
 {\overline \Psi} _{{\rm L}}^{ia} 
 t_{{\rm R}}^a\right] \left[ 
 M_0{\tilde \Phi }^i+\sqrt{G}\bar 
 t_{{\rm R}}^b{\Psi }_{{\rm L} }^{ib}
 \right] 
\nonumber \\
& = & i {\overline \Psi} {\partial \llap /} \Psi
- M_0 \sqrt{G} \left[ {\overline \Psi}_L {\tilde \Phi} t_R
+ \bar t_R {\tilde \Phi}^{\dagger} \Psi_L \right]
- M_0^2 {\tilde \Phi}^{\dagger} {\tilde \Phi} + \cdots \ , 
\label{TSM2}
\end{eqnarray}
\begin{equation}
\mbox{where:}\quad 
\tilde \Phi =i\tau _2\Phi ^{\dagger T} \ ,\qquad
\Phi =\frac 1{\sqrt{2}}\left( 
\begin{array}{c}
\sqrt{2}{\cal {G}}^{(+)} \\ {\cal H}+i{\cal {G}}^{(0)}
\end{array}
\right) \ ,\qquad 
{\cal {G}}^{(\pm)}= \frac{1}{\sqrt{2}}
 ({\cal {G}}^{(1)} \pm i {\cal {G}}^{(2)}) \ ,
\label{TSM2not}
\end{equation}
and the dots in (\ref{TSM2}) denote the terms containing
EW gauge bosons and gluons.
Addition of such a term changes the generating functional
only by a source-independent factor
\cite{Kugo76,Kikkawa76},
and is therefore assumed to lead to physics equivalent to 
(\ref{TSM}).\footnote{
This is true in the present case. However, in other types of
NJLVL models, introduction of one auxiliary isodoublet
${\Phi}$ may not always lead to a physics equivalent to
those four-fermion NJLVL interactions, 
as pointed out by~\citeasnoun{Dudas93}.
Various ans\"atze for the scalar sector
constrain us in fact to various specific 
condensation scenarios. Only the
scenario with the lowest value of the energy density 
(the vacuum) should materialize 
(cf. discussion in Sec.~\ref{SPTAQ2}).}
Here, ${\cal H}$, ${\cal G}^{(0)}$, ${\cal G}^{(1)}$ and 
${\cal G}^{(2)}$ are the Higgs and the three real Nambu--Goldstone 
components of the auxiliary complex isodoublet field $\Phi $, 
and $M_0$ is an unspecified bare mass parameter (usually taken:
$M_0 $$\sim$$ {\Lambda}$) for $\Phi $ at 
$\mu $$\sim$$ \Lambda $.  
Physical results will turn out to be independent of the
specific value of $M_0$.
These auxiliary fields become 
through quantum effects the physical bound state Higgs and 
the ``scalar'' longitudinal components of EW bosons
at energies $\mu$$>$${\Lambda}$. 
Equations of motion for the
scalars in (\ref{TSM2}) indeed reveal the composite structure  
(\ref{HD1})
\begin{equation}
\Phi = \frac{ \sqrt{G} }{ 2 M_0 } \left(
\begin{array}{c}
\bar b^a (1+\gamma_5) t^a \\
- \bar t^a (1+ \gamma_5) t^a
\end{array}
\right) = \frac{\sqrt{G}}{M_0} ( i {\tau}_2 ) \left( {\bar t}_R
{\Psi}_L \right)^{\dagger T} \ \Rightarrow \ {\tilde \Phi} =
- \frac{\sqrt{G}}{M_0} \left( {\bar t}_R {\Psi}_L \right)
\ .
\label{HD2}
\end{equation}

\subsection{Quark loop approximation}
\label{MRGE2}
\begin{figure}[htb]
\mbox{}
\vskip4cm\relax\noindent\hskip-.8cm\relax
\includegraphics{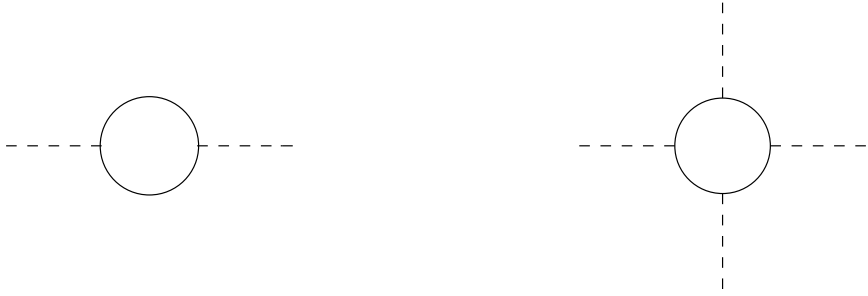} \vskip0.4cm
\caption{\footnotesize The relevant graphs contributing to
induced terms of composite scalars in quark loop
approximation. Full lines represent the top quark,
and dashed lines the (composite) scalars.}
\label{rmd1f}
\end{figure}

To illustrate how the truncated TSM 
transforms into the minimal Standard
Model (MSM) as the cutoff $\mu$ decreases below
${\Lambda}$, it is instructive to consider the
quark loop approximation, i.e., ignore
loops of particles other than quarks, as well as
radiative contributions of the composite
scalar sector (``feedback'' effects). The approximation
amounts to truncating the corresponding 
$1/N_{{\rm c}}$ expansions to the leading-$N_{{\rm c}}$ 
terms and without QCD, where
$N_{{\rm c}} $$=$$ 3$ is the number of quark colors.
Integrating out the heavy quark components at one-loop level
in energy interval $[\mu, \Lambda]$ (cf.~Fig.~\ref{rmd1f}), 
ignoring terms $\sim$$1$ when compared with
$\ln ( {\Lambda}^{\!2}/\mu^2)$, and ignoring masses
of particles,\footnote{
Based on the 
assumption that $m/\Lambda$, $m/\mu $$\ll$$ 1$, where
$m$ is a typical particle mass.
As pointed out by~\citeasnoun{Bandoetal90},
this is not correct for the bare mass of the
scalar doublet $M_0 $$\sim$$ \Lambda$, and calculation
has to be modified. Modification
does not affect the leading-$N_{{\rm c}}$ quark loop
approximation, but it does affect the full analysis
by RGE's. See Sec.~\ref{CCVAMF3} for details.} 
leads to additional (induced) terms in the 
Lagrangian density:
\begin{eqnarray}
{\cal {L}}^{(\mu)} &= & 
i {\overline {\Psi}}{\partial \llap /} {\Psi}
- M_0 \sqrt{G} 
\left[ {\overline \Psi}_L {\tilde \Phi} t_R + \mbox{ h.c.}
\right] + \triangle {\cal {L}}_{{\rm gauge}}
\nonumber \\
& & + Z_{\Phi}(\Lambda;\mu) 
(D_{\nu} \Phi)^{\dagger} D^{\nu} \Phi
- M_{\Phi}^2(\Lambda;\mu) \Phi^{\dagger} \Phi
- \frac{ \lambda(\Lambda;\mu) }{2} 
\left( \Phi^{\dagger} \Phi \right)^2
+ \cdots \ .
\label{Lmu}
\end{eqnarray}
The dots represent the old terms with gauge bosons,
$\triangle {\cal {L}}_{{\rm gauge}}$ quark
loop corrections to gauge coupling constants, 
$D_{\nu}$ are the covariant derivatives.
In (\ref{Lmu}), the fields are still those from the
theory with the $\Lambda$ cutoff (\ref{TSM2}). 
$\Lambda$- and $\mu$-dependent parts of the two- and 
four-leg Green functions\footnote{
Green functions with more external legs are ``finite''
(${\Lambda}$- and $\mu$-independent for
$\Lambda,$ $\mu$$\gg$$ E_{{\rm ew}}$).}
of Fig.~\ref{rmd1f} give
\begin{eqnarray}
Z_{\Phi}(\Lambda;\mu) & = & 
\frac{ N_{{\rm c}} M_0^2 G }{ (4 \pi)^2 } \ln \left(
 \frac{ {\Lambda}^{\!2} }{ \mu^2 } \right) \ ; \qquad
M_{\Phi}^2(\Lambda;\mu) = 
M_0^2 - \frac{ 2 N_{{\rm c}} M_0^2 G }{ (4 \pi)^2 }
\left( {\Lambda}^{\!2} - \mu^2 \right) \ ;
\label{Lmunot1}
\\
\lambda(\Lambda;\mu) & = &  
\frac{ 2 N_{{\rm c}} (M_0^2 G )^2 }{ (4 \pi)^2 } \ln \left(
 \frac{ {\Lambda}^{\!2} }{ \mu^2 } \right) \ .
\label{Lmunot2}
\end{eqnarray}
Lagrangian density (\ref{Lmu}) can be brought into the canonical
form of the MSM by rescaling the scalar field
${\Phi} = \Phi^{(\mu)}/\sqrt{Z_{\Phi}(\Lambda;\mu) }$
\begin{eqnarray}
{\cal {L}}^{(\mu)} & = & 
i {\overline {\Psi}}{\partial \llap /} {\Psi}  
- g_t(\mu) \left[ {\overline \Psi}_L 
{\tilde \Phi}^{(\mu)} t_R + \mbox{ h.c.}
\right] + \triangle {\cal {L}}_{{\rm gauge}}
\nonumber \\
& & + (D_{\nu} \Phi^{(\mu)})^{\dagger} D^{\nu} \Phi^{(\mu)}
- m_{\Phi}^2(\mu) \Phi^{(\mu) \dagger} \Phi^{(\mu)} 
- \frac{ \lambda (\mu) }{2} 
\left( \Phi^{(\mu) \dagger} \Phi^{(\mu)} \right)^2
+ \cdots \ ,
\label{Lmunor}
\end{eqnarray}
\begin{eqnarray}
\mbox{where: } \qquad
g_t(\mu) & = &
\frac{ M_0 \sqrt{G}}{ \sqrt{Z_{\Phi}(\Lambda;\mu)} }   
= \frac{ (4 \pi) }{ \sqrt{N_{{\rm c}}} } 
\frac{1}{ \sqrt{ \ln({\Lambda}^{\!2}/\mu^2) } } \ ,
\label{Lmunornot1}
\\
m_{\Phi}^2 (\mu) & = & 
\frac{ M_{\Phi}^2 (\Lambda;\mu) }{ Z_{\Phi}(\Lambda;\mu) }
= \left[ \frac{8 \pi^2}{N_{{\rm c}} G} - 
({\Lambda}^{\!2} -\mu^2) \right]
\frac{2}{\ln({\Lambda}^{\!2}/\mu^2)} \ ,
\label{Lmunornot2}
\\
\lambda(\mu) & = & 
\frac{ \lambda(\Lambda;\mu) }{ Z_{\Phi}^2 (\Lambda;\mu) }
= \frac{ 32 \pi^2 }{ N_{{\rm c}} \ln({\Lambda}^{\!2}/\mu^2) } \ .
\label{Lmunornot3}
\end{eqnarray}
These parameters are to be 
interpreted as the running parameters of the 
MSM at the ``probe'' (cutoff) energy $E$$=$$\mu$ in quark loop
approximation,\footnote{
In quark loop approximation, quark fields don't
evolve with energy $\mu$: 
${\Psi}^{(\mu)} $$=$$ {\Psi}^{(\Lambda)}$ ($=$${\Psi}$).}
and in addition they are now interdependent. 
Here we see explicitly that in these physical
parameters the initial arbitrary scaling mass $M_0$ cancels out.
For $\mu $$\ll$$ \Lambda$, 
they can be applied in either the unbroken 
($m_{\Phi}^2 $$>$$ 0$) or broken phase ($m_{\Phi}^2 $$<$$ 0$).
The latter requires fine-tuning the
parameter $G$ close to $G_{{\rm crit.}} $$=$$
(8 \pi^2)/(N_{{\rm c}} {\Lambda}^{\!2})$: 
$ G $$ \stackrel{>}{\approx}$$ 
(8 \pi^2)/(N_{{\rm c}} {\Lambda}^{\!2})$. 
In Sec.~\ref{MDSPS}, we will see that
this is equivalent to turning on a 
nonzero solution $m_t $$\not=$$ 0$
of the so called gap equation in quark loop approximation.
Further, if we assume that
$\mu$$\sim$$ E_{{\rm ew}}$$\ll$$ \Lambda$,
and that we have full
DEWSB due to this composite Higgs only
[$m^2_{\Phi}(E_{{\rm ew}}) $$<$$ 0$],
then the low energy Higgs potential $V( \Phi )^{(\mu)}$
in (\ref{Lmunor}) has the minimum at
$\langle \Phi^{(\mu)} \rangle_0$$=$$ (0,v/\sqrt{2})^T$. 
Here, $v$ is the full EW vacuum expectation value (VEV):
$v $$=$$ 246$ GeV. At that point we have 
$m_{\Phi}^2 $$=$$ -$$ \lambda v^2/2$.
Then (\ref{Lmunornot1}) implies
\begin{eqnarray}
 m^2_t &=& \frac{g^2_t(\mu\!=\!m_t) v^2}{ 2} = \frac{8 \pi^2 v^2}
{N_{{\rm c}} \ln({\Lambda}^{\!2}/m_t^2) } 
\quad \Rightarrow \
v^2 = m_t^2 \frac{N_{{\rm c}}}{ 8 \pi^2}
\ln \left( \frac{{\Lambda}^{\!2}}{m_t^2} \right) \ .
\label{mt}
\\
m_H^2 &=& 
\frac{ \partial^2 V ( {\cal {H}}; 
{\cal {G}}^{(j)}\!=\!0 )^{(\mu=m_H)} }
{\partial {\cal {H}}^2} {\Big |}_{ {\cal {H}} = v }
= \lambda (\mu\!=\!m_H) v^2 =  
\frac{ 32 \pi^2 v^2}{ N_{{\rm c}} \ln({\Lambda}^{\!2}/m_H^2) } \ .
\label{mh}
\end{eqnarray}
Comparison of (\ref{mt}) and (\ref{mh}) yields 
for the Higgs and the top quark mass the relation
$m_H$$\approx$$2 m_t$,
valid for $\Lambda $$\gg$$ m_t,$$ m_H$.
This is a result of Nambu and Jona-Lasinio (1961)
and of Vaks and Larkin (1961).
In Sec.~\ref{MDSPS} we will see that this 
is in fact the Pagels--Stokar relation in the quark loop 
approximation. Note that in this approximation $m_t^2(\mu)
$$=$$ g_t^2(\mu) \langle \Phi^{(\mu)0} \rangle_0^2 =
M_0^2 G \langle \Phi^{(\Lambda)0} \rangle_0^2$
is not evolving with $\mu$. 
For $m_t $$=$$ 175$ GeV, (\ref{mt}) yields
$\Lambda $$\approx$$ 3.6\!\cdot\!10^{13}$ GeV.

\subsection{Renormalization group equations plus compositeness
conditions}
\label{MRGE3}
Bardeen, Hill and Lindner (BHL) (1990) performed a calculation
in the truncated TSM (\ref{TSM})\footnote{
We will sometimes refer to the truncated TSM 
framework as the minimal (${\bar t t}$ condensation) framework,
in accordance with terminology used in the literature.
Strictly speaking, the minimal framework is more general --
it is not restricted to the picture 
of four-quark interactions.}
by using a method involving perturbative RGE's
plus a compositeness condition (at scale ${\Lambda}$).
The latter condition was motivated by the quark loop
approximation described in the 
previous Section \ref{MRGE2}.\footnote{
The quark loop arguments of Sec.~\ref{MRGE2} were
presented for the truncated TSM first by BHL.}
This calculation relies on the following assumption:
The model behaves as the MSM (with certain additional relations)
at energies $\mu $$<$$ \mu_{\ast}$, where $\mu_{\ast}$ is 
not far below the compositeness scale $\Lambda$:
$\ln(\Lambda/\mu_{\ast}) $$\approx$$ 1$. We can
intuitively expect that this assumption 
implicitly necessitates large $\Lambda$ and that the binding of the
quark constituents into the composite scalar is very tight.
It is expected that under such circumstances the quark
loop approximation (\ref{Lmunor})-(\ref{Lmunornot3}) describes,
at least qualitatively, the evolving of the physical parameters
$g_t(\mu)$, $m_{\Phi}^2(\mu)$ and $\lambda(\mu)$ in the short
interval $[\ln \mu_{\ast}, \ln \Lambda]$. In other words,
since $\ln(\Lambda/\mu_{\ast}) $$\approx$$ 1$, the approximate
boundary conditions at $\mu$$=$$\mu_{\ast}$ for
evolution of $g_t$ and $\lambda$ in the large interval 
$[\ln E_{{\rm ew}}, \ln \mu_{\ast}]$ 
can be read off from (\ref{Lmunornot1}), (\ref{Lmunornot3}):
\begin{equation}
g_t(\mu) {\big |}_{\mu=\mu_{\ast}} \gg 1 \ ; \qquad
\frac{\lambda(\mu)}{g_t^2(\mu)}{\Big |}_{\mu = \mu_{\ast}} \sim 1 
\qquad \mbox{for: } {\mu}_{\ast} < {\Lambda}, \
\ln \left( \frac{ {\Lambda} }{ {\mu}_{\ast} }  \right) \approx 1 \ .
\label{RGEbc}
\end{equation}
BHL then applied these boundary conditions
to one-loop RGE's of the MSM
\begin{eqnarray}
16 \pi^2 \frac{d g_t(\mu)}{d \ln \mu}& =&
\left[ (N_{{\rm c}} + \frac{3}{2} ) g_t^2(\mu) 
- 3 \frac{(N_{{\rm c}}^2-1)}{N_{{\rm c}}} g_3^2(\mu) 
- \frac{9}{4} g_2^2(\mu) 
- \frac{17}{12} g_1^2(\mu) \right] g_t(\mu) \ ,
\label{RGEgt}
\\
16 \pi^2 \frac{d \lambda(\mu)}{d \ln \mu}& =&
- 4 N_{{\rm c}} g_t^4(\mu) + 4 N_{{\rm c}} \lambda(\mu) g_t^2(\mu) 
+12 {\lambda}^{\!2}(\mu) - {\cal {A}}(\mu) \lambda(\mu) 
+ {\cal {B}}(\mu) \ ,
\label{RGEl}
\end{eqnarray}
where $g_3$, $g_2$ and $g_1$ are the usual coupling parameters of 
$SU(3)_{{\rm c}}$, $SU(2)_L$ and $U(1)_Y$,
respectively, satisfying their own one-loop RGE's
\begin{eqnarray}
16 \pi^2 \frac{d g_j(\mu)}{d\ln \mu}& =& - C_j g_j^3(\mu) \ ,
\label{gjs1}
\\
C_3 = \frac{1}{3} \left( 11 N_{{\rm c}} - 2 n_q
\right) \ , \quad 
C_2 &=& \frac{43}{6} - \frac{2}{3} n_q \ , \quad
C_1 = -\frac{1}{6} - \frac{10}{9} n_q \ .
\label{gjs2}
\end{eqnarray}
Here, $n_q$ is number of effective quark
flavors (for $\mu $$>$$ m_t$: $n_q$$=$$ 6$),
and $N_{{\rm c}}$ number of colors ($N_{{\rm c}} $$=$$ 3$).
Expressions ${\cal {A}}$ and ${\cal {B}}$ in (\ref{RGEl})
are:
\begin{equation}
{\cal {A}} = 9 g_2^2 + 3 g_1^2 \ , \qquad
{\cal {B}} = \frac{9}{4} g_2^4 + \frac{3}{2} g_2^2 g_1^2
+ \frac{3}{4} g_1^4 \ .
\label{calAB}
\end{equation}
BHL used for low energy $g_j$'s values:
$g_3^2(M_Z)$$\approx$$1.44$, 
$g_2^2(M_Z)$$\approx$$0.446$ and
$g_1^2(M_Z)$$\approx$$0.127$. They ignored contributions
of $g_b$ to the evolution of $g_t$ and $\lambda$.
Solutions of MSM RGE's (\ref{RGEgt})-(\ref{RGEl}) 
with compositeness boundary conditions
(\ref{RGEbc}) gave the renormalized masses
\begin{equation}
m_t^{{\rm ren.}} = 
 g_t(\mu\!=\!m_t^{{\rm ren.}}) v /{\sqrt{2}} \ ,
\qquad m_H^{{\rm ren.}} = 
\lambda^{1/2}(\mu\!=\!m_H^{{\rm ren.}}) v \ ,
\label{mtmhphy}
\end{equation}
which, for very large
$\Lambda $$\approx$$ \mu_{\ast} $$>$$ 10^8$ GeV, turned out to be
rather stable against variations of boundary
conditions (\ref{RGEbc}). Specifically, predicted
$m_t^{{\rm ren.}}$ and
$m_H^{{\rm ren.}}$ are rather insensitive if:
\begin{itemize}
\item $g_t^2(\mu_{\ast})/(4 \pi)$ is varied between
$\infty$ and $1$, and $\Lambda$ is large 
($\Lambda $$\sim$$ \mu_{\ast} $$>$$ 10^8$ GeV);
\item Compositeness scale $\Lambda$ (or: ${\mu}_{\ast}$;
${\mu}_{\ast} $$\sim$$ {\Lambda}$) is varied on logarithmic
scale by quantities of order $1$, 
and $\Lambda$ is large ($\Lambda $$\sim$$ 
\mu_{\ast} $$>$$ 10^8$ GeV).
\end{itemize}
This is known as infrared fixed-point
behavior. This feature makes 
the application of the RGE approach
an important contribution to the ${\bar t} t$
condensation program. It has its origin in the presence of 
QCD contributions on the right of
RGE (\ref{RGEgt}), and was discussed earlier
\cite{Chang74,Cabibboetal79,PendletonRoss81,Hill81,HillLeungRao85}
in a context independent of condensation.
This behavior can be seen explicitly in Table~\ref{tabl1} which shows
results of BHL (1990), as well as those in the more primitive
quark loop approximation.
\begin{table}[ht]
\vspace{0.3cm}
\par
\begin{center}
\begin{tabular}{l c c c c c c c c c}
$\Lambda $$\approx$$ {\mu}_{\ast}$ [GeV] &
$10^{19}$ & $10^{17}$ & $10^{15}$ & $10^{13}$ & $10^{11}$ & 
$10^9$ & $10^7$ & $10^5$ & $10^4$ \\
\hline \hline
$m_t$ [GeV] (quark loop) & 
143 & 153 & 165 & 179.5 & 200 & 228 & 276 & 378 & 519 \\
$m_H$ [GeV] (quark loop) & 
289.5 &309 & 333 & 364 & 406 &468 & 571.5& 814.5&1235\\
\hline
$m_t(m_t)$ [GeV] RGE &
218 & 223 & 229 & 237 & 248 & 264 & 293 & 360 & 455 \\
$m_H(m_H)$ [GeV] RGE &
239 & 246 & 256 & 268 & 285 & 310 & 354 & 455 & 605 \\
\end{tabular}
\end{center}
\caption{\footnotesize Predicted $m_t$ and $m_H$ in
quark loop approximation [cf.~Eqs.~(\ref{mt})-(\ref{mh})], 
and by the full one-loop RGE approach of BHL.}
\label{tabl1}
\end{table}

As seen in Table~\ref{tabl1}, the (one-loop) RGE results of
BHL gave too high a mass $m_t^{\rm ren.}$.
The larger the $\Lambda$, the smaller the $m_t^{\rm ren.}$.
For $\Lambda $$\sim$$ E_{{\rm Planck}}$ ($\sim$$ 10^{19}$ GeV),
they obtained $m_t^{{\rm ren.}} $$\approx$$ 218$ GeV,
substantially higher than the measured $m_t^{{\rm phys.}}
$$\approx$$ 170-180$ GeV
\cite{Abeetal95,Adachietal95}. 
Consideration of the two-loop RGE's for $g_t$ and 
$\lambda$ doesn't change these results significantly -- 
it only increases $m_t^{{\rm ren.}}$ further
by a few GeV, thus slightly exacerbating the problem
\cite{Lavoura92,Veldhuis92}.

As an interesting point, we mention that solution (\ref{Lmunornot1})
for $g_t(\mu)$, obtained in quark loop approximation, can be
obtained also directly from RGE (\ref{RGEgt}) by keeping on the
right of the equation only the term $N_{{\rm c}} g_t^3(\mu)$ 
and applying the boundary condition $g_t(\Lambda) $$=$$ \infty$. 
Thus, in retrospect, we see that it is the leading-$N_{{\rm c}}$ 
Yukawa term $N_{{\rm c}} g_t^3(\mu)$
on the right of RGE (\ref{RGEgt}) which represents 
the quark loop effects on the evolution of $g_t(\mu)$.

\subsection{Marciano's approach}
\label{MRGE4}
Independently of BHL (1990),
Marciano (1989, 1990) introduced his own version
of the RGE approach to ${\bar t} t$ condensation. 
He investigated the behavior of
$k_t(\mu) $$\equiv$$ g_t^2(\mu)/(4 \pi)$.
He included in his calculation the one-loop QCD contribution,
and neglected the (small) contributions of the EW gauge bosons.
The one-loop RGE of the MSM is then
\begin{equation}
\mu \frac{d k_t (\mu)}{d \mu} = 
\frac{1}{2 \pi} \left( N_{{\rm c}} + \frac{3}{2} \right) k_t^2(\mu)
- \frac{4}{\pi} {\alpha}_3(\mu) k_t(\mu) \ ,
\label{marc2}
\end{equation}
where $N_{{\rm c}}$$=$$ 3$ was used throughout. 
Since in the massless version of this theory only the gauge coupling
${\alpha}_3$ exists as an independent parameter, Marciano
made the assumption that $k_t$ is a function of 
${\alpha}_3$ only: $k_t $$=$$ k_t({\alpha}_3(\mu))$. Therefore:
\begin{equation}
\mu \frac{d k_t(\mu)}{d \mu} =  \left[ \mu \frac{d}{d \mu}
{\alpha}_3 \right] \left[ \frac{d k_t}{d {\alpha}_3} \right] \ .
\label{marc2b}
\end{equation}
This is a version of
the assumption that the compositeness (condensation)
at a scale $\Lambda$ causes a reduction of 
coupling parameters in the low energy theory.
A free constant of integration $C$ appearing in the thus
obtained formula for the family of solutions for $k_t(\mu)$
[cf.~(\ref{marc4}) below] is then assumed to be determined
solely by the scale $\Lambda$ of the new physics: $C $$=$$ C(\Lambda)$. 
It is in this way that, in addition to
the standard MSM evolution of $k_t$ (\ref{marc2}), 
certain asymptotic conditions
at $\mu $$=$$ \Lambda $$\gg$$ E_{{\rm ew}}$ were imposed
in order to take into account indirectly the effects of
a new underlying physics responsible for the ${\bar t} t$
condensation. The leading-$N_{{\rm c}}$ case,
when $N_{{\rm c}}$$ +$$ 1.5 $$\mapsto$$ N_{{\rm c}} $$\equiv$$ 3$
in (\ref{marc2}), was interpreted as a case of
a less tightly bound ${\bar t} t$ scalar which does not
contribute any ``feedback'' effects to its own binding.
The full one-loop case, as given by (\ref{marc2}), was
interpreted as the case of a tightly bound (pointlike)
${\bar t} t$ condensate. Most of the investigation was
focused on the latter case. In addition to
MSM RGE for $k_t $$\equiv$$ g_t^2/(4 \pi)$ and
assumption (\ref{marc2b}) of the reduction of coupling
parameters, Marciano used one-loop RGE for ${\alpha}_3$
and consequently arrived at the general 
family of solutions\footnote{
In leading log approximation, all procedures {\em must\/}
give (\ref{marc4}). The constant of integration $C$
depends on the underlying dynamics. Any prescription that
does not give the form (\ref{marc4}) is lacking in that
it misses leading logs by some approximation.}
\begin{equation}
k_t(\mu)  =  \frac{2}{9} \frac{ {\alpha}_3^{8/7}(\mu) }
{ {\alpha}_3^{1/7}(\mu) - C } \ , 
\quad \mbox{with: } \
{\alpha}_3^{-1}(\mu)  =  {\alpha}_3^{-1}(m_t)
+ \frac{7}{2 \pi} \ln (\mu/m_t) \ .
\label{marc4}
\end{equation}
Here, $m_t$ is the renormalized mass $m_t(m_t)$, $\mu $$\geq$$ m_t$,
and $C $$=$$ C(\Lambda)$ is the previously mentioned
arbitrary constant of integration to be determined
by assuming certain behavior on $k_t(\mu)$ in the
asymptotic region of the onset of new physics
(i.e., for large $\mu $$\sim$$ \Lambda$). Specification of $C$ then
determines also $m_t $$=$$ v \sqrt{2 \pi k_t(m_t)}$,
where $v $$\approx$$ 246$ GeV is the VEV. In
the first work (Marciano, 1989),
the solution $C $$=$$ 0$ ($\Rightarrow k_t $$=$$ 2 {\alpha}_3/9$)
was proposed, leading to $m_t $$\approx$$ 98$ GeV. This solution
corresponds formally to the solution of the BHL approach with
${\Lambda} $$=$$ \infty$ 
and the effects of EW gauge bosons neglected. 
In the second work (Marciano, 1990),
two other choices for $C$ were
advocated. First choice was $C $$=$$ {\alpha}_3^{1/7}(\Lambda)$,
where $\Lambda$ was the energy of the onset of new
physics responsible for the condensation, thus leading to
\begin{equation}
k_t(\mu)  \left( \equiv \frac{g_t^2(\mu)}{4 \pi} \right)
=  \frac{2}{9} \frac{ {\alpha}_3^{8/7}(\mu) }
{ \left[ {\alpha}_3^{1/7}(\mu) - {\alpha}_3^{1/7}(\Lambda) \right] } \ . 
\label{marc5a}  
\end{equation}
This is the solution with 
the boundary condition $k_t(\mu)\!\to\! \infty$ 
when ${\mu}\!\to\!{\Lambda}$, 
just like the boundary condition (\ref{RGEbc}) of the BHL approach
where it was motivated by the results of the diagrammatic
approach in quark loop approximation 
(\ref{Lmunornot1})-(\ref{Lmunornot3}) of the truncated
TSM model (\ref{TSM}).
However, Marciano offered an alternative motivation
for this asymptotic behavior of $k_t$, namely by employing
a variant of the Pagels--Stokar (PS) relation
\cite{PagelsStokar79}
[cf.~(\ref{PSch})-(\ref{PSne})]
with a variable (``running'') lower integration bound
\begin{equation}
M_W^2(\mu) 
\left[ \equiv \frac{{\alpha}_2(\mu) m_t^2(\mu)}{2 k_t(\mu)}
\equiv  \pi \alpha_2(\mu) v^2(\mu) \right] 
\simeq \frac{3}{8 \pi} \int_{\mu^2}^{{\Lambda}^{\!2}} 
d {\bar p}^2 {\bar p}^2
{\alpha}_2(\bar p) \frac{ {\Sigma}_t^2( {\bar p}^2 )}
{ \left[ {\bar p}^2 + {\Sigma}_t^2( {\bar p}^2 ) \right]^2 } \ ,
\label{marc5b}
\end{equation}
where ${\Sigma}_t( {\bar p}^2 )$ is the top quark self-energy
(the dynamical ``running'' mass) and the UV cutoff $\Lambda$
indicates the new underlying physics. From this relation it is seen
that $M_W(\mu)\!\to\! 0$ when ${\mu}\!\to\!{\Lambda}$, and therefore 
this indicates $k_t(\mu)\!\to\! \infty$ when
${\mu}\!\to\!{\Lambda}$, i.e., the solution (\ref{marc5a}).
However, Marciano also argued that this asymptotic boundary
condition is not realistic and can sometimes,
particularly for lower cutoffs  $\Lambda $$<$$ 10^8$ GeV,
be even misleading, because it extends the consideration
of the one-loop RGE (\ref{marc2}) [or: (\ref{RGEgt}) of the
full approach by BHL] into a highly nonperturbative region
where perturbative unitarity bounds are also violated.
Therefore, he used for the boundary condition perturbative
unitarity bounds
\cite{MarcianoValenciaWillenbrock89}
as an indicator of the onset of
new underlying physics. Specifically, he used unitarity bounds
originating from consideration of the 
${\bar t} t \to {\bar t} t$ scattering:
$k_t(\mu) $$\leq$$ 2/3$ $\Rightarrow \ k_t(\Lambda)$$=$$ 2/3$.
This led him to a value of $C(\Lambda)$ (second choice)
somewhat smaller than the value 
$C({\Lambda}) $$=$${\alpha}_3^{1/7}(\Lambda)$ [first choice,
(\ref{marc5a})]. 
The differences from the BHL approach arising from this
boundary condition were substantial for lower
cutoffs $\Lambda $$<$$ 10^8$ GeV, due to the disappearance
of the infrared fixed-point behavior of the
RGE's for these cutoffs. 
Marciano also took subsequently into account 
the gluon cloud corrections 
\begin{equation}
m_t^{{\rm phys.}} \approx m_t(\mu\!=\!m_t)
\left[1 + \frac{4}{3 \pi} {\alpha}_3(\mu\!=\!m_t) \right]
\approx 1.047 \times m_t(m_t) \ ,
\label{marc7}
\end{equation}
and obtained for $\Lambda$$ \simeq$$ 10^{15}$-$10^{19}$ GeV
predictions 
$m_t^{{\rm phys.}}$$\simeq$$ 214$-$202$ GeV, 
about 16-17 GeV below the
BHL results of Table~\ref{tabl1}. However, he pointed out
that only the difference of about 4 GeV stems
from using the boundary condition $k_t({\Lambda})
$$=$$ 2/3$, and the rest from the different input parameters
[he used ${\alpha}_3(M_W)$$=$$0.107$, while BHL used
${\alpha}_3(M_Z) $$=$$ 0.115$] and from his neglecting
the contributions of the EW gauge bosons.

It should be stressed that the 
RGE approach is applicable to any high energy realization
of strong attraction which may result in quark-antiquark
condensation, not just to the NJLVL four-quark frameworks.
This was apparently first pointed out by Marciano
(1989, 1990). For example, 
the compositeness condition (\ref{RGEbc}), 
although motivated in the work of BHL by quark loop
approximation within an NJLVL-type of picture
[TSM (\ref{TSM})], can be regarded as independent of
such a specific model.

\subsection{Other related works}
\label{MRGE5}

Another approach not referring to any particular
realization of the strong attraction was given
by~\citeasnoun{PaschosZakharov91}.
They, at a leading-$N_{\rm c}$
level, used only the fine-tuning condition
$m_t $$\ll$$ \Lambda$ and symmetries of the low energy theory to
show that $M_H $$\approx$$ 2 m_t$.

Fishbane, Norton and Truong (1992)
and subsequently Fishbane and Norton (1993),
employing perturbative RGE methods,
investigated the infrared fixed-point structure of 
an NJLVL model with an internal $SU(N)$ ``color'' symmetry,
with fermions $f$ (one flavor) and without gauge bosons.
Their model involves a composite scalar
$\sigma $$\sim$$ {\bar f} f$. They concluded
the model supports the hypothesis
that, for fine-tuning cases of DSB
(${\mu} $$\sim$$ E_{{\rm ew}} $$\ll$$ {\Lambda}$),
NJLVL-type of models give definite ${\Lambda}$-independent 
predictions for $M_{\sigma}/m_f$, i.e.,
that the BHL-type of approach is predictive in such cases.

Cooper and Perez-Mercader (1991)
investigated the connection between results obtained
when compositeness conditions are imposed 
(in analogy with BHL) and 
when an assumption of coupling parameter reduction 
is made in a theory (partially in analogy with
Marciano). They investigated specifically
the linear four-dimensional ${\sigma}$ model with $N$
``colors'' which contains a Yukawa parameter $g_f$ 
and the four-scalar self-coupling parameter $\lambda$.
In the first approach, compositeness conditions 
$Z_3 $$=$$ Z_4 $$=$$ 0$ at $\mu $$=$$ \Lambda$ for
these two parameters were imposed. In the second approach,
the assumption $\lambda $$=$$ f(g_f^2(\mu))$ was made and
RGE's were applied (cf.~also approach by Marciano,
Sec.~\ref{MRGE4}). The analysis was done in
the fermion loop (leading-$N$) approximation, and
the authors demonstrated that both approaches consistently give
the relationship between the two parameters:
$\lambda(E_{{\rm ew}})/
g_f^2(E_{{\rm ew}}) $$=$$ {\alpha}$ for 
$\Lambda $$\gg$$ E_{{\rm ew}}$, where ${\alpha}$ is
a ${\Lambda}$-independent constant. The first approach gave
a more specific result: ${\alpha} $$\equiv$$ 2$.

Recently, Aoki {\em et al.\/} (1997; Aoki, 1997)
proposed a systematic approximation scheme for
the Wilsonian nonperturbative renormalization group
(NPRG) approach to the DSB.
The approximation is based on the local
potential approximation (LPA), in which the
effective action $S(\phi; \Lambda)_{{\rm eff}}$
of a physical system with a field $\phi$ (generic notation)
and a finite ultraviolet cutoff $\Lambda$ is written
in terms of a local potential $V(\phi)$:
$S_{{\rm eff}}$$=$$\int\!d^4 x 
[V(\phi)\!+\!\partial_{\mu} \phi \partial^{\mu} \phi/2 ]$.
The authors expand $V(\phi)$ in powers of
a (conveniently defined) polynomial of $\phi$.
The method appears to be promising because
it describes the RG flow of the system nonperturbatively
in the entire energy range, and the flow
can be systematically refined within the LPA frame. 
In contrast to the previously described method
of perturbative RGE's with compositeness boundary condition, 
the approach does not rely on the
infrared fixed point behavior of RGE's (its applicability is
hence not restricted to large ${\Lambda} $$\stackrel{>}{\sim}$$
10^8$ GeV), and it deals with four-fermion
interaction terms directly.
Therefore, the method has some similarity with the
method of Dyson--Schwinger (DS) and
Pagels-Stokar (PS) equations (cf.~the next
Section). Aoki {\em et al.\/} emphasize that,
unlike the DS$+$PS approach, their NPRG approach 
is gauge independent and does not
lead to divergent series.
The method has not yet been applied to frameworks
involving ${\bar t} t$ condensation.

\subsection{Conclusions}
\label{MRGE6}
The minimal ${\bar t} t$ condensation framework is
usually (but not necessarily) discussed in terms 
of an effective model with NJLVL
four-quark interactions (\ref{TSM}) involving
$t_R$, $t_L$, and $b_L$, called the (truncated)
top-mode standard model (TSM).
The minimal framework is based on the assumption that
top quark condensation alone is responsible 
simultaneously for $m_t$ and for the {\em full\/} (dynamical)
EWSB -- i.e., ${\bar t}t $$\sim$$ {\cal {H}}$, where
${\cal {H}}$ is the only Higgs particle,
and $\langle {\cal {H}} \rangle_0$ at low energies equals the
full EW VEV $v $$\approx$$ 246$ GeV needed to reproduce
$M_W$ and $M_Z$. The resulting effective
theory at low energies is essentially the MSM.

The method of perturbative RGE's in this
minimal framework considers renormalization
effects $(\delta g_t)_{{\rm ren.}}$ and
$(\delta \lambda)_{{\rm ren.}}$,
by applying compositeness boundary conditions (\ref{RGEbc})
at a scale $\mu $$=$$ \mu_{\ast} ($$\sim$$ \Lambda)$  
to RGE's of the MSM for the Yukawa 
and scalar self-coupling parameters ($g_t,$$\lambda$). 
These compositeness conditions reflect a strong attraction
in the top quark sector, and were motivated by the behavior of
$g_t(\mu)$ and $\lambda(\mu)$ at high energies $\mu $$\sim$$ 
\Lambda$ in quark loop approximation of the truncated TSM 
model [cf.~(\ref{Lmunornot1}), (\ref{Lmunornot3})]. 
However, the results are basically independent of
any particular realization of the strong attractive sector
of the top quark. 
An implicit assumption is that
the model starts behaving as (a special version of) the MSM 
as we come down from $\Lambda$ to energies
${\mu}_{\ast}$ which are still close
to $\Lambda$: $ \ln ( {\Lambda}/{\mu}_{\ast} ) $$\approx$$ 1$.
Stated differently, binding of the
quark constituents in the composite scalars is very tight,
in contrast to low energy QCD.
For very large compositeness scales
($\Lambda $$\stackrel{>}{\sim}$$ 10^8$ GeV),
predictions of this approach turn out to be rather
insensitive to the details of the actual condensation
mechanism. It is
this infrared fixed-point behavior of the RGE's
that makes the application of the method an important contribution 
to the ${\bar t} t$ condensation program.
Compositeness conditions (\ref{RGEbc}) should be regarded only as a
crude qualitative estimate of the dynamics in the strong
condensation sector. Therefore, the method can
be trusted only for high ${\Lambda}$'s --
in the minimal framework (\ref{TSM}): 
$\Lambda $$\stackrel{>}{\sim}$$ 10^8$ GeV.

Nonperturbative renormalization group
(NPRG) approaches appear to be very promising for
lower ${\Lambda}$'s. However, they have not yet been applied to
${\bar t} t$ condensation.

\section{The method of Dyson--Schwinger and Pagels-Stokar equations}
\label{MDSPS}
\setcounter{equation}{0}

\subsection{General top-mode standard model}
\label{MDSPS1}
We now turn to a different computational approach, 
which was first applied to top quark condensation by 
Miransky, Tanabashi and Yamawaki (MTY) (1989a, 1989b),\footnote{ 
Independently of Nambu's proposal (Nambu, 1988).}
and subsequently by King and Mannan (KM) (1990, 1991a).

MTY began by introducing a rather general TSM containing
$SU(3)_c\!\times\!SU(2)_L\!\times\!U(1)_Y$ invariant 
four-quark NJLVL interactions
\begin{eqnarray}
{\cal {L}}_{4q}^{(\Lambda)} & = & 
\frac{4 \pi^2}{N_{{\rm c}} {\Lambda}^{\!2}}
{\Big [}
g^{(1)}_{\alpha \alpha^{\prime}; \beta \beta^{\prime}}
\left( {\overline \Psi}_L^{\alpha i} 
\Psi_R^{\alpha^{\prime} j} \right)
\left( {\overline \Psi}_R^{\beta j} 
\Psi_L^{\beta^{\prime} i} \right)
+ g^{(2)}_{\alpha \alpha^{\prime} ; \beta \beta^{\prime}}
\left( {\overline \Psi}_L^{\alpha i} 
\Psi_R^{\alpha^{\prime} j} \right)
( i \tau_2)^{i k} (i \tau_2)^{j \ell}
\left( {\overline \Psi}_L^{\beta k} 
\Psi_R^{\beta^{\prime} \ell} \right)
\nonumber \\
& &  + g^{(3)}_{\alpha \alpha^{\prime}; \beta \beta^{\prime}}
\left( {\overline \Psi}_L^{\alpha i} 
\Psi_R^{\alpha^{\prime} j} \right)
( \tau_3)^{j k}
\left( {\overline \Psi}_R^{\beta k} 
\Psi_L^{\beta^{\prime} i} \right) {\Big ]}
+ \mbox{ h.c.} \ .
\label{genTSM}
\end{eqnarray}
Coefficients $g^{(k)}_{\alpha \alpha^{\prime}; 
\beta \beta^{\prime}}$ ($k$$=$$1,2,3$) 
are ``bare'' dimensionless ${\cal {O}}(1)$ four-quark parameters
in the above theory with $\Lambda$ cutoff;
($\alpha,$$\alpha',$$\beta,$$\beta'$) and ($i,$$j,$$k,$$l$) are
family and isospin indices, respectively. 
E.g., $\Psi_R^{3 1} $$=$$ t_R$,
$\Psi_R^{3 2} $$=$$ b_R$, etc. 
The color indices, omitted in (\ref{genTSM}),
are distributed the same way
as in the truncated TSM (\ref{TSM}), thus leading
to the $N_{{\rm c}}$-enhancement of quark loop induced parameters,
as in (\ref{Lmunot1})-(\ref{Lmunot2}). 
We return to this point later and show why 
$g^{(k)}_{\alpha \alpha^{\prime}; \beta \beta^{\prime}}
$$=$$ {\cal {O}}(N_{{\rm c}}^0)$.

The four-quark terms involving the
third generation quarks induce 
heavy quark masses ($m_t$, and possibly $m_b$) 
as well as DEWSB -- thus they are of particular interest
\begin{eqnarray}
{\cal {L}}_{4q}^{(\Lambda){\rm 3rd}} & = & 
\frac{8 \pi^2}{N_{{\rm c}} {\Lambda}^{\!2}}
{\Big \{} 
\kappa_t \left( {\overline \Psi}_L^{ia} t_R^a \right)
\left( \bar t_R^b \Psi_L^{ib} \right) +
\kappa_b \left( {\overline \Psi}_L^{ia} b_R^a \right)
\left( \bar b_R^b \Psi_L^{ib} \right) 
\nonumber\\
&&+ 2 \kappa_{tb} \left[
\left( \bar t_L^a t_R^a \right)
\left( \bar b_L^b b_R^b \right) +
\left( \bar t_R^a t_L^a \right)
\left( \bar b_R^b b_L^b \right) -
\left( \bar t_L^a b_R^a \right)
\left( \bar b_L^b t_R^b \right) -
\left( \bar t_R^a b_L^a \right)
\left( \bar b_R^b t_L^b \right) 
\right] {\Big \}} \ ,
\label{3gTSM}
\end{eqnarray}
where $\kappa_t $$=$$ (g^{(1)}$$+$$g^{(3)})_{33;33}$,
$\kappa_b $$=$$ (g^{(1)}-g^{(3)})_{33;33}$,
and $\kappa_{tb} $$=$$ g^{(2)}_{33;33}$. Color indices $a$, $b$
are explicitly written in (\ref{3gTSM}). 
The first term here is the truncated
TSM (\ref{TSM}). We will call the framework of (\ref{3gTSM}) 
the general TSM, although not just ${\bar t}t$
but also ${\bar b}b$ condensation may occur in this framework 
(if ${\kappa}_b $$>$$ {\kappa}_{{\rm crit.}}$).
Without loss of generality, we can regard 
$g^{(j)}_{33;33}$ ($j$$=$$1,3$) to be real.
In addition, $\kappa_{tb} $$=$$ g^{(2)}_{33;33}$
is also assumed to be real. All these parameters are
``bare'' parameters of a theory with a finite
UV cutoff ${\Lambda}$: ${\kappa}_t $$\equiv$
$ {\kappa}_t({\mu}$$=$${\Lambda})$, etc.

Later, in Sec.~\ref{EWESG} (Subsec.~\ref{EWESG12}), it will be
argued that the NJLVL Lagrangian density (\ref{3gTSM})
in general leads at low energies to an effective SM
with two composite Higgs doublets, as long as the
four-quark parameters are near or above the critical
values.

\subsection{Dyson--Schwinger integral equations}
\label{MDSPS2}

In contrast to BHL,
the authors MTY and KM investigated directly
the actual condensation mechanism leading to ${\bar t} t$
condensation.
The mathematical formalism employed in their approach is
the Dyson--Schwinger (DS) integral equation for the
dynamical mass functions $\Sigma_t(p^2)$
and $\Sigma_b(p^2)$ of the top and bottom quarks,
and the Pagels--Stokar (PS)
relation for the decay constants $F_{\pi}$
of the electroweak Nambu--Goldstone bosons
(Pagels and Stokar, 1979). 
In this approach, the propagator $S_i(p)$ of quark $i$
($i$$=$$t,b$) in momentum space is
\begin{equation}
S^{-1}_i(p) = (-i) Z^{-1}_i(p^2) \left[ 
{p \llap /} - {\Sigma}_i(p^2) \right] \ ,
\label{qprop}
\end{equation}
where $Z_i(p^2)$ is the quark field renormalization constant.
For the Landau gauge in ladder approximation,
$Z_i(p^2)$$=$$1$. In the improved
ladder approximation\footnote{
Ladder approximation in this context means that only the 
leading-$N_{{\rm c}}$
contributions are calculated, i.e.,
quark loop and one-gluon-exchange QCD contributions, 
and that QCD loop contributions
are calculated by using a nonrunning QCD coupling
parameter, e.g., ${\alpha}_3({\bar p}) $$\equiv$$ {\alpha}_3(m_t)
$$\approx$$ 0.11$. Improved ladder approximation
uses a running ${\alpha}_3({\bar p})$ instead.}
discussed below, it can be shown that
$Z_i(p^2)$$ \to$$ 1$ for $|p^2| $$\gg$$ E^2_{{\rm ew}}$;
thus, in actual calculations $Z_i(p^2) $$\equiv$$ 1$ is used.

\begin{figure}[htb]
\mbox{}
\vskip3.5cm\relax\noindent\hskip-.5cm\relax
\includegraphics{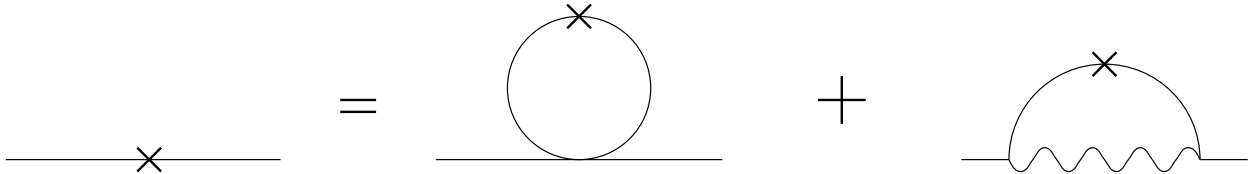} \vskip0.4cm
\caption{\footnotesize Graphical illustration of
the gap (DS) equation for a 
dynamically generated heavy quark mass. The cross represents
insertion of the dynamical ``running'' 
quark self-energy $\Sigma_i({\bar p}^2)$
($i$$=$$t$ or $b$).
The first term on the right represents the four-quark interaction,
and the second the gluon exchange (wavy line).}
\label{rmd2f}
\end{figure}

DS equation was applied
at the leading-$N_{{\rm c}}$ level, including QCD in 
the Landau gauge with one-loop running
$\alpha_3(p)$, i.e., 
in improved ladder approximation. The DS equation is
essentially a variational version of the usual gap equation
for the dynamically induced mass of the heavy quarks. 
The general formalism leading to DS equation in the context of
nonperturbative DSB involves the
Cornwall-Jackiw-Tomboulis (CJT) effective action 
(effective potential) $V_{{\rm CJT}}[\Sigma]$
\cite{CornwallJackiwTomboulis74}. 
It is a functional of the 
dynamical ``running'' mass ${\Sigma}(p^2)$,
the DS equation is obtained by requiring that the
functional (variational) derivative of
$V_{{\rm CJT}}[\Sigma]$ be zero:
$\delta V_{{\rm CJT}}[\Sigma]/\delta \Sigma(p^2) $$=$$ 0$.
The analogous relation for the case of
a nonrunning dynamical mass $\Sigma(p^2) $$=$$ m$ is usually
called a gap equation.
A detailed derivation, based on CJT formalism,
of the ladder DS equation within the context of TSM
was given by~\citeasnoun{KondoTanabashiYamawaki93}, 
and this derivation can be
applied also to the case of the improved ladder approximation.
A graphical representation of the DS equation
is given in Fig.~\ref{rmd2f}. 
After angular
integration in the Wick-rotated Euclidean space and introducing
a cutoff $\Lambda$ for the (Euclidean) quark and gluonic momenta,
the improved ladder DS equations take in the $t$-$b$ decoupled
case (${\kappa}_{tb}$$=$$0$) the form:
\begin{equation}
{\Sigma}_i({\bar p}^2) = \frac{{\kappa}_i}{{\Lambda}^{\!2}}
\int_0^{{\Lambda}^{\!2}} \frac{d{\bar q}^2 {\bar q}^2 
{\Sigma}_i({\bar q}^2)}{{\bar q}^2+ {\Sigma}_i^2({\bar q}^2)}
+ \int_0^{{\Lambda}^{\!2}} 
\frac{d{\bar q}^2 {\bar q}^2 {\Sigma}_i({\bar q}^2)}
{{\bar q}^2+ {\Sigma}_i^2({\bar q}^2)}
K({\bar p}^2;{\bar q}^2) \ , \quad (i=t \ \mbox{or} \ b) \ .
\label{DSmty} 
\end{equation}
Bars over momenta ${\bar p}$ and ${\bar q}$ 
mean that they are in the Euclidean metric.
Function $K({\bar p}^2;{\bar q}^2)$ originating from
gluon propagator in the Landau gauge is:
\begin{equation}
 K({\bar p}^2;{\bar q}^2) = 
\frac{\lambda_{{\rm qcd}}(\max({\bar p}^2;{\bar q}^2))}
          {\max({\bar p}^2;{\bar q}^2)} \ , \qquad 
\lambda_{{\rm qcd}}({\bar p}^2) = 
\frac{3 (N_{{\rm c}}^2 - 1 )}{8 \pi N_{{\rm c}}} 
\alpha_3({\bar p}) \ .
\label{Kqcd}
\end{equation}
QCD parameter 
$\alpha_3({\bar p}) $$\equiv$$ g^2_3({\bar p})/(4 \pi)$ 
follows from one-loop RGE 
(\ref{gjs1})-(\ref{gjs2}) ($j$$=$$3$)
\begin{equation}
\alpha_3({\bar p}) =
\alpha_3(\mu_{{\rm IR}}) 
\ ({\bar p}^2 \leq \mu^2_{{\rm IR}}) \ ; \quad
\alpha_3({\bar p}) =
\frac{12 \pi}{(11 N_{{\rm c}} - 2 n_q)}  
\frac{1}{\ln [{\bar p}^2/{\Lambda}^{\!2}_{{\rm QCD}} (n_q)]} 
\ ({\bar p}^2 \geq \mu^2_{{\rm IR}})  \ . 
\label{alph3}
\end{equation}
Here, $n_q$ is number of effective flavors of quarks
(for $|{\bar p}|$$>$$m_t$: $n_q$$=$$6$).
For $N_{{\rm c}}$$=$$3$ and 
$\alpha_3(180 \mbox{ GeV})$$=$$ 0.11$, 
we have ${\Lambda}_{{\rm QCD}} (n_q$$=$$6) $$\approx$$ 51$ MeV.
An infrared cutoff $\mu_{{\rm IR}}
$$\sim$$ \Lambda_{{\rm QCD}}$ was introduced to avoid
spurious divergence in the infrared. Details of the IR
cutoff do not influence appreciably the results of the analysis.
Nonzero solution of (\ref{DSmty}) exists if 
${\kappa}_i $$>$$ {\kappa}_{{\rm crit.}}$.
  
In the quark loop approximation 
($\lambda_{{\rm qcd}} $$=$$ 0$), the DS equation
(\ref{DSmty}) has one nonzero solution $\Sigma_i$$=$$m^{(0)}_i$
if ${\kappa}_i $$\stackrel{>}{\approx}$$ 1$$=$
${\kappa}^{(0)}_{{\rm crit.}}$,
and this solution is not evolving with ${\bar p}^2$:
\begin{equation}
{\Sigma}_i({\bar p}^2) = m_i^{(0)} \ , \qquad
 {\kappa}_i = \left[ 1 - (m^{(0)2}_i/{\Lambda}^{\!2})
\ln \left( {\Lambda}^{\!2}/m^{(0)2}_i+1 \right) \right]^{-1}
\ \mbox{for } {\kappa}_i > 1 \quad (i=t,b) \ .
\label{qlgap}
\end{equation}
Relations (\ref{qlgap}) are usually referred to as gap equations
in quark loop (bubble) approximation.
Here we see that ${\kappa}_i $$\sim$$ 1 $$=$
$ {\cal {O}}(N_{{\rm c}}^0)$.
Therefore, the motivation for the way ${\kappa}_t$, ${\kappa}_b$ and
${\kappa}_{tb}$ were normalized 
in (\ref{3gTSM}) can now be understood.
From (\ref{Kqcd})-(\ref{alph3}) we see that 
$K({\bar p}^2,{\bar q}^2) $$\propto$$
\lambda_{{\rm qcd}} $$=$$ {\cal {O}}(N_{{\rm c}}^0)$.
Since also $\kappa_i $$=$$ {\cal {O}}(N_{{\rm c}}^0)$ 
as we have just argued,
we see in retrospect that the (improved) ladder QCD contribution
on the right of Eq.~(\ref{DSmty}) 
is formally of leading-$N_{{\rm c}}$
order, just as the numerically dominant quark loop contribution
is. The effect of this QCD contribution is that the
dynamical mass solution ${\Sigma}_i({\bar p}^2)$ becomes 
``running'' -- it is weakly dependent on Euclidean 
momentum ${\bar p}$, with the asymptotic solution
\begin{equation}
{\Sigma}_i({\bar p}^2) \approx m_i \left[
\frac{\alpha_3({\bar p})}{\alpha_3(m_i)} \right]^{c_m}
= m_i
\left[ \frac{ \ln \left( m_i^2/\Lambda_{{\rm QCD}}^{\!2} \right) }
{ \ln \left( {\bar p}^2/\Lambda_{{\rm QCD}}^{\!2} \right) } 
\right]^{c_m} \ ,
\quad \mbox{for:  }   m_i^2 \ll {\bar p}^2 \ll {\Lambda}^{\!2} \ , 
\label{asymS}
\end{equation}
where $m_i$ is the low energy (``renormalized'') mass of the
quark $i$ ($i$$=$$t,b$) and 
\begin{equation}
 c_m= \frac{9 (N_{{\rm c}}^2-1)}
{2 N_{{\rm c}} (11 N_{{\rm c}} - 2 n_q)} \quad ( = \frac{4}{7}
\quad \mbox{for } N_{{\rm c}}=3, n_q=6) \ .
\label{asymcm}
\end{equation}

We will comment more extensively on the effect of QCD term
in (\ref{DSmty}) on ${\Sigma}_i({\bar p}^2)$ 
in Secs. \ref{CCVAMF1}-\ref{CCVAMF2} where
comparisons with the RGE approach are discussed.

\subsection{Generalized Pagels--Stokar relations}
\label{MDSPS3}
An essential aspect of the analysis by MTY and 
KM was the use of generalized Pagels--Stokar relations 
for the decay constants $F_{\pi^0}$ and $F_{\pi^{\pm}}$
of the composite Nambu--Goldstone bosons (NGB's).
If the heavy quark condensation
in the general TSM (\ref{3gTSM}) is assumed to
be responsible for all or most of the 
(dynamical) EWSB, then 
$F_{\pi^0} $$\approx$$ F_{\pi^{\pm}} $$\approx$$
v $$=$$ 246$ GeV appear
in mass formulas for EW gauge bosons\footnote{For models where
the top (plus bottom) quark condensation
is responsible only for a minor part of the DEWSB 
($F_{\pi^0}$, $F_{\pi^{\pm}} $$\ll$$ v$), see Sec.~\ref{RMUP6}.}
\begin{equation}
M_W^2 = \frac{g_2^2}{4} F^2_{\pi^{\pm}} \ ; \qquad
M_Z^2 \cos^2 \theta_W =
\frac{g_2^2}{4} F^2_{\pi^0} \ .
\label{MWMZ}
\end{equation}
PS relations relate these decay constants with
mass functions of the heavy quarks and thus provide
a correspondence between gauge boson and heavy quark masses. 
These relations are
an approximation to a sum rule that originates
from the amplitudes of the Bethe--Salpeter (BS) equation for the
bound states of the NGB's in 
the leading-$N_{{\rm c}}$ (improved ladder) approximation.
In fact, PS relations are obtained from BS amplitudes
in the weak coupling limit 
$\lambda_{{\rm qcd}}$$\to$$ 0$ 
when DS equations (\ref{DSmty}) and $\Sigma_i$
derived from them are kept unchanged. It turns out that this weak
coupling approximation changes $F_{\pi}^2$'s 
by less than $3$\%, 
as discussed 
by~\citeasnoun{Aokietal90} 
(cf.~also Aoki, 1991). Generalized PS relations
(generalized from the QCD-like to the general TSM case), connecting
$F_{\pi^{\pm}}$ and $F_{\pi^0}$ with
dynamical masses ${\Sigma}_i$ in Euclidean metric, were
first given by MTY
(1989a, 1989b)
\begin{eqnarray}
F^2_{\pi^{\pm}} & = &
\frac{N_{{\rm c}}}{8 \pi^2} 
\int_0^{{\Lambda}^{\!2}} d {\bar p}^2 {\bar p}^2
\left[ {\bar p}^2+{\Sigma}_t^2({\bar p}^2) \right]^{-1}
\left[ {\bar p}^2+{\Sigma}_b^2({\bar p}^2) \right]^{-1}
{\Bigg \{} 
\left( 1 - \frac{{\bar p}^2}{4} \frac{d}{d {\bar p}^2} \right)
\left( {\Sigma}_t^2({\bar p}^2) 
+ {\Sigma}_b^2({\bar p}^2) \right) +
\nonumber\\
&& + \frac{{\bar p}^2}{2}
\left( {\Sigma}_t^2({\bar p}^2) - {\Sigma}_b^2({\bar p}^2) \right)
{\Big [}
\left( {\bar p}^2+{\Sigma}_t^2({\bar p}^2) \right)^{-1}
\left(1 + \frac{d}{d {\bar p}^2 } {\Sigma}_t^2({\bar p}^2) \right) 
\nonumber\\
 && - \left( {\bar p}^2+{\Sigma}_b^2({\bar p}^2) \right)^{-1}
\left(1 + \frac{d}{d {\bar p}^2} {\Sigma}_b^2({\bar p}^2) \right) 
{\Big ]} {\Bigg \}} \ ,
\label{PSch}
\end{eqnarray}
\begin{eqnarray}
F^2_{\pi^0} & = &
\frac{N_{{\rm c}}}{8 \pi^2} 
\int_0^{{\Lambda}^{\!2}} d{\bar p}^2 {\bar p}^2
{\Bigg \{}
\left[ {\bar p}^2+{\Sigma}_t^2({\bar p}^2) \right]^{-2}
\left[ {\Sigma}_t^2({\bar p}^2) 
- \frac{{\bar p}^2}{4} \frac{d}{d {\bar p}^2}
{\Sigma}_t^2({\bar p}^2) \right] +
\nonumber\\
&& +\left[ {\bar p}^2+{\Sigma}_b^2({\bar p}^2) \right]^{-2}
\left[ {\Sigma}_b^2({\bar p}^2) 
- \frac{{\bar p}^2}{4} \frac{d}{d {\bar p}^2}
{\Sigma}_b^2({\bar p}^2) \right] {\Bigg \}} \ .
\label{PSne}
\end{eqnarray}
Since TSM contains no explicit
custodial symmetry $SU(2)_V$ of the MSM Higgs sector, 
one might worry that the electroweak 
$\rho$$ \equiv$$ M_W^2/(M_Z^2 \cos^2 \theta_W)$ parameter
becomes phenomenologically 
unacceptable in the case of maximal isospin
violation in the heavy quark sector, i.e., in the case of
${\kappa}_t $$>$$ {\kappa}_{{\rm crit.}} $$>$$
{\kappa}_b$ when ${\Sigma}_t({\bar p}^2) $$\not=$$ 0$ and
${\Sigma}_b({\bar p}^2) $$=$$ 0$. However, it can be seen that
PS relations (\ref{PSch})-(\ref{PSne}) reproduce
approximately the same suppressed $\delta \rho \equiv (\rho-1)$ as
the MSM with protective custodial symmetry. For example,
if we assume that the dynamical mass ${\Sigma}_t$$=$$m_t$ is
not running [i.e., quark loop approximation, 
$\lambda_{{\rm qcd}} $$=$$ 0$, of
DS equation (\ref{DSmty})] and ${\Sigma}_b$$=$$0$, 
we obtain from the above PS relations
\begin{equation}
F^2_{\pi^{\pm}}(\mbox{q.l.}) = 
\frac{N_{{\rm c}}}{8 \pi^2} m_t^2
\left[ \ln \frac{{\Lambda}^{\!2}}{m_t^2} + \frac{1}{2} \right] \ ,
\quad 
F^2_{\pi^0}(\mbox{q.l.}) = \frac{N_{{\rm c}}}{8 \pi^2} m_t^2
\ln \frac{{\Lambda}^{\!2}}{m_t^2} \ .
\label{PSql}
\end{equation}
Here, ``(q.l.)'' denotes quark loop approximation.
These relations lead to
\begin{equation}
{\delta \rho}^{{\rm (q.l.)}} 
= \left( \frac{M_W^2}
{M_Z^2 \cos^2 \theta_W} \right)^{{\rm (q.l.)}} - 1
=  \frac{ F^2_{\pi^{\pm}}(\mbox{q.l.}) }
{ F^2_{\pi^{0}}(\mbox{q.l.}) } -1 =
\frac{N_{{\rm c}} m_t^2}{16 \pi^2 F^2_{\pi^{0}}(\mbox{q.l.}) }
= \frac{1}{2 \ln ( {\Lambda}^{\!2}/m_t^2)} \ .
\label{rho}
\end{equation}
Therefore, ${\delta \rho}^{{\rm (q.l.)}}$$\ll$$1$ for 
large ${\Lambda}$.
We see that the familiar expression
\cite{Veltman77}
of the MSM for $\delta \rho$ is reproduced:
$\delta \rho $$\approx$$ G_F \sqrt{2} N_{{\rm c}} m_t^2/(4 \pi)^2$ 
(note: $G_F \sqrt{2} $$=$$ v^{-2} $$\approx$$ F_{\pi}^{-2}$). 
Even the maximal isospin violation in the TSM framework
does not raise a problem of phenomenologically unacceptable
$\delta \rho$. As a matter of fact, 
custodial $SU(2)_V$ symmetry is implicitly contained in this
condensation mechanism, as suggested by (\ref{rho}).

\subsection{Results of the approach}
\label{MDSPS4}
In the quark loop approximation (no QCD), 
MTY and BHL approaches give the same results,
written in Table~\ref{tabl1} in the
line ``$m_t [\mbox{GeV}] \mbox{ (quark loop)}$''
(cf.~also discussion in Sec.~\ref{CCVAMF2}).
However, DS equations (\ref{DSmty}) contain
also contributions of QCD loops, and the latter
change quark loop predictions of Table~\ref{tabl1}
substantially, as seen explicitly from Table~\ref{tabl2}.
Solutions in Table~\ref{tabl2}, 
in the line ``$m_t $$=$$ {\Sigma}_t(0)$'',
are for the simplest case of 
${\kappa}_b $$<$$ {\kappa}_{{\rm crit.}}$ and
${\kappa}_{tb}$$=$$0$ 
(${\Sigma}_b $$=$$ 0$, i.e., truncated TSM).
\begin{table}[ht]
\vspace{0.3cm}
\par
\begin{center}
\begin{tabular}{ l c c c c c c c c c }
$\Lambda$ [GeV] &
$10^{19}$ & $10^{17}$ & $10^{15}$ & $10^{13}$ & $10^{11}$ & 
$10^9$ & $10^7$ & $10^5$ & $10^4$ \\
\hline \hline
$m_t$ [GeV] (quark loop) & 
143 & 153 & 165 & 179.5 & 200 & 228 & 276 & 378 & 519 \\
\hline
$m_t$$=$${\Sigma}_t(0)$ [GeV] & 
253 & 259 & 268 & 279 & 293 & 316 & 354 & 446 & 591 \\ 
\hline
$m_t(m_t)$ [GeV] (RGE) &
218 & 223 & 229 & 237 & 248 & 264 & 293 & 360 & 455 \\
\end{tabular}
\end{center}
\caption{\footnotesize Predicted 
${\Sigma}_t({\bar p}^2$$=$$0) $$\approx$$m_t^{\mbox{\tiny phys.}}$ 
in the improved ladder approximation of the $\mbox{DS}+
\mbox{PS}$ approach. The values are taken from
King and Mannan (1990);
they used QCD parameters:
$\mu_{\mbox{\tiny IR}}$$=$$0.30$ GeV,
$\alpha_3(\mu_{\mbox{\tiny IR}})$$\approx$$ 3.93$, ${\alpha}_3(M_Z)
$$=$$ 0.127$ [$\Rightarrow$
${\Lambda}_{\mbox{\tiny QCD}}($$ n_q$$=$$6) $$\approx$$ 70$ MeV].
For comparison, predictions in quark loop approximation,
and of the RGE approach of BHL are also given.}
\label{tabl2}
\end{table}
To obtain these results, 
the authors proceeded in the following way:
\begin{itemize}
\item
At each chosen cutoff $\Lambda$,
solutions ${\Sigma}_t({\bar p}^2)$ of the DS equation (\ref{DSmty})
were found by iterative calculation, for various choices of
four-quark coupling parameters ${\kappa}_t$.
\item These solutions were 
then used in PS relations (\ref{PSch})-(\ref{PSne})
(with ${\Sigma}_b$$=$$0$).
\item Demanding that decay constants $F_{\pi^0}$
and $F_{\pi^{\pm}}$
be approximately equal to the full EW VEV 
$v $$\approx$$ 246$ GeV led to fine-tuning
${\kappa}_t$ to values near (slightly above) the critical value
${\kappa}_{{\rm crit.}} $$\approx$$  1$$ -$$ 
\lambda_{{\rm qcd}}(\Lambda)$, thus resulting
in predictions $m_t$$\approx$$ {\Sigma}_t(0)$ 
given in Table~\ref{tabl2}. 
\end{itemize}
It turned out that the critical value satisfies:
$1\!-\!{\kappa}_{{\rm crit.}}\!\sim\!
\lambda_{{\rm qcd}}({\Lambda}^{\!2})\!\sim 
1/\ln({\Lambda}^{\!2}/\Lambda_{{\rm QCD}}^{\!2})$.
In this context, Mahanta (1990) pointed out, 
within the ladder DS approach (3.4)
($i$=$t$), that the fine-tuning of the four-quark
parameter ${\kappa}_t$ to 
${\kappa}_{\mbox{\scriptsize crit.}}$
for the hierarchy $m_t \ll {\Lambda}$
gets softened substantially by the presence of
(QCD) gauge interaction. He showed,
in an additional approximation of nonrunning
${\alpha}_3$, that the fine-tuning changes 
from $\sim$$m_t^2/{\Lambda}^{\!2}$ 
[no gauge interaction; cf.~(3.8)] to
$\sim$$(m_t^2/{\Lambda}^{\!2})^b$,
where $b\!=\!\sqrt{
1\!-\!{\alpha}_3/{\alpha}_3^{\mbox{\scriptsize crit.}}}
< 1$. Here, ${\alpha}_3^{\mbox{\scriptsize crit.}}$ 
is the critical value of ${\alpha}_3$ for the 
pure QCD-induced DSB: 
${\alpha}_3^{\mbox{\scriptsize crit.}}$$=$$
2 \pi N_{\mbox{\scriptsize c}}/[3 (N_{\mbox{\scriptsize c}}^2 -1)]$.

KM (1991a)
numerically investigated
also cases ${\kappa}_b $$>$$ {\kappa}_{{\rm crit.}}$,
and ${\kappa}_{tb} $$\not=$$ 0$.
One aspect of the case
${\kappa}_{tb} $$\not=$$ 0$ 
(and ${\kappa}_b$$<$${\kappa}_{{\rm crit.}}$)
was pointed out by
MTY (1989a, 1989b):
${\kappa}_{tb}$-term leads to a nonzero
$m_b$ via the ``feed-down''
effect coming from\footnote{
At the leading-$N_{{\rm c}}$ level:
$\langle {\bar t} t \rangle_0 
 = - N_{{\rm c}}/(4 \pi^2)
\int_0^{\Lambda^2} d {\bar q}^2 {\bar q}^2 m_t({\bar q}^2) 
[ {\bar q}^2 + m_t^2({\bar q}^2) ]^{-1}$. }
the $\langle {\bar t} t \rangle_0$
expectation value: $m_b $$\approx$$ -$$ (8 \pi^2) {\kappa}_{tb}
\langle {\bar t} t \rangle_0 / (N_{{\rm c}} {\Lambda}^{\!2}) 
$$\approx$$ 2 {\kappa}_{tb} m_t$.
The $\kappa_{tb}$-term,
unlike the other two, breaks $U(1)_{\gamma_5}$ invariance
[$q\!\mapsto\! \exp(i \alpha \gamma_5) q$]. Therefore,
if ${\kappa}_{tb}$$=$$0$, the Lagrangian is invariant
under $U(1)_{\gamma_5}$. This invariance plays the role
of the Peccei--Quinn (PQ) symmetry
\cite{PecceiQuinn77}
which is explicitly broken only by the color anomaly
(strong Adler-Bell-Jackiw anomaly). Because of that
explicit breaking, the resulting 
Nambu--Goldstone boson, called axion,
acquires a mass (it would be massless if the breaking were
spontaneous). This mass is 
estimated (see Tanabashi, 1992; Miransky, 1993)
to about $2$-$3$ MeV,
for $m_t $$\approx$$ 180$ GeV. However, such a visible axion is
phenomenologically unacceptable
\cite{Peccei89}. 
Therefore,
${\kappa}_{tb} $$\not=$$ 0$ must be assumed, and this cross term
can be adjusted to give an acceptable axion mass.

To summarize, the use of the
improved ladder DS equations and PS relations
(MTY, 1989a, 1989b; KM, 1990, 1991a)
gave higher predictions for $m_t$
than the RGE approach by BHL (1990). Even for
very high $\Lambda $$\sim$$ E_{{\rm Planck}}
$$\sim$$ 10^{19}$ GeV the predicted mass is $m_t $$\approx$$ 253$
GeV, i.e., $35$ GeV higher than the one-loop RGE prediction,
and $70$-$80$ GeV higher than the experimental value.
Further, as shown by KM (1990),
when the dynamical structure (propagator) of a heavy gauge boson, 
whose exchange is assumed to generate the 
effective ${\kappa}_t$-four-quark
contact term, is also taken into account, the predicted
$m_t$ remains practically unchanged. However, the effective
critical coupling parameter ${\kappa}_t $$\approx$$
{\kappa}_{{\rm crit.}}$ needed for condensation
increases in this case, offsetting
the smaller gauge boson propagator
[$\approx$$ 1/({\bar q}^2$$ +$$ {\Lambda}^{\!2})$$ < $$1/
{\Lambda}^{\!2}$].
In Sec.~\ref{CCVAMF}, we discuss 
connections, analogies and differences 
between the DS$+$PS and RGE approaches, 
as well as compare these two
methods with the ``hard mass'' effective
potential method. It should be stressed that all approaches
must give increasingly similar results when increasingly
similar assumptions are used.

\subsection{Other related approaches}
\label{MDSPS5}
Gribov (1994) derived a relation similar to
the PS equation (\ref{PSne}).
He investigated the polarization operator of the
$SU(2)_L$ bosons
within a leading-$N_{{\rm c}}$ framework.
His framework was based only on the assumption that the
Landau pole $\lambda$ of the 
$U(1)_Y$ coupling parameter is responsible
for the composite structure of the Higgs ($\lambda $$=$$ \Lambda$).
In his picture, the Higgs is predominantly a ${\bar t} t$
condensate, with admixtures of condensates of other fermions
proportional to their masses. Incidentally, Gribov also
derived in his leading-$N_{{\rm c}}$ framework the following
remarkable PS-type of relation for the Higgs ({\em not\/} $W$) mass 
\begin{equation}
M_H^2 = \frac{N_{{\rm c}}}{2 v^2 \pi^2}
\int_{m_t^2}^{{\Lambda}^{\!2}} \frac{d {\bar q}^2}
{{\bar q}^2} m_t^4({\bar q}^2) \ ,
\label{Gribov1}
\end{equation}
which can be apparently applied also in the TSM-type of scenarios.

The question of formal renormalizability of gauged
NJLVL models with nonrunning gauge coupling constant 
has been investigated by Bardeen, Leung and Love (1989)
and by Kondo, Tanabashi and Yamawaki (1993). They 
showed that in the ladder approximation of the DS approach
such models are renormalizable.
This means that they were
able to construct a well-defined algorithm to express
low energy physical predictions as finite quantities even
in the limit $\Lambda\!\to\! \infty$. The conclusion has been
additionally confirmed by the work of~\citeasnoun{Kondoetal94}
in which the flow of (renormalized) Yukawa parameter and mass
parameter has been investigated in the corresponding theory
with composite Higgs and gauge bosons. One possible drawback
of these proofs was that they were confined to the case of
a nonrunning gauge coupling and zero scalar self-coupling
parameters. On the other hand, 
Yamawaki (1991)
and later also 
Kondo, Shuto and Yamawaki (1991)
investigated explicit solutions of the DS equation
for ${\Sigma}_t({\bar p}^2)$ and their behavior
in the asymptotic region for the case of
a running gauge coupling parameter -- they showed that such (QCD) 
gauge interactions are crucial for the 
behavior in the asymptotic region.
Moreover, the decay constant $F_{\pi}$ as
determined by the PS relation then remains finite even
when ${\Lambda} $$\to$$ \infty$. In this sense, they argued that
QCD makes the NJLVL framework formally renormalizable.
Krasnikov (1993)
investigated the framework in $4$$-$${\epsilon}$ dimensions,
by including the scalar self-interactions, and argued
that it remains formally renormalizable also in this
case. Later, Harada {\em et al.\/} (1994)
and Kugo (1996)
discussed rigorously the relation between the gauged
NJLVL model on the one hand,
and the system of gauge bosons and (composite) Higgs
with renormalized Yukawa
parameter $y$ on the other hand, in the general
case of a running gauge coupling and scalar quartic 
self-coupling parameter $\lambda$,
for a finite compositeness scale (cutoff) $\Lambda$. 
This allowed them to take the continuum limit 
$\Lambda $$\to$$ \infty$ (${\epsilon} $$\to$$ 0$),
and to show with RGE methods that the two models
(gauged NJLVL, and gauge-Higgs-Yukawa) become equivalent
and nontrivial in this limit when renormalized
parameters $(y,\lambda)$ take
a specific critical value in the parameter plane. Therefore,
gauged NJLVL models are formally renormalizable.

One of the main arguments against the use of 
(ladder) DS equations
in studying nonperturbative physics has been
the gauge dependence of the resulting critical coupling
parameter(s) and of the dynamical mass(es)
$m $$=$$ {\Sigma}({\bar p}^2 $$=$$ m^2)$. While it has been
assumed by many that investigation
of the (improved) ladder DS equation gives the most reasonable 
results in the Landau gauge, arguments supporting this
assumption appeared only relatively recently
(Kondo, 1992; Curtis and Pennington, 1993; and references therein).
Curtis and Pennington worked within the strongly coupled
QED and in the general covariant $R_{\xi}$ gauge.
They showed that application of a nonperturbative {\em ansatz\/}
for the fermion-photon vertex, satisfying the Ward-Takahashi
and the (derivative) Ward identity, 
leads to values of ${\alpha}_{{\rm crit.}}$
and $m^{{\rm dyn.}}$ which are only weakly gauge dependent 
and close to the corresponding results of calculations in the 
usual ladder (``rainbow'') approximation in 
the Landau gauge.\footnote{
Ladder (``rainbow'') approximation
means here above all that fermion-boson vertex
in the single-photon-exchange picture of DS equation
is taken to be that of the tree level:
${\Gamma}^{\mu}(k,p)\!\mapsto\! {\gamma}^{\mu}$.
An improvement of such an approximation along
this ``rainbow'' approach (i.e., using tree level vertices) 
would be calculation of higher, e.g.~next-to-leading in
$1/N_{{\rm c}}$, contributions to DS equation 
-- cf.~Sec. \ref{NTLEE}. Such an expansion is systematic and
gauge noninvariant, in contrast to the above approach which
is nearly gauge invariant and nonunique (nonsystematic)
due to application of an {\em ansatz\/}.}

Recently, Blumhofer and Manus (1998) proposed 
treating the DS equation in a drastically different way
-- in Minkowski instead of Euclidean metric. The
main motivation behind this approach lies in the
possibly insurmountable difficulty to analytically
continue in the general case the Euclidean dynamical quark mass
${\Sigma}_q({\bar p}^2)$ to the time-like on-shell 
region ${\bar p}^2\!\mapsto\! -p^2 $$=$$ -m_q^2$. They worked in
the framework of single exchange of 
strongly coupled gauge boson (ladder approximation), 
using the Landau gauge and the linear approximation for
the denominator stemming from the quark propagator
[$(k^2- {\Sigma}_q(k^2))^{-1} \mapsto (k^2-m_q^2)^{-1}$]
in the DS equation.
When they applied the unsubtracted dispersion relation, 
valid for the {\em regular} solution 
${\Sigma}_q(p^2) $$\propto$$ 1/p^2$ [in the time-like region, 
${\Sigma}_q(p^2)$ is in general complex], 
to the DS equation, they obtained a one-dimensional
integral equation for $I(p^2) $$\equiv$$ \mbox{Im}[
{\Sigma}_q(p^2)]$. After introducing asymptotically
free running of the gauge coupling ${\alpha}(p^2) $$\propto$$
(\ln p^2/{\Lambda}_{{\rm IR}}^2)^{-1}$,
they were able to solve the mentioned integral equation,
even when mass $M$ of the exchanged gauge boson
was zero. The authors pointed out that the results agree 
in the asymptotic region $p^2 $$\gg$$ M^2$
with the corresponding regular solutions of the Euclidean
metric approach.
(Analytic continuation ${\bar p}^2
\mapsto - {\bar p}^2 $$<$$ 0$ for
${\bar p}^2 $$\gg$$ M^2$, where $M$ is effectively
the cutoff, can be performed easily in the
asymptotic region since the solution there is
explicitly known.)
The authors further indicated
how to use the {\em subtracted} dispersion relations
to obtain the {\em irregular} solution
[for QCD: ${\Sigma}_q(p^2)_{{\rm irreg.}}
$$\propto$$ (\ln p^2/{\Lambda}_{{\rm IR}})^{-8/14}$]
which corresponds to the actual running quark mass $m_q(p^2)$
[cf.~(\ref{mtQCD}) for RGE-running $m_q(p^2)$ in QCD].

We defer the discussion of works connected with investigations
of DS (gap) equations beyond the (improved) ladder approximation
to Sec.~\ref{NTLEE}.

\section{Next-to-leading (NTL) effects in $1/N_{\rm c}$ expansion}
\label{NTLEE}
\setcounter{equation}{0}

\subsection{Effective potential as a function of a ``hard mass''}
\label{NTLEE1}  
\begin{figure}[htb]
\mbox{}
\vskip3.cm\relax\noindent\hskip1.2cm\relax
\includegraphics{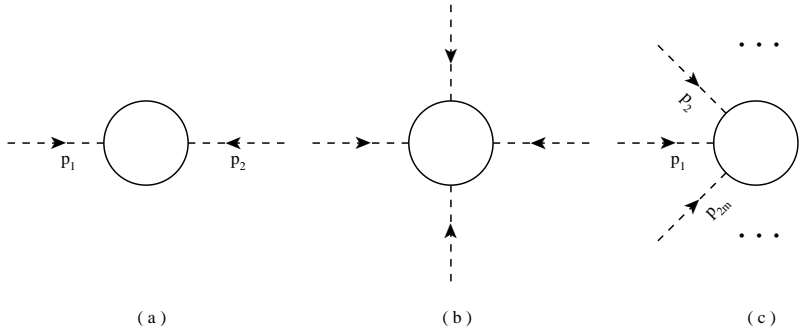} \vskip0.5cm
\caption{\footnotesize One-loop 1-PI diagrams
giving the relevant 1-PI Green functions 
${\tilde {\Gamma}}_{H}^{(2m;1)}(p_1, \ldots, p_{2m})$.
The latter, for zero external momenta,
enable us to form an infinite power series in ${\cal {H}}_0$
leading to the leading-$N_{{\rm c}}$
quark loop part of the effective potential 
$V_{{\rm eff}}({\cal {H}}_0)$. Full lines
represent massless top quarks, and dotted external lines
the nondynamical Higgs. The couplings used
are those of the TSM Lagrangian density (\ref{TSM2}).}
\label{rmd3f}
\end{figure}

With a view toward studying the next-to-leading (NTL) contributions
to the condensation mechanism, we present here
a third approach.
It makes use of an auxiliary field method to calculate
the effective potential\footnote{
The method of $V_{{\rm eff}}$ to investigate spontaneous
symmetry breaking was introduced by 
Coleman and E.~Weinberg (1973).
S.~Weinberg (1973) 
showed that
cancellation of tadpole contributions in the vacuum
(in the Landau gauge) is
equivalent to the minimization of $V_{{\rm eff}}$.
Path-integral method
to evaluate $V_{{\rm eff}}$ was worked out by 
Jackiw (1974).}
$V_{{\rm eff}}$ as a function of
a ``hard mass'' (nonrunning top quark mass) $\sigma$.
At the minimum of $V_{{\rm eff}}$ (vacuum), we have:
$\langle {\hat \sigma} \rangle_0\!=\!m_t({\Lambda})$,
where $m_t({\Lambda})$ is the dynamical bare mass. The
mass $m_t^{{\rm ren.}}$ can then be obtained
by including renormalization effects in the energy
interval below ${\Lambda}$. The approach is an alternative
to solving the DS integral equation for 
${\Sigma}_t({\bar p}^2)$, and it
allows in a systematic way inclusion of the 
NTL contributions.\footnote{
The Bethe--Salpeter (BS) equation
for the Nambu--Goldstone bosons (NGB's) 
at the NTL level (or an NTL analog of the
PS relation of the MTY approach) has not yet been investigated. 
Thus, we cannot yet make predictions at the NTL level
for $m_t^{{\rm ren.}}$ as a function of the 
cutoffs $\Lambda$ alone.}
The approach has been applied in a series of papers
\cite{CveticPaschosVlachos96,CveticVlachos96,Cvetic97}, 
at the NTL level, specifically for the minimal
framework of Eqs.~(\ref{TSM2})-(\ref{TSM2not}).
$V_{{\rm eff}}$ is calculated as a function of
$\langle {\hat {\cal H}} \rangle\!=\!{\cal H}_0$
and $\langle {\hat {\cal G}}^{(j)} 
\rangle\!=\!{\cal G}^{(j)}_0$ via the path integration
formula\footnote{
In the vacuum we have: $\langle {\hat \sigma}
\rangle_0\!=\!\sqrt{G} M_0 \langle {\hat {\cal H}} \rangle_0/
\sqrt{2}\!=\!m_t({\Lambda})$ and $\langle {\hat {\cal G}}^{(j)}
\rangle_0\!=\!0$}
\cite{FukudaKyriakopoulos75}
\begin{eqnarray}
\lefteqn{
\exp \left[ -\Omega V_{{\rm eff}}({\cal{H}}_0,
{\cal{G}}_0^{(j)} )  \right]
=\mbox{const.} \times
\int \prod_{j=0}^2 \left[ {\cal{D}}{\cal{G}}^{(j)}
\delta \left( \int d^4 {\bar y} {\cal{G}}^{(j)}({\bar y})
-\Omega {\cal{G}}_0^{(j)}  \right)
\right] \times }
\nonumber \\
&&
 \int {\cal{D}}{\cal{H}}
\delta \left( \int d^4 {\bar y} 
{\cal{H}} ({\bar y})-\Omega {\cal{H}}_0 \right)
\int {\cal{D}} {\bar \Psi} {\cal{D}} \Psi
\exp \left[ +\int d^4 {\bar x} {\cal{L}}({\bar x}) \right] \ .
\label{pathint1}
\end{eqnarray}
Euclidean metric is used, $\hbar $$=$$ 1$,
$\Omega\!=\!\int d^4 {\bar x}$ (formally infinite),
${\cal {L}}({\bar x})$ is the Euclidean 
version of density
(\ref{TSM2}) with cutoff $\Lambda$. Path integration
over the quark degrees of freedom (d.o.f.'s) can be done exactly
because it involves Gaussian integrals. Path integration over the
scalar d.o.f.'s can be performed subsequently in the sense of 
the $1/N_{{\rm c}}$ expansion up to the NTL order, and
even an analogous integration over the QCD (gluonic) d.o.f.'s can be
included. Details are given in the
work of Cveti\v{c} (1997), where
proper time cutoff and a Pauli-Villars regularization cutoff
were used for the fermionic momenta, 
within the proper time regularization approach, as well as
the covariant spherical cutoff (${\Lambda}_{{\rm f}}$). 
For the bosonic
(scalar) momenta, which appear at the NTL level,
spherical covariant cutoff 
(${\Lambda}_{{\rm b}}$) had to be used
(${\Lambda}_{{\rm f}}\!\sim\!{\Lambda}_{{\rm b}}\!\sim\!{
\Lambda}$).
In the following, we will give results only for the case of
spherical covariant cutoffs. $V_{{\rm eff}}$
can be expressed conveniently by dimensionless parameters
\begin{eqnarray}
\Xi_{{\rm eff}}({\varepsilon}^2; a;
{\Lambda}^{\!2}_{{\rm b}}/{\Lambda}^{\!2}_{{\rm f}})&  \equiv&
8\pi ^2 V_{{\rm eff}}/
(N_{{\rm c}}\Lambda_{{\rm f}}^{\!4})=
\Xi ^{(0)}+\frac 1{N_{{\rm c}}}\Xi^{(1)}
+ {\cal {O}}(\frac{1}{N^2_{{\rm c}}}) \ .  
\label{defka2}
\\
{\varepsilon}^2 &\equiv&
\frac{\sigma_0^2}{{\Lambda}^{\!2}_{{\rm f}}} \equiv
\frac{G M_0^2}{2 {\Lambda}^{\!2}_{{\rm f}}}
\langle {\cal{H}} \rangle_0^2\ , \qquad
a \equiv \frac{ G N_{{\rm c}}
{\Lambda}^{\!2}_{{\rm f}} }{ 8\pi^2 } \ 
\left( \equiv {\kappa}_t \right) \ .
\label{defka1}
\end{eqnarray}
Parameter $a$ is in fact
${\kappa}_t$ of Sec.~\ref{MDSPS} 
[cf.~(\ref{3gTSM})]. 
The leading-$N_{{\rm c}}$ part $\Xi^{(0)}$ is made up
of the quark loop (q.l.) and of the QCD (gluonic)
contributions\footnote{The QCD integrals were regulated
by the proper time cutoffs
${\tau}_{{\rm f}} $$\geq$$ 1/{\Lambda}^{\!2}_{{\rm f}}$ 
and ${\tau}_{{\rm b}} $$\geq$$ 1/{\Lambda }_{{\rm b}}^2$.} 
\begin{eqnarray}
{\Xi} ^{(0)}_{{\rm q.l.}}\left( {\varepsilon}^2;a\right) &=&
\frac{{\varepsilon}^2}a
 -\int_0^1d{\bar k}^2{\bar k}^2
\ln \left( 1+\frac{{\varepsilon}^2}{{\bar k}^2}\right) \ ,  
\label{Xi0}
\\
\Xi^{(0)}_{{\rm gl}}
\left( {\varepsilon}^2;
{\Lambda }_{{\rm b}}^2/{\Lambda}
_{{\rm f}}^2 \right) 
&=& \frac{(N_{{\rm c}}^2-1)}{4 N_{{\rm c}}} 
\int_0^{{\Lambda }_{{\rm b}}^2/{\Lambda}
_{{\rm f}}^2}
d{\bar p}^2{\bar p}^2
\ln \left[ 1-a_{{\rm gl}}
 {\cal {J}}_{{\rm gl}}\left( {\bar p}^2;
 {\varepsilon}^2\right)
\right] \ , 
\label{Xigl}
\\
\Xi^{(0)}& = &\Xi^{(0)}_{{\rm q.l.}} +
\Xi^{(0)}_{{\rm gl}} \ . 
\end{eqnarray}
Here, $a_{{\rm gl}}\!\approx\!3 {\alpha}_3(m_t^{{\rm phys.}})/
\pi\!\approx\!0.105$. 
The (QCD) gluon part (\ref{Xigl}) has its origin in exchanges of the
gluons (in the Landau gauge). We note that $\Xi_{{\rm gl}}^{(0)}$
is really leading-$N_{{\rm c}}$,
i.e., ${\Xi}^{(0)}_{{\rm gl}}$$\sim$$N_{{\rm c}}^0$, 
because ${\alpha}_3$$\sim$$
N_{{\rm c}}^{-1}$ as seen from (\ref{alph3})
and thus: $\ln [ 1 $$-$$ a_{{\rm gl}} 
{\cal {J}}_{{\rm gl}} ] \propto
- a_{{\rm gl}} {\cal {J}}_{{\rm gl}}
\propto a_{{\rm gl}} \propto {\alpha}_3
\sim N_{{\rm c}}^{-1}$.
The NTL part $\Xi^{(1)}$ comes from exchanges of (composite) scalars:
\begin{eqnarray}
\frac{1}{N_{{\rm c}}} {\Xi}^{(1)}\left( {\varepsilon}^2; a;
{\Lambda }_{{\rm b}}^2/
{\Lambda }_{{\rm f}}^2 \right) &=&
\frac{1}{4 N_{{\rm c}}} \int_0^{{\Lambda }_{{\rm b}}^2/
{\Lambda }_{{\rm f}}^2} 
 d\bar p^2 \bar p^2 
  \sum_{{\rm X}} A_{{\rm X}} \ln \left[ 
  1-a {\cal J}_{{\rm X}} 
\left( {\bar p}^2; {\varepsilon}^2\right) \right] \ .  
\label{Xi1}
\end{eqnarray}
Fermionic (${\bar k}$)
and bosonic (${\bar p}$) momenta were rescaled:
${\bar k}_{{\rm new}} $$=$$ {\bar k}_{{\rm old}}/{\Lambda}_{{\rm f}}$,
${\bar p}_{{\rm new}} $$=$$ {\bar p}_{{\rm old}}/{\Lambda}_{{\rm f}}$.
X $=$ H (Higgs), Gn, Gch (neutral and charged NGB's);
respective multiplicity factors are: $A_{{\rm X}} $$=$$ 1, 1, 2$.
${\cal {J}}_{{\rm X}}$ and ${\cal {J}}_{{\rm gl}}$ 
represent loop contributions of $t$
and $b$ to the dimensionless two-leg Green functions 
for scalars and gluons.\footnote{
They are represented by Fig.~\ref{rmd3f}(a), where external 
momenta are in this case nonzero, and $t$
in the loop is regarded as already having acquired the
(dynamical) mass $m_t(\Lambda) $$=$$ 
{\varepsilon} {\Lambda}_{{\rm f}}$, while $m_b$$=$$0$.}
We refer to the mentioned references for details.

\begin{figure}[htb]
\mbox{}
\vskip5.9cm\relax\noindent\hskip1.0cm\relax
\includegraphics{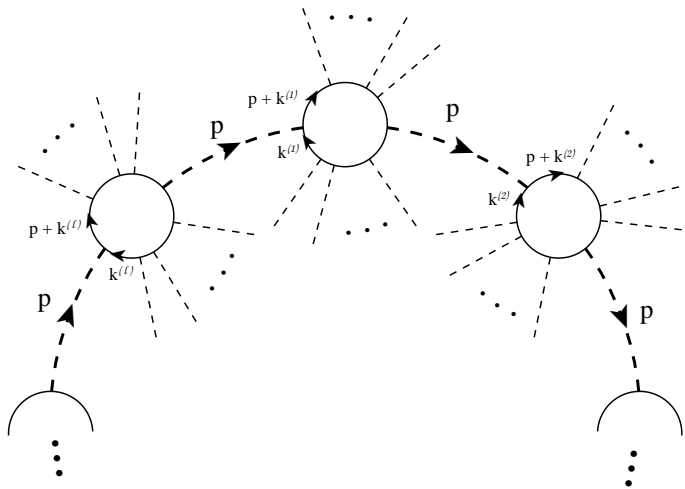} \vskip-0.1cm
\caption{\footnotesize The higher loop 1-PI diagrams
which contribute to 1-PI Green functions at the NTL level
in $1/N_{{\rm c}}$ expansion. Full lines represent the 
massless top and bottom quarks, and
dotted lines either the (nondynamical)
scalars or gluons. The couplings used are those 
of the TSM Lagrangian density (\ref{TSM2}).}
\label{rmd4f}
\end{figure}

Alternatively, $\Xi^{(0)}_{{\rm q.l.}}$,
$\Xi^{(0)}_{{\rm gl}}$ and $\Xi^{(1)}$ can be
rederived diagrammatically by summing up terms corresponding
to the one-particle irreducible (1-PI) 
Green functions depicted in Figs.~\ref{rmd3f} 
(for $\Xi^{(0)}_{{\rm q.l.}}$)
and \ref{rmd4f} (for $\Xi^{(1)}$ and $\Xi^{(0)}_{{\rm gl}}$).

\subsection{Gap equation and mass renormalization at NTL level}
\label{NTLEE2}
Minimizing the part $\Xi^{(0)}_{{\rm q.l.}}$
(of $\Xi_{{\rm eff}}$) leads to the familiar gap equation in the
leading-$N_{{\rm c}}$ approximation without QCD
(quark loop approximation) -- relating the
cutoff ${\Lambda}_{{\rm f}}$, the four-quark
coupling strength $G$ ($\Leftrightarrow$$a$) and the 
leading-$N_{{\rm c}}$ (q.l.) mass $m_t^{(0)}$ 
\begin{equation}
\frac{\partial \Xi ^{(0)}_{{\rm q.l.}}
\left( {\varepsilon}^2;a\right) }
{\partial {\varepsilon}^2}{\Bigg |}_{{\varepsilon}^2=
{\varepsilon}_0^2}= \left[ \frac{1}{a} - 1
+ {\varepsilon}^2 \ln \left( {\varepsilon}^{-2} + 1 \right)
\right]{\Bigg |}_{{\varepsilon}^2 = {\varepsilon}_0^2} = 0 \ , 
\label{gaplead1}
\end{equation}
\begin{equation}
\Rightarrow \quad a \quad 
\left( \equiv \frac{GN_{{\rm c}} 
{\Lambda}^{\!2}_{{\rm f}}}{8\pi ^2} \right)
=\left[ 
1-{\varepsilon}_0^2\ln \left( {\varepsilon}_0^{-2}+1\right)
\right]^{-1} \quad 
\left( \sim 1 \right) \ ,\qquad 
{\varepsilon}_0^2 \equiv
\left( m_t^{(0)}/{\Lambda}_{{\rm f}} \right)^2 \ .  
\label{gaplead}
\end{equation}
This coincides with the quark-loop-approximated solution
(\ref{qlgap}) of DS equation, as it should [there are
no renormalization corrections to $m_t^{(0)}$ at the
quark loop level, i.e., $m_t^{(0)}({\bar p}^2)$ is
standing]. 
Eq.~(\ref{gaplead}) shows that 
$a$$>$$1$ and $a\!\sim\!1$ ($\sim\!N_{{\rm c}}^0$).
The NTL information connecting the bare mass 
$m_t({\Lambda}_{{\rm f}})$, the cutoff 
${\Lambda}_{{\rm f}}$ and the parameter $G$ (or: $a$), 
can be obtained by minimizing $\Xi_{{\rm eff}}$ of
(\ref{defka2})-(\ref{Xi1})
\begin{eqnarray}
\lefteqn{
\frac{ {\partial } {\Xi}_{{\rm eff}}
\left( {\varepsilon}^2 ; a ; 
{\Lambda}^{\!2}_{{\rm b}}/{\Lambda}^{\!2}_{{\rm f}} \right) }{
{\partial} {\varepsilon}^2 }{\Bigg |}_{ {\varepsilon}^2 =
{\varepsilon}^2_{{\rm gap}} } =
\frac{1}{a} - \left[ 1 - {\varepsilon}^2_{{\rm gap}}
\ln \left( {\varepsilon}^{-2}_{{\rm gap}} + 1 \right) \right]
} \nonumber\\
& &
- a_{{\rm gl}} \frac{(N_{{\rm c}}^2-1)}{4 N_{{\rm c}}}
\int_0^{{\Lambda}^{\!2}_{{\rm b}}/
{\Lambda}^{\!2}_{{\rm f}}} d{\bar p}^2 {\bar p}^2
\frac{ \partial {\cal {J}}_{{\rm gl}} 
( {\bar p}^2 ; {\varepsilon}^2 ) }{ \partial {\varepsilon}^2 }
\left[ 1 - a_{{\rm gl}} {\cal {J}}_{{\rm gl}}
( {\bar p}^2 ; {\varepsilon}^2 )  \right]^{-1}
{\Bigg |}_{ {\varepsilon}^2 = {\varepsilon}^2_{{\rm gap}} }
\nonumber\\
& &
 - \frac{1}{ 4 N_{{\rm c}} }
\int_0^{ {\Lambda}^{\!2}_{{\rm b}}/
 {\Lambda}^{\!2}_{{\rm f}} }
d {\bar p}^2 {\bar p}^2 
{\Bigg \{ } \sum_{{\rm X}} A_{{\rm X}}
\frac{ {\partial} {\cal{J}}_{{\rm X}} }
{ {\partial} {\varepsilon}^2 }
\left[ 1 - {\varepsilon}^2 
\ln \left( {\varepsilon}^{-2} + 1 \right)
- {\cal{J}}_{{\rm X}} \left( {\bar p}^2; {\varepsilon}^2 
\right) \right]^{-1}
{\Bigg \} }{\Bigg |}_{ {\varepsilon}^2 = 
{\varepsilon}^2_{{\rm gap}} } = 0 .
\label{NTLgapS2}
\end{eqnarray}
where ${\varepsilon}^2_{{\rm gap}}(a;r)$$=$$
m_t^2(\Lambda)/{\Lambda}^{\!2}_{{\rm f}}$, with
$a$ and $r$$\equiv$${\Lambda}_{{\rm b}}/{\Lambda}_{{\rm f}}$
the input parameters.
Since this gap equation includes NTL contributions, we have
formally: ${\varepsilon}^2_{{\rm gap}} $$=$$
{\varepsilon}^2_0 [ 1 $$+$$ {\cal {O}}(1/N_{{\rm c}})]$. 
An important point should be mentioned.
Denominators in the above integrands of the NTL part
are propagators of the composite scalars in the 
leading-$N_{{\rm c}}$ approximation. Formally,
as implied by (\ref{Xi1}), they should be written as
$(a/2) [ 1$$ -$$ a {\cal{J}}_{{\rm X}} 
({\bar p}^2; {\varepsilon}^2) ]^{-1}$ 
(X $=$ H, Gn, Gch). However, they contain singularities
(poles) in the interval $0 $$<$$ {\bar p}^2 $$\leq$$ 1$,
because $1/N_{{\rm c}}$ expansion
does not automatically respect 
the Goldstone theorem (masslessness
of NGB's). This theorem is ensured in the
NTL gap equation (\ref{NTLgapS2}) by
replacing the mentioned propagators in 
${\partial} {\Xi}^{(1)}/ {\partial} {\varepsilon}^2$ by
$(1/2) [1 $$-$$ {\varepsilon}^2 \ln ( {\varepsilon}^{-2} $$+$$ 1 )
$$-$$ {\cal{J}}_{{\rm X}} 
({\bar p}^2; {\varepsilon}^2) ]^{-1}$. The
difference between the inverse of the old and of the latter 
expression is $2 {\partial} {\Xi}^{(0)}_{{\rm q.l.}}
/ {\partial} {\varepsilon}^2$ and
is hence $\sim\!1/{N_{{\rm c}}}$
for the values of ${\varepsilon}^2$
near the new minimum: ${\varepsilon}^2 $$\approx$$
{\varepsilon}^2_{{\rm gap}} $$=$$
{\varepsilon}_0^2 [ 1 $$+$$ {\cal {O}}(1/N_{{\rm c}}) ]$. Thus,
this changed the NTL expression
${N_{{\rm c}}}^{-1} {\partial} {\Xi}^{(1)}/ 
{\partial} {\varepsilon}^2$
formally by a next-to-next-to-leading order 
($\!\sim\!1/{N_{{\rm c}}}^2$),
resulting in a fully legitimate modification of 
the NTL gap equation. 
This change makes now all three propagators
nonsingular in the interval $0 $$<$$ {\bar p}^2 $$\leq$$ 1$,
the integrals in (\ref{NTLgapS2}) are finite and well defined.
The Goldstone theorem is now explicitly respected at
the NTL level of the gap equation,\footnote{
For more discussion on these points see
Cveti\v c (1997).
Note that an earlier work (Cveti\v c, Paschos
and Vlachos, 1996) didn't include the composite NGB's, and 
the integrands stemming from the composite Higgs exchanges
were not regularized; as a result, for small ${\varepsilon}^2
$$\ll$$ {\varepsilon}_0^2$ a singularity was encountered, but
that didn't affect the basic conclusions of the paper.} 
because
$ [1 $$-$$ {\varepsilon}^2 \ln ( {\varepsilon}^{-2}$$+$$ 1 )
 $$-$$ {\cal{J}}_{{\rm X}} 
({\bar p}^2; {\varepsilon}^2) ] \propto {\bar p}^2$
for X $=$ Gn, Gch, when ${\bar p}^2 $$\to$$ 0$. Similar problems
were discussed also by~\citeasnoun{Nikolovetal96}
in an $SU(2)$-invariant NJLVL model which they
regarded as a model of low energy QCD.

\begin{figure}[htb]
\mbox{}
\vskip4.4cm\relax\noindent\hskip1.4cm\relax
\includegraphics{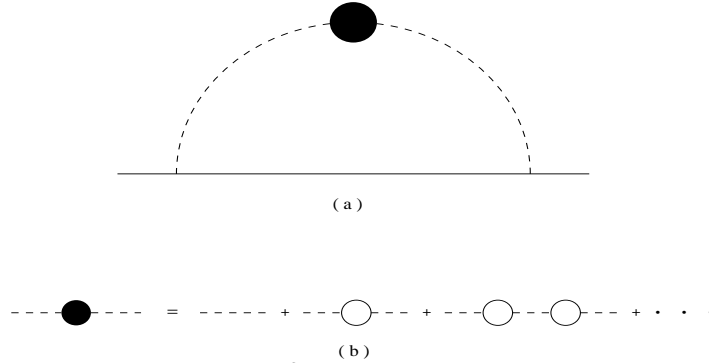} \vskip-0.2cm
\caption{\footnotesize The 1-PI diagrams which give
${\delta}({\varepsilon}^2)_{{\rm ren.}}$. The $t$ quarks
in loops have mass $m_t(\Lambda)
$$\equiv$$ {\Lambda}_{\rm f} {\varepsilon}_{\rm gap}$
as determined by the NTL gap equation (\ref{NTLgapS2}).
Dashed line is the propagator of either the (nondynamical)
scalar or of gluon. According to Fig.~(b), the black blob
makes the propagator of scalars dynamical.}
\label{rmd5f}
\end{figure}

The renormalization 
${\varepsilon}_{{\rm gap}}(a;r) $$\mapsto$$
{\varepsilon}_{{\rm ren.}}(a;r)$ allows one to obtain values
of energies where the new physics responsible for
the condensation is expected to set in:
$\Lambda $$\sim$$ {\Lambda}_{{\rm f}}$
$ = $$m_t^{{\rm ren.}}/
{\varepsilon}_{{\rm ren.}}(a;r)$$ =$$ {\Lambda}_{{\rm f}}(a; r)$,
where $a$ and $r$$ \equiv$
$ {\Lambda}_{{\rm b}}/{\Lambda}_{{\rm f}}$
are the input parameters, and 
$m_t^{{\rm ren.}}$$\approx$$ 170$-$180$ GeV. The
QCD contribution is formally leading-$N_{{\rm c}}$
\begin{equation}
\delta ({\varepsilon}^2)_{{\rm  ren.}}
^{{\rm QCD}} \approx 
\frac{ 9 (N_{{\rm c}}^2\!-\!1) }{ 
2 N_{{\rm c}} (11 N_{{\rm c}}\!-\!12) }
\frac{1}{ \ln ( m_t/{\Lambda}_{{\rm QCD}} ) } 
{\varepsilon}^2
\left[ 
\ln \left( {\varepsilon}^{-2} \right) 
\!+\!\ln (r^2)\!+\!0.256 
\!+\!{\cal {O}}({\varepsilon}^2)
\right] ,
\label{kapglr}
\end{equation}
where ${\varepsilon}$$=$${\varepsilon}_{{\rm gap}}(a;r)$ [solution
of the NTL gap Eq.~(\ref{NTLgapS2})].
Contribution of the (composite) scalars is NTL, but at the
cutoffs ${\Lambda}$$\sim$$1$ TeV it is
numerically larger than (\ref{kapglr})
\begin{equation}
\delta ( {\varepsilon}^2 )_{{\rm ren.}}
^{{\rm (NTL)}} =
\kappa_{1{\rm r}}/N_{{\rm c}}  =
\left(  \kappa _{1{\rm r}}^{{\rm (H)}}
+\kappa _{1{\rm r}}^{{\rm (Gn)}}
+\kappa _{1{\rm r}}^{{\rm (Gch)}}
\right)/N_{{\rm c}}  \ .
\label{kappar}
\end{equation}
For details on integrals ${\kappa}_{1{\rm r}}^{{\rm (X)}}$,
including the needed analytical continuation to the
on-mass-shell values of the external quark momentum in
Fig.~\ref{rmd5f}(a) 
(${\bar q}^2 $$=$$ -q^2 $$=$$ - {\varepsilon}^2_{\rm gap}
$$<$$ 0$), we refer to the mentioned literature.
Just as in (\ref{NTLgapS2}),
the dynamical scalar propagators in these integrals 
are modified to ensure the Goldstone theorem.

\subsection{Numerical results: leading-$N_{\rm c}$ 
vs.~NTL gap equation}
\label{NTLEE4}
The NTL contributions in the
gap equation (\ref{NTLgapS2})
represent mostly the so called ``feedback effects'' of
the composite scalars on its own binding.
The basic conclusion following
from the numerical results is:
these NTL contributions are strong and,
for given values of the four-quark coupling parameter $a$ 
($\equiv$$ {\kappa}_t$) and of the ratio 
$r\!\equiv\!{\Lambda}_{{\rm b}}/{\Lambda}_{{\rm f}}$ 
($<\!1$),\footnote{
Diagrams of Fig.~\ref{rmd4f} suggest that
${\bar p}_{{\rm max}}^2
$$\leq$$ {\bar k}_{{\rm max}}^2$, implying
the input values ${\Lambda }_{{\rm b}}/
{\Lambda}_{{\rm f}} $$\stackrel{<}{\sim}$$ 1$.}
change the leading-$N_{{\rm c}}$ predictions for the bare mass
$m_t({\Lambda}_{{\rm f}})$ so 
drastically that $1/N_{{\rm c}}$ expansion
may lose predictability unless the cutoffs are low
${\Lambda}_{{\rm f}} $$\sim$$ {\Lambda}_{{\rm b}} 
$$\stackrel{<}{\sim}$$ 1$ TeV.
This is illustrated in Table~\ref{tabl3}. For given values of
$r\!\equiv\!{\Lambda}_{{\rm b}}/{\Lambda}_{{\rm f}}
\!=\!1/\sqrt{2},\!1$, the other input parameter
$a$ was adjusted so that ${\varepsilon}_{{\rm gap}}^2(a;r)/
{\varepsilon}_0^2(a)\!\equiv\!(m_t({\Lambda})/m_t^{(0)})^2$
attained values $1/4,\!1/3,\!1/2$, where
${\varepsilon}_{{\rm gap}}(a;r)$ solves the
NTL gap equation ({\ref{NTLgapS2}), and ${\varepsilon}_0(a)$
the leading-$N_c$ quark loop\footnote{
Although QCD contributions are formally
part of the leading-$N_{{\rm c}}$ contributions,
we consider them numerically
as part of NTL effects 
-- because QCD contributions turn out to
be here even smaller than the NTL ``feedback'' contributions
of the composite scalars.}
gap equation (\ref{gaplead}). 
\begin{table}[ht]
\vspace{0.3cm}
\par
\begin{center}
\begin{tabular}{ c l c c c c }
$a$ &
${\Lambda}_{{\rm b}}/
 {\Lambda}_{{\rm f}}$ & 
$ m_t({\Lambda})/m_t^{(0)}  $ & 
$ m_t^{{\rm ren.}}/m_t^{(0)}  $ &
${\Lambda}_{{\rm f}}$ & 
${\Lambda}_{{\rm b}}$ \\
\hline\hline
1.441 & 1/$\sqrt{2}$  & 0.500 ($=$$\sqrt{1/4}$) & 0.448 
& 1.037 & 0.733 \\
1.520 & 1/$\sqrt{2}$  & 0.577 ($=$$\sqrt{1/3}$) & 0.515 
& 0.816 & 0.577 \\
1.841 & 1/$\sqrt{2}$  & 0.707 ($=$$\sqrt{1/2}$) & 0.624 
& 0.506 & 0.358 \\
\hline
1.959 & 1 & 0.500 ($=$$\sqrt{1/4}$) & 0.397 & 0.737 & 0.737 \\
2.292 & 1 & 0.577 ($=$$\sqrt{1/3}$) & 0.450 & 0.548 & 0.548 \\
\end{tabular}
\end{center}
\vspace{-0.1cm}
\caption {\footnotesize
Quark (fermionic) and bosonic cutoffs ${\Lambda}_{{\rm f}}$
and ${\Lambda}_{{\rm b}}$ (in TeV), obtained for given
$r\!\equiv\!{\Lambda}_{{\rm b}}/{\Lambda}_{{\rm f}}\!=
\!1/\sqrt{2},\!1$ and with imposing
the requirement that the ratio 
${\varepsilon}^2_{\rm gap}/{\varepsilon}_0^2
$$=$$ (m_t({\Lambda})/m_t^{(0)})^2$
not be smaller than: $1/4$, $1/3$, $1/2$
($a$ was adjusted accordingly).
The corresponding ratios $m_t^{{\rm ren.}}/m_t^{(0)}$
are also given.
Covariant spherical (S) cutoffs were employed. The value
$m_t^{{\rm ren.}} $$=$$ 180 \mbox{ GeV}$ was used
(Abe {\em et al.\/}, 1995;
Adachi {\em et al.\/}, 1995), to obtain
${\Lambda}_{{\rm f}}(a;r)\!=\!m_t^{{\rm ren.}}/
{\varepsilon}_{{\rm ren.}}(a;r)$.}
\label{tabl3}
\end{table}

In Table~\ref{tabl4} the resulting
cutoffs are given when we allow a highly nonperturbative
behavior of $1/N_{{\rm c}}$ expansion: 
${\varepsilon}^{\mbox{\tiny (NTL)}}_{{\rm gap}}/
{\varepsilon}_0 $$=$$ 0.05$. The cutoffs are then higher
($\sim\!10$ TeV). However, these cases
are speculative since the meaning of $1/N_{{\rm c}}$
expansion is then in serious danger. Such results could
make sense if the next-to-next-to-leading (NNTL)
terms of the gap equation (which were not calculated)
are very small and provoke only a small relative
change of the obtained small ${\varepsilon}^2_{{\rm gap}}$
of the NTL gap equation (\ref{NTLgapS2}).
While this scenario seems to be unlikely, it
cannot be completely ruled out since the strong
``feedback'' effects might be represented
almost exclusively by the NTL contributions.
\begin{table}[ht]
\vspace{0.3cm}
\par
\begin{center}
\begin{tabular}{ c l c c c c }
$a$ &
${\Lambda}_{{\rm b}}/
 {\Lambda}_{{\rm f}}$ & 
$ m_t({\Lambda})/m_t^{(0)} $ & 
$ m_t^{{\rm ren.}}/m_t^{(0)}$ &
${\Lambda}_{{\rm f}}$ & 
${\Lambda}_{{\rm b}}$ \\
\hline\hline
1.303 & $1/\sqrt{2}$ & 0.050 & 0.046 & 12.7 & 9.0 \\
1.549 & 1           & 0.050 & 0.041 & 9.8 & 9.8 \\
\end{tabular}
\end{center}
\vspace{-0.1cm}
\caption {\footnotesize 
Results for the cutoffs (in TeV)
in the case of a highly nonperturbative
behavior in the sense of $1/N_{\rm c}$ expansion: 
$m_t(\Lambda)\!=\!0.05 m_t^{(0)}\!\ll\!m_t^{(0)}$. The
results are given for two choices of the cutoff ratios
${\Lambda}_{\rm b}/{\Lambda}_{\rm f}$: $1$, $1/\sqrt{2}$. 
The value $m_t^{\rm ren.}\!=\!180$ GeV was used.}
\label{tabl4}
\end{table}

The contributions of the pure 
(i.e., transverse) components of 
the electroweak gauge bosons were not
included in the discussed NTL calculations. 
There are indications that these contributions are
numerically not very important. The RGE approach of
BHL indicates that these contributions, at least
those to the mass renormalization, are small --
because of relatively small $SU(2)_L$$\times$$U(1)_Y$
gauge coupling parameters.

\subsection{Small-${\varepsilon}$ (large-$\Lambda/m_t$) 
expansion of the NTL gap equation}
\label{NTLEE5}
Despite the just mentioned reservations 
about small-${\varepsilon}^2_{{\rm gap}}$ 
(large-$\Lambda$) solutions,
it is instructive to investigate the latter 
to discern the nature of the phase transition
between the symmetric and the broken phase
at the NTL level.\footnote{
The author acknowledges that W.~A.~Bardeen had
suggested small-${\varepsilon}^2$ expansion 
at the NTL level for this reason, and that
E.~A.~Paschos had conveyed to him this idea of
Bardeen. The results of this
Section have not appeared in the literature.} 

The small-${\varepsilon}^2$ NTL gap equation is
obtained by performing in the relevant integrals
expansions for small ${\varepsilon}^2$ and small
${\bar p}^2 $$\equiv$$ {\bar q}^2_{{\rm boson}}/
{\Lambda}^{\!2}_{{\rm f}}$. Specifically, such expansions
in the case of the proper-time cutoff (PTC) for fermionic
momenta give the gap equation\footnote{
No QCD was included.
For details on the integrals of the NTL gap equation 
in the PTC case, see Ref. Cveti\v{c} (1997) -- Eqs.~(41),
(39), (B.4), (B.20), (B.21) there.} 
\begin{equation}
\frac{\partial {\Xi}_{{\rm eff}}}
{\partial {\varepsilon}^2} \equiv
\left[ \frac{1}{a} - \frac{1}{a_{{\rm crit.}} }
+ {\kappa}({\varepsilon}^2 \ln {\varepsilon}^{-2}) +
{\cal {O}}\left( \frac{1}{N_{{\rm c}}} {\varepsilon}^2 
\ln \ln {\varepsilon}^{-2} \right) 
\right] = 0 \ ,
\label{NTLgapsmall}
\end{equation}
with the NTL solution
${\varepsilon}^2 $$=$$ {\varepsilon}^2_{{\rm gap}}(a;r)
$$\equiv$$ m_t^2(\Lambda)/{\Lambda}^{\!2}_{{\rm f}} $$\ll$$ 1$,
and
\begin{equation}
a^{-1}_{\rm crit.}(r) = 1 - {\tilde a}_1^{-1}(r)/N_{\rm c}  
\ , \qquad
{\kappa}(r) = 1 + {\tilde {\kappa}}_1(r)/N_{\rm c}  \ .
\label{acritNTL}
\end{equation}
Here, ${\tilde a}_1^{-1},\!{\tilde {\kappa}}_1(r)\!\sim\!1$;
${\tilde a}_1^{-1}$ is independent 
and ${\tilde {\kappa}}_1$ is very tamely dependent on
${\varepsilon}^2$ ($\sim\!\ln \ln {\varepsilon}^{-2}$, see
below). For $r$ ($\equiv\!{\Lambda}_{{\rm b}}/{\Lambda}_{{\rm f}}$)
$=\!1$, we have:
\begin{eqnarray}
{\tilde a}_1^{-1} &\approx& 1.79202 \ \Rightarrow
\ a^{-1}_{{\rm crit.}} \approx 1\!-\!(1.79202)/N_{{\rm c}} 
\approx 0.403 \ \left( \Rightarrow
\  a_{{\rm crit.}} \approx 2.5 \right) \ ,
\label{a1tilde}
\\
{\tilde {\kappa}}_1 &=& ( 3 \ln \ln \varepsilon^{-2} - C) \ ,
\qquad 8.7 \stackrel{<}{\sim} C \stackrel{<}{\sim} 10.8 
\qquad (\mbox{for $r$$=$$1$; PTC}) \ .
\label{kappa1tilde}
\end{eqnarray}
For $m_t(\Lambda) $$\sim$$ 10^2$ GeV and
$10^{10}$GeV $\stackrel{<}{\sim}$$ {\Lambda}_{{\rm f}}$
$\stackrel{<}{\sim}$ $10^{16}$ GeV 
(${\Lambda}_{{\rm b}}$$=$${\Lambda}_{{\rm f}}$), we then obtain:
$0$$\stackrel{<}{\sim}$${\tilde {\kappa}}_1$$ \stackrel{<}{\sim}$$4$
$\Rightarrow$
$1$$<$${\kappa}$$< 2.4$.
The bounds for the constant $C$ 
can be systematically made narrower.
When $r\!<\!1$, smaller ${\tilde a}_1^{-1}(r)$
and $a_{\rm crit.}(r)$ are obtained
($r\!\downarrow\!0$ $\Rightarrow$ $a_{\rm crit.}\!\downarrow\!1$).
Qualitatively similar results are expected when other 
quark momentum regularizarion prescriptions are employed
and/or QCD contributions are included (the latter decrease
$a_{\rm crit.}$).

Ensuring the respect of the Goldstone theorem at the NTL
level was crucial for obtaining the above results.
Comparing (\ref{NTLgapsmall}) and (\ref{gaplead1}),
we see that the NTL contributions for small ${\varepsilon}^2$
do not change the structure of the leading-$N_{{\rm c}}$
(quark loop) gap equation.\footnote{
$a_{\rm crit.}\!=\!1\!=\!{\kappa}$ at the q.l.
level in the case of covariant spherical
cutoff and in the PTC case.}
Thus, the phase transition remains second order.
This means: when $a\!\downarrow\!a_{\rm crit.}(r)$
continuously, then ${\varepsilon}^2_{{\rm gap}}(a;
r)\!\downarrow\!0$ continuously, too. For 
$a\!\leq\!a_{\rm crit.}$, we have 
${\varepsilon}_{{\rm gap}}\!=\!0$ [$m_t({\Lambda})\!=\!0$].
Also the question of the mass renormalization in the
small-${\varepsilon}^2_{{\rm gap}}$ (large-${\Lambda}$)
limit should be investigated, probably under inclusion
of the (MSM) RGE's.

Comparing the results of this Section with those of the
previous, we can express the latter in this way: For given
values of parameters of the new physics ($a$ and
$\Lambda $$\sim$$ {\Lambda}_{{\rm f}} $$\sim$$ {\Lambda}_{{\rm b}}$), 
and if $\Lambda $$\gg$$E_{{\rm ew}}$, we can probably predict
with the $1/N_{{\rm c}}$ approach 
reasonably well and in a systematic
way the critical parameters $a_{{\rm crit.}}$
and ${\kappa}$ [cf.~(\ref{acritNTL})
and (\ref{a1tilde})-(\ref{kappa1tilde})]. However,
predictions for ${\varepsilon}_{{\rm gap}}
$$\equiv$$ m_t(\Lambda)/{\Lambda}_{{\rm f}}$ 
and ${\varepsilon}_{\mbox
{\scriptsize ren.}} $$\equiv$$ m_t^{{\rm ren.}}/
{\Lambda}_{{\rm f}}$,
($\Rightarrow\!m_t^{{\rm ren.}}$)
appear to be in serious
danger unless $\Lambda $$\stackrel{<}{\sim}$$ 1$ TeV, i.e.,
unless we abandon the case of fine-tuning $a $$\approx$$
a_{{\rm crit.}}(r)$. For example, 
from the last line of Table~\ref{tabl4} 
it follows that for high cutoffs 
${\Lambda}_{{\rm b}} $$=$$ {\Lambda}_{{\rm f}} $$=$$ 9.8$ GeV, 
we have to choose $a $$=$$ 1.549$ and\footnote{
Covariant spherical cutoffs were used there, and QCD contributions
included; $a\!=\!1.549$ is a bit above, but apparently close to 
$a_{{\rm crit.}}$ in this case.}
and the leading-$N_{{\rm c}}$ (quark loop) prediction $m_t^{(0)}$
is in this case drastically higher than $m_t^{{\rm ren.}}$:
$m_t^{(0)}\!\approx\!24 m_t^{{\rm ren.}}\!\approx\!4.4$ TeV.
This illustrates drastically the gravity of the fine-tuning
problem within the $1/N_{{\rm c}}$ expansion approach.
We will return briefly to questions concerning
$1/N_{{\rm c}}$ expansion in Sec.~\ref{SPTAQ3}.

\subsection{Other works on NTL effects 
in quark condensation mechanism}
\label{NTLEE6}
In addition to the forementioned series of works,
other authors have also investigated NTL 
contributions in NJLVL frameworks without
gauge bosons: 
\citeasnoun{Nikolovetal96},
\citeasnoun{Dmitrasinovicetal95},
Hands, Koci\'c and Kogut (1991, 1993),
Gracey (1993, 1994),
\citeasnoun{Derkachovetal93},
\citeasnoun{Akama96}, 
\citeasnoun{LurieTupper93},
\citeasnoun{Dyatlov96},
and~\citeasnoun{PreparataXue96}.

Nikolov {\em et al.\/} (1996)
calculated NTL contributions with an effective action formalism, 
Dmitra\v sinovi\'c {\em et al.\/} (1995)
with diagrammatic methods. Both groups 
investigated an $SU(2)$ symmetric NJLVL
model and regarded it as a framework of
low energy QCD,\footnote{
$SU(2)$ and $SU(3)$ versions of NJLVL frameworks
as models of low energy QCD were reviewed by Klevansky (1992).} 
where cutoffs ${\Lambda}_{{\rm b}} $$\sim$$ 
{\Lambda}_{{\rm f}} $$\sim$$ 1$ GeV 
are relatively close to the
constituent (dynamical) mass of the light quarks
$m_q $$\approx$$ 0.2$-$0.4$ GeV.
They took care that their formalism
respected the Goldstone theorem. They
calculated at the NTL level (meson loop level)
the gap equation which connects the NJLVL four-quark
coupling parameter with $m_q$ and
the cutoffs ${\Lambda}_{{\rm f}}$, ${\Lambda}_{{\rm b}}$.
In addition, they calculated at the NTL level
the pion decay constant $f_{\pi}$ as
a function of $m_q$ and of cutoffs. The latter relation
can be viewed as a sum rule for
an NTL version of the Bethe--Salpeter pion bound state 
wave function, i.e., an NTL extension of the Pagels--Stokar (PS)
relation (\ref{PSql}) for a nonrunning constituent mass $m_q$.
In the chiral limit, this relation gave the
Goldberger-Treiman relation $m_q $$=$$ g_{\pi q q} f_{\pi}$,
where $g_{\pi q q}$ is the pion-quark coupling parameter.
Their NTL relation for $f_{\pi}$, just like
PS relations (\ref{PSch})-(\ref{PSql}), does not
involve the NJLVL four-quark coupling constant
$a$$=$${\kappa}_q$ ($\Leftrightarrow$$G$). Hence, it can be
regarded independently of the
NTL gap equation. Taking the experimental
value $f_{\pi} $$\approx$$ 93$ MeV, 
they used only this relation 
$f_{\pi} $$=$$ f_{\pi}(m_q; {\Lambda}_{{\rm f}}; 
{\Lambda}_{{\rm b}}/{\Lambda}_{{\rm f}})$ 
to find numerically, among other things,
relations $m_q $$=$$ m_q({\Lambda}_{{\rm f}})$
at fixed ratios ${\Lambda}_{{\rm b}}/{\Lambda}_{{\rm f}}$, 
as well as (cf.~Dmitra\v sinovi\'c {\em et al.\/})
relations $m_q $$=$$ m_q({\Lambda}_{{\rm b}}/{\Lambda}_{{\rm f}})$
for fixed values of ${\Lambda}_{{\rm f}}$. The latter relations
show that $m_q$ increases when NTL effects
are included (at a fixed value of ${\Lambda}_{{\rm f}}$),
as pointed out by Dmitra\v sinovi\'c {\em et al.\/}
(cf.~Fig.~12 of that reference). For example,
for ${\Lambda}_{{\rm f}} $$=$$ 0.8$ GeV, $m_q$
can increase from $0.25$ GeV (leading-$N_{{\rm c}}$
result, i.e., for 
${\Lambda}_{{\rm b}}/{\Lambda}_{{\rm f}} $$=$$ 0$)
to $0.4$ GeV (NTL result, for 
${\Lambda}_{{\rm b}}/{\Lambda}_{{\rm f}} $$\approx$$ 1$).

The works of 
Dmitra\v sinovi\'c {\em et al.\/}
and of Nikolov {\em et al.\/},
although applied to low energy QCD, have 
relevance for studies of ${\bar t} t$
condensation at the NTL level. In particular,
they may be relevant for a derivation of an
NTL sum rule analog of the (leading-$N_{{\rm c}}$) PS relations
described in Sec.~\ref{MDSPS}. Such a sum rule
would allow us to predict $m_t^{{\rm dyn.}}$  
at NTL level as a function of the cutoffs only.
It could alternatively be derived by
first considering the Bethe--Salpeter (BS) equation
for the NGB bound states at the NTL level.\footnote{
The problem of incorporating systematically NTL effects
in the BS equation for NGB's has to our knowledge not been
investigated in the literature.} 
However, applying
in the chiral limit blindly the NTL relation for $f_{\pi}$, 
obtained by the two groups, to ${\bar t} t$ condensation
may be under circumstances somewhat misleading.
Such an application would consist of
replacing $f_{\pi} $$\approx$$ 93$ MeV 
by the NGB decay constant 
$f_{\pi}$$=$$ \sqrt{2}\,F_{\pi} $$=$$ v\,\sqrt{2} $$\approx$
$ 350$ GeV
and $\Lambda $$\sim$$1 {\rm GeV} \mapsto
\Lambda$$\stackrel{>}{\sim}$$1 {\rm TeV}$.
There are at least two reasons for caution. Firstly, the model
these authors describe is an $SU(2)\!\times\!SU(3)_c$-symmetric
NJLVL model, while for the ${\bar t} t$ condensation
the TSM models (\ref{TSM}) and (\ref{3gTSM}) are of relevance.
The latter are NJLVL models with 
$SU(2)_L\!\times\!U(1)_Y\!\times\!SU(3)_c$ symmetry.
Secondly, the two groups
did not consider effects of mass
renormalization $m_q(\Lambda) $$\mapsto$$ m_q(m_q)$, or equivalently,
the influence of the RGE evolution $m_q({\bar p}^2)$
on the value of $f_{\pi}$. 
The latter effects may be important in
${\bar t} t$ condensation frameworks, especially if
cutoffs are above $1$ TeV. This contrasts with low energy QCD. 
Further, besides the NJLVL four-quark interactions,
gluon exchanges may play an important role in predictions of
PS relations for $m_t^{{\rm phys.}}$
as a function of cutoffs $\Lambda$,
at least when ${\Lambda}$'s are large, as seen in Sec.~\ref{MDSPS}. 
NTL gap equation effects, discussed in
previous parts of the present Section~\ref{NTLEE},
indicate that cutoffs should be low [$\sim$$1$ TeV]
if $1/N_{{\rm c}}$ expansion for the
bare mass solution 
$m_t^2({\Lambda}_{{\rm f}})/{\Lambda}^{\!2}_{{\rm f}}$$\equiv$$
{\varepsilon}^2_{{\rm gap}}
(a; {\Lambda}_{{\rm b}}/{\Lambda}_{{\rm f}})$
is not to be endangered. For thus low
cutoffs, $f_{\pi}$-relation of
Dmitra\v sinovi\'c {\em et al.\/},
recalculated for the TSM model (\ref{TSM}),
could give us at least an estimate of how NTL contributions 
change the predicted mass $m_t^{{\rm phys.}}$ as a function 
of cutoffs ${\Lambda}_{{\rm f}}$ and ${\Lambda}_{{\rm b}}$.

Authors 
Hands, Koci\'c and Kogut (1991, 1993),
Gracey (1993, 1994)
and Derkachov {\em et al.\/} (1993) 
calculated for various
dimensions ($d $$\leq$$ 4$) NTL contributions to critical
exponents of the fields at the fixed points, i.e., at
locations where $\beta$ function has a nontrivial zero. 
The implications of these works for physical predictions of
four-dimensional NJLVL models with finite cutoff are
not clear and would deserve investigation.

Akama (1996)
investigated the NTL contributions
in the truncated TSM,
by considering compositeness condition for the
composite scalars. This condition says that
the renormalization constants of the composite scalars
and of their self-interactions are zero.
Further, 
Luri\'e and Tupper (1993)
had earlier investigated compositeness condition
by taking into account some of the NTL effects.
Akama, as well as Luri\'e and Tupper, concluded
that compositeness condition implies that
NTL contributions to the corresponding
physical quantities for $N_{{\rm c}}$$=$$3$ are larger
than the leading-$N_{{\rm c}}$ contributions, and
indicate that $1/N_{{\rm c}}$ expansion is in trouble.
Approaches used by these three authors are
similar in spirit, but not identical, to the approach
of BHL (1990). TSM was treated as a renormalizable
Yukawa type model without gauge bosons plus the
compositeness condition. The implicit assumption was
that the cutoffs are large:
$\ln (\Lambda/E_{{\rm ew}}) $$\gg$$ 1$.
Thus, their results are consistent with
conclusions of papers 
described in previous parts of this Sec.~\ref{NTLEE}
(Cveti\v c {\em et al.\/}, 1996, 1997):
${\bar t} t$ condensation
can be described by $1/N_{{\rm c}}$ expansion
as long as $\Lambda $$\stackrel{<}{\sim}$$ 1$ TeV,
and this expansion may be endangered for
${\Lambda} $$>$$ 1$ TeV.

Dyatlov (1996) investigated NTL contributions to quark masses
in an almost ``flavor-democratic'' 
[$U(1)^n\!\times\!U(1)^n\!\times\!SU(N_{{\rm c}})$] 
NJLVL model. It was
based partly on his phenomenological framework for dynamical
generation of quark mass hierarchy and CKM mixing (Dyatlov, 1992,
1993). He ensured validity of the Goldstone theorem
at the NTL level. However, it appears that he accounted at the 
NTL level only for the effective renormalization contributions
to the quark masses, i.e.,
contributions corresponding to the (1-PI) diagrams
of Fig.~\ref{rmd5f}(a). The tadpole-type (1-PR) NTL diagrams
in the gap equation which would correspond to the
derivative $\partial \Xi^{(1)}/\partial \langle {\cal {H}}
\rangle_0^2/N_{{\rm c}}$ [cf.~(\ref{Xi1}) and 
(\ref{NTLgapS2})] were apparently not included in his analysis 
(diagrams corresponding to $\Xi^{(1)}/N_{{\rm c}}$ 
are in Fig.~\ref{rmd4f}).

Preparata and Xue (1996) studied 
analytically ${\bar q} q$ condensation
in an NJLVL framework on a lattice with finite
lattice constant $a_{{\rm Planck}} $$\sim$$
10^{-33}$ cm. They employed a lattice version of
leading-$N_{{\rm c}}$ (quark loop)
gap equation. They then calculated the resulting
vacuum energy ${\triangle E}_{{\rm vac.}}$,
via the effective Wilson action over the ground state, 
as a function of
the number of quark generations $N_{{\rm g}}$
acquiring nonzero dynamical masses. When they included
in ${\triangle E}_{{\rm vac.}}$ also the
(NTL) contributions of composite (pseudo)scalars, they
were able to show that the energetically preferred
configuration is that with $N_{{\rm g}}$$=$$1$.
Thus, there is only one (third) quark generation with
dynamical masses. The
conclusion depends critically on inclusion of
NTL contributions to ${\triangle E}_{{\rm vac.}}$.
It appears that their results would become
more consistent if they included the NTL contributions
also in their gap equation. An intriguing
feature of their analysis is disappearance of
(composite) Higgs from low energy spectrum
($M_{H} $$\sim$$ E_{{\rm Planck}}$).

In the framework of technicolor
theories (TC), Appelquist, Lane and Mahanta (1988),
Mahanta (1989), and Kamli and Ross (1992) investigated
corrections beyond the (improved) ladder approximation in the
variational version of the gap equation (i.e., the DS equation).
Appelquist {\em et al.\/} (1988) performed a two-loop
calculation for the dynamical fermion mass ${\Sigma}({\bar p}^2)$
in the nonrunning limit of 
${\alpha}_{{\rm TC}}$, using the linearized form of 
the DS equation [i.e., ${\bar p}^2 $$\gg$$ {\Sigma}^2({\bar p}^2)$
was taken for large $|{\bar p}| $$>$$ {\Lambda}_{{\rm TC}}
$$\sim$$ E_{{\rm ew}}$]. They restricted themselves
to two specific gauges (Landau gauge $\xi $$=$$ 0$; and
$\xi $$=$$ -3$). They showed semianalytically that the two-loop
corrections ($\sim$$ {\alpha}^2_{{\rm TC}}$)
result under such assumptions in negative corrections to
${\alpha}_{{\rm TC}}^{{\rm crit.}}$
not larger than $20 \%$ [cf.~(\ref{a1tilde}) for the
TSM, where the NTL corrections to analogous
four-quark parameter $a_{{\rm crit.}}$
are positive and about $150 \%$]. 
Hence, they argued that
ladder (one-loop) results in TC theories can in general be 
trusted. Mahanta (1989) employed
the same approximation of linearized DS and nonrunning
${\alpha}_{{\rm TC}}$, and worked in 
the Landau gauge. He showed
that ${\Sigma}({\bar p}^2)$$\propto$$|{\bar p}|^{-1}$ 
to all orders in ${\alpha}_{{\rm TC}}$
when ${\alpha}_{{\rm TC}} $$=$$
{\alpha}_{{\rm TC}}^{{\rm crit.}}$.
Thus, the anomalous dimension ${\gamma}_m$ of the
technifermion condensate\footnote{
${\gamma}_m({\alpha}_{\mbox{\tiny TC}})$ is defined by:
$d \ln \langle {\overline {\Psi}}^{(\mu)}_{\mbox{\tiny TC}}
{\Psi}^{(\mu)}_{\mbox{\tiny TC}} \rangle_0 / d \ln \mu
\equiv {\gamma}_m({\alpha}_{\mbox{\tiny TC}}(\mu))$. }
$\langle {\overline {\Psi}}_{{\rm TC}}
{\Psi}_{{\rm TC}} \rangle_0$
is relatively large ${\gamma}( 
{\alpha}_{{\rm TC}}^{{\rm crit.}})
$$=$$ 1$ [since ${\Sigma}({\bar p}^2) \propto
|{\bar p}|^{{\gamma}_m -2}$]. 
This implies that the FCNC-suppressing
hierarchy of the extended TC (${\Lambda}_{{\rm ETC}}
$$\gg$$ {\Lambda}_{{\rm TC}}$) is not just an
artifact of the ladder approximation. 

Kamli and Ross (1992),
on the other hand, investigated numerically gauge dependence
of the one- plus two-loop contributions to the DS equations in TC
theories. They didn't invoke any further approximations (in contrast to
Appelquist {\em et al.\/} and Mahanta). They worked 
in the general covariant $R_{\xi}$ gauge and showed that
$F_{\pi}$ (obtained by using the PS relation which is 
{\em leading\/}-$N_{{\rm c}}$) is only
moderately dependent on the $\xi$ parameter, but the
technifermion condensate
$\langle {\overline {\Psi}}_{{\rm TC}}
{\Psi}_{{\rm TC}} \rangle_0$
depends drastically on $\xi$. Hence, they argued
that the predictability of perturbative expansions
in powers of ${\alpha}_{{\rm TC}}$ in
the DS approach is doubtful.

It should be stressed that conclusions of the
works of Appelquist {\em et al.\/}, Mahanta, and
Kamli and Ross, do not apply to the TSM-type of
models of ${\bar t} t$ condensation. These authors
investigated exchanges of {\em massless\/}
(techni)gluons, and did not derive
the NTL terms of $1/N_{{\rm c}}$
($1/N_{{\rm TC}}$) expansion but rather
NTL terms of expansion in powers of ${\alpha}$
(${\alpha}_{{\rm TC}}$).
The TSM-type of interactions (\ref{TSM}) 
and (\ref{3gTSM}) are usually assumed to arise from exchange
of {\em massive\/} particles (e.g., massive gauge bosons
with mass $M $$\sim$$ {\Lambda} $$\stackrel{>}{\sim}$$ 1$ TeV).
They lead to a tightly bound $\langle {\bar t} t \rangle_0$
condensate (${\gamma}_m $$\simeq$$ 2$), in contrast to
the more weakly bound
$\langle {\overline {\Psi}}_{{\rm TC}}
{\Psi}_{{\rm TC}} \rangle_0$
(${\gamma}_m $$\simeq$$ 1$) -- cf.~also discussion
in Subsec.~\ref{SPTAQ10}.

Smith, Jain and McKay (1995)
investigated DS equations for quarks within the
minimal SM (MSM) in the ladder approximation.
Interestingly, they
found out that exchanges of (elementary) scalars
contribute to the heavy quark
self-energies ${\Sigma}_q({\bar p}^2)$ more strongly
than exchanges of gluons do, especially
for large $|p| $$>$$ m_q $($\sim$$ 10^2$ GeV). 
This would indirectly
suggest that in TSM-type of models exchanges of
tightly bound composite scalars (Higgs $\sim$$ {\bar t}t$, etc.)
are more important than other (gluon) exchanges.
This was explicitly confirmed in works described 
in previous parts of this Section 
(\ref{NTLEE4} -- \ref{NTLEE5}).

\section{Comparisons of and comments on various approaches 
in the minimal framework}
\label{CCVAMF}
\setcounter{equation}{0}

\subsection{Initial remarks}
\label{CCVAMF0}
The following two Subsections concentrate basically
on some relations between the two dominant computational
approaches for ${\bar t} t$ condensation mechanism: the
RGE approach (BHL; Sec.~\ref{MRGE})
and the DS$+$PS approach in the leading-$N_{{\rm c}}$ approximation 
(MTY; Sec.~\ref{MDSPS}).
Several similarities and differences are pointed out.
We should stress that various computational approaches
must be equivalent in principle. Differences arise due to
approximations made, as is also suggested by the
results discussed in this Section. Advantages and drawbacks
of the two methods have been mentioned in 
Secs.~\ref{MRGE} and \ref{MDSPS}. To reiterate briefly,
the RGE approch is simpler, it properly sums up leading log
effects, and accounts for a part of the effects beyond the
leading-$N_{{\rm c}}$. The actual condensation mechanism is
indirectly accounted for by a compositeness boundary condition.
The predictions rely heavily on the infrared fixed-point behavior
of RGE's, and are thus reliable for high compositeness
scales ${\Lambda}$$\stackrel{>}{\sim}$$10^8$ GeV.
DS$+$PS approach is more complicated, it has been worked
out only at the leading-$N_{{\rm c}}$ level, and
suffers from gauge dependence.
In principle, it deals with the condensation mechanism (e.g., 
NJLVL terms) directly and is applicable also for low ${\Lambda}$'s.

\subsection{Preliminaries -- RGE solutions in closed form}
\label{CCVAMF1}
To compare the RGE with the DS$+$PS approach 
(in the leading-$N_{{\rm c}}$ approximation),
it is useful to write down first an analytic approximate
formula for the one-loop running Yukawa parameter
$g_t(\mu)$ and mass $m_t( \mu)$ 
in the minimal Standard Model (MSM). The approximation
consists of neglecting the small contributions
of the transverse components of the EW gauge bosons 
($g_1\!\mapsto\!0$ and $g_2\!\mapsto\!0$). Using one-loop RGE
solution (\ref{alph3}) for $\alpha_3(\mu)$
(for $\mu $$\geq$$ m_t^{{\rm phys.}}$), RGE (\ref{RGEgt})
for the Yukawa parameter $g_t(\mu)$ can be solved explicitly
\begin{eqnarray}
g_t^2(\mu) &=& 
\frac{(A-1)}{A} \frac{2 (4\pi)^2}{(N_{{\rm c}} + 1.5)}
{\vartheta}_s(\mu)^A
\left\{ \left[ {\vartheta}_s(\mu)^{A-1} 
- {\vartheta}_s(m_t)^{A-1} \right]
+ c(m_t) \right\}^{-1} \ ,
\label{gtmu}
\\
\mbox{where: } 
\vartheta_s(\mu) &\equiv&
\frac{\alpha_3(\mu)}{4\pi} 
\frac{3 (N_{{\rm c}}^2 - 1)}{2 N_{{\rm c}}}
= \frac{ A }{ 2 \ln(\mu^2/{\Lambda}^{\!2}_{{\rm QCD}}) }
\quad (\mbox{for } \mu > m_t^{{\rm phys.}} ) \ ,
\label{vartheta}
\\
A(N_{{\rm c}})& =&  
\frac{ 9 (N_{{\rm c}}^2\!-\!1) }{ N_{{\rm c}} 
(11 N_{{\rm c}}\!-\!12) }\!\sim\!N_{{\rm c}}^0 \ , \quad
c(m_t) = \frac{(A\!-\!1)}{A} \frac{2 (4\pi)^2}
{(N_{{\rm c}}\!+\!1.5)}
\frac{ {\vartheta}_s(m_t)^A }{ g_t^2(m_t) } \ ,
\label{Avarthcmt}
\end{eqnarray}
and ${\Lambda}_{{\rm QCD}}\!\approx\!51$ MeV.
Inserting this $g^2_t(\mu)$ into the one-loop MSM RGE 
for the evolution of the VEV $v(\mu)$
\cite{Arasonetal92}:
\begin{equation}
16 \pi^2 \frac{ d \ln v(\mu)}{d \ln \mu} = 
- N_{{\rm c}} g_t^2(\mu)
+ \frac{9}{4}g_2^2(\mu) + \frac{3}{4}g_1^2(\mu) \ ,
\label{RGEv}
\end{equation}
setting again $g_1,\!g_2$$\mapsto$$0$, we obtain
\begin{equation}
v^2(\mu) \equiv \langle H^{(\mu)} \rangle_0^2
= v^2(m_t)
\left\{ 1 -  \left[ {\vartheta}_s(m_t)^{A-1} 
- {\vartheta}_s(\mu)^{A-1} \right]/c(m_t) \right\}^{ 
\frac{ N_{{\rm c}} }{ N_{{\rm c}}+1.5 } } \ ,
\label{vmu}
\end{equation}
where $c(m_t)$ is given in (\ref{Avarthcmt}). Combining 
(\ref{gtmu}) and (\ref{vmu}), we obtain 
$m_t^2(\mu)\!\equiv\!g_t^2(\mu) v^2(\mu)/2$
\begin{equation}
m_t^2(\mu) = m_t^2 
\left[ \frac{ \ln( \mu/\Lambda_{{\rm QCD}} ) }
{ \ln( m_t/\Lambda_{{\rm QCD}} ) } \right]^{-A}
\left\{  \frac{ c(m_t) }
{ c(m_t) - \left[ {\vartheta}_s(m_t)^{A-1} 
- {\vartheta}_s(\mu)^{A-1} \right] }
\right\}^{\left( 1 - 
\frac{ N_{{\rm c}} }{ N_{{\rm c}}+1.5 } 
\right) } \ .
\label{mtmu}
\end{equation}
The mass $m_t$ denotes throughout here 
$m_t(m_t)$ (renormalized mass), constant $A$ and $c(m_t)$ 
are given in (\ref{Avarthcmt}), and QCD parameter ${\vartheta}_s$ 
in (\ref{vartheta}).

\subsection{Comparisons at the leading-$N_{\rm c}$ level}
\label{CCVAMF2}
When only quark loops are taken into account in
RGE for $g_t$ (\ref{RGEgt}), the only term kept on
the right of that RGE is $N_{{\rm c}} g_t^3(\mu)$
because this is the term corresponding to the
quark loop diagram of Fig.~\ref{rmd6f}.
\begin{figure}[htb]
\mbox{}
\vskip2.4cm\relax\noindent\hskip.0cm\relax
\includegraphics{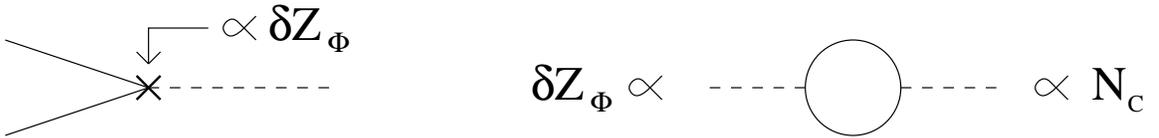} \vskip0.5cm
\caption{\footnotesize Loop contribution of $t$
to evolution of $g_t$ -- it is
the origin of the term $N_{{\rm c}} g_t^3(\mu)$ on the right
of MSM RGE (\ref{RGEgt}). Full lines represent
$t$, and dashed ones (neutral) scalars. ${\delta} Z_{\Phi}$
is change in the scalar field renormalization when the cutoff
is changed by $\delta \mu$. ${\delta} Z_{\Phi}$ is caused by the 
$t$-loop effect and is hence $\propto$$N_{{\rm c}}$.}
\label{rmd6f}
\end{figure}
Therefore, RGE structure (\ref{RGEgt}) implies that
the quark loop approximation (q.l.a.) is leading-$N_{{\rm c}}$
with no QCD (and $g_1,$$g_2$$\mapsto$$0$).
Mass $m_t(\mu)$ is in this case RGE-nonrunning, as
seen from RGE solution (\ref{mtmu}) -- by
putting there $A$$ \mapsto$$ 0$ (to exclude QCD) and
replacing $(N_{{\rm c}}\! +\!1.5) \!\mapsto\! N_{{\rm c}}$.
Hence, the RGE approach gives
in quark loop approximation the same behavior for
$m_t(\mu)$ as the approach with the Dyson--Schwinger (DS) 
equation in quark loop approximation (\ref{qlgap}).
Also the approach with the ``hard mass '' effective potential
and with subsequent mass renormalization (Sec.~\ref{NTLEE})
gave mass renormalization
effects (when QCD was switched off) only at the NTL
level of $1/N_{{\rm c}}$ expansion (\ref{kappar}). 
Hence, all three approaches give at the level of the q.l.a.
consistently nonrunning dynamical $m_t$.
 
Further, in the q.l.a. the
Pagels--Stokar (PS) relation gave the same relation 
(\ref{PSql}) between
the condensation cutoff $\Lambda$, $m_t$ and 
$F_{\pi}$ as did the RGE approach. Namely, keeping in
RGE (\ref{RGEgt}) only quark loop terms
[$g_i$$\mapsto$$0$ ($i$$=$$1,2,3$)
and $(N_{{\rm c}}$$ +$$ 1.5) $$\mapsto $$N_{{\rm c}}$] 
and taking into
account the compositeness condition $g_t(\Lambda) $$=$$ \infty$,
we arrive precisely at the 
quark loop effective action solution
(\ref{Lmunornot1}), and the latter is in turn identical with
(\ref{PSql}) of the DS$+$PS approach in the q.l.a.
once we impose the minimal framework condition
for the full DEWSB: $F_{\pi}$$ =$$ v$($\approx$$246$ GeV).
The effective potential approach of Sec.~\ref{NTLEE}, however, 
did not contain any relations analogous to the PS equation, 
and thus cannot be compared with it.

When including, in addition to quark loops,
also the (one-gluon-exchange) QCD contributions
in the DS$+$PS approach, we remain formally in the
leading-$N_{{\rm c}}$ approximation, as argued
in Sec.~\ref{MDSPS}. The same is true in the
RGE approach. Namely, the QCD term on the right
of RGE for $g_t$ (\ref{RGEgt}) 
is of the same leading-$N_{{\rm c}}$ 
order as the quark loop term $N_{{\rm c}} g_t^3$, 
because $g^2_3$$ =$$ {\cal {O}}(1/N_{{\rm c}})$ 
[cf.~(\ref{alph3})], and
$g_t^2$$=$${\cal {O}}(1/N_{{\rm c}})$ [the latter can be
seen by solving RGE (\ref{RGEgt}) explicitly in
quark loop approximation between the Landau
pole $\Lambda$ and $\mu$, giving exactly
(\ref{Lmunornot1})].
One-loop RGE result (\ref{mtmu}) in this approximation
[$(N_{{\rm c}}$$ +$$ 1.5) $$\mapsto$$ N_{{\rm c}}$;
and $A$$=$$8/7$ to include QCD] gives the QCD-run
mass $m_t(\mu)$
\begin{equation}
m_t^2(\mu) = m_t^2(m_t)
\left[ \frac{ \ln( \mu/\Lambda_{{\rm QCD}} ) }
{ \ln( m_t/\Lambda_{{\rm QCD}} ) } \right]^{-8/7} \ ,
\label{mtQCD}
\end{equation}
agreeing with the solution of the nonperturbative DS equation 
(\ref{asymS})-(\ref{asymcm}) in the asymptotic $\mu$-region 
$m_t(m_t) $$\ll$$ \mu $$\ll$$ \Lambda$.
Why does the {\it perturbative\/} leading-$N_{{\rm c}}$
RGE solution (\ref{mtQCD}) agree with solution
(\ref{asymS}) of the {\em nonperturbative\/}
leading-$N_{{\rm c}}$ DS approach (\ref{DSmty})?
A way to explain this:
in the leading-$N_{{\rm c}}$ DS approach (\ref{DSmty}),
the truly nonperturbative physics is contained
in the four-quark contribution ($\propto$$ {\kappa}_t \!\sim\! 1$)
which is not directly responsible for the running of $m_t(\mu)$
-- only the weaker QCD contribution causes ${\Sigma}_t$$\equiv$$
m_t$ to (slowly) run.
In the approach of Sec.~\ref{NTLEE},
QCD mass renormalization effects are given in (\ref{kapglr}) --
they are perturbative and can be reproduced to the leading order
in $\ln({\varepsilon}^{-2})$ and in ${\alpha}_3$
from RGE solution (\ref{mtQCD}):
\begin{equation}
\frac{m_t^2}{{\Lambda}^{\!2}} - 
\frac{m_t^2(\Lambda)}{{\Lambda}^{\!2}}
= \frac{m_t^2}{{\Lambda}^{\!2}} \left[ 
\frac{A}{ \ln (m_t^2/{\Lambda}^{\!2}_{{\rm QCD}} ) }
\ln ({\Lambda}^{\!2}/m_t^2) + \cdots \right] \ ,
\label{mtQCDexp}
\end{equation}
where $m_t$$=$$m_t(m_t)$ is the renormalized mass 
and $A$$=$$8/7$ for $N_{{\rm c}} $$=$$ 3$. The
dots represent corrections smaller than 50\%
(25\%) when ${\Lambda} $$<$$ 5$ TeV (${\Lambda} $$\leq$$ 1$ TeV).
Using (\ref{Avarthcmt}) for $A$,
we see that the dominant term in the above RGE expansion
is equal to the dominant $\ln({\varepsilon}^{-2})$ term in
(\ref{kapglr}), if we set there ${\varepsilon}
$$=$$ m_t^{{\rm ren.}}/\Lambda$ and 
${\Lambda}_{{\rm b}} $$=$$ {\Lambda}_{{\rm f}} $$=$$ \Lambda$.
Therefore, all three approaches at the leading-$N_{{\rm c}}$
level (i.e., quark loop $+$ QCD) give similar
behavior for $m_t(\mu)$. 

Further, at the leading-$N_{{\rm c}}$ level,
the RGE approach and the PS equation give approximately the same
relation between the VEV, $m_t$ and the cutoff $\Lambda$.
In the RGE approach of Sec.~\ref{MRGE}, this relation is
implicitly given by (\ref{gtmu}),
when, in addition, the RGE compositeness condition 
is used, i.e., $g_t({\Lambda}) $$=$$ \infty$
$\Rightarrow 
[ {\vartheta}_s(\Lambda)^{A-1}\!-\!{\vartheta}_s(m_t)^{A-1} ]\!+\!
c(m_t)\!=\!0 \Rightarrow$
\begin{equation}
g_t^2(m_t) \left( \equiv \frac{2 m_t^2}{v^2} \right)
= \frac{(A-1)}{A} \frac{2 (4\pi)^2}{(N_{{\rm c}} + 1.5)}
\frac{ {\vartheta}_s(m_t)^A }{ \left[
{\vartheta}_s(m_t)^{A-1} - 
{\vartheta}_s(\Lambda)^{A-1} \right] } \ .
\label{vmtLb}
\end{equation}
This analytic expression was first obtained by 
Marciano (1990),
in the realistic case 
$N_{{\rm c}}$$=$$3$ and 
$n_q$$=$$6$ ($A$$=$$8/7$), cf.~Eq.~(\ref{marc5a}). 
It was also reproduced by Gribov (1994) within his
leading-$N_{{\rm c}}$ framework (cf. Sec.~\ref{MDSPS5}).
In the DS$+$PS approach of 
Sec.~\ref{MDSPS}, relation (\ref{vmtLb})
can be obtained the following way: take for
the dynamical mass ${\Sigma}_t({\bar p}^2) $$=$$ 
m_t(|{\bar p}|)$ the asymptotic solution 
(\ref{asymS})-(\ref{asymcm})
of the DS integral equation (\ref{DSmty}) [equal to 
the RGE solution (\ref{mtQCD})] and insert it in the PS relation
(\ref{PSne}). There, neglect the term with
$d \Sigma_t/ d {\bar p}^2$ due to the slow
${\bar p}^2$-dependence, and set ${\Sigma}_b$$=$$0$. Neglect 
contributions from ${\bar p}^2 $$\stackrel{<}{\sim}$$ m_t^2$,
i.e., approximate
$[{\bar p}^2 $$+$$ {\Sigma}_t^2({\bar p}^2) ]^{-2}$
$\approx$$[{\bar p}^2]^{-2}$. The resulting integrand can be
analytically integrated
\begin{eqnarray}
F^2_{\pi}& \approx& 
\frac{N_{{\rm c}}}{8\pi^2} \int_{m_t^2}^{{\Lambda}^{\!2}}
d {\bar p}^2 \frac{ {\Sigma}_t^2({\bar p}^2) }{ {\bar p}^2 }
=  \frac{N_{{\rm c}} m_t^2}{8 \pi^2} 
{\int}_{m_t^2}^{{\Lambda}^{\!2}}
\frac{ d {\bar p}^2 }{ {\bar p}^2 } 
\left[ \frac{ \ln( {\bar p}^2/{\Lambda}^{\!2}_{{\rm QCD}} ) }
{ \ln( m^2_t/{\Lambda}^{\!2}_{{\rm QCD}} ) } \right]^{-A} 
\nonumber\\
& = & 
\frac{N_{{\rm c}} m_t^2}{16 \pi^2}  \frac{A}{ (A-1) } 
\frac{ \left[
{\vartheta}_s(m_t)^{A-1} - {\vartheta}_s(\Lambda)^{A-1} \right] }
{{\vartheta}_s(m_t)^A } \ .
\label{vmtLbPS}
\end{eqnarray}
This is now equivalent to the RGE result (\ref{vmtLb})
at the leading-$N_{{\rm c}}$ level 
[when $(N_{{\rm c}}\!+\!1.5)\!\mapsto\! N_{{\rm c}}$
in (\ref{vmtLb})], if keeping in mind that $F_{\pi} $$=$$ v$.
We note that this equivalence proof at the leading-$N_{{\rm c}}$ 
level is valid only when cutoff $\Lambda$ is high:
$\ln(\Lambda/m_t) $$\gg$$ 1$. Anyway, the
full RGE approach of BHL (1990) has predictive power
only for large $\Lambda $$\stackrel{>}{\sim}$$ 10^{8}$ GeV.

The discussed
approximate equivalence of the RGE approach of BHL (1990)
and of PS $+$ DS approach of MTY (1989a, 1989b),
at the leading-$N_{{\rm c}}$ level for which the latter method
had been carried out, was shown 
analytically by
Blumhofer, Dawid and Lindner (1995)
and reemphasized by
Yamawaki (1995).
It was shown earlier to hold quite well also numerically by
Barrios and Mahanta (1991)
-- they obtained
good agreement (up to a few per cent)
of predicted masses $m_t$ of both
approaches at the leading-$N_{{\rm c}}$ level for almost all cutoffs,
even for cutoffs as low as 100 TeV. However, we recall that
the RGE approach at the leading-$N_{{\rm c}}$ level doesn't
show infrared fixed-point behavior at such low cutoffs.

In fact, it turns out that for 
$\Lambda $$\stackrel{<}{\sim}$$ 10^8$ GeV
the leading-$N_{{\rm c}}$ RGE predictions for $m_t(m_t)$ 
decrease considerably (by several per cent) if the
compositeness condition is taken to be $k_t $$\equiv$$
g_t^2(\Lambda)/(4 \pi)\!=\!1$, $2/3$ or $1/3$, 
instead of $\infty$. This effect is more pronounced
at lower ${\Lambda}$'s.
The results of the DS$+$PS approach, given
in Table~\ref{tabl2} 
[$m_t $$\approx$$ {\Sigma}_t(0)$], do not suffer from
the mentioned ambiguity problems of the RGE approach
stemming from the choice of various boundary
values $g_t(\Lambda)$, not even for low ${\Lambda}$'s.
This is so because DS$+$PS
actually treats directly the TSM physics (\ref{TSM}) or 
(\ref{3gTSM}) responsible for the condensation.
In contrast to the RGE approach, the
DS$+$PS in the minimal framework predicts, for any given ${\Lambda}$,
not just values $m_t(\mu)$ (and $m_t^{{\rm phys.}}$), but also
the value of the dimensionless four-quark parameter
$a $$\equiv$$ G {\Lambda}^{\!2} N_{\rm c}/(8 \pi^2)$ of the TSM
(\ref{TSM}) [or: ${\kappa}_t$ of (\ref{3gTSM})].
However, we should keep in mind that the DS$+$PS
approach suffers from the problem of gauge noninvariance
(cf. Sec.~\ref{MDSPS5}) and the difficulties of
extending it beyond the leading-$N_{{\rm c}}$ level
(cf. Sec.~\ref{NTLEE}).

We haven't discussed yet the mass $M_H$ of the composite Higgs
in the comparison of the RGE and DS$+$PS approaches.
Chesterman, King and Ross (1991)
investigated relation between predictions 
of the RGE and DS approaches for the mass $M_H$
in the leading-$N_{{\rm c}}$ approximation. They inserted
solution ${\Sigma}_t({\bar p}^2)$ of the (improved lattice) 
DS equation of the truncated TSM (\ref{TSM})
into the Bethe--Salpeter (BS) amplitude for 
${\bar t} t $$\to$$ {\bar t}t$ scattering. In this amplitude,
they included dominant QCD corrections in the Landau gauge
and searched for poles ${\bar p}^2\!=\!-X^2$ in the scalar channel,
identifying $X$ with the physical mass $M_H$.
As a consistency check, they showed that 
the pseudoscalar channel gives a pole at ${\bar p}^2\!=\!0$.
QCD corrections pushed down the prediction $M_H $$=$$ 2 m_t$ of
quark loop approximation. They compared the results
with those of the BHL RGE approach at the leading-$N_{{\rm c}}$
level, i.e., with the solutions $m_t^2(m_t) $$=$$ g_t^2(m_t) v^2/2$
and $M^2_H\!=\! v^2 \lambda(\mu$$=$$ M_H)$ stemming from the
leading-$N_{{\rm c}}$ version of MSM RGE's (\ref{RGEgt}) and
(\ref{RGEl})
\begin{eqnarray}
16 \pi^2 \frac{ d g_t(\mu) }{ d \ln\mu } &=&
\left[ N_{{\rm c}} g_t^2(\mu) 
- 3 \frac{(N_{{\rm c}}^2-1)}{N_{{\rm c}}} g_3^2(\mu) \right]
g_t(\mu) \ ,
\label{RGEgtlnc}
\\
16 \pi^2 \frac{d \lambda(\mu)}{d \ln \mu} &=&
- 4 N_{{\rm c}} g_t^4(\mu) + 4 N_{{\rm c}} \lambda(\mu) g_t^2(\mu) \ ,
\label{RGEllnc}
\end{eqnarray}
where $N_{{\rm c}}$$=$$3$ is the actual number of colors, 
$g_3^2 $$=$$ 4 \pi {\alpha}_3$ is the one-loop solution
(\ref{alph3}), and the usual BHL compositeness boundary conditions
(\ref{RGEbc}) are taken at $\mu$$=$$\mu_{\ast} $$\approx$$ \Lambda$.
Chesterman {\em et al.\/} 
considered only cutoffs $\Lambda $$\leq$$
10^{7}$ GeV, because at larger values they ran into numerical
instabilities due to subtractions and cancellations
of very large terms $\sim\!{\Lambda}^{\!2}$. They obtained 
values for $M_H$ which differ from those of 
the leading-$N_{{\rm c}}$ BHL approach
by less than 10\% 
for $10^5$GeV $\leq$$ \Lambda $$\leq$$ 10^7$ GeV,
and for $\Lambda $$=$$ 10^4$ GeV they are lower by about 15\%.
These  results are given in Table~\ref{tabl5}.
\begin{table}[ht]
\vspace{0.1cm}
\par
\begin{center}
\begin{tabular}{l c c c c }
$\Lambda$ [GeV] &
$10^4$ & $10^5$ & $10^6$ & $10^7$ \\
\hline \hline
$m_t $$=$$ {\Sigma}_t(0)$ (DS$+$PS, l.-$N_{{\rm c}}$) & 
594 & 448 & 391 & 357 \\
$M_H$ (BS, l.-$N_{{\rm c}}$) & 
1062 & 769 & 634 & 617 \\
\hline
$m_t(m_t)$ (RGE, l.-$N_{{\rm c}}$) &
577 & 446 & 391 & 358 \\
$M_H$ (RGE, l.-$N_{{\rm c}}$) &
1246 & 827 & 676 & 592 \\
$M_H$ (Gribov) &
1068 & 778 & 655 & 578 \\
\end{tabular}
\end{center}
\vspace{-0.1cm}
\caption{\footnotesize Predicted masses $m_t$ and $M_H$ in the 
leading-$N_{{\rm c}}$ approximation
(quark loop $+$ QCD). Entries of the 
first two lines are from the DS$+$PS approach, and the
Bethe--Salpeter (BS) amplitude of
Chesterman, King and Ross (1991), and the
other two lines contain results of the 
leading-$N_{{\rm c}}$ RGE approach of
(\ref{RGEgtlnc})-(\ref{RGEllnc}). 
The results were taken from the mentioned work of
Chesterman {\em et al.\/} where apparently
${\Lambda}_{{\rm QCD}} $$\approx$$ 0.122$ [$\Rightarrow
{\alpha}_3(M_Z) $$\approx$$ 0.136$] was used.
For comparison, the corresponding values of $M_H$ as
determined by Gribov's leading-$N_{{\rm c}}$
sum rule (\ref{Gribov1}), (\ref{Gribov2}), were also included.
All entries are in GeV.}
\label{tabl5}
\end{table}
For comparison, the corresponding values of $M_H$
as obtained from Gribov's leading-$N_{{\rm c}}$ sum rule
(\ref{Gribov1}) were also included in the Table:
\begin{equation}
M_H^2 = \frac{7}{3 \pi^2} \frac{m_t^4}{v^2}
\ln \left( \frac{m_t}{ {\Lambda}_{{\rm QCD}} } \right)
{\Bigg \{} 1 - \left[ 
\frac{ \ln ( m_t/{\Lambda}_{{\rm QCD}} ) }
{ \ln ( {\Lambda}/{\Lambda}_{{\rm QCD}} ) } \right]^{9/7}
{\Bigg \}} \ ,
\label{Gribov2}
\end{equation}
where shorthand notation $m_t(m_t) $$\equiv$$ m_t$ is used.

\subsection{Mass dependent RGE approach}
\label{CCVAMF3}
After the paper by Bardeen, Hill and Lindner (BHL, 1990)
had appeared,
another group
\cite{Bandoetal90}
pointed out that
the approach by BHL has a theoretical and practical
deficiency because it employed the usual RGE's 
(\ref{RGEgt})-(\ref{RGEl}) of the
$\overline{MS}$ scheme which is mass-independent.
Bando {\em et al.\/} (1990) started with the cutoff dependent
Lagrangian density (\ref{Lmunor}) of the MSM.
They included $g_b(\mu)$,
however, their discussion applies also
to the truncated case $g_b\!\mapsto\!0$,
this case we illustrate here. Compositeness
of the scalar doublet $\Phi^{(\mu)}$ in (\ref{Lmunor})
can be discerned at a large energy $\Lambda$ analogously
to the discussion in Sec.~\ref{MRGE}, 
Eqs.~(\ref{TSM})-(\ref{Lmunornot3}),
where it was discussed in
the quark loop approximation. In general,
a dynamical scalar isodoublet ${\Phi}^{(\mu)}$ is
composite up to a large $\Lambda$ 
[$\ln ( \Lambda / E_{{\rm ew}} ) $$\gg$$ 1$] where it
decomposes into constituents, if an
auxiliary (nondynamical) scalar isodoublet $\Phi$ can
be defined as ${\Phi} = {\Phi}^{(\mu)}/\sqrt{Z_{\Phi}(\Lambda;\mu)}
\equiv {\Phi}^{(\mu)}/\sqrt{{\epsilon}(\Lambda;\mu)}$, with
${\epsilon}(\Lambda;\mu) $$\equiv$$ Z_{\Phi}(\Lambda;\mu)$ 
having the two properties: 
\begin{itemize}
\item In the relation
${\Phi}^{(\mu_1)} \equiv {\Phi}^{(\mu_2)} 
\sqrt{ Z_{\Phi}(\Lambda; \mu_1) } / 
\sqrt{ Z_{\Phi}(\Lambda; \mu_2) }$, the ratio of square roots is
for $\mu_1,$ $\mu_2 $$\ll$$ \Lambda$ independent of $\Lambda$,
leading to the usual MSM renormalization prescription.
\item $Z_{\Phi}(\Lambda; \mu) \to 0$ 
when ${\mu}\!\to\!{\Lambda}$; this is
the usual compositeness condition, resulting in the disappearance
of the dynamical kinetic term of the scalar degrees of freedom
when the UV cutoff $\mu$ for the Lagrangian density
approaches the energy $\Lambda$ 
where the composite scalar is supposed
to fall apart into its constituent parts (\ref{HD2}).
\end{itemize}
Also the quark fields in (\ref{Lmunor}) are, in general,
energy (cutoff) dependent: ${\Psi}\!\mapsto\! {\Psi}^{(\mu)}$
(in the quark loop approximation they are not). Hence:
${\Psi}^{(\mu_1)}\!=\! {\Psi}^{(\mu_2)} 
\sqrt{ Z_{\Psi}(\mu_1) } / \sqrt{ Z_{\Psi}(\mu_2) }$.
The quark fields appearing in the 
truncated TSM (\ref{TSM})-(\ref{TSM2})
at $\mu $$=$$ \Lambda$, where the scalars decompose, are then
${\Psi}^{(\Lambda)} \!=\! {\Psi}^{(\mu)}
\sqrt{ Z_{\Psi}(\Lambda) } / \sqrt{ Z_{\Psi}(\mu) }$, 
with still existing kinetic terms -- quarks 
remain dynamical, don't ``decompose'' at ${\Lambda}$. 
The Lagrangian density at $\mu $$<$$ \Lambda$ is
\begin{eqnarray}
{\cal {L}}^{(\mu)} & = & i 
{\overline {\Psi}}^{(\mu)}{ {D} \llap {/} } {\Psi}^{(\mu)}
+ (D_{\nu} \Phi^{(\mu)})^{\dagger} D^{\nu} \Phi^{(\mu)}
- m_{\Phi}^2(\mu) \Phi^{(\mu) \dagger} \Phi^{(\mu)} 
- \frac{ \lambda (\mu) }{2} 
\left( \Phi^{(\mu) \dagger} \Phi^{(\mu)} \right)^2
\nonumber\\
&& 
- g_t(\mu) \left[ {\overline \Psi}_L^{(\mu)} 
{\tilde \Phi}^{(\mu)} t_R^{(\mu)} + \mbox{ h.c.}
\right] + \triangle {\cal {L}}^{(\mu)}_{{\rm gauge}} \ ,
\label{Lmuban}
\end{eqnarray}
where ${\Psi}^{T} $$=$$ (t,b)$, and 
${\tilde \Phi}^{(\mu)} $$=$$ i {\tau}_2
{\Phi}^{(\mu) \dagger T}$. $D_{\nu}$ and ${ {D} \llap {/} }$ 
are the usual covariant derivatives, and 
$\triangle {\cal {L}}^{(\mu)}_{{\rm gauge}}$
contains terms without the scalars and quarks.
This density at $\mu$
can then be written in terms of the auxiliary (nondynamical)
scalar isodoublet $\Phi$
\begin{eqnarray}
{\cal {L}}^{(\mu)} & = & i 
\left[ Z_{\Psi}(\mu)/Z_{\Psi}(\Lambda) \right]
{\overline {\Psi}}^{(\Lambda)}{ {D} \llap {/} } {\Psi}^{(\Lambda)}
+ {\epsilon}(\Lambda; \mu) (D_{\nu} \Phi)^{\dagger} D^{\nu} \Phi
- {\tilde m}_{\Phi}^2 \Phi^{\dagger} \Phi
- \frac{ {\tilde \lambda} }{2} 
\left( \Phi^{\dagger} \Phi \right)^2
\nonumber\\
&& 
- {\tilde g}_t \left[ Z_{\Psi}(\mu)/Z_{\Psi}(\Lambda) \right]
\left[ {\overline \Psi}_L^{(\Lambda)} 
{\tilde \Phi} t_R^{(\Lambda)} + \mbox{ h.c.}
\right] + \triangle {\cal {L}}^{(\mu)}_{{\rm gauge}} \ ,
\label{LmuLbban}
\end{eqnarray}
Here, quantities ${\tilde m}_{\Phi}$, ${\tilde \lambda}$
and ${\tilde g}_t$, are finite, independent of $\mu$. The
running parameters of (\ref{Lmuban}) are related to them through
\begin{equation}
g_t(\mu) = {\tilde g}_t {\epsilon}^{-1/2}(\Lambda; \mu) \ ,
\quad m_{\Phi}^2(\mu) = 
{\tilde m}_{\Phi}^2 {\epsilon}^{-1}(\Lambda; \mu) \ ,
\quad \lambda(\mu) = {\tilde \lambda} 
{\epsilon}^{-2}(\Lambda;\mu) \ ,
\label{muTSMcon}
\end{equation}
where we keep in mind that ${\epsilon}(\Lambda;\mu)\!\to\! 0$ when
${\mu}\!\to\!{\Lambda}$, as a consequence of compositeness. 
At $\mu $$=$$ \Lambda$
(${\epsilon}$$=$$0$) and for ${\tilde {\lambda}}$$=$$0$, 
the above Lagrangian density reduces to the
truncated TSM (\ref{TSM2}) which is equivalent
to the NJLVL-type of interaction (\ref{TSM}). Therefore,
the special case ${\tilde {\lambda}}$$=$$0$
represents the case of NJLVL-type of ${\bar t} t$ condensation
at high $\Lambda$ [$\ln (\Lambda/E_{{\rm ew}}) $$\gg$$ 1$],
and ${\tilde {\lambda}} $$\not=$$ 0$ represents some other
more general framework of ${\bar t} t$ condensation at
high $\Lambda$.

The problem pointed out by Bando {\em et al.\/} (1990)
is that $m_{\Phi}^2(\mu)$ becomes 
large when $g_t(\mu)$ becomes large for
${\mu}\!\to\!{\Lambda}$, as seen explicitly in (\ref{muTSMcon}). 
In fact, when $g_t(\mu)\!\to\! \infty$ 
for ${\mu}\!\to\!{\Lambda}$, then
$m_{\Phi}^2(\mu)\!\to\! \infty$ in such a way that the ratio
$[m_{\Phi}^2(\mu)/(\mu^2 g_t^2(\mu))]$ remains finite, equal to 
$[{\tilde m}_{\Phi}^2/({\Lambda}^{\!2} {\tilde g}_t^2)]$ in the limit. 
Therefore, the effect of the scalar exchange of Fig.~\ref{rmd7f}(b)
which resulted in the NTL term $1.5 g_t^3$
on the left of RGE (\ref{RGEgt}), should be screened out
for $\mu $$\sim$$ \Lambda$ because of 
the decoupling due to the large
mass $m_{\Phi}(\mu)$. This is not taken into account
in $\overline{MS}$ RGE (\ref{RGEgt}). 
\begin{figure}[htb]
\mbox{}
\vskip2.4cm\relax\noindent\hskip0.cm\relax
\includegraphics{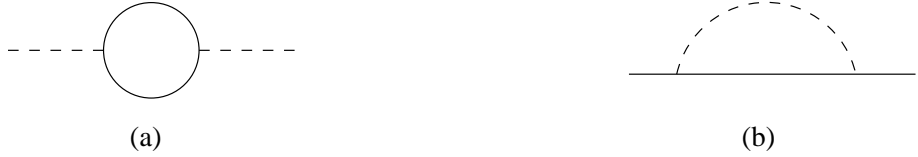} \vskip0.5cm
\caption{\footnotesize Feynman diagrams responsible for the 
terms on the left of RGE (\ref{RGEgt}) and (\ref{RGEgtban}):
(a) the leading-$N_{{\rm c}}$ term 
$N_{{\rm c}} g_t^3$ ($N_{{\rm c}} g_t^2$); 
(b) the next-to-leading term $1.5 g_t^3$ ($1.5 g_t^2 f$). 
Full lines are top quarks, dashed ones are (neutral) components of
the composite isodoublet scalar $\Phi^{(\mu)}$.}
\label{rmd7f}
\end{figure}
Therefore, Bando {\em et al.\/} used
the mass-dependent renormalization devised 
by~\citeasnoun{GeorgiPolitzer76} 
in which contributions of the
propagators of heavy particles are suppressed. The
scalar exchange term was then suppressed by
a factor $f(c)\!\sim\! c \!+\! {\cal {O}}(c^2)$, 
where $c$$ \equiv$$ {\mu}^2/m_{\Phi}^2(\mu)$$ \to$$ 0$
when $\mu$$ \to$$ \Lambda$.
The modified one-loop RGE for $g_t$ then reads
\begin{eqnarray}
8 \pi^2 \frac{ d \ln g^2_t(\mu)}{d \ln \mu} &=&
 N_{{\rm c}} g_t^2(\mu) + 1.5 g_t^2(\mu) f(c(\mu)) - G_t(\mu)  \ ,
\label{RGEgtban}
\\
\mbox{where } \qquad G_t(\mu) &=& 
  3 \frac{(N_{{\rm c}}^2-1)}{N_{{\rm c}}} g_3^2(\mu) 
+ \frac{9}{4} g_2^2(\mu) + \frac{17}{12} g_1^2(\mu) \ .
\label{Gt}
\end{eqnarray}
In order to see from another perspective that a
shielding factor $f(c(\mu))$ is needed in (\ref{RGEgtban}),
we write down also RGE for the quantity 
$c(\mu) \!=\! \mu^2/m_{\Phi}^2(\mu)$
\begin{equation}
8 \pi^2 \frac{ d \ln c(\mu) }{ d \ln \mu } =
 16 \pi^2 - (1 + c(\mu)) N_{{\rm c}} g_t^2(\mu)  \ .
\label{RGEm}
\end{equation}
Diagram of Fig.~\ref{rmd7f}(a) is the origin of this RGE and 
of the leading-$N_{{\rm c}}$ term $N_{{\rm c}} g_t^2$ in
(\ref{RGEgtban}) (cf.~also Fig.~\ref{rmd6f}).
Incidentally, (\ref{RGEm}) tells us that $m_{\Phi}^2(\mu)
\!\to\! {\infty}$ when ${\mu}\!\to\! {\Lambda}$, because
$g_t^2(\mu) $$\sim$$ [ \ln ( {\Lambda}/{\mu} ) ]^{-1}$ for
${\mu} $$\sim$$ {\Lambda}$ by (\ref{RGEgtban}). Adding
RGE's (\ref{RGEm}) and (\ref{RGEgtban}), leads to
\begin{equation}
8 \pi^2 \frac{ d \ln \left[ m_{\Phi}^2(\mu)/g_t^2(\mu) \right] }
{d \ln \mu} = \frac{\mu^2}{m_{\Phi}^2(\mu)} N_{{\rm c}} g_t^2(\mu)
- 1.5 g_t^2(\mu) f(c(\mu)) + G_t(\mu) \ .
\label{RGErat}
\end{equation}
From here we can see that the ${\overline{MS}}$ case 
[$f(c) $$\equiv$$ 1$] implies:
$m_{\Phi}^2(\mu)/g_t^2(\mu)\!\to\! 0$ when ${\mu}\!\to\!{\Lambda}$.
This is so because: 
(a) $g^2_t(\mu)\!\propto\![\ln(\Lambda/\mu)]^{-1}$
for $\mu $$\sim$$ \Lambda$; (b) ${\mu}^2/m_{\Phi}^2(\mu)\!\to\! 0$
when ${\mu}\!\to\! {\Lambda}$; (c) the second term 
on the right of (\ref{RGErat})
consequently becomes dominant for $\mu $$\sim$$ \Lambda$. However, 
from compositeness conditions (\ref{muTSMcon}) we should 
have $m_{\Phi}^2(\mu)/g_t^2(\mu)\!\to\! 
{\tilde m}_{\Phi}^2/{\tilde g}^2_t$,
which is finite nonzero.\footnote{
In the truncated TSM (\ref{TSM2}),
${\tilde m}_{\Phi} $$=$$ M_0$ and 
${\tilde g}_t $$=$$ M_0 \sqrt{G}$, thus
the ratio is $1/G$ (finite nonzero).}
This contradiction shows from another
perspective that RGE (\ref{RGEgt}) of a mass-independent 
renormalization scheme is at $\mu $$\simeq$$ \Lambda$ not consistent 
with condensation scenario. On the other hand, if
$f(c(\mu)) $$\sim$$ c(\mu) $$\equiv$$ \mu^2/
m_{\Phi}^2(\mu)\!\to\! 0$
when ${\mu}\!\to\!{\Lambda}$, it is straightforward to check that
(\ref{RGErat}) gives a finite nonzero ratio
$m_{\Phi}^2(\mu)/g_t^2(\mu)$ when ${\mu}\!\to\! {\Lambda}$, 
in accordance with compositeness conditions (\ref{muTSMcon}).

The precise behavior of the modification
factor $f(c)$ depends on the
choice of renormalization conditions for the scalar and
the top quark propagators. For a specific choice, 
Bando {\em et al.\/} (1990)
obtained values of $m_t(m_t)$ which were, for
$\Lambda $$=$$ 10^{15}$ and $10^{10}$ GeV, higher than those of
BHL by about 2.2\% and 1.6\%, respectively. However,
for lower cutoff $\Lambda $$=$$ 10^4$ GeV, they obtained
$m_t(m_t) $$=$$ 405$ GeV, while BHL value in this case
is $455$ GeV. Therefore, it appears that the
scalar mass decoupling effect is numerically important
only for low cutoffs $\Lambda$, while for high cutoffs
it is marginal -- a consequence of the
infrared fixed-point behavior of the RGE's for
high cutoffs.

\subsection{Possible effects of higher than 6-dimensional operators}
\label{CCVAMF4}
Soon after the appearance of the first works on
${\bar t} t$ condensation within the 
NJLVL models, several authors
\cite{Suzuki90a,Bastero-GilPerez-Mercader90,MannanKing91,Hasenfratzetal91}
questioned the validity of
such a framework. They argued that four-quark effective interactions
other than those of the NJLVL-type
(\ref{TSM}) and (\ref{3gTSM}), i.e., terms with dimension higher
than six also contribute to the ${\bar t} t$ condensation.
They argued that predictions for mass ratios
and/or masses of the top quark $m_t$ 
and of the composite Higgs boson $M_H$ can
in principle acquire any value, if only the coupling parameters
of the higher than 6-dimensional four-quark terms are
chosen to be strong enough. 
Thus, this generalized ${\bar t}t$ condensation mechanism
becomes in principle as unpredictive as the SM. 
While Suzuki, and Hasenfratz {\it et al.\/}, argued in the
quark loop approximation, Mannan and King extended Suzuki's
work by including the (improved) ladder QCD contributions
in the DS$+$PS approach, and Veldhuis (1992) investigated two-loop
RGE's using the quark loop results of Suzuki as UV boundary
condition. Hasenfratz {\it et al.\/} emphasized their point
additionally by the title of their paper
(``The equivalence of the top quark condensate and the
elementary Higgs field''). Their paper had an immediate and strong
impact in the ${\bar t}t$ condensation community.  
Further, a paper by Zinn-Justin (1991)
argued similarly that large classes of composite models
are not more predictive than those with explicit (elementary)
bosons. To illustrate these claims, he considered
the extended Gross-Neveu model near the four dimensions.

Other authors
\cite{Bardeen90,Hill91,Lindner93} 
subsequently argued against the
conclusions of 
Hasenfratz {\em et al.\/}. 
One of the
principal arguments is very transparent: the mentioned 
additional coupling parameters are not arbitrary,
but are determined by the underlying physics at
energies $E $$\stackrel{>}{\sim}$$ \Lambda$. These
parameters were defined as dimensionless, via multiplication
with an appropriate power of ${\Lambda}$. Since
the compositeness scale ${\Lambda}$ is also the
UV cutoff of the theory, it is the natural energy scale
of the theory, and therefore it can be naively expected that the
mentioned dimensionless parameters are $\sim$$1$.
For large cutoffs, when $\ln(\Lambda/m_t) $$\gg$$ 1$,
such parameters then modify
predictions of the original (NJLVL) framework very little.

Strictly speaking, the effects of the additional operators
can be calculated only when we know the underlying physics. 
Hill (1991) 
calculated approximately one of such parameters
($d$ in the notation by Hasenfratz {\it et al.\/};
he called it $\chi$) in his renormalizable topcolor
model of the underlying physics. Exchange of massive colorons
($M_B $$=$$ \Lambda$) is responsible for the TSM interaction term
(\ref{TSM}) of the NJLVL type. However, the box diagrams
containing colorons contribute at
the leading-$N_{{\rm c}}$ to an 8-dimensional four-quark operator
and consequently to the $\chi$ parameter.
His calculation
gives $\chi $$\approx$$ 1/8 $$=$$ 0.125$,
which is even substantially smaller than 1. 
In the block-spin renormalization group approach,
even for cutoffs as low as 1 TeV this would imply corrections to
the masses $m_t$ and $M_H$ of only $\sim\!10$\%,
and for significantly higher cutoffs the corrections
would be negligible. Hill thus expressed the belief that
conclusions of Hasenfratz {\em et al.\/} (1991)
are not relevant in most cases of realistic theories of underlying
physics leading to ${\bar t}t$ condensation.

\subsection{Deficiencies of the minimal framework --
motivation for extensions}
\label{CCVAMF5}

The minimal ${\bar t} t$ condensation framework
represents an elegant solution to several problems
that would otherwise be solved within the conventional
MSM in a rather {\em ad hoc\/} manner -- it explains,
in a dynamical way, simultaneously the mass of 
by far the heaviest known fermion (together with its
Yukawa parameter) and the scalar sector leading to the
EWSB. However, various approximate methods of
calculation (perturbative RGE's, DS$+$BS in
leading-$N_{{\rm c}}$) consistently indicate
that the phenomenologically acceptable values
$\langle {\cal H} \rangle_0 $$\equiv$$ v $$\approx$$ 246$ GeV
and $m_t $$\approx$$ 175$ GeV cannot be simultaneously 
produced with this mechanism in the minimal framework.
When the EW VEV $v$ is adjusted to $246$ GeV,
$m_t^{{\rm dyn.}}$ is too high by at least $20 \%$, even when
compositeness scale ${\Lambda}$ is very high
($\sim$$E_{{\rm Planck}} $).
An indication of this problem can be seen qualitatively 
already by inspecting the quark-loop-approximated
PS relation (\ref{PSql}) 
-- when requiring $F_{\pi^0}$$=$$v$ ($=246$ GeV),
phenomenologically acceptable values of
$m_t$ in this crude approximation can be obtained
only for very high ${\Lambda} $$\sim$$ 10^{13}$ GeV.
Therefore, there appears to be a need to relieve
${\bar t} t$ condensate from
the burden of being responsible for the entire EWSB,
i.e., to allow $F_{\pi^0}$$<$$v$. 
This can be achieved by either introducing
dynamical DEWSB frameworks with
at least one additional condensate (e.g., ${\bar f} f$,
where $f$ is a fermion different from $t$),
or by having an elementary Higgs in addition to the
composite ${\bar t}t$ scalar. Both scalars together
can then lead to the correct EW VEV $v $$\approx$$ 246$
GeV, while keeping
$m_t^{{\rm dyn.}}$ at the value
$\approx$$175$ GeV, and there is an additional 
possibility to reduce substantially ${\Lambda}$, 
thus even eliminating any need for fine-tuning.

This almost automatically leads to extended frameworks of
DEWSB. In Secs. \ref{EWESG} and \ref{EGSG} we discuss 
some such {\em effective\/}
extensions of the (effective) minimal framework
which don't involve or involve, respectively,
an enlargement of symmetries.
In Sec.~\ref{RMUP}, on the other hand, we discuss
some {\em renormalizable\/} models of underlying physics
which, at lower energies, lead in general to
extended effective frameworks of DEWSB and
${\bar t} t$ condensation.

\section{Extensions without enlarging the symmetry group}
\label{EWESG}
\setcounter{equation}{0}

\subsection{Composite two-Higgs-doublet (2HD) scenarios}
\label{EWESG1}
\subsubsection{A general framework with more than one family}
\label{EWESG11}

The basic idea that a
NJLVL framework can lead to a composite Higgs
via ${\bar t}t$ condensation can be extended in a
straightforward manner to scenarios with two composite
Higgs doublets (composite 2HD). 
Suzuki (1990b)
investigated this
possibility for the general NJLVL scenario with the SM symmetry
$SU(3)_c\!\times\!SU(2)_L\!\times\!U(1)_Y$, as written for example
in (\ref{genTSM}). First he relabeled biquark combinations 
${\bar q}_{\alpha} q_{\beta}$ with a
single ``pair'' index ${\cal {A}}$$=$$({\alpha},{\beta})$
\begin{equation}
{\cal {L}}_{4q}^{(\Lambda)} =  \sum_{{\cal {A}},{\cal {B}}=1}^{2 n^2}
G_{{\cal {A}} {\cal {B}}}J^{\dagger}_{{\cal {A}}} J_{{\cal {B}}} ,
\quad \mbox{where: } \ J_{{\cal {A}}} = \left\{
\begin{array}{ll}
{\overline {\Psi}}_L^{{\alpha}a} u_R^{{\beta}a} & ({\cal {A}} 
= 1, \ldots, n^2) 
\\
{\overline {\Psi}}_L^{c{\alpha}a} d_R^{c{\beta}a} & ({\cal {A}} 
= n^2\!+\!1, \ldots, 2n^2) 
\end{array}
\right.  \ .
\label{relab}
\end{equation}
As in (\ref{genTSM}) and (\ref{3gTSM}), ${\Psi}^{{\alpha}} $$=$$ 
(u^{\alpha},d^{\alpha})^T$, and 
$({\alpha},{\beta})$ and $a$ are family and color indices, 
respectively. ${\Psi}^{c{\alpha}} $$=$$ 
(d^{c{\alpha}},-u^{c{\alpha}})^T$ are the charge
conjugated fields of family ${\alpha}$. 
Then the $2n^2\!\times\!2n^2$ Hermitian
coupling matrix $G_{ {\cal {A}} {\cal {B}} }$ was diagonalized
by a unitary matrix $K$ to obtain
${\cal {L}}_{4q}^{(\Lambda)}$$=$$G_D J_D^{\dagger} J_D$,
with $J_D$$=$$K_{D {\cal {A}}} J_{{\cal {A}}}$, where
$D$ and ${\cal {A}}$ run from $1$ to $2 n^2$,
and $G_D$ are the (real) eigenvalues of 
$G_{{\cal {A}}{\cal {B}}}$.
Suzuki argued that one composite 
Higgs doublet is generated in the
eigenchannel with the largest $G_D$ (say: $G_{D=1}$), i.e., 
in the channel with the strongest attractive force. 
The mass eigenstate quarks are then obtained
by diagonalizing the neutral component of 
$J_{D=1} = \sum K_{1 {\cal {A}}}
J_{{\cal {A}}}$ by using unitary rotations for the $L$ and $R$
components of the up-type and down-type quarks separately -- 
a procedure analogous to that in the MSM. 
In this way, the four-quark terms
of the most attractive channel don't contribute any flavor-changing
neutral currents (FCNC's), in accordance with experimental evidence.
Suzuki then argued that, in this general framework, the second
most attractive channel (corresponding to, say $D$$=$$2$) would lead
analogously to the generation of the second composite Higgs doublet.
This doublet, in the described general NJLVL framework, 
would generally give nonzero FCNC's in the
basis of mass eigenstates of quarks. The author then discussed
in detail the conditions under which the composite two-Higgs-doublet
scenario results in suppressed FCNC's, 
using the quark loop (bubble) approximation
for gap equations and assuming $E_{{\rm ew}}/\Lambda $$\ll$$ 1$.

\subsubsection{The four-quark interaction
picture vs.~the composite 2HDSM(II) picture}
\label{EWESG12}

It can be shown that the general NJLVL-type of TSM Lagrangian
density for the third quark generation (\ref{3gTSM}) leads 
at $\mu $$\ll$$ {\Lambda}$ under certain conditions
to an effective standard model with two Higgs doublets
of type II -- 2HDSM(II). 
This is the model in which one Higgs isodoublet (${\Phi}_u$) is
responsible for the generation of $m_t$, and
the other (${\Phi}_d$) for the generation of $m_b$
\cite{DonoghueLi79,HallWise81}:
\begin{equation}
{\cal {L}}_{{\rm Yukawa}}^{(\mu)} =
- \left[ {\overline \Psi}_L^{(\mu)} 
{\tilde \Phi}^{(\mu)}_u t_R^{(\mu)} + 
\mbox{ h.c.} \right] 
- \left[ {\overline \Psi}_L^{(\mu)} 
{\Phi}^{(\mu)}_d b_R^{(\mu)}  + 
\mbox{ h.c.} \right] \ .
\label{2HDSMII}
\end{equation}
To see this, let us start with the 
following ``2HDSM(II)-type'' of
Lagrangian density at the 
compositeness scale ${\mu} = 
{\Lambda}$, with ${\Phi}_u$, ${\Phi}_d$ being
auxiliary scalar isodoublet fields:
\begin{equation}
{\cal {L}}^{(\Lambda)} = - {\mu}_u^2 {\Phi}_u^{\dagger} {\Phi}_u
- {\mu}_d^2 {\Phi}_d^{\dagger} {\Phi}_d
- {\mu}_{ud}^2 \left( {\Phi}_u^{\dagger} 
{\Phi}_d + \mbox{ h.c.} \right)
- \left[ {\overline \Psi}_L {\tilde \Phi}_u t_R 
 +  {\overline \Psi}_L {\Phi}_d b_R  + \mbox{ h.c.} \right] \ ,
\label{2HDSMIIaux}
\end{equation}
where ${\Psi}_L $$\equiv$$ (t_L,b_L)^T$ and
${\tilde \Phi} $$\equiv$$ i {\tau}_2 {\Phi}^{\dagger T}$. 
We omit the color indices and the UV energy 
cutoff superscripts $(\Lambda)$.
The Yukawa coupling parameters are normalized to
$1$ for convenience, by appropriate redefinition of 
${\Phi}_u$, ${\Phi}_d$.
If the mass parameters above satisfy certain constraints, 
the auxiliary isodoublets become dynamical -- 
they develop kinetic and
self-interaction terms and VEV's
at ${\mu} $$<$$ {\Lambda}$, thus leading to dynamical generation of
$m_t$ and $m_b$. The resulting low energy
theory is a composite 2HDSM(II).
To show equivalence of (\ref{2HDSMIIaux}) with
the NJLVL density (\ref{3gTSM}) of Sec.~\ref{MDSPS},
we employ equations of motion for the
auxiliary fields, leading to
\begin{eqnarray}
{\tilde \Phi}_u \equiv i {\tau}_2 {\Phi}_u^{\dagger T} 
& = & \left[ -\!{\mu}_d^2 \left( {\bar t}_R {\Psi}_L 
\right)\!+\!{\mu}_{ud}^2 (i {\tau}_2)
\left( {\bar b}_R {\Psi}_L \right)^{\dagger T} \right]/{\cal {A}} ,
\label{tildePhiu}
\\ 
{\Phi}_d & = & \left[ -\!{\mu}_{u}^2 \left( {\bar b}_R {\Psi}_L 
\right)\!-\!{\mu}_{ud}^2 (i {\tau}_2) 
\left( {\bar t}_R {\Psi}_L \right)^{\dagger T} \right]/{\cal {A}} ,
\ \mbox{with: } \
{\cal {A}}\!=\!\left[ 
{\mu}_u^2 {\mu}_d^2\!-\!({\mu}_{ud}^2)^2 \right] .
\label{Phid}
\end{eqnarray}
In (\ref{tildePhiu})-(\ref{Phid}), ${\tau}_2$ is the second Pauli
matrix acting on isospin components, and
$({\bar t}_R {\Psi}_L)^{\dagger T}$ 
$ =$$({\bar t}_L t_R, {\bar b}_L t_R)^T$. Inserting
(\ref{tildePhiu})-(\ref{Phid}) into (\ref{2HDSMIIaux})
leads to an equivalent Lagrangian density, which is of the
NJLVL-type (i.e., four-quark interactions without derivatives)
\begin{eqnarray}
{\cal {L}}^{(\Lambda)} & = & 
{\Big \{} {\mu}_d^2 \left( {\overline \Psi}_L^{ia} t_R^a \right)
\left( {\bar t}_R^b {\Psi}_L^{ib} \right) +
{\mu}_u^2 \left( {\overline \Psi}_L^{ia} b_R^a \right)
\left( {\bar b}_R^b {\Psi}_L^{ib} \right) +
\nonumber\\
&&+ (- {\mu}_{ud}^2 ) 
\left[ ( {\bar t}_R^a t_L^a )( {\bar b}_R^b b_L^b )
-({\bar t}_R^a b_L^a ) ({\bar b}_R^b t_L^b) + \mbox{ h.c.} \right] 
{\Big \}}/{\cal {A}} \ ,
\label{2HDNJLVL}
\end{eqnarray}
where the isospin ($i$) and color indices ($a, b$)
are explicitly written.
This density has the form (\ref{3gTSM}) of Sec.~\ref{MDSPS}.
Comparison gives relations between the parameters of
(\ref{3gTSM}) and (\ref{2HDSMIIaux})
\begin{equation}
{\kappa}_t = \frac{ {\cal {B}} }{ {\mu}_u^2 } \ , \quad
{\kappa}_b = \frac{ {\cal {B}} }{ {\mu}_d^2 } \ , \quad
2 {\kappa}_{tb} = 
- \frac{ {\cal {B}} {\mu}_{ud}^2 }{ {\mu}_u^2 {\mu}_d^2 } 
\quad \mbox{where: } \ 
{\cal {B}} = \frac{ N_{{\rm c}} {\Lambda}^{\!2} }{ 8 \pi^2 } 
\frac{ {\mu}_u^2 {\mu}_d^2 } 
{ \left[ {\mu}_u^2 {\mu}_d^2 - ({\mu}_{ud}^2)^2 \right] } \ .
\label{kapmu}
\end{equation}
From the above formulas we see that 
\begin{equation}
(8 \pi^2)^2 [ {\kappa}_t {\kappa}_b - 4 {\kappa}_{tb}^2 ]
= ( N_{{\rm c}} {\Lambda}^{\!2} )^2
\left[ {\mu}_u^2 {\mu}_d^2 - ({\mu}_{ud}^2)^2 \right]^{-1}
\left( \not= 0 \right) \ .
\label{not1HD}
\end{equation}
This shows that the NJLVL-type of framework (\ref{3gTSM})
can lead to a composite dynamical 2HDSM(II)
only as long as ${\kappa}_t {\kappa}_b - 4 {\kappa}_{tb}^2 \not= 0$.
When ${\kappa}_t {\kappa}_b - 4 {\kappa}_{tb}^2 = 0$
[note: ${\kappa}_t $$\not=$$ 0$ 
and ${\kappa}_b $$=$$ {\kappa}_{tb}$$=$$0$
is the case of truncated TSM (\ref{TSM})],
this can lead only to an
effective model with {\em one\/} (composite) scalar isodoublet --
a composite MSM (minimal framework).

Eqs.~(\ref{tildePhiu})-(\ref{Phid}) imply
that the generalized TSM model (\ref{3gTSM})
discussed in Sec.~\ref{MDSPS} leads to two dynamical composite scalar
isodoublets made up of $t$ and $b$ quarks and with
nonzero VEV's, as long as
parameters ${\kappa}_t$, ${\kappa}_b$ and
${\kappa}_{tb}$ satisfy some criticality bounds.
In the framework (\ref{3gTSM}),
conditions for the dynamical generation of heavy 
$m_t $$\sim$$ 10^2$ GeV and 
light $m_b $$\approx$$ 5$ GeV
were investigated numerically
by King and Mannan (1991a)
who employed the DS$+$PS approach (quark loop plus QCD contributions)
-- cf.~discussion in Sec.~\ref{MDSPS}. This approach
does not deal directly with the picture of dynamically generated
Higgs doublets of 2HDSM(II) or MSM,
but only with the dynamically generated masses $m_t$
and $m_b$. Results of King and Mannan show that 
the heavy $m_t^{{\rm dyn.}}$ (i.e., a corresponding
$\langle {\Phi}_u \rangle_0\!\not=\!0$
in the present picture) is generated when ${\kappa}_t$ is at least a 
bit above a critical value ${\kappa}_{{\rm crit.}} $$\sim$$ 1$. 
The latter value is slightly
dependent on ${\kappa}_b$ and ${\kappa}_{tb}$, and both latter
parameters satisfy: $0 $$\leq$$ {\kappa}_b $$<$$ {\kappa}_t$ and
$0 $$\leq$$ {\kappa}_{tb} $$\ll$$ {\kappa}_t$. 
Further, a smaller $m_b^{{\rm dyn.}}$ 
(i.e., $\langle {\Phi}_d \rangle_0\!\not=\!0$ in the present picture)
is generated when, for a given ${\kappa}_b $$<$$ {\kappa}_t$,
the mixing parameter ${\kappa}_{tb}$ acquires certain small positive
value $\sim$$ 10^{-2}$.

If the $U(1)_{{\gamma}_5}$-violating four-quark term 
in (\ref{3gTSM}) is taken to be zero [${\kappa}_{tb} $$=$$ 0$, hence
by (\ref{kapmu}): ${\mu}_{ud}^2 $$=$$0$], we see from 
(\ref{tildePhiu})-(\ref{Phid}) that the first (second) isodoublet has
the neutral scalar component made up entirely of the
${\bar t} t$ (${\bar b} b$) condensate. Therefore,
in this case the possible dynamically generated nonzero VEV's 
lead to $m_t^{{\rm dyn.}}$ ($m_b^{{\rm dyn.}}$) 
coming exclusively from the
${\bar t t}$ (${\bar b} b$) condensation. 
The DS integral equations
(\ref{DSmty}) in Sec.~\ref{MDSPS} in fact describe 
just this dynamical 2HDSM(II) scenario. 
King and Mannan (1991a) found out that, when
${\kappa}_{tb}$$=$$0$, {\em both\/} ${\kappa}_t$
and ${\kappa}_b$ have to be at least a bit above a
certain common critical value
${\kappa}_{{\rm crit.}} $$\sim$$ 1$
for the generation of {\em both\/} nonzero\footnote{
We note that, when 
${\kappa}_t,\!{\kappa}_b$$<$${\kappa}_{{\rm crit.}}$,
King and Mannan (1991a)
still obtained nonzero
$m_t^{{\rm dyn.}} $$=$$ m_b^{{\rm dyn.}} $
$\approx$$ 0.3$ GeV. This small
value originates solely from QCD gluon exchange effects.}
$m_t^{{\rm dyn.}}$ and $m_b^{{\rm dyn.}}$ -- in the 
present picture this case means
the generation of $\langle {\Phi}_u \rangle_0$,
$\langle {\Phi}_d \rangle_0\!\not=\!0$ when ${\mu}_{tb}^2 $$=$$ 0$. 
In the subcase ${\kappa}_t $$>$$
{\kappa}_{{\rm crit.}} $$>$$ {\kappa}_b$
and ${\kappa}_{tb}$$=$$0$, we have 
$m_t $$\sim$$ 10^2$ GeV and $m_b $$=$$ 0$, and equivalently:
$\langle {\Phi}_u \rangle_0 $$\not=$$ 0$ and
$\langle {\Phi}_d \rangle_0 $$=$$ 0$. In this subcase,
a question appears as to the physical picture
of composite isodoublets -- are there two dynamical composite
isodoublets or only one at low energies
($\mu $$\ll$$ {\Lambda}$)? The works
discussed in Sec.~\ref{MDSPS} did not address this question,
since the DS$+$PS formalism employed there
doesn't contain composite isodoublets explicitly.
We argue here that in such a case, the relevant four-quark
coupling ${\kappa}_b$, being too weak to lead to a nonzero
VEV $\langle {\Phi}_d \rangle_0$, appears in general to
be also too weak to bind the composite auxiliary field
${\Phi}_d $$\sim$$ {\bar b}_R {\Psi}_L$ into an existing and
detectable
dynamical field ${\Phi}_d^{(\mu)} $$\sim$$ {\bar b}_R {\Psi}_L$
at lower energies ${\mu} $$<$$ {\Lambda}$. We can understand this
claim at the level of the quark loop approximation also this
way: We replace in formulas 
(\ref{Lmu})-(\ref{Lmunornot3}) of Sec.~\ref{MRGE} the field
$t_R$ by $b_R$, the parameter 
${\kappa}_t \equiv G {N_{{\rm c}}} {\Lambda}^{\!2}/(8 {\pi}^2)$
by ${\kappa}_b$, and the auxiliary field ${\tilde \Phi} $$\equiv$$ 
{\tilde \Phi}_u$ by the auxiliary field ${\Phi}_d$. Then
those formulas describe the ${\Phi}_d $$\sim$$ {\bar b}_R {\Psi}_L$
condensation sector at the quark loop level in our subcase.
However, for ${\kappa}_b$ {\em substantially\/} smaller than
${\kappa}_{{\rm crit.}} $$=$$ 1$ 
($0 $$<$$ 1$$-$$ {\kappa}_b $$\sim$$ 1$), relation (\ref{Lmunornot2})
implies that the mass of the presumably
dynamical zero-VEV isodoublet ${\Phi}_d^{(\mu)}$ at $\mu $$\sim$$
E_{{\rm ew}}$ becomes as large as $\sim$$ {\Lambda}$:
\begin{equation}
m_{{\Phi}_d}^2(\mu) = \left[ \left( \frac{1}{ {\kappa}_b } -1 \right)
{\Lambda}^{\!2} + {\mu}^2 \right] 
\frac{2}{ \ln ( {\Lambda}^{\!2}/{\mu}^2 ) } \ 
\sim \frac{{\Lambda}^{\!2}}{ \ln ( {\Lambda}^{\!2}/{\mu}^2 ) } \quad
\left[ \mbox{roughly } \sim {\Lambda}^{\!2} \right] \ .
\label{kapbbelow}
\end{equation}
Despite the fact that quark loops induce kinetic terms for {\em both\/}
${\Phi}_u$ and ${\Phi}_d$ at ${\mu} $$<$$ {\Lambda}$ 
[cf.~$Z_{\Phi}({\Lambda};{\mu})$ in (\ref{Lmu})-(\ref{Lmunot1}), where
${\Phi}$ is now ${\Phi}_u$ or ${\Phi}_d$],
having $m_{ {\Phi}_d} $$\sim$$ {\Lambda}$ means that ${\Phi}_d^{(\mu)}
$$\sim$$ {\bar b}_R {\Psi}_L$ is undetectable since the theory
has by definition an UV cutoff at 
$\sim$$ {\Lambda}$. From (\ref{kapbbelow})
we see that this is true
as long as the value of ${\kappa}_b$ is below and not very close
to the critical value ${\kappa}_{{\rm crit.}}$
(${\kappa}_{{\rm crit.}}$$=$$1$ 
in the quark loop approximation).

When there is some small mixing (${\kappa}_{tb} $$\not=$$ 0$,
hence ${\mu}_{ud}^2 $$\not=$$ 0$), we see from
(\ref{tildePhiu})-(\ref{Phid}) that
the first (second) neutral scalar, responsible for
the mass $m_t$ ($m_b$), has a small
admixture of ${\bar b} b$ (${\bar t} t$).
The latter phenomenon is the ``feed-down'' mechanism mentioned
already by King and Mannan (1991a) in their DS$+$PS approach
(cf.~Sec.~\ref{MDSPS}): ${\bar t} t$-condensate VEV
$\langle {\bar t} t \rangle_0$, being largely
responsible for large dynamical $m_t$, gives a (small) nonzero
value $m_b$ to the bottom quark through ${\kappa}_{tb}$-mixing
term interaction.

Harada and Kitazawa (1991)
investigated the question
of whether the composite two-Higgs-doublet (2HD) scenario can lead to
a breaking of $U(1)_{{\rm em}}$, i.e.,
whether the problem of $\langle {\bar t} b \rangle_0 $$\not=$$ 0$ or
$\langle {\bar b} t \rangle_0 $$\not=$$ 0$ can occur. 
They considered
the third generation of quarks, thus investigating the composite
2HD scenario of the general TSM (\ref{3gTSM}). 
As shown above, this can lead,
when ${\kappa}_t {\kappa}_b
$$\not=$$ 4 {\kappa}_{tb}^2$, to the composite 2HDSM(II).
They then constructed
the 2HD effective potential in the quark loop approximation.
The gap equations were obtained by minimizing it, and
they found out that the resulting VEV's of the
charged components of the Higgs doublets must be zero
if ${\kappa}_t $$\not=$$ {\kappa}_b$. For 
${\kappa}_t $$=$$ {\kappa}_b$, they argued that inclusion of the
isospin-breaking $U(1)_Y$ interaction is of crucial
importance to see whether the quark condensates align also
in this case in a $U(1)_{{\rm em}}$-conserving
manner.

We implicitly assumed in 
formulas (\ref{3gTSM}) and (\ref{2HDSMIIaux}) that
the mixing parameter ${\kappa}_{tb}$, and hence
also ${\mu}_{ud}^2$, are real. Harada and Kitazawa (1991) also
considered such a case. They claimed that we always have
the freedom of redefining phases of the quark fields
in such a way that these mixing parameters
can be made real. However, we wish to stress that
this claim may not be true -- at least it should be
investigated in detail. Namely, if we assumed that
${\kappa}_{tb}$ and hence ${\mu}_{ud}$ are
complex, we could end up with a composite 2HDSM(II) 
where the two dynamically generated VEV's may acquire 
substantially different phases, thus
possibly leading to a CP violation in the dynamical scalar
sector.\footnote{
Cf.~also the work of Andrianov {\em et al.\/}
(1996), summarized toward the end of this 
Section \ref{EWESG1}.}

\subsubsection{RGE analyses of the composite 2HDSM(II)}
\label{EWESG13}

Luty (1990)
investigated the composite 2HDSM(II) scenario
within the same NJLVL framework
of TSM, by going beyond the quark loop approximation.
He employed the one-loop RGE approach, in analogy with
the original 1HD TSM approach of Bardeen, Hill and Lindner
(BHL, 1990).
It turned out that the UV boundary conditions (compositeness
conditions) for the Yukawa parameters $g_t(\mu)$
and $g_b(\mu)$ were completely analogous to those of
the RGE approach in the case of the 
minimal framework of ${\bar t}t$ condensation
(\ref{RGEbc}):
\begin{equation}
 g_t(\mu) \to \infty \ , \quad g_b(\mu) \to \infty \ ,
\qquad \mbox{when } \ {\mu}\!\to\!{\Lambda} \ .
\label{2HDRGEbc}
\end{equation}
In addition,
Luty investigated one-loop RGE's governing the composite scalar
sector (scalar self-interaction parameters $\lambda_i$, $i=1,
\ldots,4$) and the
resulting masses of the composite neutral and charged scalars and
(neutral) pseudoscalar. Since RGE's for the
Yukawa couplings $g_t$ and $g_b$ are at the one-loop level
decoupled from those of the scalar self-couplings, it suffices
for our limited purposes of illustration to consider only the former
\begin{eqnarray}
16 \pi^2 \frac{d g_t(\mu)}{d\ln \mu} & = &
\left[ (N_{{\rm c}} + \frac{3}{2}) g_t^2(\mu) + 
\frac{1}{2} g_b^2(\mu) - G_t(\mu) \right] g_t(\mu) \ ,
\label{RGE2HD1}
\\
16 \pi^2 \frac{d g_b(\mu)}{d\ln \mu} & = &
\left[ (N_{{\rm c}} + \frac{3}{2}) g_b^2(\mu) + 
\frac{1}{2} g_t^2(\mu) - G_t(\mu) + g_1^2(\mu) \right] g_b(\mu) \ ,
\label{RGE2HD2}
\end{eqnarray}
where $G_t(\mu)$ contains gauge coupling parameters $g_j$ and is
given in (\ref{Gt}).
Parameters $g_j$ satisfy RGE's (\ref{gjs1}), 
but with slightly changed constants $C_j$ (for $j$$=$$1,2$) from those
of the MSM (\ref{gjs2}), due to the presence of two Higgs doublets
\begin{equation}
N_H = 1 \ \mapsto \ N_H=2 \qquad \Longrightarrow \quad
C_1 = -\frac{1}{3} - \frac{10}{9} n_q \ , \quad
C_2 = 7 - \frac{2}{3} n_q \ .
\label{2HDgjs2}
\end{equation}
Again, $n_q$ is number of effective quark flavors
(for $\mu $$>$$ m_t$: $n_q$$=$$6$).
It turns out that at the Landau poles (${\mu}$$\to$${\Lambda}$),
where both $g_t$ and $g_b$ formally diverge, we have either
$g_b/g_t$$\to$$ 0$ or $g_b/g_t $$\to$$ \infty$,
depending on whether at $E$$=$$E_{{\rm ew}}$ the ratio
$r $$\equiv$$ g_b/g_t$ is smaller or larger than a critical value 
$r_{{\rm crit.}}(m_t) $$=$$ 1 $$-$$ \delta(m_t)$, 
respectively.\footnote{
Here, ``($m_t$)'' means that $r_{{\rm crit.}}$
and ${\delta}$ are functions 
of the renormalized mass $m_t = m_t(E$$=$$m_t)$.}
Here, $\delta$ is a very small positive number, nonzero due to
the different $g_1^2$-terms in RGE's (\ref{RGE2HD1})-(\ref{RGE2HD2}).
The difference in $g_1^2$-terms originates
from the fact that $t$ and $b$ have
different hypercharges $Y$. E.g., for $m_t(m_t) $$=$$ 250$ GeV,
numerical analysis gives: $\delta $$\approx$$ 6\!\times\!10^{-3}$.
Stated differently, for any given $m_t$, there is
a critical ratio of VEV's $\tan\!\beta_{{\rm crit.}}(m_t)
$$=$$ (v_u/v_d)_{{\rm crit.}} $$\approx$$ 
(m_t/m_b) r_{{\rm crit.}}$ below which 
$g_b/g_t$ at $\mu $$=$$ \Lambda$ becomes zero and above which
it becomes infinity.
These arguments suggest that the full compositeness
in this framework,
i.e., the compositeness of both ${\Phi}_u$ and ${\Phi}_d$,
can be achieved only if $g_t(\Lambda) $$\approx$$ g_b(\Lambda)$, 
in which case
then $g_t(\mu) $$\approx$$ g_b(\mu)$ for any $\mu $$\leq$$ \Lambda$
by (\ref{RGE2HD1})-(\ref{RGE2HD2}).

However, there is also another argument
leading to the requirement 
$g_t(\mu) $$\approx$$ g_b(\mu)$ for $\mu $$\leq$$ \Lambda$.
We denote the two composite doublet fields as ${\Phi}_u^{(\mu)}$
and ${\Phi}_d^{(\mu)}$, where $\mu$ is any chosen
upper energy cutoff of the 2HDSM(II). RGE's for the
running renormalization ``constants'' $Z_j(\mu)^{1/2} =
{\Phi}_j^{(\mu)}/{\Phi}_j^{(E_{\mbox{\tiny ew}})}$ ($j$$=$$u,d$)
can be derived in a straightforward manner by regarding
diagrams of Fig.~\ref{rmd7f}(a) of Sec.~\ref{NTLEE}
(with particles in the
loop being either the electroweak gauge bosons, or top quarks,
or bottom quarks)
\begin{eqnarray}
8 \pi^2 \frac{ d \ln Z_u(\mu)}{d \ln \mu}  & = &
- N_{{\rm c}} g_t^2(\mu) + \frac{3}{4} g_1^2(\mu) 
+ \frac{9}{4} g_2^2(\mu) \ ,
\label{2HDRGEVEV1}
\\
8 \pi^2 \frac{ d \ln Z_d(\mu)}{d \ln \mu}  & = &
- N_{{\rm c}} g_b^2(\mu) + \frac{3}{4} g_1^2(\mu) 
+ \frac{9}{4} g_2^2(\mu) \ .
\label{2HDRGEVEV2}
\end{eqnarray}
$Z_j(\mu)^{1/2}$ are proportional to the
corresponding running VEV's $v_j(\mu)$, and in 
(\ref{2HDRGEVEV1})-(\ref{2HDRGEVEV2}) we can replace:\footnote{
We denote $v_j(\mu $$=$$ E_{\mbox{\tiny ew}})$ simply as $v_j$.
We will use $E_{\mbox{\tiny ew}} $$=$$ M_Z$.} 
$Z_j(\mu)$$\mapsto$$ v_j^2(\mu)$ ($j$$=$$u,d$). Since
compositeness of the scalar isodoublets ${\Phi}_j^{(\mu)}$
requires their disappearance at $\mu $$=$$ \Lambda$,
this means: $Z_j(\mu)$$\to$$ 0$ when ${\mu}$$\to$${\Lambda}$. However, 
for any chosen $m_t$ [$\equiv$$ m_t(m_t)$], the following results
from numerical analysis:
when the ratio $g_b/g_t$ at $E_{{\rm ew}}$
is not fine-tuned to $r_{{\rm crit.}}(m_t)
$$=$$ 1$$-$$ \delta(m_t)$, or equivalently,
when $\tan\!\beta $$\equiv$$ v_u/v_d$ is not fine-tuned to the
corresponding $\tan\!\beta_{{\rm crit.}}(m_t)$,
we end up either with
$Z_u(\mu)$$ \ll$$ 1$ and $Z_d(\mu) $$\sim$$ 1$, or
$Z_u(\mu) $$\sim$$ 1$ and $Z_d(\mu)$$ \ll$$ 1$,
for $\mu $$\sim$$ \Lambda$ ($\equiv$$ {\Lambda}_{{\rm pole}}$). 
Hence,
one of the scalar isodoublets is not composite in such cases.
For example, if we set $m_t(m_t)\!=\!250$ GeV, then
$\tan\!\beta_{{\rm crit.}}$$\approx$$ 81.85$.

\begin{table}
\vspace{0.1cm}
\par
\begin{center}
\begin{tabular}{ l c c c c c c c c c }
${\Lambda} \rightarrow$ & & & & & & & & & \\ 
\cline{1-1} 
$k_t = k_b \downarrow$ & \raisebox{1.5ex}[-1.5ex]{$10^{19}$} &
\raisebox{1.5ex}[-1.5ex]{$10^{17}$} &
\raisebox{1.5ex}[-1.5ex]{$10^{15}$} &
\raisebox{1.5ex}[-1.5ex]{$10^{13}$} &
\raisebox{1.5ex}[-1.5ex]{$10^{11}$} &
\raisebox{1.5ex}[-1.5ex]{$10^{9}$} &
\raisebox{1.5ex}[-1.5ex]{$10^{7}$} &
\raisebox{1.5ex}[-1.5ex]{$10^{5}$} &
\raisebox{1.5ex}[-1.5ex]{$10^{4}$} \\
\hline\hline
 & 205.4 & 208.9 & 213.3 & 219.1 & 226.9 & 238.2 & 255.8 & 
288.5 & 319.0 \\
 \raisebox{2.ex}[-1.5ex]{2/3} & [66.68] & [67.94] & 
[69.50] & [71.50] &
[74.18] & [77.95] & [83.77] & [94.12] & [103.28] \\
\hline
 & 205.9 & 209.5 & 214.1 & 222.1 & 228.3 & 240.3 & 259.5 & 
296.6 & 333.7 \\
 \raisebox{2.ex}[-1.5ex]{1} & [66.85] & [68.14] & [69.76] & 
[71.84] & [74.65] & [78.66] & [84.96] & [96.60] & [107.47] \\
\hline
 & 206.7 & 210.5 & 215.5 & 221.9 & 230.9 & 244.3 & 266.7 & 
313.4 & 367.0 \\
 \raisebox{2.ex}[-1.5ex]{10} & [67.15] & [68.51]& [70.23]&[72.47]&
[75.53] & [79.99] & [87.24] & [101.61] & [116.49] \\
\end{tabular}
\end{center}
\vspace{-0.1cm}
\caption {\footnotesize Mass $m_t(m_t)$ in the
composite 2HDSM(II) as a function of the 
cutoff ${\mu}_{\ast}$$=$$\Lambda$ 
and the chosen value $k_t$ 
[$\equiv$$ g_t^2(\Lambda)/(4 \pi)$] $=$$k_b$
[$\equiv$$ g_b^2(\Lambda)/(4 \pi)$]. Masses and energies are
in GeV. In brackets $[ \ldots ]$,  
the corresponding VEV ratios $\tan\!\beta $$=$$ v_u/v_d$
(at $E$$=$$M_Z$) are given. Used values: ${\alpha}_3(M_Z)$$=$$0.120$
[$\Rightarrow\!{\alpha}_3(180{\rm GeV})$$=$$0.110$];
$m_b(m_b)$$=$$4.3$ GeV [$\Rightarrow\!m_b(M_Z)$$\approx$$3.175$ GeV];
quark flavors number $n_q$$=$$6$ was taken, for $\mu$$\geq$$M_Z$.}
\label{tabl6}
\end{table}
Thus, we are led to the conclusion that
RGE compositeness conditions, i.e.,  
$Z_j(\mu)$$\to$$ 0$ ($j $$=$$ u,d$)
[or: $g_j(\mu)$$\to$$ \infty$ ($j$$=$$t,b$)] 
as ${\mu}$$\to$${\Lambda}$,
can only be fulfilled when $g_t(\Lambda) $$\approx$$ g_b(\Lambda)$,
thus resulting in $g_t(E_{{\rm ew}}) $$\approx$$
g_b(E_{{\rm ew}})$.
This is the reason why Luty (1990)
used implicitly for one-loop
RGE's (\ref{RGE2HD1})-(\ref{RGE2HD2}) the boundary condition 
$g_t(\Lambda) $$=$$ g_b(\Lambda)$ ($\gg$$ 1$). 
This implied that 
$\tan\!\beta $$=$$ v_u/v_d$ of the 2HDSM(II) must be very large:
$\tan\!\beta $$\approx$$ m_t(M_Z)/m_b(M_Z) $$\sim$$10^2$. 
This is also
explicitly seen in Table~\ref{tabl6}, where we have chosen
as compositeness condition for one-loop RGE's
(\ref{RGE2HD1})-(\ref{RGE2HD2}): 
$g^2_t(\Lambda)/(4 \pi) $$=$$ g^2_b(\Lambda)/(4 \pi)
$$=$$ 1/3$, $2/3$, $1$ and $10$ ($=$``$\infty$''). Results
of Luty for $m_t(m_t)$ are similar to those in
the last line of Table~\ref{tabl6}.

Table~\ref{tabl6} shows that the RGE approach to ${\bar t}t$
(and ${\bar b}b$) condensation in the 2HDSM(II) framework
leads to phenomenologically unacceptable (too high) values of 
$m_t(m_t)$: $m_t(m_t) $$\stackrel{>}{\approx}$ 
$206$ GeV for
${\Lambda} $$\stackrel{<}{\sim}$$ E_{{\rm Planck}}$,
similar to the BHL result (BHL, 1990) for the
minimal framework (cf.~Table~\ref{tabl1}).
Namely, since $g_t $$\approx$$ g_b$,
RGE in (\ref{RGE2HD1}) looks almost like
RGE (\ref{RGEgt}) of the minimal framework,
with $(N_{{\rm c}}\!+\! 1.5)$$\mapsto$$ (N_{{\rm c}}\!+\! 2)$ 
and slight changes in evolution of $g_1$ and $g_2$, while
compositeness condition $g_t(\Lambda) $$=$$ \infty$
is the same as in the minimal framework. 
Thus, for a given $\Lambda$, $g_t(E_{{\rm ew}})$
is somewhat below but close to the value in the
minimal framework. Also the corresponding VEV's in the two frameworks
are $v $$\approx$$ v_u$, due
to $v_u/v_d $$\approx$$ m_t/m_b $$\gg$$ 1$ and
the relation $v^2_u $$+$$ v^2_d $$=$$ v^2$.
Therefore, the resulting $m_t(m_t)$ in the two
frameworks, for a given $\Lambda$, turn out to have values
close to each other.

Composite 2HDSM(II) framework was also investigated by 
Froggatt, Knowles and Moorhouse (1990, 1992) who
performed an RGE analysis similar to that of Luty (1990).
In one of the discussed scenarios, ${\Phi}_d$ couples
also to ${\tau}_R$. The compositeness conditions were
$g_f^2(\mu_{\ast})/\pi $$\sim$$ 1$
($f=t,b,{\tau}$) and ${\lambda}_i(\mu_{\ast})^2/\pi $$\sim$$ 1$ 
($i=1,\ldots,5$) at $\mu_{\ast} $$=$$ 10^{14}$-$10^ {15}$ GeV.
The resulting infrared fixed-point behavior led them to
predictions of dynamical masses of $t$, $b$ and ${\tau}$
($m_b/m_{\tau} $$\approx$$ 2.6$) and of the composite scalars.
In their first work (1990), unlike Luty, they did not include
${\Phi}_u$-${\Phi}_d$ mixing term [cf.~(\ref{2HDSMIIaux})]
and consequently obtained a very light -- phenomenologically
unacceptable -- composite neutral scalar:
$M_{\eta^0} $$\stackrel{<}{\approx}$$ m_b$. In the overlapping
region of parameter space their results (1990) agree
with those of Luty. Later (1992) they included the mixing term,
thus raising in general $M_{\eta^0}$ to an acceptable level.
In particular, they proposed a decoupling scenario, in which the
scalar mass term parameters $\mu_u^2(\mu)$ and $\mu_d^2(\mu)$
at energies $\mu $$\sim$$ E_{{\rm ew}}$
are fine-tuned to values $\sim$$ E_{{\rm ew}}$
in such a way that $\sqrt{v_u^2+v_d^2}
$$=$$ v$ $\approx$$ 246$ GeV, and the
mixing mass term parameter is not fine-tuned:
${\mu}^2_{ud}(E_{{\rm ew}}) $$\sim$$ 
{\Lambda}^{\!2}$$\gg$$ E_{{\rm ew}}$
(${\Lambda}$ is the compositeness scale).\footnote{
All the mass term parameters ${\mu}_u^2(\mu)$, ${\mu}_d^2(\mu)$,
${\mu}_{ud}^2(\mu)$ at $\mu $$\sim$$ E_{\mbox{\tiny ew}}$ are in
general roughly $\sim$$ {\Lambda}^{\!2}$ unless fine-tuned --
cf.~$m_{\Phi}^2(\mu)$ in (\ref{Lmunornot2}) for the
case of {\em one\/} composite Higgs doublet.}
Then, only one composite (neutral) scalar $h^{(0)}$ retains
a small mass $M_{h^0} $$\sim$$ E_{{\rm ew}}$,
while the other four
acquire masses $\sim$$ {\Lambda}$ and
decouple from the EW physics. Thus,
the low energy physics behaves as the MSM,
RGE's having the MSM form for $\mu $$<$$ {\Lambda}$, but
$v_u$ can be substantially smaller than
$v $$\approx$$ 246$ GeV. In one of the discussed cases (case B), 
they set $v_u/v_d $$\approx$$ 1$, thus $v_u $$\approx$$ v/\sqrt{2}$,
and consequently obtained a lower $m_t $$\approx$$ 163$ GeV.
However, we previously argued that
abandoning the condition
$v_u/v_d $$\approx$$ m_t(M_Z)/m_b(M_Z) $$\gg$$ 1$
and replacing it by a smaller $v_u/v_d$ implies
abandoning the condition
$Z_d({\mu})$$\to$$ 0$ when $\mu$$\to$${\Lambda}_{{\rm pole}}$
[if $Z_u({\mu})$$\to$$ 0$ remains]. Therefore,
${\Phi}_d$ is not (fully) composite in this scenario. 
This is the price to be paid to decrease $m_t^ {{\rm dyn.}}$ 
to acceptable values $160$-$180$ GeV. 
The mass $m_t^{{\rm dyn.}}$ can be decreased
only if the composite top quark sector is not made (almost)
fully responsible for the EWSB 
($v_u $$\not\approx$$ v$ $=$$ 246$ GeV).\footnote{ 
We will see later that this is also a feature of the
fully renormalizable TC2 frameworks (cf.~Sec.~RMUP6).
A somewhat different mechanism
of decreasing $m_t$ works in a ${\bar t} t$ condensation
framework within supersymmetric models
(cf.~Sec.~\ref{EGSG1}).} 
Froggatt, Knowles and Moorhouse
included in their work (1992) also an extensive analysis
of radiative corrections of the scalar sector to
the vacuum polarizations of vector bosons. They
showed that these corrections are very close to those
in the MSM, because contributions of the four
additional heavy scalars (with masses $M $$\sim$$ {\Lambda}$)
cancel as these scalars have almost degenerate masses
($|\triangle M| $$\ll$$ E_{{\rm ew}}$).

Mahanta (1992) discussed a possible CP violation effect
in the composite 2HDSM(II), which arises when the
four-quark parameter ${\kappa}_{tb}$ in (\ref{3gTSM})
[or equivalently, ${\mu}_{ud}^2$ in (\ref{2HDNJLVL})] is
complex.\footnote{ 
Of course, ${\cal {L}}^{(\Lambda)}$ is kept
Hermitian, i.e., the corresponding ${\kappa}_{tb}$-terms 
in (\ref{3gTSM}) read in this case: $2 {\kappa}_{tb}
[ ({\bar t}_L^a t_R^a)({\bar b}_L^b b_R^b) -
({\bar t}_L^a b_R^a)({\bar b}_L^b t_R^b)] + \mbox{h.c.}$;
and the ${\mu}_{ud}^2$-terms in (\ref{2HDNJLVL}) accordingly.}
CP violation in this framework
stems from the mixing of
neutral CP-even and CP-odd composite scalars at low energies,
and is manifested through ($t$-loop-induced) electric dipole 
moment $d_e$ of the electron. Mahanta employed essentially
the one-loop RGE approach in quark loop (bubble) approximation
to construct from ${\cal {L}}^{(\Lambda)}$ of (\ref{2HDNJLVL}),
with ${\mu}_{ud}^2$ being complex, the effective
density ${\cal {L}}^{(\mu)}$ of 2HDSM(II) at
low energies $\mu $$\sim$$ v$, and he was subsequently able to impose
an upper bound on a parameter $\propto$${\kappa}_{tb}$
by requiring that $d_e$ not exceed the phenomenological upper
bound.

In this context, we mention another work
\cite{CveticKim93}
in which (one-loop) RGE evolution of Yukawa coupling parameters of
the second and third quark generation in the 2HDSM(II)
was investigated,
with the purpose of seeing whether the model
allows flavor-democratic (FD) structures at high energies.
Indeed, the authors found that, in contrast to the MSM,
the Yukawa matrices in the up-type and the down-type
sectors move toward the flavor-democratic (FD) form\footnote{
A matrix has the FD form when
all its elements are equal.}
in a weak basis
when the energy increases. Also the CKM mixing matrix moves
toward the identity matrix as the energy increases.
The chosen values of the VEV ratio were:
$v_u/v_d = 0.5, 1.0$ and $5$. Therefore, at
the pole energies ${\mu} $$=$$ {\Lambda}$ the ratio 
$g_b/g_t$ was zero, although both $g_t$ and $g_b$ diverged there.
Equivalently, $Z_u(\Lambda) $$=$$ 0$ and $Z_d(\Lambda) $$\sim$$ 1$.
Thus, the model can be interpreted
as having a composite Higgs doublet 
${\tilde \Phi}_u $$\sim$$ {\bar t}_R {\Psi}_L$
(with compositeness scale ${\Lambda}$) and another
Higgs doublet ${\Phi}_d$ of unspecified nature.
The mass $m_t$$\approx$$180$ GeV is then of completely,
and the EWSB of partially dynamical origin
($v$$\equiv$$\sqrt{v_u^2+v_d^2}$$>v_u$).
Interestingly, for low values of the VEV ratio
$v_u/v_d $$=$$ 0.5$, the pole (compositeness) energy
is as low as $\sim$$ 10$ TeV, for $m_t$ 
($=$$m_t^{{\rm dyn.}}$) $\approx$$ 180$ GeV. 

\subsubsection{Explaining large isospin breaking
$m_t $$\gg$$ m_b$ in composite 2HDSM(II)}
\label{EWESG14}

The problem of explaining by underlying physics
the large isospin breaking ($m_t $$\gg$$ m_b $$\not=$$ 0$)
in the composite 2HDSM(II) frameworks was discussed by
Bando, Kugo and Suehiro (1991),
Bando, Kugo, Maekawa and Nakano (1991),
Nagoshi, Nakanishi and Tanaka (1991),
and King and Suzuki (1992).
Discussion in the previous Subsec.~\ref{EWESG13} showed that, 
in the fully composite 2HDSM(II), the problem of
explaining the mass hierarchy $m_t $$\gg$$ m_b$
can be traded for the problem of explaining the VEV
hierarchy $v_u $$\gg$$ v_d$.
In one of the schemes discussed by these authors, 
NJLVL interactions (\ref{3gTSM}) 
with bare parameters 
${\kappa}_{tb}(\Lambda) $$=$$ 0$ and ${\kappa}_t(\Lambda)$$=$$ 
{\kappa}_b(\Lambda)$ were investigated. 
Such interactions can originate from
exchange of heavy gauge bosons which have equal coupling
strengths to $t$ and $b$.
Since $t$ and $b$ have different hypercharges, 
the weak hypercharge boson exchange
causes the critical values ${\kappa}_t^{{\rm (crit.)}}$
and ${\kappa}_b^{{\rm (crit.)}}$, and the effective
values ${\kappa}_t^{{\rm (eff.)}}$ and ${\kappa}_b^{{\rm (eff.)}}$
in the gap equation, to differ slightly:\footnote{
The bare effective coupling ${\kappa}^{\mbox{\tiny (eff.)}}$
($\not=$$ {\kappa}$) incorporates $U(1)_Y$ gauge boson exchange 
contributions from the entire energy interval $[ E_{\mbox{\tiny ew}},
\Lambda]$, while the original bare coupling ${\kappa}$ doesn't.}
\begin{equation}
{\kappa}_t^{{\rm (eff.)}}\!(\Lambda) -
{\kappa}_b^{{\rm (eff.)}}\!(\Lambda) =
{\kappa}_b^{{\rm (crit.)}} -
{\kappa}_t^{{\rm (crit.)}} = g_1^2(\Lambda)/(16 \pi^2) \ ,
\label{isosp1}
\end{equation}
where $g_1(E)$ is the running $U(1)_Y$ gauge coupling parameter.
They showed in the quark loop approximation that 
the difference of the VEV's 
becomes consequently enhanced
\begin{equation}
v_u^2 - v_d^2 \approx 2 N_{{\rm c}} g_1^2(\Lambda) {\Lambda}^{\!2}/
(16 \pi^2)^2 \ ,
\label{isosp2}
\end{equation}
where $\sqrt{v_u^2 + v_d^2}\!=\!v $$\approx$$ 246$ GeV
and $\Lambda $$\gg$$ E_{{\rm ew}}$.
This result is expected to hold qualitatively 
also beyond the quark loop approximation. 
Relation (\ref{isosp2}) suggests that $v_u $$\gg$$ v_d$, i.e.,
$v_u/v_d $$\gg$$ 1$. Keeping in mind that
$g_t $$\approx$$ g_b$ (cf.~Subsec.~\ref{EWESG13}),
this explains why a very tiny 
$U(1)_Y$ splitting of the effective
parameters at a large scale $\Lambda$ can lead to
$m_t^{{\rm phys.}} $$\gg$$ m_b^{{\rm phys.}}$.

Large isospin breaking was investigated indirectly 
within the generalized TSM model (\ref{3gTSM}) by
King and Mannan (1991a).
No direct reference was made to the fact that this
was a composite 2HDSM(II), since the employed
DS$+$PS formalism (cf.~Sec.~\ref{MDSPS}) doesn't
use composite scalar fields. Their calculations include
quark loop and QCD contributions, but not
those of $U(1)_Y$. Therefore, their bare four-quark
parameters ${\kappa}_i({\Lambda})$ should be 
regarded as ${\kappa}_i^{{\rm (eff.)}}\!({\Lambda})$ 
with $U(1)_Y$ contributions already contained in them.
For very large\footnote{
For thus large ${\Lambda}$, (\ref{isosp2}) is not applicable.}
$\Lambda $$=$$ 10^{15}$ GeV, already a tiny $U(1)_Y$ splitting
${\kappa}_t^{{\rm (eff.)}} $$-$$ {\kappa}_b^{{\rm (eff.)}} $
$<$$ 10^ {-6}$ 
resulted in $m_t^{{\rm dyn.}} $$\gg$$ m_b^{{\rm dyn.}}$.
However, 
Miransky, Tanabashi and Yamawaki (MTY, 1989a), 
within the DS$+$PS approach,
proposed that the cross term 
$\propto$${\kappa}_{tb}(\Lambda)$ in (\ref{3gTSM}) 
be also nonzero, leading to a ``feed-down'' effect --
cf.~beginning of Sec.~\ref{MDSPS2}.
Mass $m_t^{{\rm dyn.}}$
then feeds down through this term to yield 
$m_b $$\sim$$ {\kappa}_{tb}(\Lambda) m_t $$\ll$$ m_t$. 
King and Mannan subsequently showed numerically that
in such a case there is no need to fine-tune 
${\kappa}_b^{{\rm (eff.)}}\!(\Lambda)$
close to ${\kappa}_t^{{\rm (eff.)}}\!(\Lambda)$,
and that then 
${\kappa}_{tb}^{{\rm (eff)}}\!(\Lambda)$$\sim$$ 10^{-2}$.

Andrianov, Andrianov and Yudichev (1996)
investigated a framework in which there are two composite
Higgs doublets arising from a generalized version of NJLVL
interactions which include 
higher dimensional terms with two and four derivative
insertions. For a special configuration of four-quark
constants, they obtained a dynamically generated CP violation
reflected in a complex ratio of the two 
dynamically generated VEV's. It appears that
in their scenario both Higgs doublets are responsible for the
dynamical generation of $m_t$ and $m_b$,
i.e., theirs is a composite 2HDSM of ``type III''.

Partly related to this context is the work of 
Fr\"ohlich and Lavoura (1991)
who investigated the case
when NJLVL interactions (\ref{3gTSM})
satisfy relation ${\kappa}_t {\kappa}_b $$=$$
4 {\kappa}_{tb}^2$. The resulting Lagrangian density 
can then be rewritten with {\em one\/} single
auxiliary isodoublet, thus leading at $E $$\ll$$ \Lambda$
to an effective MSM.
In contrast to the truncated TSM (\ref{TSM}),
their framework leads to a nonzero $m_b^{{\rm dyn.}}$, 
in addition to $m_t^{{\rm dyn.}}$. 
They calculated in quark
loop approximation and obtained an interesting relation
$M_H^2$$\approx$$ 4 (m_t^2$$-$$m_b^2)$, if $m_b $$\ll$$ m_t$. 
They assumed hierarchy
$m_b $$\ll$$ m_t$, while noting that $m_b$
is in principle arbitrary in this framework.
Mass $m_t$ acquired values close to those of the truncated
TSM (\ref{TSM}) when $m_b $$\ll$$ m_t$.

\subsection{Two Higgs doublets -- one elementary and one composite}
\label{EWESG2}
Clague and Ross (1991)
investigated a scenario in which multiple exchanges of an 
elementary Higgs between ${\bar t}$ and $t$
lead to ${\bar t} t$ condensation (to a composite doublet
${\Phi}_{{\rm comp.}}$) and to nonzero $m_t^{{\rm dyn.}}$.
This Higgs has a strong Yukawa coupling parameter $h_t$ to $t_R$:
$h_t^2(E)/(4 \pi) $$\stackrel{>}{\sim}$$ 1$,
for $E $$\sim$$ E_{{\rm ew}}$. In the scenario, the 
$\langle {\bar t} t \rangle_0$ VEV induces
the VEV $v $$=$$ \langle (\phi_3)_{{\rm el.}} \rangle_0$
$\approx$$ 246$ GeV of the 
elementary Higgs doublet ${\Phi}_{{\rm el.}}$.
The authors calculated in the quark loop (bubble)
approximation, thus not dealing directly with composite scalars.
Mass eigenstates of scalars were mixtures of components of 
both isodoublets ${\Phi}_{{\rm el.}}$ and
${\Phi}_{{\rm comp.}}$, with mass eigenvalues
$M_s^{(1)} $$\sim$$ m_t$, $M_s^{(2)} $$\sim$$ m_{\phi}$ ($\gg $$m_t$). 
Here, $m_{\phi}^2 $$>$$ 0$ is the original mass
parameter of ${\Phi}_{{\rm el.}}$
The authors argued that such a ${\bar t}t$ binding,
parametrized by an NJLVL interaction of the TSM form
(\ref{TSM}), results in the dependence of four-quark
form factor $G$ on the momentum transfer $q^2$:
$G(q^2)$$ =$$ H/(q^2-b m_{\phi}^2)$, where $H,$$b$$\sim$$1$
($H$$>$$0$). The effective cutoff scale is 
${\Lambda}$$\sim$$ m_{\phi}$. For the case $G($$q^2$$=$$0)$$ >$$ 0$
they obtained only slightly reduced $m_t$ in comparison with
the minimal framework in the corresponding approximation, 
i.e., too high $m_t$. When $G($$q^2$$=$$0)$$ <$$0$ (i.e., $b$$>$$0$), 
$m_t$ could be substantially lowered 
(and $m_{\phi}$ as low as $E_{{\rm ew}}$).
However, their modified gap equation implies that
the average $\langle G(q^2) \rangle$ over the relevant
momenta must be positive. Therefore, it is unclear how the
(otherwise interesting) case of
$G(q^2) $$<$$0$ for $q^2 $$\stackrel{<}{\sim}$$ m_{\phi}^2$
($\sim$$ {\Lambda}^{\!2}$) can be reconciled with
a motivation for working with effective models
- namely, that the momenta above the cutoff don't
affect appreciably the low energy predictions of the model.

The work of Clague and Ross has some similarity with
the original ``bootstrap'' idea of 
Nambu (1989).
In this early work, which stimulated much of the subsequent 
research activity on ${\bar t}t$ condensation, Nambu
proposed that BCS mechanism
\cite{BardeenCooperSchrieffer57}, 
and its low energy
effective description through Ginzburg-Landau-type
of Hamiltonian, be incorporated into relativistic
field theories. He showed that this possibility does
exist and leads to theories which can contain composite
scalars and have a structure
he called ``relativistic quasi-supersymmetry.''
He proposed two scenarios:
\begin{itemize}
\item
A hierarchical chain
of dynamical symmetry breaking (``tumbling'')
-- composite scalars
could mediate between massive fermions forces
attractive enough to cause another symmetry breaking,
and so on repeatedly down to the electroweak scales.
\item
Self-sustaining (``bootstrap'') mechanism --
composite scalars mediate forces between fermions
leading to four-fermion interactions strong
enough to explain the condensation (composite scalars
responsible for their own existence). 
The dynamics should be self-consistent in the sense
that it wouldn't be necessary to look for any new physics
(substratum) at a high energy generating the four-fermion
interactions. This would imply insensitivity
of the electroweak model to the UV cutoff $\Lambda$, and therefore
the ${\Lambda}^{\!2}$-tadpole effects for the masses $m_f$ should
cancel, and preferably even $\ln \Lambda$-terms.\footnote{
For earlier discussions of these points, without the
condensation interpretation, see Decker and Pestieau (1979, 1989),
and Veltman (1981).}
\end{itemize}

The main difference between the
picture of Nambu and that of Clague and Ross is that
the latter assumed that the scalar
responsible for binding together ${\bar t}$ and
$t$ is not the generated ${\bar t}t$ condensate itself, 
while Nambu did (``bootstrap'' assumption).

In this context, we mention that already
Tanabashi, Yamawaki and Kondo (1990)
observed
that a strong Yukawa coupling parameter 
$h_t$ ($\equiv\!g_t$) of an elementary scalar
to the top quark may generate ${\bar t} t$ condensate
and that the VEV $\langle {\bar t} t \rangle_0$ induces
a VEV $v$ for ${\Phi}_{{\rm el.}}$.

An approach somewhat different from that of Clague and
Ross was later taken by 
Del\'epine, G\'erard and Gonz\'alez Felipe (1996).
They assumed that the strong Yukawa parameter $g_t$
of ${\Phi}_{{\rm el.}}$ doublet
induces an effective TSM-type interaction
(\ref{TSM}); the four-quark parameter $G$ was taken
to be independent of the momentum
transfer.\footnote{
This appears to be contrary to what they suggest in their
Eqs.~(3) and (4).} 
This led them to an additional ${\Phi}_{{\rm comp.}}$ doublet
of the TSM-type (${\bar t} t$). Both ${\Phi}_{{\rm el.}}$
and ${\Phi}_{{\rm comp.}}$ couple to $t_R$.
They applied formalism of effective potential
$V_{{\rm eff}}$ (cf.~Sec.~\ref{NTLEE}),
constructing $V_{{\rm eff}}$ as a function of
both VEV's, in the quark loop approximation
and using simple spherical cutoff $\Lambda$.
Their formalism allows {\em both\/} scalar doublets 
to develop appreciable VEV's -- 
this is apparently a major difference from the
scenario of Clague and Ross. 
Neutral scalar mass eigenstates
are linear combinations of the neutral scalar components
of ${\Phi}_{{\rm el.}}$ and ${\Phi}_{{\rm comp.}}$.
Taking $g_t^2/(4 \pi) $$\simeq$$ 1$ and $m_t $$=$$ 174$ GeV
resulted in the lightest Higgs being almost entirely
composite, and with a mass possibly
as small as 80 GeV, if $\Lambda $$\sim$$ 1$ TeV.
Masses of EW gauge bosons could
be reproduced even for thus low ${\Lambda}$'s, because
{\em both\/} ${\Phi}_{{\rm el.}}$ and ${\Phi}_{{\rm comp.}}$
contribute to the EWSB.
Their Lagrangians for ${\Phi}_{{\rm el.}}$ and 
for ${\Phi}_{{\rm comp.}}$ have identical form. 
Formally it is not possible
to distinguish in their formalism between the two isodoublets
in the quark loop approximation of $V_{{\rm eff}}$ 
(cf.~Fig.~\ref{rmd3f} in Sec.~\ref{NTLEE}) where the composite
nature is not felt.
Going beyond this approximation (i.e., to the
next-to-leading terms -- NTL) in calculating $V_{{\rm eff}}$
would make the difference between
${\Phi}_{{\rm el.}}$ and ${\Phi}_{{\rm comp.}}$
also formally manifest, because the diagrams needed to calculate
the NTL terms include the scalar propagators which
are dynamical and nondynamical, respectively (cf.~Fig.~\ref{rmd4f}
where they are nondynamical).
Further, since in their framework
both isodoublets contribute to $m_t$,
this would lead at low
energies to an effective 2HDSM of ``type III''.
Yukawa sector of such a model has no discrete symmetry, and thus
suppression of the FCNC's is not automatically ensured
when including the lighter quark generations.
Thus, the framework of Del\'epine {\em et al.\/}, while giving
interesting results at quark loop level,
raises also several interesting questions which
deserve further investigation.

\subsection{Colored composite scalars}
\label{EWESG3}
To obtain phenomenologically acceptable prediction for
$m_t$ in models involving ${\bar t} t$ condensation,
several groups of authors have proposed NJLVL
interactions which, while respecting the SM 
gauge symmetry, lead to
(${\bar t} t$-dominated) condensation into composite scalars
with {\em color\/}. Such interactions must differ from those of
(\ref{TSM}), (\ref{genTSM}), (\ref{3gTSM}) -- the
latter lead to formation of only {\em colorless\/} 
composite scalars (for ${\kappa}_t $$>$$ {\kappa}_{{\rm crit.}}$).

Babu and Mohapatra (1991)
proposed an NJLVL model leading to
a ${\bar t} t$-dominated composite scalar sector with
one usual isodoublet $\Phi$ and two color-triplet isoscalars
$\omega_1$ and $\omega_2$.
At energies $\mu $$<$$ \Lambda$, scalars
become dynamical through quantum effects
\begin{equation}
{\cal {L}}^{(\mu)}_{{\rm Y.}}  = 
h_t(\mu) {\bar Q}_L^a {\tilde \Phi} t_R^a + 
f_1(\mu) Q_L^{aT} C^{-1} i {\tau}_2 Q_L^b {\omega}_1^d
{\epsilon}_{abd} +
f_2(\mu) t_R^{aT} C^{-1} b_R^b {\omega}_2^d {\epsilon}_{abd} +
\mbox{ h.c.} 
\label{babu3}
\end{equation}
Here, ($a,\!b,\!d$) are color indices; isospin indices and
superscripts ``$(\mu)$'' were suppressed.
The authors then studied 
RGE's for Yukawa parameters $h_t, f_1, f_2$, by imposing
compositeness conditions $h_t, f_1, f_2$$\to$$ \infty$ when
${\mu}$$\to$${\Lambda}$. Only $\Phi$ develops a VEV
($\approx$$ 246$ GeV), colored scalars don't 
because the vacuum is colorless.
Thus, $m_t^{{\rm dyn.}} $$=$$ m_t(m_t)$ is determined only by 
$h_t(m_t)$. RGE for $h_t$ is different from
RGE (\ref{RGEgt}) of the MSM -- 
$f_1$ and $f_2$ give positive contributions
to the slope $d h_t/d \ln \mu$. This results in
lower values of $m_t(m_t) $$=$$ h_t(m_t) v/\sqrt{2}$
than in the minimal framework. Acceptable values 
$m_t(m_t) $$\approx$$ 170$ GeV can be obtained,
for $\Lambda $$\sim$$ 10^{17}$$-$$10^{19}$ GeV. 
Mass $m_b$ ($\ll$$ m_t$) gets induced 
out of $m_t^{{\rm dyn.}}$ 
as a one-loop radiative effect depicted in Fig.~\ref{rmd8f}.
\begin{figure}[htb]
\mbox{}
\vskip2.4cm\relax\noindent\hskip0.cm\relax
\includegraphics{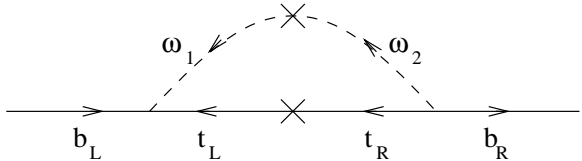} \vskip0.cm
\caption{\footnotesize Generation of $m_b$ 
out of $m_t^{{\rm dyn.}}$, via mixing of the
composite isosinglet color-triplet scalars $\omega_1$ and
$\omega_2$. The diagram is reproduction of Fig.~1 of the
work of Babu and Mohapatra (1991).}
\label{rmd8f}
\end{figure}
They calculated also the mass $M_H$ of the colorless Higgs,
by considering the scalar potential of $\Phi$, ${\omega}_1$
and ${\omega}_2$, and investigating one-loop RGE's for the
self-coupling parameters $\lambda_i$ appearing in the potential.
Compositeness conditions were $\lambda_i$$\to$$ \infty$
when ${\mu}$$\to$${\Lambda}$. The resulting $M_H$
were also lower than those of the minimal
framework -- e.g., $M_H $$\approx$$ 166$, $158$ GeV
for $\Lambda $$\sim$$ 10^{17}$, $10^{19}$ GeV, respectively. 
They subsequently included
also four-fermion interactions involving 
the third generation of leptons 
(no $\nu_R$), leading to two additional
composite color-triplet isosinglet scalars $\chi_1$ and $\chi_2$ and
to a modified RGE flow for $h_t$. The mass, $m_t(m_t)$,
was reduced: $m_t(m_t) $$\approx$$ 153$-$155$ GeV for
$\Lambda$$ =$$ 10^{17}$-$10^{19}$ GeV;
$m_t(m_t) $$\approx$$ 170$ GeV for $\Lambda $$\sim$$ 10^{11}$ GeV.

Kundu, De and Dutta-Roy (1993) 
investigated another ${\bar t} t$ condensation scenario with
colored composite scalars. The starting density was
used first by
Dai {\em et al.\/} (1992)
\begin{equation}
{\cal {L}}^{(\Lambda)}_{4q} = 
\frac{g}{{\Lambda}^{\!2}} ({\bar Q}_L^a t_R^a)({\bar t}_R^b Q_L^b) +
\frac{g^{\prime}}{{\Lambda}^{\!2}} 
({\bar Q}_L^{a} {\lambda}^{(\alpha)}_{ab} t_R^b)
({\bar t}_R^{d} {\lambda}^{(\alpha)}_{de} Q_L^e) \ ,
\label{Dai1}
\end{equation}
where $(a,b,d,e)$ are color indices, ${\lambda}^{(\alpha)}$
(${\alpha}=1,$$\ldots,$$8$) are the $SU(3)_c$ Gell-Mann matrices
satisfying $\mbox{tr}({\lambda}^{(\alpha)} {\lambda}^{(\beta)})
$$=$$ 2 {\delta}_{{\alpha} {\beta}}$, and
$Q_L $$=$$ (t_L, b_L)^T$. Isospin indices are omitted
for simplicity. The first term is the truncated TSM
(\ref{TSM}). The second term is also invariant under the
SM gauge group. If strong enough, it generates
a composite color-octet isodoublet scalar $\chi$ by
${\bar t} t$-dominated condensation, 
as pointed out by 
Kundu {\em et al.\/} (1993).

Dai {\em et al.\/} (1992) 
calculated $m_t$ in this model by employing the ladder
Dyson--Schwinger (DS) integral equation, and Bethe--Salpeter
(BS) equation for the Nambu--Goldstone boson (instead of the 
Pagels--Stokar equation) for calculating the 
decay constant $F_{\pi} $$=$$ v$. The formalism of the approach 
contains explicitly dynamical masses of quarks, 
but not composite scalars - hence
Dai {\em et al.\/} didn't refer to the latter.
They argued that the $g^{\prime}$-term in (\ref{Dai1}) 
does not contribute in the ladder approximation to the DS
equation, but they included it in the BS equation.
For large $g^{\prime}$
($g^{\prime} $$=$$ 3 g$)\footnote{
We note that $g^{\prime} $$=$$ 3 g/2$ corresponds to
the case when the origin of 
(\ref{Dai1}) is an exchange of massive ($M $$\sim$$ {\Lambda}$)
colorless vector bosons between the quarks --
cf.~identity (\ref{Fierz2}) in Sec.~\ref{RMUP2}.} 
they obtained values of $m_t$ 
lower than those in the truncated
TSM model in the DS$+$PS approach 
(cf.~Sec.~\ref{MDSPS}, Table~\ref{tabl2})
by $30$--$40$ GeV for $\Lambda $$\sim$$ 10^{11}$--$10^{19}$ GeV.
Nonetheless, for $g^{\prime} $$=$$ 3 g$, $m_t$'s were
still above 200 GeV, for
$\Lambda $$\stackrel{<}{\sim}$$ E_{{\rm Planck}}$.
Since they used the ladder DS, their results
at high $\Lambda$'s seem to correspond partly
to the RGE approach in leading-$N_{{\rm c}}$
(and with gauge couplings $g_1,\!g_2 $$\mapsto$$ 0$).
Hence, it can be expected that their $m_t$'s are
higher than those of the model in the RGE approach. 

Indeed, Kundu, De and Dutta-Roy (1993)
recalculated $m_t$ in this model by using the RGE approach. 
For $g^{\prime} $$>$$g^{\prime}_{{\rm crit.}}$, 
they argued that the composite scalar sector
contains beside the usual isodoublet $\Phi$ also
a color-octet isodoublet field $\chi$ (with zero VEV). 
At low energies, $t_R$ couples to both fields 
with Yukawa parameters $g_t$ and $g^{\prime}_t$,
respectively. RGE for $g_t(\mu)$, modified by the
presence of $g^{\prime}_t(\mu)$, gives an appreciably higher
slope $d g_t/d \ln \mu$ than in the 
MSM case (\ref{RGEgt}). They investigated the infrared
fixed point of one-loop RGE's for $g_t$ and $g^{\prime}_t$,
in conjunction with RGE's for the scalar self-coupling
parameters (with $g_1,\!g_2\! \mapsto\! 0$).
Demanding in addition that the scalar potential
have meaningful finite minima, they obtained the fixed
point solution $m_t^{{\rm (f.p.)}} $$\approx$$
145$ GeV, and masses of scalars
$M_{\chi^0} $$\approx$$ 135$ GeV, $M_{\chi^+} $$\approx$$ 74$ GeV
and $M_H $$\approx$$ 209$ GeV. In analogy with the minimal
framework, it appears that this $m_t^{{\rm (f.p.)}}$ 
value represents solution of 
RGE's to an accuracy of a few per cent
for compositeness 
scales ${\Lambda} $$\stackrel{>}{\sim}$$ 10^{10}$ GeV.
The obtained value for $m_t$
is substantially lower than those
obtained by Dai {\em et al.\/} (1992), because of the
difference of approaches used.

Emergence of colored scalars with relatively low
masses $\sim$$E_{{\rm ew}}$ may pose a problem 
when confronted with experimental evidence. Babu and
Mohapatra didn't address this question.
Kundu {\em et al.\/}
argued that the question is not alarming because
$M_{\chi}$'s are higher than masses of
light hadrons (thus not influencing light hadron phenomenology)
and that physical objects involving ${\chi}$
will be colorless.

\subsection{Other structures of composite scalars}
\label{EWESG4}
Arata and Konishi (1992)
introduced a ${\bar t} t$ condensation model containing the
isodoublet ${\tilde \Phi}_i $$\sim$$ {\bar t}_R^a Q_L^{ia}$ 
and an isotriplet ${\chi}_{ij} $$=$$ {\chi}_{ji}
$$\sim$$ ({\bar t}_R^a Q_L^{ia}) ({\bar t}_R^b Q_L^{jb})$.
They started with
\begin{equation}
{\cal {L}}^{(\Lambda)}_{4q}  =  G 
({\bar Q}_L^i t_R)({\bar t}_R Q_L^i)
+ G^{\prime} ({\bar Q}_L^i t_R)({\bar Q}_L^j t_R)
({\bar t}_R Q_L^i)({\bar t}_R Q_L^j)  ,
\label{Arata1}
\end{equation}
where: $Q_L $$=$$ (t_L,b_L)^T$; $i,\!j$ are isospin indices;
color indices (of the TSM-type) were suppressed.
The first term is the truncated TSM,
for $G {N_{{\rm c}}} {\Lambda}^{\!2}/(8 {\pi}^2) $$>$$ 
{\kappa}_{{\rm crit.}}$ it leads to
the emergence of ${\tilde \Phi}$.
The eight-quark term leads, for strong enough
$G^{\prime} {\Lambda}^8$, to the emergence of 
${\chi}$. The main motivation for introducing this
term, and hence $\chi$,
lied in the fact that values $m_t(m_t) $$\approx$$ 220$-$230$ GeV 
of the minimal framework 
(for $\Lambda $$\stackrel{>}{\sim}$$ 10^{15}$ GeV) 
were irreconcilable with the experimental
restrictions on the electroweak $\delta \rho$
parameter within the MSM ($m_t$ had not yet been
measured at that time). The extended framework allowed 
$\langle {\chi}_{11} \rangle_0 $$\not=$$ 0$, resulting in a
reduction of the effective $\delta \rho$.
If $m_t(m_t) $$\approx$$ 220$-$230$ GeV and the VEV ratio
$\eta $$\equiv$$ \langle {\chi}_{11} \rangle_0/
\langle {\tilde \Phi}_1 \rangle_0 $$\sim$$ 10^{-1}$,
acceptable values of $\delta \rho$ were obtained.
The authors then used these values of $\eta$ in 
one-loop RGE investigation
of the scalar self-couplings, concluding that
at least one neutral scalar mass eigenstate 
[a lin.~combination of
$\mbox{Re}( {\tilde \Phi}_1 )$ and $\mbox{Re}({\chi}_{11})$],
with $M_H $$<$$ 300$ GeV, couples to quarks as in the MSM.
Other mass eigenstates turned out to couple to
quarks only weakly. Evolution of $g_t(\mu)$ 
was almost the same as in the minimal framework, as was
the VEV $\langle {\tilde \Phi}_1 \rangle_0$,
resulting in nowadays unacceptable mass $m_t(m_t) $$>$$ 210$ GeV.

Achiman and Davidson (1991)
investigated, within the NJLVL framework (\ref{3gTSM}) involving
${\bar t} t$ condensation,
the possibility that the Peccei--Quinn (PQ) 
spontaneous symmetry breaking 
(Peccei and Quinn, 1977)
is dynamical. They identified the global axial $U(1)_{\gamma_5}$ 
symmetry\footnote{
Fermions transform under $U(1)_{\gamma_5}$:
$f\!\mapsto\! \exp(i {\alpha} {\gamma}_5) f$, i .e.,
$f_L\!\mapsto\! e^{-i {\alpha}} f_L$,
$f_R\!\mapsto\! e^{+i {\alpha}} f_R$.}
of the first two terms in (\ref{3gTSM})
with the color-anomalous PQ symmetry 
group $U(1)_{{\rm PQ}}$. In (\ref{3gTSM}), the third term
($\propto$${\kappa}_{tb}$) violates this symmetry
explicitly. In their framework, this term is generated
dynamically at an intermediate scale $E_{{\rm PQ}}$:
$E_{{\rm ew}} $$\sim$$ 10^2 \mbox{ GeV}$ $\ll$$ 
E_{{\rm PQ}} $$\sim$$ 10^{11} \mbox{ GeV}$ $\ll$
${\Lambda}$, where $\Lambda$ is
${\bar t} t$ condensation scale.
They proposed that the condensate responsible for this PQ
DSB be made up of third generation right-handed
neutrinos:\footnote{
${\bar \nu}^c_R$ is a shorthand notation for:
${\bar \nu}^c_R $$\equiv$$  {\overline { ( {\nu_R} )^c }}$.}
$\Phi $$\sim$$
{\tilde \nu}_R {\nu}_R $$\equiv$$ {\bar \nu}^c_R {\nu}_R$.
Since this condensate is not invariant under $U(1)_{{\gamma}_5}$
[${\Phi}\!\mapsto\! e^{i2 {\alpha}} {\Phi}$
when $f\!\mapsto\! \exp(i {\alpha} {\gamma}_5) f$],
a nonzero VEV
$\langle {\Phi} \rangle_0 $$\not=$$ 0$ breaks
PQ symmetry dynamically. The see-saw mechanism
yields heavy right-handed third generation neutrinos
(with Majorana mass $M $$\sim$$ E_{{\rm PQ}}$)
and very light left-handed third-generation neutrinos
and axions ($m_{\nu_L} $$\sim$$ m_{{\rm axion}}$).

Kaus and Meshkov (1990)
proposed a flavor-democratic NJLVL model
\begin{equation}
{\cal {L}}^{(\Lambda)}_{4f} = - G \sum_{i,j,k,l} {\bar \psi}_i
{\psi}_j {\bar \psi}_k {\psi}_l \ ,
\label{Kaus1}
\end{equation}
where the sum is over ``ur quarks'' of a given electric
charge (i.e., either up-type or down-type).
The motivation was to explain masses and mass hierarchies
of fermions in a semi-phenomenological way.
If the sum $({\psi}_1\!+\!{\psi}_2\!+\! {\psi}_3)$ in the up-type
sector is identified as the top
quark ${\Psi}_t$, then the resulting interaction
is $\propto$$| {\bar \Psi}_t {\Psi}_t |^2$. This yields
a TSM-type of model which, for large enough $G$,
leads to ${\bar t}t$ condensation and
$m_t^{{\rm dyn.}} $$\not=$$ 0$.
The other two up-type quarks remain massless. This
leads to a $3\!\times\!3$ up-type quark mass matrix
which, in a specific basis, has a completely
flavor-democratic (FD) form  (all nine elements
equal). The ${\bar t} t$ condensation represents, 
in that basis, breaking
of the original global $U(3)_L\!\times\!U(3)_R$ symmetry
(masslessness) to the $S(3)_L\!\times\!S(3)_R$
symmetry, where $S(3)$ is the group of permutations of
three elements. The masses $m_u,\!m_c $$\not=$$0$
were explained phenomenologically via explicit
breaking of the $S(3)_L\!\times\!S(3)_R$.
The down-type quark mass matrix was
constructed analogously and the hierarchy of the weak
mixing (CKM) angles was taken into account
in the procedure. 

Lindner and L\"ust (1991)
proposed an explanation of how to obtain
acceptable lower $m_t^{{\rm dyn.}}$ without enlarging 
the SM symmetry of the
NJLVL framework and without introducing an enlarged
{\em low mass\/} composite scalar sector. They argued that
the minimal framework naturally allows a larger spectrum involving
additional composite scalars and vectors, made up of ${\bar t}$,
$t$ and $b_L$, and with masses $M $$>$$ 
{\cal {O}}(E_{{\rm ew}})$. For example,
the additional scalars ${\tilde H}^{(i)}$ can be radial excitations
of the ``ground'' state $H $$\sim$$ {\bar t}t$.
They allowed the existence of
colorless composite excited states 
$\tilde H^{(i)}$, $\tilde B^{(j)}$ and
$\tilde W^{(k)}$ -- scalar isodoublets, vector isosinglets,
and vector isotriplets, respectively. Although their
RGE analysis doesn't rely on any specific choice of
the effective terms at the compositeness scale
$\Lambda$, they motivated the model by various scalar and vector
NJLVL terms. The authors then constructed RGE's for Yukawa
parameters $g_t$ and ${\tilde g}^{(i)}_t$
(${\tilde g}^{(i)}_t$ corresponds to the coupling of
${\tilde H}^{(i)}$ to $t_R$) and for the
gauge coupling parameters $g_1,\!g_2,\!g_3,\!{\tilde g}^{(j)}_B$
and ${\tilde g}^{(k)}_W$. At $\mu $$=$$ \Lambda$
they imposed compositeness conditions:
$g_t,\!{\tilde g}^{(i)}_t,\!{\tilde g}^{(j)}_B, 
\!{\tilde g}^{(k)}_W \!\to\! \infty$.
In the RGE's they took into account threshold effects
of the heavy composite particles --
by switching off these degrees of freedom (d.o.f.'s) 
of mass $M$ for energies ${\mu} $$<$$ M$.
Although $M $$\gg$$ E_{{\rm ew}}$,
the presence of these d.o.f.'s in the RGE for $g_t$
nonetheless influences evolution of $g_t$ substantially,
leading to values $m_t(m_t) $
different from those of the minimal framework.
In particular, emergence of excited
scalars, i.e., of their Yukawa parameters ${\tilde g}^{(i)}_t$,
increases the slope $d g_t^2/d \ln \mu$ and thus
leads to smaller values of $g_t(E_{{\rm ew}})$ and
$m_t(m_t)$ than in the minimal TSM framework.
The authors chose a form for the mass spectrum of the excited
particles which is equidistant on the log scale
and extends through the interval $[M_1, \Lambda)$, 
where $M_1$ ($\gg\!E_{{\rm ew}}$)
is the mass of the lowest excited state.
At the thresholds, they apparently took:
${\tilde g}_t^{(i)}(\mu) $$=$$ {\tilde g}_t^{(i+1)}(\mu)$
[$\equiv$$ {\tilde g}_t(\mu)$] at $\mu $$=$$ M_{i+1}$
($i\!\geq\!1$). 
For ${\tilde g}_t(\Lambda)/g_t(\Lambda)$
they chose values $2$, $1$, $0.5$. 
The choice ${\tilde g}_t(\Lambda)/g_t(\Lambda) $$=$$ 2$
gave the lowest values for $m_t$.
However, as they argued, this choice represents very
strongly coupled excited scalars and is not likely
within NJLVL-type of dynamical scenarios. 
They obtained the lowest
$m_t$'s when the excited composite states
were scalars only.\footnote{
When including composite vector (gauge) fields,
their framework implied also compositeness of
(the transverse components of) $Z$, $\gamma$ and $W^{\pm}$:
$g_1({\Lambda})$, $g_2({\Lambda}) $$=$$ {\infty}$.
NJLVL frameworks leading to effective MSM
with fully composite EW gauge bosons were investigated by
Terazawa, Akama and Chikashige (1976, 1977),
D.~Kahana and S.~Kahana (1991, 1995), 
and by
Akama and Hattori (1997).}
In such cases,
for the choice ${\tilde g}_t(\Lambda)/g_t(\Lambda) $$=$$ 1$,
$M_1 $$=$$ 10^2 M_W$, and for the number of excited scalars
${\tilde N}_H $$=$$ 5$,$1$,
acceptable values $m_t(m_t) $$\approx$$ 167$ GeV 
($m_t^{{\rm phys.}} $$\approx$$ 175$ GeV)
were obtained at $\Lambda $$\sim$$ 10^9$ GeV,$10^{19}$ GeV,
respectively. For much heavier
excitations $M_1 $$=$$ 10^{10} M_W$,
the value $m_t(m_t) $$\approx$$ 167$ GeV
can be reached for $\Lambda $$<$$ E_{{\rm Planck}}$
only if ${\tilde N}_H $$>$$ 5$. 
If another realistic choice 
${\tilde g}_t(\Lambda)/g_t(\Lambda)$$=$$0.5$
is made (weakly coupled excited scalar states),
$m_t$ becomes higher: $m_t(m_t) $$\stackrel{>}{\sim}$$
190$ GeV, for ${\Lambda} $$<$$ {\Lambda}_{{\rm GUT}}$ and
${\tilde N}_H $$\leq$$ 5$ and $M_1$$\geq$$ 10^2 M_W$.

\subsection{Condensation including the fourth generation}
\label{EWESG5}
The existence of a heavy fourth quark generation 
($t^{\prime}, b^{\prime}$) has been proposed 
and/or assumed by many authors, among others by: 
Marciano (1989, 1990); 
Bardeen, Hill and Lindner (1990); 
King (1990);
Barrios and Mahanta (1991); 
Chesterman, King and Ross (1991).
The purpose was to avoid the too high $m_t$
of the minimal framework. Usually, in such 
extended frameworks the fourth quark masses were considered
to be heavy and (almost) degenerate:
$m_{t^{\prime}} $$\approx$$ m_{b^{\prime}} $$>$$ m_t$. 
Approximate degeneracy was needed to keep
value of the electroweak $\delta \rho$ parameter
acceptably low. Heaviness was needed to
have the fourth generation primarily
responsible for the DEWSB (for $v $$\approx$$ 246$ GeV),
thus making $m_t$ less constrained from below
and adjustable to the experimental value 
$m_t(m_t) $$\approx$$ 170$ GeV. Some
of the results are given in Table~\ref{tabl7},
and were obtained by the RGE approach.
Implicitly, it was assumed that
the fourth generation leptons have masses
substantially lower than $m_{t^{\prime}}$.
\begin{table}
\vspace{0.3cm}
\par
\begin{center}
\begin{tabular}{l c c c c c c }
$\Lambda$ [GeV] &
$10^{19}$ & $10^{15}$ & $10^{11}$ & 
$10^7$ & $10^4$ & $2 \times 10^3$ \\
\hline \hline
$m_{t^{\prime}} $$=$$ m_{b^{\prime}}$ &
218 & 229 & 248 & 293 & 455 & $\sim$$ 1000$ \\
$m_H$ &
239 & 256 & 285 & 354 & 605 & $\sim$$ 2000$ \\
\end{tabular}
\end{center}
\vspace{-0.1cm}
\caption{\footnotesize Predicted renormalized masses (in GeV)
of the degenerate fourth generation quarks and of the
(composite) Higgs. The results are taken from 
BHL (1990)
and King (1995).}
\label{tabl7}
\end{table}

Hill, Luty and Paschos (1991)
considered a model in
which the {\em leptonic\/} sector plays a crucial role in the
fourth generation condensation. The model was partly motivated
by the LEP$+$SLC limits on the mass of the fourth generation
neutrino eigenstate: $m_{\nu_4} $$>$$ M_Z/2$. 
Further, the
$\delta \rho$ constraints imply an almost degeneracy of the
heavy fourth generation fermions:
$m_{t^{\prime}} $$\approx$$ m_{b^{\prime}}$ and
$m_{\nu_4} $$\approx$$ m_{\ell_4}$, where $\ell_4$ is the charged
fourth generation lepton.
Hence, if there are four generations, then the masses of 
the fourth generation mass eigenstates 
$t^{\prime},\!b^{\prime},\!\nu_4$ and $\ell_4$ 
are by at least a factor $\sim$$10^2$ higher than 
those of other fermions except the top quark.
The authors therefore regarded that these four
heavy fermions play the central role in their own
dynamical mass generation. 
Another motivation was to have a viable scenario
with the cutoff $\Lambda$ as low as $\sim$$1$ TeV, thus allowing
direct probes of the underlying physics by future LHC experiments.
However, the top quark was not playing a direct role
in their dynamical framework -- its mass was fixed to be
$m_t(m_t)\!\approx\!130$ GeV by choosing the value of its
Yukawa parameter accordingly in their RGE analysis. The
compositeness conditions were applied to the Yukawa
parameters of the fourth generation fermions and to
those of the Majorana sector, but not to the top quark
Yukawa parameter. Although the dynamical framework of
these authors is rich and interesting, discussing it in
more detail would take us
beyond the limited scope of this review article
(-- frameworks involving ${\bar t} t$ condensation).

\subsection{Including the leptonic sector with 
third generation only}
\label{EWESG6}
Martin (1991)
proposed a scenario in which
the third generation of quarks {\em and\/} the third
generation of leptons participate in the condensation.
The third generation neutrino sector included
also a right-handed component $\nu_R$ with a large Majorana
mass term $\propto$$ M_{\mbox{\tiny M}}$.
He assumed the hierarchy: 
$E_{{\rm ew}} $$\ll$$ M_{\mbox{\tiny M}} $$\ll$$ \Lambda$,
where ${\Lambda}$ is the compositeness scale.
The Higgs turned out to be a combination of third generation 
quark and lepton condensates:
${\tilde H}\!\sim\!k_t ({\bar t}_R^a 
Q_L^a)\!+\!k_{\nu} ({\bar \nu}_R L_L)$.
In contrast to
Hill, Luty and Paschos (1991),
Martin didn't devise any specific dynamical mechanism for 
generation of $M_{\mbox{\tiny M}}{\bar \nu}_R^c {\nu}_R$,
he just assumed its presence.
For the third generation Dirac mass 
$m_{\nu,\mbox{\tiny D}}$,
one needs an additional hierarchy
$m_{\nu,\mbox{\tiny D}} $$\sim$$ 
E_{{\rm ew}}$$\ll$$ M_{\mbox{\tiny M}}$, 
since then the see-saw mechanism 
(Gell-Mann, Ramond and Slansky, 1979; 
Yanagida, 1979)
can yield an acceptably
light eigenmass $m_{\nu}^{(2)} $$\approx$$ 
m^2_{\nu,\mbox{\tiny D}}/M_{\mbox{\tiny M}} $$\ll$$
E_{{\rm ew}}$.
The inclusion of threshold effects in the RGE's
then simplifies because 
${\nu}_R $$\approx$$ {\nu}^{(1)}$,
where ${\nu}^{(1)}$ is the heavier mass eigenstate
with mass $m_{\nu}^{(1)} $$\approx$$ M_{\mbox{\tiny M}}$.
The heavy ${\nu}_R$ contributes to the evolution of the 
Dirac Yukawa parameters $g_t(\mu)$ and 
$g^{\mbox{\tiny D}}_{\nu}(\mu)$
in the energy interval
$[M_{\mbox{\tiny M}},\Lambda]$, by
increasing there the slopes $d g_t(\mu)/d \ln \mu$
and $d g^{\mbox{\tiny D}}_{\nu}(\mu)/d \ln \mu$.
RGE compositeness conditions for ${\tilde H}$ were:
\begin{equation}
g_t(\mu) \to \infty \ , \quad 
g_{\nu}^{\mbox{\tiny D}}(\mu) \to \infty
\quad \mbox{when: } {\mu}\!\to\!{\Lambda} \ , \quad \mbox{and: }
\ \frac{ k_{\nu} }{ k_t } \approx \lim_{\mu\!\to\!{\Lambda}} 
\frac{ g_{\nu}^{\mbox{\tiny D}}(\mu) }{ g_t(\mu) } \not= 0 \ 
(\sim 1) \ .
\label{Mart6}
\end{equation}
Renormalized Dirac masses 
$m_t(m_t)$$=$$g_t(m_t) v/{\sqrt{2}}$ and
$m_{\nu,\mbox{\tiny D}}$$=$$g^{\mbox{\tiny D}}_{\nu}
(M_Z) v/{\sqrt{2}}$
were obtained from the RGE's
under inclusion of (\ref{Mart6}).
Both were $\sim$$E_{{\rm ew}}$, the hierarchy
$m_{\nu,\mbox{\tiny D}} $$\sim$$ E_{{\rm ew}} $$\ll$$ 
M_{\mbox{\tiny M}}$
thus appears automatically.
Results for various
choices of $M_{\mbox{\tiny M}}$ and $\Lambda$ satisfying 
the assumed hierarchy
$E_{{\rm ew}} $$\ll$$ M_{\mbox{\tiny M}} $$\ll$$ \Lambda$
were presented.
Some of them far exceeded the experimental
upper bound $m_{\nu}^{(2)}$$<$$24$ MeV for tau neutrinos. 
For given values of $M_{\mbox{\tiny M}}$ and high enough
values of ${\Lambda}$, it is in general possible to adjust
the finite nonzero ratio 
$g^{\mbox{\tiny D}}_{\nu}({\Lambda})/g_t({\Lambda})$
of the compositeness condition so that
$m_t(m_t)$$\approx$$ 170$ GeV 
($m_t^{{\rm phys.}} $$\approx$$ 175$ GeV).
Such results are included in Table~\ref{tabl8}. 
\begin{table}
\vspace{0.3cm}
\par
\begin{center}
\begin{tabular}{ l l c c l }
$\Lambda$ [GeV] & $M_{\mbox{\tiny M}}$ [GeV] & 
$m_t$ [GeV] & $M_H$ [GeV]& 
$m_{\nu}^{(2)}$ [MeV] \\
\hline \hline
$10^{10}$ & $10^6$ & 170 & 235 & 87 \\
$10^{10}$ & $10^8$ & 170 & 225 & 1.7 \\
$10^{15}$ & $10^9$ & 170 & 210 & $5.1 \times 10^{-2}$ \\
$10^{15}$ & $10^{11}$&170& 208 & $7.6 \times 10^{-4}$ \\
$10^{15}$ & $10^{13}$&170& 205 & $1.5 \times 10^{-5}$ \\
\end{tabular}
\end{center}
\vspace{-0.1cm}
\caption{\footnotesize Some values of compositeness scale 
$\Lambda$ and of third generation
Majorana mass $M_{\mbox{\tiny M}}$, which allow
$m_t(m_t) $$\approx$$ 170$ GeV.
Corresponding values of Higgs mass $M_H$ and of
the light third generation neutrino eigenmass $m_{\nu}^{(2)}
$$\approx$$ m^2_{\nu,\mbox{\tiny D}}/M_{\mbox{\tiny M}}$ are also given. 
The first line is excluded by experimental bounds
$m_{\nu}^{(2)} $$<$$ 24$ MeV.
The numbers were read off or
deduced from graphs of the work of Martin (1991).}
\label{tabl8}
\end{table}
The values of the Higgs masses 
were obtained by considering, in addition to the forementioned RGE's,
also the one-loop RGE for the scalar
self-coupling parameter $\lambda(\mu)$ modified by the 
presence of the heavy ${\nu}_R$.
From the Table, we see that the acceptable values
$m_t(m_t) $$\approx$$ 170$ GeV are possible here without
resorting to very
high cutoff values $\Lambda $$>$$ E_{{\rm Planck}}
$$\sim$$ 10^{19}$ GeV. In fact, values $\Lambda $$\stackrel{>}{\sim}$$
10^8$ GeV appear to be acceptable, while 
much lower values lead to an unacceptable 
$m_{\nu}^{(2)} $$>$$ 24$ MeV. When $\ln M_{\mbox{\tiny M}}$
is rather close to $\ln {\Lambda}$, the needed ratio
$g^{\mbox{\tiny D}}_{\nu}({\Lambda})/g_t({\Lambda})$
($=\!k_{\nu}/k_t$) must be taken larger to attain the low value
$m_t(m_t)\!=\!170$ GeV. Then, ${\tilde H}$ has
an appreciable admixture of the lepton condensate
${\bar \nu}_R L_L$.

Thus, the described scenario gives acceptable
values of $m_t^{{\rm dyn.}}$ and $m_{\nu}^{(2)}$, 
for compositeness scales ${\Lambda}$ between 
$10^8$ GeV and $E_{{\rm Planck}}$.
In comparison 
with the minimal framwork, at least two additional relatively
free parameters 
[$M_{\mbox{\tiny M}}$ and 
$g_{\nu}^{\mbox{\tiny D}}({\Lambda})/g_t({\Lambda})$]\footnote{
The minimal framework 
has only one free parameter (${\Lambda}$).}
appear in the framework, thus making predictions of the model
dependent on them, too. 

\section{Enlarging the symmetry or the gauge symmetry group}
\label{EGSG}
\setcounter{equation}{0}

\subsection{Top quark condensation in supersymmetry}
\label{EGSG1}
Top quark condensation within the
minimal supersymmetric (MSSM) framework was first proposed by
Bardeen, Clark and Love (1990)
The main motivation was to avoid fine-tuning of the
relevant four-quark parameter of the non-supersymmetric
framework for large ${\Lambda}$, in view of the fact that
SUSY models are free from ${\Lambda}^{\!2}$-terms in the two-point
Green functions of the Higgs superfields. This is
a consequence of cancellation between boson and corresponding
fermion loops. 
The authors used an 
$SU(3)\!\times\!SU(2)\!\times\!U(1)$-invariant 
softly broken NJLVL-type of interaction
\cite{BuchmuellerLove82,BuchmuellerEllwanger84}
at a compositeness scale ${\Lambda}$.
The extra SUSY particles were assumed to decouple at
a SUSY soft-breaking scale $\triangle_S $$<$$ {\Lambda}$.
The model becomes at scales $\mu $$<$$ {\Lambda}$
(${\mu} $$>$$ {\triangle}_S$) a minimal SUSY model (MSSM),
but with both Higgs superfield multiplets being composite, 
made up of pairs of top and bottom quark SUSY chiral multiplets. 

Couplings of these Higgs
bosons to the fermion chiral multiplets 
in the mentioned energy intervals are analogous
to the non-SUSY 2HDSM(II) discussed in 
Sec.~\ref{EWESG1}.
However, there is one crucial difference between the
discussed composite MSSM and the composite 2HDSM(II): in the
latter, both composite Higgs doublets become dynamical
at energies ${\mu} $$<$$  {\Lambda}$, 
predominantly through quark loop
quantum effects; in the composite MSSM, the composite 
$H_2 $$\equiv$$ H_u$ develops kinetic 
energy term at ${\mu} $$<$$ {\Lambda}$ 
in a similar way as in the non-SUSY case, predominantly
through quark superfield loops, 
while the composite $H_1 $$\equiv$$ H_d$
already has a canonical kinetic energy term at the NJLVL-level
(at ${\mu} $$=$$ {\Lambda}$). Comparing this with the
discussion of the composite 2HDSM(II) in Subsec.~\ref{EWESG13}, this
implies: in SUSY RGE's 
with the third generation quarks, the usual ``non-SUSY''
compositeness conditions are taken only for 
$H_2 $$\equiv$$ H_u$: 
$Z_{u}(\mu)\!\to\!0$ when ${\mu}\!\to\!{\Lambda}$,
thus $g_t({\mu})\!\to\!{\infty}$ when ${\mu}\!\to\!{\Lambda}$.
There is no such compositeness boundary condition for
$Z_{d}(\mu)$, or equivalently, for $g_b({\mu})$ and
$g_{\tau}({\mu})$, in contrast to the fully composite
non-SUSY 2HDSM(II).  

Bardeen, Clark and Love
ignored the fermion multiplet masses other
than $m_t$, and derived the gap equation at the 
quark-multiplet-loop (bubble) level
\begin{equation}
\left(a_S \equiv \right) 
\frac{G N_{{\rm c}} {\triangle}_S^2 }{ 8{\pi}^2 }
= \left[ \left( 1 + \frac{m_t^2}{{\triangle}_S^2} \right)
\ln \left( \frac{ {\Lambda}^{\!2} }{ 
{\triangle}_S^2 + m_t^2 } \right)
- \frac{ m_t^2 }{ {\triangle}_S^2 } \ln \left(
\frac{ {\Lambda}^{\!2} }{ m_t^2 } \right) \right]^{-1} \ \ 
\left( \sim 1 \right) \ .
\label{SUSYgaplead}
\end{equation}
$G$ is the coupling strength of the SUSY NJLVL
four-quark interaction, analogous to $G$ of the non-SUSY
case (\ref{TSM}). Comparing (\ref{SUSYgaplead}) with
its non-SUSY counterpart (\ref{gaplead}), we see that
the low SUSY-breaking scale ${\triangle}_S $$\sim$$
1$ TeV (${\triangle}_S $$\ll$$ {\Lambda}$) ameliorates the familiar
fine-tuning of high-${\Lambda}$ case -- there is no
${\Lambda}^{\!2}$-dependence any more. 
However, (\ref{SUSYgaplead})
implies $G $$\sim$$ {\triangle}_S^{-2}$
($\gg$$ {\Lambda}^{-2}$), which may be somewhat unnatural 
since the physics responsible for the
four-quark interaction (and hence for $G$) appears at
scales ${\mu} $$\stackrel{>}{\sim}$$ {\Lambda}$ 
($\gg$$ {\triangle}_S$).
Thus, composite SUSY framework does not explain
the hierarchy ${\triangle}_S/{\Lambda} $$\ll$$ 1$. The latter is
imposed by hand.

In analogy with the non-SUSY framework, the
MSSM RGE analysis, with the corresponding compositeness
boundary condition, represents an improvement of the
quark-multiplet-loop aproximation. In the interval 
$[{\triangle}_S, {\Lambda}]$, the one-loop MSSM RGE for
$g_t$ is
\begin{equation}
16 \pi^2 \frac{ d g_t(\mu) }{d \ln \mu } =
\left[ 6 g_t^2(\mu) - \frac{16}{3} g_3^2(\mu) - 3 g_2^2(\mu) -
\frac{13}{9} g_1^2(\mu) \right] g_t(\mu) \ ,
\label{SUSYRGEgt}
\end{equation}
and the compositeness condition is
$g_t({\Lambda}) $$=$$ {\infty}$.
Gauge coupling parameters $g_j$ satisfy the MSSM-modified
RGE's when compared with the MSM RGE's (\ref{gjs1})-(\ref{gjs2}):
$C_j$$\mapsto$$ {\tilde C}_j$; ${\tilde C}_3$$=$$3$, 
${\tilde C}_2$$=$$-1$, ${\tilde C}_1$$=$$-11$.
Evolution below ${\triangle}_S$ for the Yukawa and gauge coupling
parameters is governed either by MSM RGE's
or by RGE's of the composite 2HDSM(II),
depending on values of
mass parameters in the scalar potential.
At $\mu$$=$${\triangle}_S$, continuity of the 
quark masses $m_q(\mu)$ [and of $g_j(\mu)$'s] has to
be implemented.

Bardeen, Clark and Love (1990) included in their work 
one-loop RGE analysis, and chose the VEV ratio
$\tan\!\beta $($\equiv$$v_u/v_d $)$=$$ \infty$
($\Rightarrow v_d $$\approx$$ 0$, $v_u $$\approx$$ v $).
The lowest $m_t$ [$\equiv$$ m_t(m_t)$], 
for a given value of ${\triangle}_S$,
was obtained at the highest ${\Lambda} $$\sim$$ 
E_{{\rm Planck}} $$\sim$$ 10^{19}$ GeV.
For ${\triangle}_S $$\sim$$ 1$ TeV and 
${\Lambda} $$\sim$$ 10^{19}$ GeV,
$m_t $$\approx$$ 196$ GeV was obtained.
When ${\triangle}_S$ was increased, $m_t$
increased, too.

Subsequently, 
Bardeen {\em et al.\/} (1992)
allowed in their RGE analysis
in general a nonzero VEV 
$\langle H_1 \rangle_0$$\equiv$$v_1 $($\equiv$$ v_d$)
($\tan\!\beta $$<$${\infty}$), and included
the bottom quark chiral multiplet and its
Yukawa parameter $g_b$.
Their results for $m_t(m_t)$,
when ${\triangle}_S $$=$$ 1$ TeV, are given in Fig.~\ref{rmd9f}.
\begin{figure}[htb]
\mbox{}
\vskip7.3cm\relax\noindent\hskip2.0cm\relax
\includegraphics{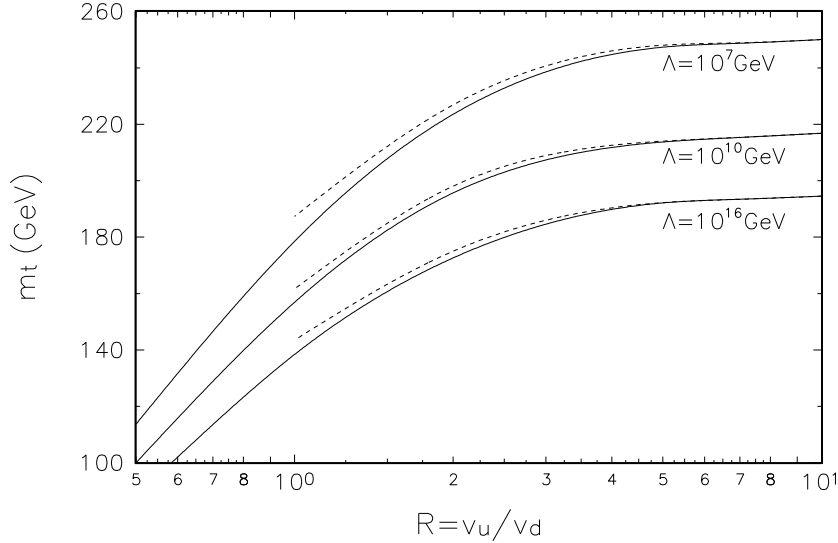} \vskip0.2cm
\caption{\footnotesize Mass $m_t(m_t)$ as a function of the
VEV ratio $R$$=$$v_u/v_d$, in the 
composite MSSM for the SUSY breaking scale
$\triangle_S$$=$$1$ TeV, and for three different values of the
compositeness scale $\Lambda$. Solid lines are for the case
when two Higgs doublets are present also at $\mu $$<$$ \triangle_S$,
and dotted lines for the case of one Higgs doublet
at ${\mu} $$<$$ {\triangle}_S$. The Figure is reproduced from
Bardeen {\em et al.\/} (1992).}
\label{rmd9f}
\end{figure}
It is seen from there that, for
$\tan\!\beta $($\equiv$$ v_u/v_d $)
$\stackrel{>}{\approx}$$ 1$, acceptable
values $m_t(m_t)$$\approx$$160$-$180$ GeV 
can be obtained if $\Lambda $$\stackrel{>}{\sim}$$
10^7$ GeV. For $\tan\!\beta $$\gg$$ 1$ 
($ \Rightarrow v_u $$\approx$$ v
$$=$$ 246$ GeV), $m_t$'s are too large.
Incidentally, it is straightforward to see
why $m_t$ decreases when $\tan\!\beta $ decreases (or when
${\Lambda}$ increases).

The main reason why the composite MSSM framework allows
acceptably low $m_t$, in contrast to the fully composite (non-SUSY)
2HDSM(II) of Subsec.~\ref{EWESG13}, 
is the freedom to change (decrease) $\tan\!\beta$ while
still keeping $m_b$$\approx$$5$ GeV.
This freedom has its origin in the fact that the SUSY
compositeness condition is represented by 
$g_t({\Lambda})$${\infty}$
and by {\em no\/} analogous requirement for $g_b$.
Therefore, $g_b(\Lambda)/g_t(\Lambda) $$=$$ 0$ 
[in fully composite 2HDSM(II), 
$g_b(\Lambda)/g_t(\Lambda) $$\gg$$ 1$]. This is so
because no compositeness condition is imposed on 
$H_1 $$\equiv$$ H_d$ at ${\mu} $$=$$ {\Lambda}$, since the
latter composite superfield already has a canonical kinetic
energy term at ${\Lambda}$ (cf.~second paragraph
of the present Sec.~\ref{EGSG1}). 
Even when $\tan\!\beta$$=$${\infty}$ ($v_d$$=$$0$)
is chosen, which is analogous to
the minimal framework, the predicted $m_t$'s
(Bardeen, Clark and Love, 1990) are lower than those of the
minimal framework. Namely, comparing
(\ref{SUSYRGEgt}) with (\ref{RGEgt}), we see that     
the extra degrees of freedom of the MSSM 
in the interval $[{\triangle}_S, {\Lambda}]$
weaken the negative contributions of QCD
to the slope $d g_t/d \ln \mu$ and enhance the positive
contributions of Yukawa interactions, thus lowering
the infrared fixed point of $g_t$.

In a later work 
\cite{BardeenCarenaPokorskiWagner94},
evolution of the $\tau$ Yukawa parameter in
the MSSM framework was included.
They showed that the $g_b$-$g_{\tau}$
Yukawa unification at $E $$\sim$$ E_{{\rm GUT}}$,
as predicted by many grand unification schemes, can be implemented
in the MSSM with acceptable $m_t(m_t)$.
They demonstrated that in such cases the value of
$m_t(m_t)$ is very near the infrared fixed-point 
(within 10 \%),
thus depending basically only on
$\tan\!\beta$$\equiv$$ v_u/v_b$. Observation that $g_b$-$g_{\tau}$
unification requirement at $E_{{\rm GUT}}$
drives $m_t(m_t)$ close to the infrared
fixed point had already been made somewhat earlier
by other authors. However, 
Bardeen, Carena {\em et al.\/} pointed out that
large $g_t$ at $E_{{\rm GUT}}$
are suggestive of the onset of nonperturbative
physics which may result
in compositeness of $H_2 $$\equiv$$ H_u$.

It was emphasized by Hill (1997)
that the top quark appears to be the only known
particle yielding a nontrivial vanishing of the
${\beta}$-function of its Yukawa coupling parameter,
i.e., that the actual top quark mass $m_t $$\sim$$ E_{{\rm ew}}$
is controlled by the RGE infrared fixed point, if the (Landau)
pole energies are very large: ${\Lambda} $$\gg$$ E_{{\rm ew}}$.
He stressed that this is the case not just in the MSSM, but also
in all those condensation scenarios which have (drastic)
fine-tuning ${\Lambda} $$\gg$$ E_{{\rm ew}}$, including the
BHL scenario of the minimal framework.

Clark, Love and ter Veldhuis (1991)
investigated\footnote{
That paper was submitted somewhat later than
the paper of Bardeen {\em et al.\/} (1992), and contains
also results showing that $m_t$ decreases to
acceptable values when $\tan\!\beta $$\equiv$$ v_u/v_d$
decreases.} 
whether the effective 10-dimensional
four-quark interactions of the Suzuki type
[cf.~Sec.~\ref{CCVAMF4}]
can change substantially predictions of the 
MSSM top condensation framework discussed above.
These authors also used an 
$SU(3)\!\times\!SU(2)\!\times\!U(1)$-invariant softly 
broken NJLVL-type of interaction
at a compositeness scale ${\Lambda}$, but added Suzuki terms.
They applied quark-multiplet-loop approximation
to this effective action in an interval 
$[{\mu}_{\ast}, {\Lambda}]$
determined by the ``perturbativity breakdown'' condition: 
$g_t^2(\mu_{\ast})/(4 \pi)$$=$$ 1$.
The ratio ${\mu}_{\ast}/{\Lambda}$ 
was dependent on ${\Lambda}$
and on Suzuki parameter $\xi$, where ${\xi}^2/{\Lambda}^4$
is the ratio of the coupling parameters of the 10- and
6-dimensional four-quark interactions.
Then they applied the usual MSSM RGE's in the
interval $[{\triangle}_S, {\mu}_{\ast}]$, taking into
account at ${\mu}_{\ast}$ the mentioned condition. 
In the low energy interval $[M_Z, {\triangle}_S]$, 
MSM RGE's were applied. 
They chose ${\triangle}_S $$\sim$$ 1$ TeV or $10$ TeV, 
${\Lambda} $$\sim$$ 10^{16}$ GeV, and
$v_u/v_d$$ = 4, 2, 1$.
They concluded that, as long as $\xi$
is not unreasonably large ($| {\xi} | $$<$$ 3$), variation
of the predicted $m_t$ changed very little
(by $5$-$6$ GeV). This result is similar to the one
in the non-SUSY framework by Suzuki (1990a), Bardeen (1990)
and Hill (1991) -- cf.~Sec.~\ref{CCVAMF4}.

Froggatt, Knowles and Moorhouse (1993)
applied in their composite MSSM framework
the compositeness condition at ${\Lambda}$ 
for $H_2 $$\equiv$$ H_u$
[$Z_u(\Lambda) $$\approx$$0$, i.e., $g_t({\Lambda})$$\gg$$1$]
{\em and\/} for $H_1 $$\equiv$$ H_d$
[$Z_d(\Lambda) $$\approx$$0$, i.e., $g_b({\Lambda})$$\gg$$1$],
and predicted $m_t $$\approx$$ 184$ GeV, when
${\Lambda} $$\sim$$ 10^{16}$ GeV.
However, motivation within the MSSM for the second
compositeness condition 
[$g_b(\Lambda)/g_t(\Lambda) $$\sim$$ 1 $$\not=$$ 0$]
in their work is unclear, in view of remarks in the 
second paragraph of this Section~\ref{EGSG1}.

Ellwanger (1991)
showed that supersymmetric nonlinear sigma models
can be the origin of the supersymmetric NJLVL-type
of interactions, where the latter lead to the DEWSB. As in the
works discussed above, the supersymmetry
was assumed to be broken explicitly, not dynamically,
at a scale ${\triangle}_S $$<$$ {\Lambda}$.
He investigated in detail a model based on the coset
space $E_8/SO(10)\!\times\!SU(3)$. The $SU(3)$ family
symmetry is broken at $E_{{\rm GUT}}$
dynamically, simultaneously with the GUT group $SO(10)$.
Masses of the
lighter quarks and leptons are then generated in this
framework through radiative corrections. 

Ellwanger's framework is an {\em effective\/} one. 
An attempt was later made by Dawid and Reznov (1996)
to construct a supersymmetric 
{\em renormalizable\/} underlying theory
leading to a supersymmetric NJLVL model and thus to ${\bar t}t$
condensation. They showed that this is improbable,
because exchange of heavy particles with mass $M_S$ ($\equiv$$
{\Lambda}$) in such a framework would not separate the scales
${\Lambda}^{\!2}$ and $G^{-1} $$\sim$$ {\triangle}_S^2$
[cf.~Eq.~(\ref{SUSYgaplead})] and would 
thus not lead to an effective MSSM at
$E $$\stackrel{<}{\sim}$$ {\Lambda}$. 
The authors proposed a framework
in which the nonrenormalizable structure is maintained up to
$E_{{\rm Planck}}$, thus arguing that in a
supersymmetric framework an underlying theory for ${\bar t} t$
condensation should be connected to supergravity.

\subsection{Dynamical left-right plus electroweak symmetry breaking}
\label{EGSG2}
Akhmedov {\em et al.\/} (1996)
studied in detail a fully dynamical scenario of 
symmetry breaking in a left-right (L-R) symmetric model.
In the model, standard fermions of the third generation and
an additional (nonstandard) fermion participate in condensation.
The model contains no elementary scalars. They started
at a certain compositeness scale $\Lambda$ ($>$$ E_{{\rm ew}}$)
with an NJLVL Lagrangian density
${\cal {L}}^{(\Lambda)}$$=$${
\cal {L}}^{(\Lambda)}_1\!+\!{\cal {L}}^{(\Lambda)}_2$ 
respecting gauge symmetry
$G $$\equiv$$ SU(3)_c\!\times\!SU(2)_L\!\times\!SU(
2)_R\!\times\!U(1)_{B-L}$. 
${\cal {L}}^{(\Lambda)}_1$ 
consisted of six NJLVL\footnote{
NJLVL terms are
four-fermion terms without derivatives.}
terms (and their h.c.) involving only the third generation
fermions $Q $$=$$ (t,b)^T$
and ${\Psi} $$=$$ (\nu_{\tau}, \tau)^T$.
Terms which may lead to colored condensates were not included.
${\cal {L}}^{(\Lambda)}_2$ consisted of two NJLVL terms
involving leptons ${\Psi}$ and an additional 
gauge singlet fermion $S_L $$\sim$$ (1,1,1,0)$. The latter
was introduced to have a correct dynamical symmetry breaking 
(DSB) pattern with a spontaneous breaking of parity. 
The four-fermion coupling parameters
$G_k $$=$$ 8 \pi^2 a_k / {\Lambda}^{\!2}$ 
($k=1, \ldots 8$; $a_k $$\sim$$ 1$) originate from an unspecified
underlying physics at ${\mu} $$>$$ {\Lambda}$.
If some of them ($G_2,\!G_4,\!G_5$ and $G_6$) are real,
${\cal {L}}^{(\Lambda)}$
is invariant under the discrete parity symmetry:
$Q_L\!\leftrightarrow\!Q_R$, ${\Psi}_L\!\leftrightarrow\!{\Psi}_R$,
$S_L\!\leftrightarrow\!(S_L)^c$ [$\equiv\!(S^c)_R$].

The authors argued that in a fully dynamical scenario without the
additional fermion $S_L$, we would have to have in general 
``strong enough'' four-lepton interactions of Majorana type. 
Namely, in order to arrive at a phenomenologically
required dynamical tumbling symmetry-breaking scenario $G \equiv$
$SU(3)_c\!\times\!SU(2)_L\!\times\!SU(2)_R\!\times\!U(1)_{B-L}$
$\stackrel{E_R}{\to} SU(3)_c\!\times\!SU(2)_L\!\times\!U(1)_Y$
$\stackrel{E_{\mbox{\tiny ew}}}{\to} 
SU(3)_c\!\times\!U(1)_{{\rm em}}$, 
these Majorana-type
NJLVL interactions would have to be strong enough
(critical or supercritical) to lead to a formation of two composite
Higgs triplets ${\triangle}_R $$\propto$$ 
{\Psi}^T_R C {\tau}_2 {\vec \tau} {\Psi}_R$ and
${\triangle}_L $$\propto$$ {\Psi}^T_L C {\tau}_2 {\vec \tau} {\Psi}_L$,
which transform under the full gauge group $G$ as $(1,1,3,2)$
and $(1,3,1,2)$, respectively. The first part of symmetry breaking
(L-R) in the tumbling then occurs dynamically at
a right-handed scale $E_R $$\sim$$ \langle {\triangle}_R \rangle_0$.
Incidentally, such a composite Higgs scenario would lead to
the desirable see-saw mechanism and the consequent natural
suppression of the lighter neutrino eigenmasses. However, the
presence of the two {\em composite\/} Higgs triplets would
not lead to a phenomenologically viable L-R symmetry-breaking
pattern ($\langle {\triangle}_R \rangle_0 $$>$$
\langle {\triangle}_L \rangle_0 $$\sim$$ E_{{\rm ew}}$).
The authors argued, in quark loop approximation, that 
this dynamical scenario would give 
$\langle {\triangle}_R \rangle_0 $$=$$
\langle {\triangle}_L \rangle_0$ and hence no parity violation.

On the other hand, ${\cal {L}}^{(\Lambda)}_1$ 
and ${\cal {L}}^{(\Lambda)}_2$ 
lead to formation of a composite bidoublet $\phi $$\sim$$
(1,2,2,0)$, two composite semileptonic
doublets ${\chi}_R $$\sim$$ (1,1,2,-1)$ and 
${\chi}_L $$\sim$$ (1,2,1,-1)$,
and a singlet scalar $\sigma$. The authors
showed that this scenario does allow a 
phenomenologically viable DSB pattern, with
$E_R $$\sim$$ \langle {\chi}_R^0 \rangle_0 $$\gg$$
E_{{\rm ew}} $$\sim$$ \langle \phi_{ij} \rangle_0
$$\sim$$ \langle {\chi}_L^0 \rangle_0$ and 
$\langle {\chi}_R^0 \rangle_0
$$\stackrel{>}{\sim}$$ \langle {\sigma} \rangle_0$.
At ${\mu}$$=$${\Lambda}$ these fields
are auxiliary, and the authors rewrote density
${\cal {L}}^{(\Lambda)}$$=$${
\cal {L}}^{(\Lambda)}_1\!+\!{\cal {L}}^{(\Lambda)}_2$ 
with them
\begin{eqnarray}
{\cal {L}}^{(\Lambda)}_{{\rm aux.}} & = &
- M_0^2 ( {\chi}^{\dagger}_L {\chi}_L 
+ {\chi}^{\dagger}_R {\chi}_R )
- M_1^2 tr( \phi^{\dagger} \phi) 
- \frac{M_2^2}{2} tr ( {\phi}^{\dagger}
{\tilde \phi} + \mbox{ h.c.}) 
- M_3^2 {\sigma}^{\dagger} {\sigma}
\\
&&- \left[ Y_1 {\bar Q}_L {\phi} Q_R 
+ Y_2 {\bar Q}_L {\tilde \phi} Q_R
+ Y_3 {\bar \Psi}_L {\phi} {\Psi}_R 
+ Y_4 {\bar \Psi}_L {\tilde \phi}
{\Psi}_R + \mbox{ h.c.} \right]
\\
&&- \left[ Y_5 \left( {\bar {\Psi}}_L {\chi}_L (S_L)^c + 
{\bar \Psi}_R {\chi}_R S_L \right)
+ Y_6 ( S^T_L C S_L) {\sigma}  + \mbox{ h.c.} \right] \ ,
\label{LRaux}
\end{eqnarray}
where ${\tilde \phi} $$\equiv$$ {\tau}_2 {\phi}^{\ast} {\tau}_2$. 
Equations of motion show that ${\phi}$ is a condensate
of third generation fermions,
${\chi}_L$ and ${\chi}_R$ condensates of leptons and $S_L$,
and ${\sigma}$ a condensate of $S_L$
\begin{eqnarray}
{\phi}_{ij} & \sim & {\alpha} ({\bar Q}_{Rj} Q_{Li}) +
{\beta}( {\tau}_2 {\bar Q}_L Q_R {\tau}_2)_{ij} +
{\gamma} ({\bar \Psi}_{Rj} {\Psi}_{Li} ) +
{\delta} ( {\tau}_2 {\bar \Psi}_L {\Psi}_R {\tau}_2)_{ij} \ ,
\label{bidoubl}
\\
{\chi}_L & \sim & S_L^T C {\Psi}_L \ , \quad
{\chi}_R \sim {\bar S}_L {\Psi}_R \ , \quad
{\sigma} \sim {\bar S}_L C {\bar S}^T_L \ ,
\label{doubl}
\end{eqnarray}
$C$ being the charge conjugation matrix.
In the mentioned DSB, the first
(L-R) breaking occurs due to the nonzero VEV
$v_R $$=$$ \langle {\chi}^0_R \rangle_0$,
and the second (EW) breaking due to the bidoublet VEV's
$\langle \phi_{11} \rangle_0 $$=$$ {\kappa}$ 
and $\langle \phi_{22} \rangle_0 $$=$$ {\kappa}^{\prime}$
($\langle \phi_{12} \rangle_0 $$=$$
\langle \phi_{21} \rangle_0 $$=$$ 0$).
Minimizing the Higgs potential, the authors obtained
zero values for the remaining VEV's:
$v_L $$\equiv$$ \langle {\chi}^0_L \rangle_0 $$=$$0$,
$\langle {\sigma} \rangle_0 $$=$$0$. 
They assumed hierarchy
$\Lambda $$\gg$$ v_R $$\gg$$ E_{{\rm ew}}$ ($\sim$$ {\kappa},$
${\kappa}^{\prime}$). 
In this phenomenologically viable tumbling DSB,
the breaking of $SU(2)_R$ and of the
discrete parity symmetry at a right-handed
scale $E_R $$\sim$$ v_R$ eventually drives the 
EWSB at a lower scale $\sim$$E_{{\rm ew}}$.
The model has nine input parameters: attractive 
(positive) $G_j$ ($j$$=$$1,$$\ldots,$$8$) 
and the compositeness scale $\Lambda$.
Quark and Dirac neutrino masses are determined by VEV's 
$\langle \phi_{ij} \rangle_0$
and by corresponding Yukawa parameters $Y_j$
\begin{equation}
m_t = Y_1 {\kappa} + Y_2 {\kappa}^{\prime} \ , \
m_b = Y_1 {\kappa}^{\prime} + Y_2 {\kappa} \ ; 
\qquad
m_{\nu,\mbox{\tiny D}} = 
Y_3 {\kappa} + Y_4 {\kappa}^{\prime} \ , \
m_{\tau} = Y_3 {\kappa}^{\prime} + Y_4 {\kappa} \ ,
\label{qlDm}
\end{equation}
where: ${\kappa}^2$$+$${\kappa}^{\prime 2}$$=$$ 
v^2/2 $$\approx$$ 174^2 \mbox{GeV}^2$.
The authors had to abandon the scenario with
composite triplets, and thus the usual Dirac-Majorana see-saw
mechanism, but their model possessed a modified see-saw
with two heavy Majorana neutrinos with eigenmasses 
$M $$\approx$$ Y_5 v_R$ and a light Majorana neutrino 
with eigenmass $m_{\nu}\!\approx\!2 Y_6 \langle \sigma
\rangle_0 (m_{\nu,\mbox{\tiny D}}/M)^2 $$-$$ 2 Y_5 v_L 
m_{\nu,\mbox{\tiny D}}/M$.
This mass vanishes when $v_L,$$\langle {\sigma} \rangle_0$ 
$\to$$0$, or $M$$\to$$ \infty$. First they investigated the described
tumbling DSB by the effective potential
approach in the approximation of including quark and
transverse gauge boson loops only (no next-to-leading scalar
``feedback'' contributions), and
subsequently by the RGE approach with compositeness
conditions $Z_{\phi}(\Lambda) $$=$$ Z_{\chi}(\Lambda)$
$=$$ Z_{\sigma}(\Lambda) $$=$$ 0$.
This translated into compositeness conditions for
Yukawa and scalar self-coupling parameters.
Values $m_t(m_t)$$\approx$$180$ GeV
can be achieved in this model for values of
the bidoublet VEV ratio 
${\kappa}^{\prime}/{\kappa}$$\approx$$1.3$-$4$ 
[see Fig.~\ref{rmd10f} for $m_t(m_t)$$=$$180$ GeV].
\begin{figure}[htb]
\mbox{}
\vskip6.3cm\relax\noindent\hskip2.0cm\relax
\includegraphics{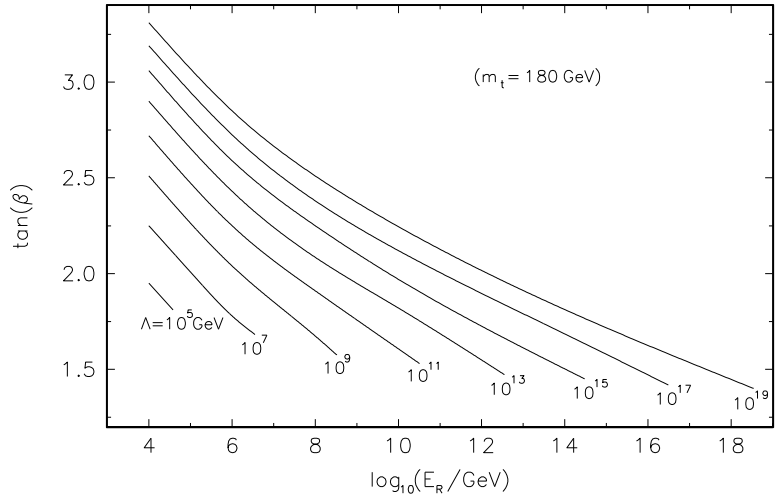} \vskip0.2cm
\caption{\footnotesize Values of 
$\tan\!\beta $$\equiv$$ {\kappa}^{\prime}/{\kappa}$ 
for $m_t(m_t)$$=$$180$ GeV and for various magnitudes
of the compositeness scale ${\Lambda}$ and the right-handed
scale $E_R$. Figure is reproduction of Fig.~9
of Akhmedov {\em et al.\/} (1996).}
\label{rmd10f}
\end{figure}
As seen from Fig.~\ref{rmd10f}, 
${\Lambda}$ can be low ($\sim$$10^2$ TeV),
and the right-handed scale $E_R $$\sim$$ v_R$
$\sim$$ 10$ TeV. For so low $\Lambda$
and $E_R$, however, the RGE approach is not reliable and gives 
only qualitative results, since infrared fixed-point 
behavior is absent. If $E_R$ and $\Lambda$ are
so low, predictions of the model, in particular
lepton flavor violation effects (e.g., $\mu$$\to$$e \gamma$)
as well as light CP-even and CP-odd neutral scalars (with masses
as low as $60$-$100$ GeV), 
could become testable in foreseeable future. 

Among the effective models of DSB in which ${\bar t} t$ 
condensate is playing a central role and which have been 
investigated so far,
the discussed model of Akhmedov {\em et al.\/} is probably
one of those with the richest phenomenological consequences.

A DSB model somewhat similar to that of the L-R
symmetric framework of Akhmedov {\em et al.\/}
was proposed by Xue (1997). It contains,
in a certain energy region between an energy
${\varepsilon}$ and the cutoff ${\Lambda}$
($E_{{\rm ew}} $$<$$ {\varepsilon} $$<$$
{\Lambda}$), three-fermion right-handed bound states
(composite fermions) $T_R $$\sim$$ ({\bar t}_R t_L) t_R$,
$B_R $$\sim$$ ({\bar b}_R b_L) b_R$. This allows
$W^{\pm}$ to have couplings 
to the right-handed composite quarks,
while at the same time formally respecting the
SM symmetry. Contributions of such
couplings to DS equations\footnote{
DS equations employed by Xue are in
quark loop approximation, with addition
of single-photon-exchange contribution; the
corresponding gluon contribution was surprisingly ignored.}
of the TSM, in the single-$W$-exchange approximation, 
lead to a coupled system for masses
${\Sigma}_t({\bar p}^2)$ and ${\Sigma}_b({\bar p}^2)$.
This system, in contrast to the usual case, does not
require the extreme fine-tuning 
$\sim$$v^2/{\Lambda}^{\!2}$ of the four-quark
parameter to its critical value, but a
less severe fine-tuning $\sim$$m_b^2/m_t^2$.

\subsection{Other enlarged symmetry groups}
\label{EGSG3}
To obtain $m_t^{{\rm dyn.}}$ lower than in the minimal framework,
Kuo, Mahanta and Park (1990)
proposed another extended symmetry group
$G$$\equiv$$ SU(3)_c\!\times\!SP(6)_L\!\times\!U(1)_Y$,
involving the flavor gauge group $SP(6)_L$.
They devised a scenario where the breaking 
$G\!\to\!G_{{\rm SM}}$
comes about in two stages: at an intermediate energy $\triangle$
($\Lambda $$\gg$$ \triangle $$\gg$$ E_{{\rm ew}}$) the $SP(6)_L$
is broken spontaneously to $SU(2)_L$
by two Higgs 14-plets $H^{(\alpha)}$ (${\alpha} $$=$$ 1,2$)
which transform under $G$ as (1,14,0). 
These $H^{(\alpha)}$'s have VEV's and masses $\sim$$\triangle$.
Details of the mechanism were left unspecified.
However, the EWSB 
$SU(2)_L\!\times\!U(1)_Y\!\to\!U(1)_{{\rm em}}$ 
was dynamical. The starting point was a $G$-invariant
NJLVL Lagrangian density
${\cal {L}}^{(\Lambda)}_{4q}\!=\!g {\bar t}_R q_L^i 
{\bar q}_L^i t_R/{\Lambda}^{\!2}\!+\!\mbox{ h.c.}$.
Here, $(q_L^i)$ quark multiplet ($i$$=$$1,$$\ldots,$$6$) 
transforms as ${\bf 6}$ under $SP(6)_L$.
The top quark $t_L$ is a linear combination of $q_L^i$'s. 
Color indices are distributed as in the TSM. 
For $g $$>$$ g_{{\rm crit.}}$, a composite Higgs
multiplet $\phi^i $$\sim$$ {\bar t}_R q_L^i$
[$\sim$$(1,6,1/2)$ under $G$] 
is dynamically generated at $\mu $$\sim$$ {\Lambda}$.
At lower $\mu $$\sim$$ \triangle$, the mentioned two
14-plets $H^{(\alpha)}$ break $SP(6)_L$ spontaneously to
$SU(2)_L$, $\phi^i$ splits into three $SU(2)_L$ doublets.
One of them (${\phi}^{(1)}$) is flavor
singlet, plays the role of the MSM Higgs doublet
at $\mu $$<$$ \triangle$, and has a low mass ($M_H $$\sim$$
E_{{\rm ew}}$). Other two doublets were
assumed heavy ($M $$\sim$$ \triangle$), thus decoupling
for $\mu $$<$$ {\triangle}$. 
After devising this scenario, the authors
investigated one-loop RGE flow for 
$g_t$ and the scalar self-interaction parameter
$\lambda$, separately in intervals $(\triangle, \Lambda)$
and $(E_{{\rm ew}}, {\triangle})$. 
In the former interval, RGE's were based on a
$G$-invariant Lagrangian and included
the $SP(6)_L$ coupling parameter $g_6$
($=$$g_2 \sqrt{3}$ at $\mu $$=$$ \triangle$ by group theory).
For $\mu $$<$$ {\triangle}$, RGE's were those of the MSM. 
They solved the RGE's, imposing the usual
compositeness conditions at $\mu $$=$$ \Lambda$. 
For $\Lambda $$\stackrel{<}{\sim}$$ 10^9$ GeV
and ${\triangle} $$\sim$$ 10^4$-$10^5$ GeV,
the resulting $m_t$ and $M_H$ were significantly
lower than those of the (BHL) minimal framework: $|\triangle m_t|
$$\stackrel{>}{\approx}$$ 13$ GeV, 
$|\triangle M_H| $$\stackrel{>}{\approx}$$ 23$ GeV. 
However, $m_t$ was still substantially above 200 GeV.
For $\Lambda $$\stackrel{>}{\sim}$$ 10^{17}$ GeV,
the results were close to those of the minimal framework,
lower by only a few GeV. Thus, $m_t(m_t) $$>$$ 210$ GeV for any
$\Lambda $$\stackrel{<}{\sim}$$ E_{{\rm Planck}}$ and
$\triangle $$\stackrel{>}{\sim}$$ 10^3$ GeV. 
This has its origin in the value 
of the infrared fixed point (IRFP) of the RGE for $g_t$
for $\mu $$>$$ \triangle$ -- it is very close to the IRFP 
value of the MSM RGE for $g_t$. 

Frampton and Yasuda (1991)
introduced an $SU(N)_X\!\times\!G_{{\rm SM}}$ model with the DEWSB,
where $N$$=$$2,$$4,$$\ldots$ is a so called sark color. 
The usual fermions are singlets under $SU(N)_X$.
Two additional colorless quarks 
$(U^a,D^a)$ ($a$$=$$1,$$\ldots,$$N$), called sarks, were introduced.
$(U^a_L, D^a_L)$ transform under $SU(N)_X\!\times\!SU(2)_L$
as (N,2), and $U^a_R$, $D^a_R$ as (N,1). In the model,
sarks play the dominant role in the DEWSB,
resulting in a degenerate dynamical mass for both sarks
and in a sark-composite Higgs isodoublet.
Mass $m_t$ is assumed to be substantially lower than
sark masses. They employed one-loop RGE analysis
similar to that of BHL, and ended up with sark masses 
in the range of 170-455 GeV and Higgs masses larger
than those of sarks by 15-40\%. 
Sark masses were lower when $\Lambda$ was higher
($\Lambda$$\stackrel{<}{\sim}$$E_{{\rm Planck}}$).
When sark color number $N$ was increased, masses
increased for high $\Lambda$'s ($\Lambda $$>$$ 10^{13}$ GeV), and
decreased for lower $\Lambda$'s. 
A major drawback of the scenario is that it doesn't
give any predictions for $m_t$ --
a drawback similar to the one in condensation scenarios
of the fourth quark generation. By introducing 
new type of quarks which play the central role
in formation of composite scalars, the
framework appears to have more resemblance to
technicolor scenarios in which condensates of new technifermions
are responsible for the DEWSB.
 
\section{Renormalizable models of underlying physics}
\label{RMUP}
\setcounter{equation}{0}

\subsection{Initial remarks}
\label{RMUP1}
Construction of renormalizable models of the underlying physics
leading to NJLVL-type of interactions believed to be
responsible for ${\bar t} t$ condensation is in general a
formidable project. It is connected with fulfilling several
requirements, among them:

(a) anomaly cancellations; 

(b) a viable spontaneous symmetry breaking (SSB) of the gauge sector
believed to be responsible for the effective four-fermion terms;
this SSB can also be dynamical;
 
(c) the effective four-fermion (NJLVL)
terms must yield the correct hierarchy of 
those fermionic dynamical masses which are predicted by
the framework (e.g., $m_t $$\gg$$ m_b$). 

One can well imagine that
constructing such models is a difficult
task. Concerning the predictability,
a drawback of such models is that they are in general
not simple, having additional unknown parameters
which can usually be adjusted so that
phenomenologically acceptable results are obtained.
But such drawbacks are characteristic of most
extensions of the minimal framework.
There is a large literature on construction of renormalizable
models involving ${\bar t} t$ condensation.
Some of the models will be described in this review
only briefly, without details.

\subsection{Gauge frameworks with additional symmetries as
factors of $U(1)$ or $SU(2)$}
\label{RMUP2}
The SM group 
$G_{{\rm SM}}$$\equiv$$SU(3)_c\!\times\!SU(2)_L\!\times\!U(1)_Y$ 
can be extended, for example, by a simple Lie group factor.
B\"onisch (1991)
investigated the
question of whether an additional
factor $U(1)$ or $SU(2)$ can lead to
NJLVL terms responsible for ${\bar t} t$ condensation.
He followed a general idea that the additional gauge
bosons acquire heavy masses $M $$\sim$$ {\Lambda}$ by some kind of 
(usually unspecified) SSB
mechanism, and that exchange of these bosons
between fermions leads at
$E $$\stackrel{<}{\sim}$$ \Lambda$ 
to effective NJLVL terms.\footnote{
King and Mannan (1990)
were apparently the first to suggest and study heavy gauge boson
exchange as a means of generating four-fermion operators; they
did it in a model-independent way (see also end of Sec. III.D).}
He considered three cases:

(1) An extra $U(1)$: 
$G$$\equiv$$SU(3)_c\!\times\!SU(2)_L\!\times\!U(1)_S\!\times\!U(1)_T$;

(2) An extra $SU(2)$ (L-R symm. models):
$G$$\equiv$$SU(3)_c\!\times\!SU(2)_L\!\times\!SU(2)_R\!\times\!U(1
)_{B-L}$;

(3) An extra gauged custodial $SU(2)_V$:
$G$$\equiv$$SU(3)_c\!\times\!SU(2)_L\!\times\!U(1)_Y\!\times\!SU(2)_V$.

He concluded that it is possible to generate in all three
cases NJLVL terms $({\bar q}_L t_R) ({\bar t}_R q_L)$
and $({\bar q}_L b_R)({\bar b}_R q_L)$ ($q$$=$$ t, b$). These terms
lead to ${\bar t} t$ and ${\bar b} b$ condensation,
respectively, if ${\kappa}_q $$>$$ {\kappa}_{{\rm crit.}}$
[$q $$=$$ t, b$; cf.~(\ref{3gTSM})].
In addition, he showed that L-R symmetric models naturally lead to
the ``feed-down'' terms 
$({\bar t}_L t_R)({\bar b}_L b_R)\!+\!\ldots$ [cf.~(\ref{3gTSM})]
which cause the
``feed-down'' generation of $m_b$ by the ${\bar t} t$ condensate.
All of these scenarios of exchange of heavy colorless gauge 
boson lead, by a Fierz rearrangement, to scalar
four-quark terms
\begin{equation}
- \frac{g_X^2}{M_X^2}
({\bar \psi}_{1L}^a {\gamma}^{\mu} {\psi}_{2L}^a)
({\bar \psi}_{3R}^b {\gamma}_{\mu} {\psi}_{4R}^b)
= + 2 \frac{g_X^2}{M_X^2} ({\bar \psi}_{1L}^a {\psi}_{4R}^b)
({\bar \psi}_{3R}^b {\psi}_{2L}^a) \ ,
\label{Boenisch1}
\end{equation}
where: $a,\!b$ are colors; ${\psi}_k$ ($k$$=$$1,$$\ldots,$$4$)
are four-component spinor fields; $M_X$ ($\sim$${\Lambda}$)
is the heavy gauge boson mass and $g_X$ its gauge
coupling parameter. He noted that
color indices are distributed here differently
than in the TSM terms (\ref{TSM}), (\ref{3gTSM}),
suggesting that only terms with $a$$=$$b$ contribute to
condensation. We will return to this point shortly.

B\"onisch in his derivation of the effective 
four-quark interactions used formulas in which 
also EW gauge bosons $W$ and $Z$ were taken to 
have nonzero masses. Thus, in such picture 
the entire symmetry breaking 
$G \to SU(3)_c\!\times\!U(1)_{{\rm em}}$
takes place before the generation of four-quark
interactions. However, at that stage, only the
breaking $G\!\to\!G_{{\rm SM}}$ should take place
(i.e., $M_W$$=$$M_Z$$=$$0$) because the four-quark terms
generated by the latter breaking are responsible for the
subsequent dynamical breaking 
$G_{{\rm SM}}\!\to\!SU(3)_c\!\times\!U(1)_{{\rm em}}$.
However, even when performing the corresponding modifications
(i.e., setting $M_W$$=$$M_Z$$=$$0$,
and changing the mixing angles correspondingly), it appears
that all the major qualitative conclusions of B\"onisch
survive, but some more quantitative conclusions change --
e.g., conclusions on when ${\bar t} t$
condensation occurs and ${\bar b}b$ doesn't, or vice-versa.

Lindner and Ross (1992)
investigated in more detail the case of gauge group
$G$$\equiv$$SU(3)_c\!\times\!SU(2)_L\!\times\!U(1)_S\!\times\!U(1)_T$.
In their framework, symmetry breaking occurs in
two stages as mentioned above. At the first stage, the
additional gauge boson 
$X_{\mu}\!\propto\!(g_S S_{\mu}$$-$$g_T T_{\mu})$ 
obtains a large mass $M_X $$=$$ g_X v^{\prime}/2
$$\sim$$ {\Lambda}$ through a Higgs mechanism,
where $v^{\prime}$ is a VEV of a remote Higgs field $F$
which can be either elementary or composite, and
$g_X$$=$$\sqrt{g_S^2+g_T^2}$. 
$B_{\mu}\!\propto\!(g_T S_{\mu}$$+$$g_S T_{\mu})$
is the usual $U(1)_Y$ field of the SM, massless at this stage. 
The authors argued, on the basis of a one-loop gap equation 
involving the propagator of the heavy $X_{\mu}$,
that the gauge coupling parameter
$g_X$ is very large: $g_X $$\gg$$ 1$, i.e.,
$g_X $$\approx$$ g_S $$\gg$$ g_T$. Therefore, the VEV 
$\langle F^0 \rangle_0 $$=$$ v^{\prime}$ 
can be much lower than $M_X $$=$$ g_X v^{\prime}/2 $$\sim$$ \Lambda$. 
This opened the possibility of nondecoupling effects when
$M_X $$\sim$$ {\Lambda}\!\to\!{\infty}$. Care was taken to ensure
cancellation of triangular anomalies.
The authors conjectured that $M_X $$\sim$$ \Lambda$
could be pushed down to $\sim$${\rm TeV}$.
However, since their scheme suggests a very strongly
coupling boson 
$X_{\mu}$ ($g_X  $$\gg$$ 1$), it is unclear whether
their one-loop gap (DS) equation has predictive
power, a point mentioned by the authors themselves.
If $\Lambda$ is low, they argued that
several composite bound states should occur, made up of
$t_L$, $t_R$ and $b_L$, among them: color-octet scalars
with possibly large decay widths
[cf. also Kundu, De and Dutta-Roy (1993), and Dai {\em et al.\/}
(1992), in Sec.~\ref{EWESG3}], as well as
vectors and axial vectors [cf.~also Lindner and
L\"ust (1991), in Sec.~\ref{EWESG4}]. They
argued that masses of the composite particles can be as low
as $\sim$$m_t$. The
model, as well as models discussed by B\"onisch
(see above), yield at energies $E $$\sim$$ \Lambda$ four-quark
terms (\ref{Boenisch1}), i.e., not
those of the TSM (\ref{TSM}), (\ref{3gTSM}).
However, we note that such interactions can be further
Fierz-transformed
\begin{equation}
- ({\bar Q}_{L}^a {\gamma}^{\mu} {Q}_{L}^a)
({\bar t}_{R}^b {\gamma}_{\mu} t_R^b)
=  2 ({\bar Q}_{L}^a t_R^b)({\bar t}_{R}^b Q_L^a)  =
\frac{2}{N_{{\rm c}}} ({\bar Q}_L^a t_R^a) ({\bar t}_R^b Q_L^b)
+ ({\bar Q}_L^{a} {\lambda}^{(\alpha)}_{ab} t_R^b)
({\bar t}_R^{c} {\lambda}^{(\alpha)}_{cd} Q_L^d) \ .
\label{Fierz2}
\end{equation}
Therefore, exchange of colorless heavy gauge boson
results in a sum of the TSM term (\ref{TSM})
and a colored sum involving Gell-Mann matrices 
${\lambda}^{(\alpha)}$ (${\alpha}$$=$$1,$$\ldots,$$8$)
employed earlier by Dai {\em et al.\/} (1992) and
Kundu, De and Dutta-Roy (1993) [cf.~Eq.~(\ref{Dai1})]
and satisfying
$\mbox{tr}( {\lambda}^{(\alpha)}{\lambda}^{(\beta)})$$=$$
2 {\delta}_{\alpha \beta}$.
The second identity in (\ref{Fierz2}) is a consequence of:
${\lambda}^{(\alpha)}_{ab} {\lambda}^{(\alpha)}_{cd}/2\!=\! 
({\delta}_{ad} {\delta}_{bc}\!-\!{\delta}_{ab} 
{\delta}_{cd}/N_{{\rm c}})$.
As mentioned in Sec.~\ref{EWESG3},
Kundu {\em et al.\/} argued that the term
involving ${\lambda}^{(\alpha)}$'s in (\ref{Fierz2}) leads to a
composite color-octet isodoublet. It is interesting that
Lindner and Ross also came to the conclusion that such composite 
states should appear in the framework
-- without employing Fierz transformations,
but rather arguments
involving a global symmetry breaking induced by the
four-quark terms [$U(6)_L\!\times\!U(3)_R\!\to\!U(3)\!\times\!U(3)$].

Since the resulting TSM term in (\ref{Fierz2}) is 
$\propto$$1/N_{{\rm c}}$, one may worry that
$1/N_{{\rm c}}$ expansion approach
of Sec.~\ref{NTLEE}, as well as related 
leading-$N_{{\rm c}}$ arguments of 
Secs.~\ref{MRGE}, \ref{MDSPS} and~\ref{CCVAMF}, 
may be in danger in such models with heavy colorless
gauge boson exchange. This is not the case, at least
as to the analysis connected to the appearance of the
colorless composites (TSM-term). Namely, the
four-quark parameter $G$ of (\ref{TSM}), 
or equivalently ${\kappa}_t$ of (\ref{3gTSM}), 
just has to be reexpressed
by the four-quark parameter $G_v $$\approx$$
- g_X^2/M_X^2$ ($M_X $$\sim$$ {\Lambda}$) of
the generated vector-vector 
$({\bar Q}_{L}^a {\gamma}^{\mu} {Q}_{L}^a)
({\bar t}_{R}^b {\gamma}_{\mu} t_R^b)$ term: 
$G $$=$$ 2|G_v|/N_{{\rm c}}$.
Then $1/N_{{\rm c}}$ analysis of the colorless composite sector
carries through unchanged. For example, leading-$N_{{\rm c}}$ 
gap equation (\ref{gaplead}) yields again condition 
$G {\Lambda}^{\!2} $$=$$ {\cal {O}}(N_{{\rm c}}^{-1})$
for ${\bar t} t$ condensation, i.e.,
$|G_v| {\Lambda}^{\!2} $$=$$ {\cal {O}}(N_{{\rm c}}^0)$.

Martin (1992a)
wrote relation (\ref{Fierz2}) in a normalized form
\begin{equation}
- ({\bar Q}_{L}^a {\gamma}^{\mu} {Q}_{L}^a)
({\bar t}_{R}^b {\gamma}_{\mu} t_R^b)
 =  2 \left[
({\bar Q}_L {\rho}^{(0)} t_R) ({\bar t}_R {\rho}^{(0)} Q_L)
+ ({\bar Q}_L {\rho}^{(\alpha)} t_R)
({\bar t}_R {\rho}^{(\alpha)} Q_L) \right] \ ,
\label{Fierz2n}
\end{equation}
where ${\rho}^{(0)}_{ab} $$=$$ {\delta}_{ab}/\sqrt{N_{{\rm c}}}$
and ${\rho}^{(\alpha)} $$=$$ {\lambda}^{(\alpha)}/\sqrt{2}$
(${\alpha}$$=$$1,\!\ldots,\!N_{{\rm c}}^2$$-$$1$)
are normalized
\begin{equation}
\mbox{tr}( {\rho}^{(\beta)} {\rho}^{(\gamma)} ) = 
{\delta}_{{\beta} {\gamma}} \ ; \qquad {\beta}, {\gamma} = 0, 1,
\ldots, N_{{\rm c}}^2-1 \ .
\label{normrho}
\end{equation}
He argued that (\ref{Fierz2n}) implies that the colorless and the
colored channels are equally attractive and could 
in principle lead not just to a colorless VEV 
$\langle {\bar t}_L t_R \rangle_0 $$\not=$$ 0$, but in general also to
colored VEV's $\langle {\bar t}_L {\lambda}^{(\alpha)} t_R
\rangle_0 $$\not=$$ 0$.
This would dynamically break $SU(3)_c$
and give gluons masses, thus contradicting experiments.
He argued that inclusion of QCD lifts this
degeneracy in favor of the first term on the right of
(\ref{Fierz2n}), because
$t_R $$\sim$$ {\bf N_{{\rm c}}}$ and ${\bar Q}_L $$\sim$$ 
{\overline {\bf N}}_{{\rm \bf c}}$ under
$SU(N_{{\rm c}})_c$ ($N_{{\rm c}} $$=$$ 3$) and because representations
${\bf 3}$ and ${\bar {\bf 3}}$ feel an attractive QCD force to
combine into a color-singlet (${\bar Q}_L t_R$) and 
a repulsive QCD force to combine into a color-octet 
(${\bar Q}_L {\lambda}^{(\alpha)} t_R$).
The latter arguments are heuristic and might
suggest, in addition to
$\langle {\bar Q}_L {\lambda}^{(\alpha)} t_R \rangle_0 $$=$$ 0$,
that ${\bar Q}_L {\lambda}^{(\alpha)} t_R$
doesn't become dynamical. 
On the other hand, Kundu, De and Dutta-Roy (1993),
as well as Lindner and Ross (1992),
argued that composite dynamical color-octet scalars
may exist, but must have zero VEV (cf.~Sec.~\ref{EWESG3} and previous
part of the present Sec.~\ref{RMUP2}). 

A variation of the described model of Lindner and Ross (1992)
was investigated by Martin (1992a)
who chose the full gauge group
$G$$\equiv$$SU(3)_c \!\times\!SU(2)_L\!\times\!U(1)_Y\!\times\!U(1)_X$,
i.e., the additional strong $U(1)_X$ in $G$ was 
orthogonal to the $U(1)_Y$ group of the SM. In
particular, he independently arrived at
conditions for triangular anomaly cancellations,
and at conditions for the colorless 
${\bar t} t$ condensation channel to be the most attractive one.
He noted that a possible mixing between
the $U(1)$'s, i.e., the case of Lindner and
Ross, does not affect anomaly cancellation or
the qualitative features of condensation.
He stressed that one
feature of the model is unfavorable for condensation:
if $U(1)_X$ is a separate (not embedded) gauge group,
perturbative methods predict that $g_X(\mu)$
decreases when energy $\mu$ decreases.

Ho\v sek (1999), working also in a framework similar
to that of Lindner and Ross, argued that spin-1
${\bar f}f$ condensates could represent dynamically
generated non-Abelian gauge bosons, and that
their appearance would solve a host of problems --
they could ``eat'' the large number of unwanted
composite Nambu--Goldstone bosons (NGB's), thus become heavy
($M $$>$$ 1$ TeV), and their presence 
in radiative corrections would modify
the otherwise for condensation unfavorable evolution of $U(1)_X$
coupling parameter to the favorable asymptotically free
evolution. He argued that various $U(1)_X$
hypercharges assigned to various fermions could
account for the hierarchies of (dynamical) fermionic
masses.

All in all, renormalizable frameworks discussed here so far
can lead in general to a rather involved low energy composite
sector, including not just the low mass scalar isodoublet
of the usual TSM effective model, but
also, among others, low mass colored scalars. This may pose some problem
in view of the fact that no colored particles of any kind have been
observed. This problem was briefly addressed by Kundu, De
and Dutta-Roy (1993) (cf.~Sec.~\ref{EWESG3}). 
Lindner and Ross briefly mentioned the possibility
that these color-octet particles might have very broad decay widths
which would make them effectively invisible. 

\subsection{Coloron (topcolor) model}
\label{RMUP3}
Hill (1991)
constructed a renormalizable framework in which, in contrast to 
the models discussed in Sec.~\ref{RMUP2}, the effective 
TSM terms (\ref{TSM}) are clearly the dominant ones
among the resulting scalar four-quark terms. 
It is motivated by the following identity 
\begin{eqnarray}
\lefteqn{
- ({\bar Q}_L^a {\gamma}^{\mu} \frac{1}{2} {\lambda}^{(\alpha)}_{ab}
Q_L^b) ( {\bar t}_R^d {\gamma}_{\mu}
\frac{1}{2} {\lambda}^{(\alpha)}_{de} t_R^e)  = 
({\bar Q}_L^a t_R^a) ({\bar t}_R^b Q_L^b) - \frac{1}{N_{{\rm c}}}
({\bar Q}_L^a t_R^b) ({\bar t}_R^b Q_L^a) }
\label{FierzH1}
\\
& = & 
 \left( N_{{\rm c}} 
- \frac{1}{N_{{\rm c}}} \right) 
( {\bar Q}_L {\rho}^{(0)} t_R)
( {\bar t}_R {\rho}^{(0)} Q_L) - \frac{1}{N_{{\rm c}}} 
({\bar Q}_L {\rho}^{(\alpha)} t_R) ({\bar t}_R {\rho}^{(\alpha)} Q_L) 
\ ,
\label{FierzH2}
\end{eqnarray}
where: $Q_L $$=$$ (t_L, b_L)^T$;
${\lambda}^{(\alpha)}$ (${\alpha}$$=$$1,$$\ldots,$$8$) are
Gell-Mann matrices; $a,\!b,\!d,\!e$ are color indices;
isospin indices were suppressed. Identity (\ref{FierzH1}) follows from 
(\ref{Boenisch1}) and from
${\lambda}^{(\alpha)}_{ab} {\lambda}^{(\alpha)}_{de}/2\!=\! 
({\delta}_{ae} {\delta}_{bd}$$-$$ {\delta}_{ab} 
{\delta}_{de}/N_{{\rm c}})$.
From (\ref{Fierz2}) follows then identity (\ref{FierzH2}),
written in the normalized form with notations as in
(\ref{Fierz2n}) and (\ref{normrho}).
We see from (\ref{FierzH2}) that the effective term 
$[- ({\bar Q}_L {\gamma}^{\mu} {\lambda}^{(\alpha)} Q_L) 
( {\bar t}_R {\gamma}_{\mu}{\lambda}^{(\alpha)} t_R)]$,
which can be induced by exchange of $SU(3)$
massive gauge bosons, really leads to a TSM-dominated
NJLVL model, thus resulting in a low mass
composite scalar isodoublet. Low mass
colored composite scalars in general
wouldn't appear, because terms
with ${\lambda}^{(\alpha)}$'s in (\ref{FierzH2})
have a negative factor (are repulsive).

Hill (1991) therefore extended $G_{{\rm SM}}$
by a factor $SU(3)$, i.e., the gauge group
at scales $\mu $$>$$ \Lambda$ was: $G $$\equiv$$
U(1)_Y\!\times\!SU(2)_L\!\times\!SU(3)_1\!\times\!SU(3)_2$. 
In the first version of such {\em topcolor\/} models (1991),
he required that $t$ behave in a
manifestly different way under $SU(3)_1\!\times\!SU(3)_2$ than
other fermions. Motivation behind this requirement
was the hierarchy $m_t $$\gg$$ m_b$. Therefore, he required that
$t_R$, $t_L$ and $b_L$ 
(relevant d.o.f.'s in ${\bar t} t$ condensation)
couple to the strong $SU(3)_2$, and the other
quarks to the weaker $SU(3)_1$. Consequently, 
he assigned to fermions the following
representations under $SU(3)_1\!\times\!SU(3)_2$: 
all right-handed quarks
except $t_R$ are $(3,1)$ while $t_R$ is $(1,3)$;
$(u,d)_L$ and $(c,s)_L$
are $(3,1)$ while $(t,b)_L$ is $(1,3)$;
all leptons are singlets [$\sim$$ (1,1)$].
He extended the quark sector in the simplest
way to ensure triangular anomaly cancellation
-- with an electroweak-singlet quark $q$ of
electric charge $-1/3$ and transforming
under $SU(3)_1\!\times\!SU(3)_2$ as: $q_R $$\sim$$ (1,3)$,
$q_L $$\sim$$ (3,1)$.

In order to provide a framework for the first stage of the
symmetry breaking [$G\!\to\!G_{{\rm SM}}$,
i.e., $SU(3)_1\!\times\!SU(3)_2\!\to\!SU(3)_c$], 
Hill introduced a scalar isosinglet Higgs field
${\Phi}^a_b$ which transforms as $(3, {\bar 3})$ under
$SU(3)_1\!\times\!SU(3)_2$. The nature of 
${\Phi}^a_b$ was left unspecified -- it
can be either elementary or composite. 
$SU(3)_1\!\times\!SU(3)_2\!\to\!SU(3)_c$ 
was then assumed to
occur at an energy $\mu $$\sim$$ \Lambda$ spontaneously through the
Higgs mechanism characterized by a development of a VEV
$\langle {\Phi}^a_b \rangle_0 = \mbox{diag}(M,M,M)$
($M $$\sim$$ \Lambda$). This VEV matrix, while breaking
$SU(3)_1\!\times\!SU(3)_2$ to the 
QCD gauge group $SU(3)_c$ with massless gluons,
gives large degenerate masses to eight other gauge bosons
(colorons) $B^{(\alpha)}_{\mu}$
\begin{equation}
m_B (= \Lambda) = M \sqrt{h_1^2 + h_2^2} = \frac{g_3}{\sin \theta
\cos \theta} M \ ,
\label{masscol}
\end{equation}
where $h_1$ and $h_2$ are gauge coupling parameters
of $SU(3)_1$ and $SU(3)_2$, respectively, and
$\theta$ is mixing angle of the
orthonormal transformation between the original massless
gauge bosons $A^{(\alpha)}_{1\mu}$ and $A^{(\alpha)}_{2\mu}$
of $SU(3)_1$ and $SU(3)_2$, on the one hand, and
massless gluons $A^{(\alpha)}_{\mu}$ and massive colorons
$B^{(\alpha)}_{\mu}$, on the other hand 
($\tan \theta $$=$$ h_1/h_2 $$\ll$$ 1$,
$h_1 \cos \theta $$=$$ g_3$, $h_2 \sin \theta $$=$$ g_3$).
QCD coupling parameter $g_3$ in (\ref{masscol}) is
$g_3(\mu)$ at ${\mu} $$\approx$$ {\Lambda}$.
Hierarchy $h_2 $$\gg$$ h_1$ ($\theta  $$\ll$$ 1$) is a reflection
of the mentioned fact that $SU(3)_2$ must be stronger
than $SU(3)_1$ so that $t$ will be favored over
other quarks in condensation.
The resulting interaction of massive colorons with
fermions at an energy $\mu$ is then
\begin{eqnarray}
{\cal {L}}^{(\mu)}(\mbox{colorons-quarks}) & = &
\frac{1}{2} g_3(\mu) \cot \theta B^{(\alpha)}_{\mu} {\Big [}
{\bar t}_R {\gamma}^{\mu} {\lambda}^{(\alpha)} t_R +
{\bar t}_L {\gamma}^{\mu} {\lambda}^{(\alpha)} t_L +
\nonumber\\
&& + {\bar b}_L {\gamma}^{\mu} {\lambda}^{(\alpha)} b_L
+ {\bar q}_R {\gamma}^{\mu} {\lambda}^{(\alpha)} q_R {\Big ]}
+ {\cal {O}}( g_3 \theta) \ ,
\label{colquar}
\end{eqnarray}
where the negligible terms $\sim$$g_3 \theta$ contain
interactions with quarks other than $t$.
At low energies $\mu $$\stackrel{<}{\sim}$$ {\Lambda} $$=$$ M_B$, 
(\ref{colquar}) leads to an effective four-quark term
\[
- \frac{g^2_3(\Lambda) \cot^2 \theta}{M_B^2}
\left( {\bar Q}_L {\gamma}^{\mu} \frac{1}{2} {\lambda}^{(\alpha)} Q_L
\right) 
\left( {\bar t}_R {\gamma}_{\mu} \frac{1}{2} {\lambda}^{(\alpha)} t_R
\right) \ , \quad \mbox{where: } \ Q_L = (t_L, b_L)^T \ . 
\]
This term, in view of (\ref{FierzH1})-(\ref{FierzH2}),
leads to the usual TSM interaction (\ref{TSM})
with $G = [g^2_3(\Lambda) \cot^2 \theta/ {\Lambda}^{\!2}]$
$[1 + {\cal {O}}(N_{{\rm c}}^{-2})]$, 
if the compositeness scale is identified naturally as 
$\Lambda $$=$$ M_B$ ($\approx$$ g_3 M/\theta$). 
Hill also verified that the additional quark
$q$ in general decouples from the low energy physics -- it obtains
a large mass $m_q $$\gg$$ m_t$ through Yukawa coupling to the
$(3, {\bar 3})$ Higgs ${\Phi}^a_b$ and therefore does not form
condensates ${\bar t}_L q_R$, ${\bar b}_L q_R$.

Martin (1992a, 1992b)
raised a point of criticism with the
original coloron model proposed by Hill (1991) and described
above. Martin noted
that in a basis of left-handed two-component Weyl fields,
the additional quark $q_L$ and 
$(b^c)_L$ transform\footnote{
The superscript $c$ denotes here that the field is charge-conjugated;
therefore: $(b^c)_L $$=$$ (b_R)^c$.} 
under the full group 
$G$$\equiv$$U(1)_Y\!\times\!SU(2)_L\!\times\!SU(3)_1\!\times\!SU(3)_2$ 
as: $q_L $$\sim$$ (-2/3, 1, 3, 1)$, 
$(b^c)_L  $$\sim$$ (+2/3, 1, {\bar 3}, 1)$.
Therefore, $q_L$ and $(b^c)_L$ form a real representation of
$G$ even before the SSB $G\!\to\!G_{{\rm SM}}$. This would allow
them in general to pair up and form large bare mass terms
(in general with mass $>$$ {\Lambda}$) and thus decouple.
In the low energy region, $q_R $$\sim$$ (-2/3, 1, 1, 3)$ 
would then play
the role of $b_R$. However, since it feels the strong $SU(3)_2$
$h_2 $$\gg$$ h_1$), just like $Q_L $$=$$( t_L, b_L)^T$ and $t_R$ do,
condensate ${\bar Q}_L q_R$ will form, alongside with
condensate ${\bar Q}_L t_R$, with equal strength.
This would give $m_t^{{\rm dyn.}} $$\approx$$ m_b^{{\rm dyn.}}$, 
rendering the scenario unacceptable.
However, Martin (1992a, 1992b) did mention that the
unfavorable decoupling of $q_L$ and $(b^c)_L$ can be avoided,
and hence the first coloron model of Hill does work, if one
assumes that the mentioned large bare mass terms are prevented
from forming, e.g., by an (unspecified) global symmetry.

Furthermore, Martin (1992a) constructed a renormalizable
model which is based on the same group $G$ as Hill's 
original coloron model 
and contains the same (elementary) $(3, {\bar 3})$
Higgs ${\Phi}^a_b$, but does not contain the 
mentioned problem of possible 
$m_b^{{\rm dyn.}} $$\approx$$ m_t^{{\rm dyn.}}$. 
To achieve this and to avoid the potential
danger of breaking $U(1)_{{\rm em}}$,
he introduced a more involved
additional spectrum of fermions, two of them
isodoublets and two of them isotriplets. Each pair
forms a real representation of $G_{{\rm SM}}$
and forms together mass terms with masses 
$\sim$$\langle {\Phi}^a_a \rangle_0 $$=$$ M$,
by coupling to ${\Phi}^a_b$. These fermions transform
under $SU(3)_1\!\times\!SU(3)_2$ either as $(3,1)$,
$({\bar 3},1)$, $(1,3)$ or $(1,{\bar 3})$. The model
contains no triangular anomalies. 
Physical $t$ (i.e., with $m_t^{{\rm dyn.}} $$\not=$$ 0$)
is a linear combination of the original
$t$ from unbroken theory and of extra fermions. Admixtures
of these extra fermions are weak when Yukawa parameters
between them and ${\Phi}^a_b$ are large ($\sim$$1$). 
Since $b_R$ doesn't participate in condensation
in this scenario, $m_b$$=$$0$. 

Later, 
Martin (1992b)
constructed, on the basis of the mentioned coloron group $G$,
a fully dynamical scenario of symmetry breaking
$G\!\to\!G_{{\rm SM}}\!\to\!U(1)_{{\rm em}}$
-- the $(3, {\bar 3})$ Higgs field ${\Phi}^a_b$ responsible
for the first part of the symmetry breaking was also composite.
He achieved this by introducing, in addition to the
SM fermions, also additional isosinglet
quarks which are $(3,3)$, $(1,{\bar 6})$ or $({\bar 6},1)$
under $SU(3)_1\!\times\!SU(3)_2$. 
In this tumbling scenario of condensation,
$(3,3)$ and $(1,{\bar 6})$ fermions
participated in forming the composite ${\Phi}^a_b$
(note: $h_2 $$\gg$$ h_1$). The other additional fermions
$({\bar 6},1)$ were present to
ensure triangular anomaly cancellations.
The scenario yielded four axion-like neutral
pseudo--Nambu--Goldstone bosons which could be
dangerously light.
Numerical calculations were not performed.
The author employed sophisticated group theoretical arguments
to explain qualitatively the tumbling scheme of DSB.
Among the arguments employed were the dynamical 
assumptions by~\citeasnoun{RabyDimopoulosSusskind80}. 
These assumptions, 
based on the approximation
of a single gauge boson exchange, lead to an algorithm
determining the ``most attractive scalar channel''
in which the condensate is supposed to appear.

It should be noted that a modification of the original
topcolor model (Hill, 1991) was later introduced by 
Hill and Parke (1994).
They performed an analysis of effects of DSB scenarios
on the production of top quarks.
The modified model was later used also by
Hill and Zhang (1995)
in analyzing $Z \to {\bar b} b$ (cf.~Subsec.~\ref{SPTAQ12}).
The modified assignment of quantum numbers to
fermions in this topcolor variant is much along the lines of the
``topcolor assisted technicolor'' framework 
(topcolor I) discussed later in Sec.~\ref{RMUP6} --
i.e., $b$ transforms now under $SU(3)_1\!\times\!SU(3)_2$ 
the same way as $t$ does:
$(t,b)_L $$\sim$$ (1,3)$; $t_R$, $b_R$$ \sim$$ (1,3)$.
Other fermions transform the same way as in the original model. 
No EW-singlet quark $q$ is introduced,
because the assignment is already anomaly-free.
As a result, the mentioned point of criticism 
by Martin (1992a), (1992b)
about the dangerous decoupling of $(b^c)_L$ and $q_L$
does not apply here.
In this model, the different $U(1)_Y$ coupling strengths
to $t_R$ and $b_R$ modify the coloron-exchange-induced
four-quark parameters so that the latter lead to
$m_t^{{\rm dyn.}} $$\not=$$ 0$ and $m_b^{{\rm dyn.}} $$=$$ 0$
(${\kappa}_t^{{\rm (eff.)}} $$>$$ {\kappa}_{{\rm crit.}}$
and ${\kappa}_b^{{\rm (eff.)}} < {\kappa}_{{\rm crit.}}$).
Therefore,
despite the fact that the coloron exchange alone
would imply in this model 
$m_t^{{\rm dyn.}} $$\approx$$ m_b^{{\rm dyn.}}$,
the weak $U(1)_Y$ interaction can cause the hierarchy
$m_t $$\gg$$ m_b$ -- 
cf.~a related discussion in Subsec.~\ref{EWESG14}.

\subsection{A renormalizable model with fine-tuning of
gauge couplings}
\label{RMUP4}
Luty (1993) 
constructed another renormalizable gauge theory leading,
via DSB at $\mu $$\sim$$ \Lambda$,
to an NJLVL model with $F$ fermion flavors and $N$ colors
\begin{equation}
{\cal {L}}^{(\Lambda)}_{4f} =  \frac{8 \pi^2}{N} 
\frac{g}{{\Lambda}^{\!2}} 
({\bar \psi}_L^{ia} {\psi}_R^{ja})
({\bar \psi}_R^{jb} {\psi}_L^{ib}) \ ,
\label{FNNJLVL}
\end{equation}
where $i,$$j$$=$$1,$$\ldots,$$F$ are flavor 
and $a,$$b$$=$$1,$$\ldots,$$N$
color indices. If $g $$>$$ g_{{\rm crit.}} $$\sim$$ 1$,
this leads to fermion condensation and
an effective $U(F)\!\times\!U(F)$ linear ${\sigma}$ model
(at $\mu $$<$$ \Lambda$) with a composite scalar ${\Phi}_{ij}$
($i,j =1, \ldots, F$). Similarly as in Hill's coloron model,
it was assumed that the above effective terms
are generated by an exchange of non-Abelian, in general
colored, gauge bosons. Luty considered the gauge group
$G \equiv $ 
$SU(N)_{UC}\!\times\! 
\left[ SU(K)\!\times\!SU(K) \right]_I\!\times\!SU(N)_{C^{\prime}}$
(made up of asymptotically free factors),
and the following fermion content including additional fermions
${\chi}_L,$${\chi}_R,$${\xi}_L,$${\xi}_R$:
${\psi}_L,$${\psi}_R$$\sim$$(N,1,1,1)(F)$;
${\chi}_L$$\sim$$  (N,K,1,1)$;
${\chi}_R$$\sim$$  (1,K,1,N)$;
${\xi}_R$$\sim$$ (N,1,K,1)$;
${\xi}_L$$\sim$$ (1,1,K,N)$.
$K$ is here an (unspecified) integer number. 
The first three group factors were considered strong, leading via
DSB to fermion condensates, while $SU(N)_{C^{\prime}}$ was
considered weak in the energy range of interest.
The reason for introduction of $SU(N)_{C^{\prime}}$ 
apparently was to ensure triangular anomaly cancellations and
to evade emergence of (too) light pseudo--Nambu--Goldstone
bosons (pseudo-NGB's).
Let ${\Lambda}_{UC}$ and ${\Lambda}_I$ be the energy scales at which
the respective gauge couplings of $SU(N)_{UC}$ and
$[SU(K)\!\times\!SU(K)]_I$ become strong enough to
trigger DSB (couplings of the two $SU(K)$'s were equally strong).
He then considered two different limiting cases:
\begin{enumerate}
\item ${\Lambda}_I $$\gg$$ {\Lambda}_{UC}$: 
In this case, due to asymptotic freedom,
$SU(N)_{UC}$ is relatively weak at $\mu $$\sim$$ {\Lambda}_I$, and
$[SU(K)\!\times\!SU(K)]_I$ interactions are strong there.
They trigger DSB and
$\langle {\bar \chi}_L {\chi}_R \rangle_0,$ 
$\langle {\bar \xi}_L {\xi}_R \rangle_0$$ \sim$$ {\Lambda}^3_I$ appear. 
This leads to the DSB 
$SU(N)_{UC}\!\times\!SU(N)_{C^{\prime}}\!\to\!SU(N)_C$. 
Of the total of $2 (N^2-1)$ NGB's,
half are absorbed by gauge bosons of the
broken sector and give them heavy masses 
$\sim$$g_{UC} {\Lambda}_I$, the other half transform in the
adjoint representation of the unbroken weak
$SU(N)_{C}$ and acquire, via loops
containing $SU(N)_C$ gauge bosons, 
lighter masses $\sim$$g_{C} {\Lambda}_I$ 
(note: $g_C $$\approx$$ g_{C^{\prime}}$). Effective theory
at $\mu $$\ll$$ g_C {\Lambda}_I$ contains NJLVL
terms suppressed by ${\Lambda}^{\!2}_I$,
and fermions $\psi$ in them remain massless.
\item ${\Lambda}_{UC} $$\gg$$ {\Lambda}_I$: 
In this case, $SU(N)_{UC}$
is strong at $\mu $$\sim$$ {\Lambda}_{UC}$, it triggers
DSB resulting in 
$\langle {\bar \psi}_L {\psi}_R$$ +$$ {\bar \chi}_L {\xi}_R
\rangle_0 $$\sim$$ {\Lambda}^3_{UC}$ and in the DSB 
$[ SU(K)\!\times\!SU(K)]_I\!\to\!SU(K)_{I^{\prime}}$.
There are $(F+K)^2$$-$$1$ (potential) NGB's.
Of them, $K^2$$-$$1$ are absorbed by gauge bosons of the
broken sector and give them masses $\sim$$g_I {\Lambda}_{UC}$,
$2 F K$ get masses $\sim$$g_{I^{\prime}} {\Lambda}_{UC}$ from
$SU(K)_{I^{\prime}}$ gauge boson loops
(note: $g_{I^{\prime}} $$\sim$$ g_I$), but $F^2$
remain massless. Fermions ${\chi}_R$ and ${\xi}_L$ 
remain massless, transforming in the fundamental
representation of the unbroken $SU(K)_{I^{\prime}}$. The latter
group becomes strong at a lower scale ${\Lambda}_{I^{\prime}}$ and
triggers $\langle {\bar \xi}_L {\chi}_R \rangle_0
$$\sim$$ {\Lambda}^3_{I^{\prime}}$. As a result, $N^2$$-$$1$ NGB's
appear and get masses $\sim$$ g_{C^{\prime}} {\Lambda}_{I^{\prime}}$
from $SU(N)_{C^{\prime}}$ gauge boson loops. 
\end{enumerate}
Let us summarize:
in the first limiting case 
(${\Lambda}_I $$\gg$$ {\Lambda}_{UC}$), the order parameter
$\langle {\bar \psi}_L {\psi}_R \rangle_0$ remains zero
and $\psi$'s are massless, while in the second case 
(${\Lambda}_{UC} $$\gg$$ {\Lambda}_I$) 
$\langle {\bar \psi}_L {\psi}_R \rangle_0$ 
is in general very large 
$\sim$${\Lambda}^3_{UC}$ and $\psi$'s acquire large\footnote{
In general $\langle {\bar \psi} {\psi} \rangle_0 $$\sim$$
m_{\psi}^{{\rm dyn.}} {\Lambda}^{\!2}$, 
where ${\Lambda}$ is the relevant compositeness
scale (in this case ${\Lambda}_{UC}$).
In the general, non-fine-tuning,
case we have $m_{\psi}^{{\rm dyn.}} $$\sim$$ {\Lambda}$.} 
$m_{\psi}^{{\rm dyn.}} $$\sim$$ {\Lambda}_{UC}$. 
Luty then assumed that transition
between the two limiting cases is continuous (second order),
i.e., VEV's of condensates and $m^{{\rm dyn.}}$
change continuously when ${\Lambda}_{UC}$ and
${\Lambda}_I$, or equivalently $g_{UC}(\Lambda)$ and $g_I(\Lambda)$,
are adjusted (chosen) accordingly. In an intermediate,
``fine-tuned'' case, when 
${\Lambda}_{UC} $$\sim$$ {\Lambda}_I $$\sim$$ {\Lambda}$,
fermions $\psi$ have then $m_{\psi}^{{\rm dyn.}} $$\ll$$ {\Lambda}$
and VEV's of their condensates are
$\langle {\bar \psi}_L {\psi}_R \rangle_0$$ \sim$ 
$m_{\psi}^{{\rm dyn.}} {\Lambda}^{\!2} $$\ll$$ {\Lambda}^3$. 

Having in mind the choice 
$m_{\psi}^{{\rm dyn.}} $$\sim$$ E_{{\rm ew}}$,
he then conjectured that such a fine-tuned DSB scenario
would lead to an effective NJLVL picture 
of strong enough four-$\psi$ interactions
(\ref{FNNJLVL}) at $\mu $$\sim$$ \Lambda$, with
a fine-tuned $g $$\stackrel{>}{\approx}$$
g_{{\rm crit.}}$. These 
terms have then their origin in exchanges of the discussed 
heavy $SU(N)_{UC}$ gauge bosons between $\psi$'s.
In this picture, (\ref{FNNJLVL}) leads via a DSB
(similar to the DEWSB) to a linear ${\sigma}$ model
with a dynamical light composite scalar ${\Phi}_{ij}$ 
($i,$$j$$=$$1,$$\ldots,$$F$)
at low energies $\mu $$<$$ \Lambda$. He then employed a 
Dyson--Schwinger (DS) gap equation for the
running $m_{\psi}^{{\rm dyn.}}(p^2) $$\equiv$$ \Sigma(p^2)$, 
based on the approximation
of one $SU(N)_{UC}$-gauge boson exchange. Using this DS equation,
he showed that a specific function of the mass $M$ ($=$$ {\Lambda}$) 
of these gauge bosons 
has to be fine-tuned to the value $1$ up to the order of
$(m_{\psi}^{{\rm dyn.}}/M)^2$ 
for this DSB to occur. He showed that,
in this DSB, light dynamical composite scalars occur,
and that decay constants of the resulting NGB's
are $f $$\sim$$ m_{\psi}^{{\rm dyn.}}$ ($\sim$$ E_{{\rm ew}}$).

If taking $F$$=$$2$,
$\psi_1 $$=$$ t$ and $\psi_2 $$=$$ b$, the model
leads to a TSM scenario
with equally strong ${\bar t} t$ and ${\bar b} b$ condensates
($\Rightarrow$ $m_t $$=$$ m_b$). However, as discussed in
Subsec.~\ref{EWESG14} and at the end 
of Sec.~\ref{RMUP3}, $U(1)_Y$ interactions can change
this situation so that $m_t $$\gg$$ m_b$.

\subsection{Models possessing simultaneously horizontal and 
vertical gauge symmetries}
\label{RMUP5}
A gauge theory with broken horizontal symmetries, in addition
to the left-right (L-R) symmetric gauge group 
(a ``vertical symmetry''), 
as an underlying physics for the DEWSB, has been
introduced by 
Nagoshi, Nakanishi and Tanaka (1991).
These authors assumed that the full gauge group contains,
in addition to the L-R symmetric group
$G_{L-R}$$\equiv$$SU(3)_c\!\times\!SU(2)_L\!\times\!SU(2)_R\!\times\!
U(1)_{B-L}$, also horizontal $SU(N)$ gauge groups with the
following horizontal
interactions over $N$ generations of leptons and quarks
($N$$=$$3$)
\begin{equation}
{\cal {L}}_{{\rm hor.}} = i f \sum_{k=1}^{N^2-1}
{\bar \psi} {\gamma}_{\mu} T_k H_k^{\mu} {\psi} \ .
\label{hor}
\end{equation}
${\psi}$ stands for any $N$-component sector of quarks or leptons
(up-type quarks, down-type quarks, charged leptons, neutrinos);
$H_k^{\mu}$ are the horizontal gauge fields;
$T_k$ are generators of $SU(N)$: $T_k $$=$$ {\lambda}^{(k)}/2$,
where ${\lambda}^{(k)}$ are Gell-Mann matrices, for $N$$=$$3$.
As seen from (\ref{hor}), horizontal interactions treat,
in a certain sense, all generations equally.
The authors assumed that horizontal symmetries are
broken by an unspecified mechanism at an energy scale
$\mu $$=$$ {\Lambda}_H$, while the L-R symmetry is left unbroken.
The resulting masses of the $N^2$$-$$1$$=$$8$ gauge bosons
were assumed to satisfy certain hierarchy, for example:
$M^2_{4,5,6,7}$ ($\sim$$ {\Lambda}^{\!2}$)
$\gg$$ M^2_{1,2,3,8}$ ($\sim$$ {\Lambda}^{\!2}_H$),
where ${\Lambda}_H$ satisfies:
$10^2 \mbox{ TeV} $$\stackrel{<}{\sim}$$ {\Lambda}_H $$\ll$
$E_{{\rm GUT}}$. The lower bound was needed
to avoid possibly too large induced 
FCNC's. This breaking of the horizontal
symmetry, and the resulting heavy gauge boson exchange
processes between fermions,
lead to a pattern of DSB of the L-R
gauge symmetry and eventually to the DEWSB. The authors
obtained one gap equation for each of the two quark
sectors, based on the exchange of horizontal
gauge bosons with mass $M_k $$\sim$$ {\Lambda}_H$ between the quarks. 
The UV cutoff in the gap equations was taken to be
${\Lambda}$($\sim$$M_{4,5,6,7}$). For the 
neutrino sector, they obtained
similarly two coupled gap equations for the Dirac ${\cal {M}}_D$
and Majorana ${\cal {M}}_M$ mass matrices. 
If the hierarchy 
${\cal {M}}^2_D $$\ll$$ {\cal {M}}^{\dagger}_M {\cal {M}}_M
$$\ll$$ {\Lambda}_H$ is satisfied, see-saw mechanism can give
phenomenologically acceptable eigenmasses of neutrinos,
and the authors suggested that such a framework would
lead to a dynamical L-R 
gauge symmetry breaking.\footnote{
The L-R DSB framework 
as constructed and investigated in detail by 
Akhmedov {\em et al.\/} (1996)
(cf.~Sec.~\ref{EGSG2}), is different; among other things,
it involves a modified see-saw mechanism.}
The gap equation gave $m_t^{{\rm dyn.}} $$\not=$$ 0$
and $m_b $$=$$0$, due to the mentioned gauge boson mass
hierarchy. However, the authors pointed
out that $m_b$ would become nonzero and
$m_b $$\ll$$ m_t$, through a kind of $m_t$-induced
``feed-down'' effect: $b$ obtains a small 
$m_b^{(0)} $$\not=$$ 0$ through mixing between $W_L$
and $W_R$ of the L-R gauge sector and the
large $m_t^{{\rm dyn.}}$, as illustrated in Fig.~\ref{rmd11f}. 
\begin{figure}[htb]
\mbox{}
\vskip3.8cm\relax\noindent\hskip.0cm\relax
\includegraphics{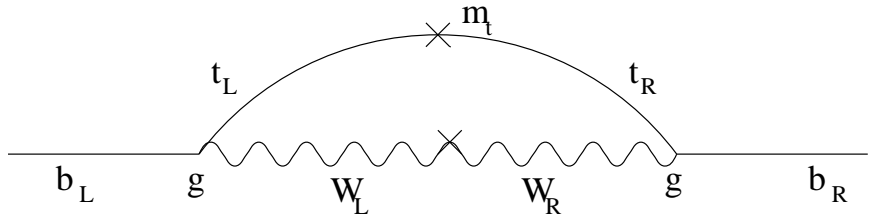} \vskip0.6cm
\caption{\footnotesize Generation of a small mass 
$m_b^{(0)}$ out of a large $m_t^{{\rm dyn.}}$ and by help of
$W_L$-$W_R$ mixing. Figure is reproduction
of Fig.~5 of Nagoshi, Nakanishi and Tanaka (1991).}
\label{rmd11f}
\end{figure}
This $m_b^{(0)}$ gets enhanced
by horizontal gauge interactions (through gap
equation) to a value $m_b $$\gg$$ m_b^{(0)}$, 
but still remains much smaller than $m_t$
as long as the mixing angle between $W_L$ and $W_R$
is small enough. Stated differently, the tiny $m_b^{(0)}$,
generated by the process of Fig.~\ref{rmd11f},
effectively increases the
four-quark parameter ${\kappa}_b$ of NJLVL interaction
(\ref{3gTSM}) from a slightly subcritical to a 
slightly supercritical value.

Nakanishi and Tanaka (1992)
subsequently applied the described framework in
a more detailed investigation of the spectrum of
the dynamical neutrino masses. The requirement of 
having three nearly massless generations of
neutrino eigenstates led them naturally to assume
the existence of higher generations of quarks and leptons.
Nagoshi, Nakanishi and Tanaka (1993), 
and later Nagoshi and Nakanishi (1995),
investigated within the described framework 
the dynamical generation of the mass hierarchy between and within
the up- and down-type quark sectors. 
The role of the vertical gauge interactions was played
this time by QED or $U(1)_Y$. This interaction played
an important role in generating mass hierarchy between
the up- and down-type sectors, while the mass hierarchy
within the sectors was explained by a suitable breaking 
pattern of horizontal gauge symmetries.
In a simplified scheme, they showed that
generated flavor mixings can involve CP violation.
The authors pointed out that two factors are crucial for
dynamical generation of flavor mixings in their framework,
i.e., for generation of CKM mixing matrix elements through 
the DEWSB: 
(a) radiative corrections from vertical gauge interactions;
(b) non-commutativity of those generators of horizontal
 symmetries which are associated with the lighter horizontal
gauge bosons.

Another type of renormalizable models 
involving horizontal symmetries and leading to the DEWSB
was constructed by King (1992),
later further investigated by Elliott and King (1992)
and Evans, King and Ross (1993).
It is based on the horizontal
gauge symmetry group 
$G$$\equiv$$SU(3)_f\!\times\!SU(3)_L\!\times\!SU(3)_R$, 
in addition to $G_{{\rm SM}}$.
The model doesn't involve leptonic
degrees of freedom in DSB. Left-handed Weyl quarks
of the SM transform under $G$: $u_L $$\sim$$ (1,3,1)$,
$(u_R)^c $$\sim$$ (1,1,{\bar 3})$. The framework contains
additional Weyl quarks $a$, $b$, $c$ and $d$,
which transform nontrivially under $SU(3)_f$.
It was assumed that condensates $\langle a d \rangle_0 
$$\not=$$ 0$ and $\langle b c \rangle_0 $$\not=$$ 0$ form. This
symmetry breaking gave heavy masses $M_A $$>$$ M_B
$$\sim$$ \Lambda$ to the corresponding gauge bosons
of the horizontal group, and preserved global flavor symmetries 
of the SM. Therefore, the resulting effective model at 
$\mu $$\stackrel{<}{\sim}$$ \Lambda$
possesses GIM mechanism and has suppressed FCNC's even for
low ${\Lambda} $$\sim$$ 1$ TeV. 
At $\mu $$\sim$$ \Lambda$, the model is in fact the truncated
TSM (\ref{TSM}), generated by exchange of heavy gauge bosons.
Elliott and King (1992) studied dynamics of ${\bar t} t$
condensation (and DEWSB) in this framework, taking into account
the dynamical structure (momentum dependence) of propagators
of heavy gauge bosons involved in the exchange.
DS (gap) equation was investigated numerically under inclusion
of QCD in dressed ladder approximation,
and the PS relations were included in the analysis, too
(cf.~Sec.~\ref{MDSPS} on the DS$+$PS approach). The new high energy
dynamics reduced $m_t$ somewhat in comparison
with the minimal (TSM) model, but $m_t$ remained still too high.
Later, Evans, King and Ross (1993) discussed vacuum alignment
in a fully dynamical scenario of the model -- i.e.,
they discussed a previously unspecified scenario that is driving 
the formation of condensates $\langle a d \rangle_0 
$$\not=$$ 0$ and $\langle b c \rangle_0 $$\not=$$ 0$. Responsible for 
this is $SU(3)_f$ which becomes strong enough for
the formation of these condensates at a scale
$\Lambda_f $$\stackrel{>}{\sim}$$ 1$ TeV.

\subsection{Topcolor assisted technicolor (TC2)}
\label{RMUP6}
Several authors have proposed to combine the ${\bar t} t$
condensation mechanism with the technicolor (TC) and extended
technicolor (ETC) mechanism. Their purpose was to obtain a
phenomenologically acceptable value of $m_t$, i.e.,
a somewhat smaller $m_t$ than the minimal (TSM) scenario
gives, and a reduced compositeness
scale $\Lambda$ of ${\bar t} t$ condensate.
Furthermore, such combined (TC2) scenarios, being
more complex than the minimal TSM framework, may also
explain the lighter masses of fermions other than $t$,
flavor mixings, CP violation, and suppression
of the flavor-changing neutral currents (FCNC's).
It is interesting that such combined mechanisms tend to cure
simultaneously well known deficiencies
of the pure ${\bar t} t$ condensation mechanism
(or: of the pure topcolor scenario leading to
the minimal framework of ${\bar t} t$ condensation) 
and those of the TC$+$ETC mechanism:
\begin{itemize}
\item 
the minimal framework requirement that
${\bar t} t$ condensate be fully responsible for
the EWSB [i.e., the corresponding NGB's have
$F_{\pi} $$=$$ v$ ($=$$ 246$ GeV) leading to $M_W,\!M_Z$]
leads to too large $m_t$ ($>$$ 220$ GeV); minimal $m_t$
($\approx$$220$ GeV) is obtained when
${\Lambda}$ is exceedingly high ($\sim$$E_{{\rm Planck}}$);
some people regard such high ${\Lambda}$'s and the
related fine-tuning ${\kappa}_t $$\approx$$ {\kappa}_{{\rm crit.}}$ 
of the four-quark coupling parameter
of (\ref{3gTSM}) as unnatural;
\item
TC$+$ETC predict $m_t$ substantially lower
than the measured value $\approx$$175$ GeV, 
while TC can account for the full (dynamical) EWSB
and ETC for light quark and lepton masses, at low cutoffs
${\Lambda}_{{\rm TC}} $$\sim$$ 10^2$-$10^3$ GeV and
${\Lambda}_{{\rm ETC}} $$\sim$$ 10^5$ GeV. 
\end{itemize}
Furthermore, such combined frameworks (TC2) offer a possibility
to have a {\em fully dynamical\/} symmetry breaking scenario --
in which even the masses $M$$\sim$${\Lambda}$ of new heavy
gauge bosons (whose exchanges lead to TSM-type four-quark terms
responsible for ${\bar t}t$ condensation) can be explained
dynamically via DSB induced by condensation of
technifermion pairs.

A renormalizable
framework incorporating these ideas 
of ``topcolor assisted technicolor'' (TC2) 
was first constructed by
Hill (1995), 
and later investigated
by Buchalla, Burdman, Hill and Kominis (1996)
where it is called topcolor I model. 
The model builds
partly on Hill's previous idea of topcolor (coloron)
-- cf.~Sec.~\ref{RMUP3} (Hill, 1991).
At energies $\mu $$<$$ {\Lambda}_{{\rm ETC}}
$$\sim$$ 10^2$ TeV, the unbroken gauge group is assumed to be
a product of a TC group $G_{{\rm TC}}$ and
a variation of the topcolor (coloron) group 
$G_{{\rm cln}}\!\equiv\!\left[ 
SU(3)_1\!\times\!U(1)_{Y1} \right]\!\times\! 
\left[ SU(3)_2\!\times\!U(1)_{Y2} \right]$ and the SM isospin
group $SU(2)_L$. However, now the assignment of
$G_{{\rm cln}}$ quantum numbers
is different: third generation
of quarks ({\em including\/} $b_R$) preferentially couples to
the strong\footnote{
Hill (1995) used a different notation than in his
earlier work (1991). $SU(3)_1$ and $U(1)_{Y1}$ are
now the stronger groups; in the earlier work,
$SU(3)_1$ was the group with the weaker coupling.}
$SU(3)_1\!\times\!U(1)_{Y1}$,
first and second generations couple
to the weaker $SU(3)_2\!\times\!U(1)_{Y2}$. 
The model treats $t$ and $b$
equally under $SU(3)_1\!\times\!SU(3)_2$ -- this is in contrast
to the original topcolor model (Hill, 1991), and similar
as in a modified version of topcolor (Hill and Parke, 1994).
Leptons are assigned such quantum numbers under 
$G_{{\rm TC}}\!\times\!G_{{\rm cln}}$ that the model
has cancellation of triangular anomalies.
Strong $U(1)_{Y1}$ was introduced to ensure that the 
generated NJLVL terms which can lead
to ${\bar t}t$ condensate will be stronger 
(${\kappa}_t $$\stackrel{>}{\approx}$$ {\kappa}_{{\rm crit.}}$) 
than those which can lead to the
${\bar b}b$ condensate (${\kappa}_b $$\stackrel{<}{\approx}$$
{\kappa}_{{\rm crit.}}$). Stated differently, strong
$U(1)_{Y1}$ was introduced
to ensure isospin hierarchy $m_t $$\gg$$ m_b$.
In this context, we recall that quarks 
$t_L,\!b_L,\!t_R,\!b_R$ couple to the $U(1)_{Y1}$ gauge boson
with strengths $q_1 Y/2$, where $q_1$ is the strong $U(1)_{Y1}$
coupling parameter and $Y$ the usual EW hypercharge:
$Y$$=$$+1/3$ for $t_L$ and $b_L$; 
$Y=$$+4/3,\!-2/3$, for $t_R$, $b_R$,
respectively. Quarks of the first two generations couple
analogously to $U(1)_{Y2}$, with strengths $q_2 Y/2$
($q_2 $$\ll$$ q_1$).

In one of the scenarios proposed
by Hill, there are techniquarks $Q$ and $T$ transforming
nontrivially under subgroups $SU(3)_{TCQ}$ 
and $SU(3)_{TCT}$ of the TC gauge group $G_{{\rm TC}}
$$\equiv$$ SU(3)_{TCQ}\!\times\!SU(3)_{TCT}$, respectively,
and transforming nontrivially
under the coloron group $G_{{\rm cln}}$. They condense due to
the strong and confining TC interaction 
(${\Lambda}_{{\rm TC}} $$\sim$$ 1$ TeV).
Condensate $\langle {\bar Q} Q \rangle_0$ is 
responsible for the first stage of DSB:
$G_{{\rm cln}}\!\times\!SU(2)_L\!\to\!G_{{\rm SM}}$; 
condensate $\langle {\bar T} T \rangle_0$ for the second stage
(i.e., the DEWSB): $G_{{\rm SM}} \to
SU(3)_c\!\times\!U(1)_{{\rm em}}$.
After the first stage,
a residual global symmetry $SU(3)^{\prime}\!\times\!U(1)^{\prime}$
is left, and as a result a degenerate massive color-octet
of gauge bosons called {\em colorons\/} $B^{(\alpha)}_{\mu}$ appears,
similarly as in the earlier topcolor model (Hill, 1991).
In addition, a color-singlet heavy gauge boson
$Z^{\prime}$ appears, corresponding to
the global $U(1)^{\prime}$. Masses of these gauge bosons
satisfy the hierarchy $E_{{\rm ew}} $$\sim$$ 10^2 \mbox{ GeV}
$$<$$ M_B $$\sim$$ M_{Z^{\prime}} $$\sim$$ 10^3 \mbox{ GeV} $$<$$ 
{\Lambda}_{{\rm ETC}} $$\sim$$ 10^5 \mbox{ GeV}$.
Exchange of these bosons between the SM quarks leads to
strong effective NJLVL interactions 
at ${\Lambda} $$\sim$$ M_B $$\sim$$ M_{Z^{\prime}}$ 
($\sim$$ 1$ TeV), involving
$t_{L,R}$ and $b_{L,R}$. Due to the mentioned couplings of
these quarks to the $U(1)_{Y1}$ factor 
in the original $G_{{\rm cln}}$,
the effective four-quark parameters 
in the ${\bar b} b$ and ${\bar t} t$ channels may satisfy:
${\kappa}_b^{{\rm (eff.)}} $$<$$ {\kappa}_{{\rm crit.}} $$<$$
{\kappa}_t^{{\rm (eff.)}}$. In this case,
only $t$ condenses {\em and\/} obtains a large 
$m_t^{{\rm dyn.}} $$\not=$$ 0$;
$m_b^{{\rm dyn.}} $$=$$ 0$, although $b$
may still condense when ${\kappa}_b^{{\rm (eff.)}}$ 
is close (from below) to ${\kappa}_{{\rm crit.}}$
[cf.~(\ref{kapbbelow})].

The ${\bar t} t$ condensation in this framework is, 
to a large degree, only a spectator to the TC driven DSB (DEWSB).
This means that ${\bar t} t$ condensation contributes only
a small part to the electroweak VEV $v$, and hence only a
small part to $M_W$ and $M_Z$, in contrast to the
TC mechanism. To understand this statement in a 
bit more quantitative way, we recall Pagels--Stokar (PS) formula
in quark loop approximation (\ref{PSql}) for the decay constant
of the NGB's originating from the pure ${\bar t}t$ condensation
scenario:
\begin{equation}
F^2_{\pi^{\pm}}(\mbox{q.l.}) = 
\frac{N_{{\rm c}}}{8 \pi^2} m_t^2
\left[ \ln \frac{{\Lambda}^{\!2}}{m_t^2} + \frac{1}{2} \right] \ ,
\quad 
F^2_{\pi^0}(\mbox{q.l.}) = \frac{N_{{\rm c}}}{8 \pi^2} m_t^2
\ln \frac{{\Lambda}^{\!2}}{m_t^2} \ ,
\label{PSql2}
\end{equation}
where $m_t $$\approx$$ 175$ GeV. 
If ${\bar t} t$ condensation were entirely
responsible for the DEWSB, then 
$F_{\pi} $$\approx$$ v $$=$$ 246$ GeV.
For this we would need, as suggested by
(\ref{PSql2}), a huge ${\bar t} t$-compositeness scale
${\Lambda} $$\sim$$  10^{13}$-$10^{14}$ GeV 
(cf.~also Table~\ref{tabl2} 
in Sec.~\ref{MDSPS3}). However, since in the discussed TC2 framework
${\Lambda}$ has much lower (and hence more
attractive) values ${\Lambda} $$\sim$$ M_B $$\sim$$ M_{Z^{\prime}}$ 
$\sim$$ 1$ TeV, (\ref{PSql2}) suggests that ${\bar t} t$
condensation cannot be responsible for most of the
needed EW VEV $v $$\approx$$ 246$ GeV. Namely, for
${\Lambda} $$=$$ 1$ TeV and $m_t $$=$$ 175$ GeV, (\ref{PSql2})
yields: 
$F_{\pi}$ ($\equiv$$ f_{\pi} \sqrt{2}$ )$\approx$$64$-$68$ GeV. 
Therefore, the actual
massless NGB's, subsequently eaten by $W$'s and $Z$ 
to provide the latter with masses, are linear combinations
of composite scalars provided by the TC mechanism and those
provided by ${\bar t} t$ condensation;
the latter components are only a small
admixture in the actual eaten NGB's.
Stating this in an even more simplified (approximate) way:
the actual (composite) EW NGB's are basically provided by the TC
mechanism. 

Therefore, an additional triplet of
(uneaten) pseudo-NGB's ${\tilde \pi}^a$ occurs in the course of
${\bar t} t$ condensation. 
Techniquarks (denoted generically as $Q_i$), 
which had condensed by
strong confining TC interactions, have masses $\sim$$500$
GeV and can be neglected at the EW scale. As a result, the
dangerous TC-breaking condensates $\langle {\bar Q} t \rangle_0$
do not form. ETC interactions in this scenario (or an
elementary Higgs) generate masses of light fermions, and
give contribution ${\epsilon} m_t$ to to the full
$m_t$. This contribution is much smaller than
$m_t^{{\rm dyn.}}$$=$$(1$$-$${\epsilon})m_t$ obtained via
the described ${\bar t} t$ condensation.
Typical expected values for ${\epsilon}$
are: ${\epsilon} $$\sim$$ 10^{-2}$-$10^{-1}$. 
Mass $m_b$ originates
partly from ETC and partly from instantons in $SU(3)_1$.
Other light quarks obtain masses from ETC.
Consequently, CKM mixing matrix may be generated.

Small ETC-induced ${\epsilon} m_t$ induces 
a nonzero mass of the mentioned triplet
of pseudo-NGB's ${\tilde \pi}^a$, in a range
$m_{\tilde \pi}$$\approx$$180$-$240$ GeV,
thus ameliorating the problem of having
dangerously light scalars. Their decay
lengths are $f_{\tilde \pi}$ 
($\equiv$$ F_{\tilde \pi}/\sqrt{2}$) $\approx$$ 50$ GeV,
according to (\ref{PSql2}).
Hill called them top-pions, and argued that their
appearance is a general feature in
scenarios where ${\bar t} t$ condensation is accompanied
by some additional mechanism leading to the EWSB and to
the light quark masses
(e.g., TC$+$ETC or elementary Higgs). 

Buchalla, Burdman, Hill and Kominis (1996),
in a long paper on theory and phenomenology of topcolor
scenarios, discussed further the above 
TC2 framework which they called
topcolor I. In addition, they proposed and investigated 
also a variant of Hill's original 
coloron model (cf.~Sec.~\ref{RMUP3}), calling it topcolor II.
It has an additional (TC)
group factor $SU(3)_q$, under which only the additional
quark ($q_L$, $q_R$) transforms nontrivially -- as a triplet. 
Topcolor II is therefore based on the gauge group
$G$$\equiv$$SU(3)_q\!\times\!SU(3)_1\!\times\!SU(3)_2\!\times\!U(1)_Y
\!\times\!SU(2)_L$, where $U(1)_Y$ is the conventional
weak group which tilts
condensation to ${\bar t} t$ flavor direction. The model
is anomaly-free. Among other things,
$(c,s)_{L,R}$ are assigned different quantum numbers
under $SU(3)_1\!\times\!SU(3)_2$: 
$(c,s)_L $$\sim$$ (3,1)$ feels strong\footnote{
Buchalla {\em et al.\/} had a different notation
than Hill had in his original coloron model (1991) --
for them, $SU(3)_1$ is strong
and $SU(3)_2$ weak ($h_1 $$\gg$$ h_2$).}
$SU(3)_1$ and participates in the $t_R$-dominated
condensation resulting in a heavy $m_t^{{\rm dyn.}}$; 
$c_R,$$ s_R,$$ b_R$$ \sim$$ (1,3)$
couple only to weak $SU(3)_2$ and cannot
lead to dynamical $m_c,\!m_s,\!m_b$.
$SU(3)_q$ is confining and forms ${\bar q} q$
condensate which acts like ${\Phi}^a_b$ effective scalar,
breaking the coloron group down to QCD dynamically. The problem
of pairing of $(b^c)_L$ and $q_L$, as discussed by
Martin (1992a, 1992b) (cf.~Sec.~\ref{RMUP3}), does not occur here,
since these two quarks transform as a singlet and a triplet
under the TC group $SU(3)_q$, respectively. 
The EWSB 
$G_{{\rm SM}}\!\to\!SU(3)_c\!\times\!U(1)_{{\rm em}}$ is
assumed to occur via Higgs mechanism, by an elementary
or composite Higgs other than ${\bar t} t$.
Consequently, a rich structure of
pseudo-NGB's emerges, leading to
a rich phenomenology as investigated by these authors
(cf.~Subsec.~\ref{SPTAQ12}). 
In topcolor II, just as in topcolor I (TC2),
${\bar t} t$ condensation is just a spectator to
the (D)EWSB.

Soon after the appearance of Hill's work (1995) on
TC2, a series of papers by other authors appeared 
(Chivukula, Dobrescu and Terning, 1995;
Lane and Eichten, 1995;
Lane, 1996;
Chivukula and Terning, 1996),
discussing some problems in TC2 scenarios
and solutions to these problems.
Chivukula, Dobrescu and Terning (1995)
argued that TC2 scenarios are likely to have
a problem of incompatibility of the following
two requirements:
(a) ``naturalness'' $M_{{\rm coloron}}$$\sim$$ 1$ TeV 
to avoid the fine-tuning
(recall: $M_{{\rm coloron}}
$$\equiv$$ {\Lambda}$, where ${\Lambda}$ is
${\bar t} t$-compositeness scale);
(b) a phenomenologically acceptable 
$\delta \rho$ [$\rho $$\equiv$$ M_W^2/(M_Z^2 \cos^2 \theta_W)$].
They argued that the up-type and down-type technifermions
(their right-handed components)
are likely to have different couplings to the strong $U(1)_{Y1}$
and violate custodial symmetry, thus leading to a large
$\delta \rho$. To prevent this, $U(1)_{Y1}$ couplings
must be relatively small. However, the latter
are responsible within TC2 scenarios to bring the effective
$({\bar t} t) ({\bar t} t)$-parameter above ${\kappa}_{{\rm crit.}}$
to enable the condensation to occur [$\Rightarrow$
large $m_t^{{\rm dyn.}}\!=\!(1$$-$${\epsilon})m_t$],
while at the same time keeping (pushing) the
$({\bar b} b)({\bar b}b)$-parameter below ${\kappa}_{{\rm crit.}}$
to prevent a (large) $m_b^{{\rm dyn.}}$.
Therefore, the required weakening of $U(1)_{Y1}$,
stemming from the $\delta \rho$ restriction, results
in a fine-tuning of the ${\bar t} t$ condensation
mechanism, and consequently in a large (``unnatural'')
${\bar t} t$ condensation scale 
${\Lambda} $$\equiv$$ M_{{\rm coloron}}
$$\stackrel{>}{\approx}$$ 4.5$ TeV.

Lane and Eichten (1995)
and Lane (1996)
constructed TC2 scenarios which overcome
these problems. In particular,
in their anomaly-free scenarios, $U(1)_{Y1}$ couplings to
technifermions are isospin symmetric and preserve custodial $SU(2)$,
allowing acceptably small $\delta \rho$ and at the same
time strong $U(1)_{Y1}$ -- consequently
``naturalness'': $M_{{\rm coloron}} $$\approx$$
1$ TeV.
 
Phenomenological implications
of TC2 models will be discussed in Subsec.~\ref{SPTAQ12}.

We do not attempt to go into detail of explaining the
TC and ETC mechanism, since this would bring us
beyond the scope of this review article.
It should be stressed that earlier, other authors
\cite{KingMannan91b,MendelMiransky91,Miransky92}
had proposed the idea that ${\bar t} t$ condensation
be combined with (extended) TC to bring down~\footnote{
The idea that a full strong dynamics could lower
$m_t^{{\rm dyn.}}$ and account for $M_W$ and $M_Z$
was also mentioned by Hill (1991).}
$m_t^{{\rm dyn.}}$ and the compositeness 
scale $\Lambda$, when compared with the minimal framework.
In contrast to the renormalizable 
TC2 framework of Hill (1995) described
above, they used as a starting point
effective NJLVL terms
involving third generation quarks and TC quarks. 
In their investigation, they
employed Dyson--Schwinger equations in
the dressed ladder approximation. 
They also predicted emergence of a triplet of quark-antiquark
pseudo-NGB's (called top-pions by Hill).
King and Mannan argued that masses of these pseudo-NGB's
in their framework are $\sim$$20$ GeV,
while Mendel and Miransky obtained masses of several $100$ GeV.
Subsequently, Miransky argued that these heavy 
pseudo-NGB's could lead
to important enhancements of FCNC's in $B$ meson physics
and could be detected
at the Superconducting Super Collider (SSC).

Martin (1993)
proposed a dynamical scenario similar to 
TC2: a self-breaking TC combined with topcolor.
The usual TC is like a scaled-up QCD,
not self-breaking, gauge bosons (technigluons) of the TC
group remain massless, although their coupling to 
technifermions is strong enough to make the latter condense
and thus (in TC2 scenarios) to make 
them break the topcolor (coloron) sector 
and possibly also the SM sector of the gauge group.
However, Martin proposed that strong TC
interactions break the TC group itself and also
$G_{{\rm SM}}$ (DEWSB).
He proposed a full gauge group
$G\!\equiv\!SU(4)_{TC}\!\times\!SU(3)_{C^{\prime}} 
\!\times\!SU(3)_{C^{\prime\prime}}
\!\times\!SU(2)_L\!\times\!U(1)_{Y^{\prime}}$ with a 
complicated anomaly-free content of technifermions and other
techni-singlet fermions (in addition to SM fermions).
Three types of condensates, 
assumed to appear simultaneously at $\Lambda $$\sim$$ 1$ TeV,
are responsible for the full breaking 
$G\!\to\!SU(3)_c\!\times\!U(1)_{{\rm em}}$.
One type of condensates contains a pair of technifermions in
different TC representations, and therefore breaks TC group
$SU(4)_{TC}$ dynamically: 
$SU(4)_{TC}\!\times\!SU(3)_{C^{\prime}}\!\times\!U(1)_{Y^{\prime}}
\!\to\!SU(3)_{C^{\prime\prime\prime}}\!\times\!U(1)_Y$. 
It thus gives heavy mass to ``technigluons'' and
leads to the usual coloron model (Sec.~\ref{RMUP3}),
with $SU(3)_{C^{\prime\prime\prime}}$ being
the strong and $SU(3)_{C^{\prime\prime}}$
the weak coloron group factor. Second type of 
condensates, made up of pairs of 
TC-singlet and $SU(2)_L$-singlet fermions
(quix-antiquix pairs), then breaks 
$SU(3)_{C^{\prime\prime\prime}}\!\times\!SU(3)_{C^{\prime\prime}}$ 
to QCD $SU(3)_c$,
giving heavy mass to the eight colorons. Third type
of condensates, made up of technifermions, then breaks
$SU(2)_L\!\times\!U(1)_Y\!\to\!U(1)_{{\rm em}}$.
The ${\bar t} t$ condensate (and hence $m_t$) 
is generated by strong interactions
mediated by heavy colorons just as in 
the coloron (topcolor) model,
and is largely a ``spectator'' to the DEWSB. The
mentioned TC condensates of the third type 
are largely responsible for the DEWSB.
The model behaves similarly as topcolor I.

It should be mentioned that the broken TC helps
reduce the electroweak S parameter
(Peskin and Takeuchi, 1992), 
making it more
compatible with experimental evidence and thus avoiding
a potentially dangerous problem of too large values of S
in the usual scaled-up QCD-like
TC theories, as pointed out by 
Hill, Kennedy, Onogi and Yu (1993).

Apart from TC-scenarios, in which condensates ${\bar Q} Q$ of
technifermions are largely responsible for the DEWSB,
there exists a somewhat related dynamical scenario in which the DEWSB
is brought about by a condensate of new heavy vector
bosons: $\langle B^{(0)\dagger}_{\mu} B^{(0)\mu} \rangle_0
$$\propto$$ v^2$ (Cynolter, Lendvai and P\'ocsik, 1997,
and references therein).

\subsection{Other renormalizable scenarios 
with ${\bar t}t$ condensation}
\label{RMUP7}
Bando, Kugo and Suehiro (1991)
considered $E_6$ GUT as the underlying theory leading 
to ${\bar t} t$ condensation with a 
high compositeness scale $\Lambda $$\sim$$ E_{{\rm GUT}}
$$\sim$$ 10^{16}$ GeV.
The main motivation was to construct
a dynamical scheme which would allow in a natural way
the large isospin breaking $m_b $$\ll$$ m_t$.
The authors emphasized that $E_6$ is apparently the only gauge group
which has simultaneously the following properties:
(a) it is free of anomalies;
(b) all fermions of each generation are combined
into an irreducible representation;
(c) all the symmetry breaking can be realized by using
only those Higgs representations which give masses to fermions.
Property (c) implies that all stages in the 
breaking $E_6\!\to\!SU(3)_c\!\times\!U(1)_{{\rm em}}$
could in principle be realized dynamically. The authors
considered the breaking pattern 
$E_6\!\to\!SO(10)\!\to\!SU(5)\!\to\!G_{{\rm SM}}$, assuming that
it takes place via (unspecified) Higgs mechanisms. 
In the course of the breaking, part of $SO(10)$-spinor gauge
bosons, and fermions of those irreducible
representations in $SO(10)$ not containing SM fermions,
acquire masses $M $$\sim$$ E_{{\rm GUT}}$. 
Then a box diagram involving
these fields leads to NJLVL interactions 
($\propto$${\alpha}^2_{{\rm GUT}}/M^2$), including
terms containing simultaneously 
$t_R$ and $b_R$, i.e., terms $({\bar t}_L t_R)({\bar b}_L b_R)$,
$({\bar b}_L t_R)({\bar t}_L b_R)$, and their Hermitian conjugates.
Therefore,\footnote{
Other NJLVL terms also emerge,
e.g., those without $b_R$ ($\Rightarrow$ TSM).}
such terms lead to a ``feed-down'' effect mentioned
in Sec.~\ref{MDSPS2}: large $m_t^{{\rm dyn.}}$ 
($\Leftrightarrow$ $\langle {\bar t}_L t_R \rangle_0 $$\not=$$ 0$) 
induces a tiny $m_b^{(0)} $$\not=$$ 0$ 
(mass term $\propto$${\bar b}_L b_R$). 
Four-quark parameter ${\kappa}_b^{{\rm (eff.)}}$ of 
$({\bar b}_L b_R)({\bar b}_R b_L)$ is below 
${\kappa}_{{\rm crit.}}$ [due to $U(1)_Y$],
thus preventing a nonzero $m_b^{{\rm dyn.}}$ from emerging
from its initial zero value. 
However, the tiny ``feed-down''-induced
$m_b^{(0)}$ brings this ${\kappa}_b^{{\rm (eff.)}}$
above ${\kappa}_{{\rm crit.}}$, and solution
of the modified gap equation for the dynamical $m_b$
acquires a value $m_b $$\gg$$ m_b^{(0)}$.
This mechanism, known as Nagoshi--Nakanishi--Tanaka (NNT)
enhancement, was earlier investigated and explained
by the latter authors (Nagoshi, Nakanishi, Tanaka, 1991)
within a dynamical framework with horizontal and L-R gauge
symmetries (cf.~Sec.~\ref{RMUP5}). Also Bando {\em et al.\/},
however, can easily ensure in their $E_6$-based
framework that the obtained NNT-enhanced $m_b$ is still
small enough: $m_b $$\ll$$ m_t$.

Yoshida (1996)
performed a detailed analysis of a renormalizable
model stemming from the fine-tuning framework of Luty (1993) 
(cf.~Sec.~\ref{RMUP4})
and the topcolor frameworks of Hill (1991, 1995)
(cf.~Secs. ~\ref{RMUP3}, \ref{RMUP6}).
Yoshida considered the gauge group 
$G\!\equiv\!SU(N_A)\!\times\!SU(N_B)\!\times\!SU(2)_L\!\times\!U(1)_Y$
(with $N_B $$=$$ 3$)
as leading to a tumbling DSB scenario. He first discussed,
in analogy with Luty, two extreme
cases: ${\Lambda}_A $$\gg$$ {\Lambda}_B$ 
and ${\Lambda}_A $$\ll$$ {\Lambda}_B$.
In the first case, strong $SU(N_A)$ drives a huge
fermion condensate VEV ($\sim$${\Lambda}_A^3$)
which breaks $SU(N_B)$ dynamically and gives its gauge bosons
(``colorons'') large masses $M_B$, while gauge bosons of
$SU(N_A)$ remain massless $M_A$$=$$0$. In the second case, the roles
in the DSB scenario are simply inverted.  
He then investigated the generic case ${\Lambda}_A $$\sim$$
{\Lambda}_B$, employing formalism of the dressed ladder
Dyson--Schwinger (DS) equation\footnote{
``Dressed'' means that one-loop evolution of
$SU(N_A)$ and $SU(N_B)$ coupling parameters 
was taken into account.} 
plus the (generalized) Pagels--Stokar (PS) relations.
As Luty, he demanded
that transition between the two cases (phases) not have
discontinuity, i.e., that the phase transition be second order
-- to ensure hierarchy $m_t $$\ll$$ {\Lambda}$
($\sim$$ {\Lambda}_A $$\sim$$ {\Lambda}_B$) if also the DEWSB 
eventually occurs: 
$SU(2)_L\!\times\!U(1)_Y\!\to\!U(1)_{{\rm em}}$.
He found out that second order
phase transition occurs only if $N_A $$\geq$$ 9$ 
($N_B$$=$$3$ was taken).
After the described DSB, a broken $SU(3)$
appears, a kind of a coloron group with
gauge boson masses $M $$\sim$$ M_B$ ($\sim$$ {\Lambda}$). These
bosons then mediate at energies $\mu $$<$$ M$
four-quark interactions leading to ${\bar t} t$ condensation
and the DEWSB.

Triantaphyllou (1994),
and Blumhofer and Hutter (1997),
investigated possibilities of explaining dynamically
fermion family mass hierarchies {\em without\/} introducing
horizontal (plus vertical) gauge interactions,
in contrast to the models described in Sec.~\ref{RMUP5}.

Triantaphyllou (1994)
investigated a scenario within a TC$+$ETC framework.
In the scenario, the DS (gap) equation
allows several solutions which may correspond to the dynamical masses
of a three-generation sector of fermionic particles
(e.g., of up-type quarks). 
Each solution represents another
(local) minimum of the effective potential. This means
that fermions (of a given sector) from different generations
reside in different vacua.
CKM matrix could in
principle be calculated by considering instanton transitions
between different vacua. The familiar SM is then an effective
low energy description of this model.

Blumhofer and Hutter (1997),
on the other hand, investigated whether the DS
equation can possess such a fermion propagator
solution $i [{k \llap /}\!-\!B(k^2)]^{-1}$ which has several
(three) poles $k^2 $$=$$ m_n^2$ ($n$$=$$1,2,3$). They really found such
a possibility, with a phenomenologically acceptable
mass spectrum $m_n $$\approx$$ m_1 e^{(n-1) \alpha}$. 
A one-loop DS giving such a solution can be realized
in an unusual and peculiar theory containing a light
``hidden'' gauge boson sector with a $k^2$-dependent
coupling strength to the fermion, the origin of this 
$k^2$-dependence being predominantly mass threshold effects. 
Stated differently, the DS equation at one loop was
generated by exchange of such gauge bosons whose
masses are $\sim$$E_{{\rm ew}}$
or less, and these gauge bosons are ``hidden'' in the
sense that they manifest themselves in interactions only at the 
loop, not at the tree level. The authors showed also that
the fermion particle spectrum is stable. The three fermion
particles of a given sector (e.g., up-type quarks) have in
this scenario just {\em one\/} propagator 
$i [{k \llap /}\!-\!B(k^2) ]^{-1}$. 
The authors showed that at tree level this 
is equivalent to having the usual picture of
three fermions with three propagators 
$i [{k \llap /}\!-\!m_n]^{-1}$. Only at loop level the deviations
from the SM arise.

\section{Some phenomenological and theoretical aspects and questions}
\label{SPTAQ}
\setcounter{equation}{0}

\subsection{Phenomenological predictions}
\label{SPTAQ1}
\subsubsection{Phenomenology of general 
strong dynamics frameworks with DEWSB}
\label{SPTAQ10}
Chesterman and King (1992)
investigated how a relatively low
compositeness scale $\Lambda $$\stackrel{<}{\sim}$$ 10$ TeV
might be {\em directly\/} tested at the Superconducting Super
Collider (SSC) or CERN Large Hadron Collider (LHC).
They pointed out that the most promising channels are the
scattering processes $V_L V_L \to V^{\prime}_L V^{\prime}_L$
of the longitudinal components of EW gauge bosons ($V_L, V^{\prime}_L
= W^{\pm}_L, Z_L$). 
Such processes can be calculated by replacing 
$W^{\pm}_L$, $Z_L$ in amplitudes by the
(composite) Nambu--Goldstone bosons (NGB's)
${\cal {G}}^{(\pm)}$, ${\cal {G}}^{(0)}$. The resulting
errors are $\sim$$M_Z/M$, where
$M$$\equiv$$\sqrt{s}$ 
is the invariant mass of the gauge boson pair.
Thus, by probing such processes, compositeness of NGB's
can be tested. For example,
in the minimal (TSM) framework, composite structure of 
the NGB's is, according to (\ref{HD1}): 
${\cal {G}}^{(0)} $$\sim$$ {\bar t} {\gamma}_5 t$,
${\cal {G}}^{(+)} $$\sim$$ {\bar b} (1+ {\gamma}_5) t$.
The scalar sector of the MSM is described by 
the density ${\cal {L}}^{\sigma}(H, \pi)$ of the
Gell-Mann--Levy ${\sigma}$ model.
The authors parametrized deviations of NGB interactions 
from the MSM behavior by adding to 
${\cal {L}}^{\sigma}$ an additional nonrenormalizable
contact term
$\triangle {\cal {L}}^{\sigma}\!=\!- g^2\left[ ({\vec \pi} 
\partial^{\mu} {\vec \pi})\!+\!(H \partial^{\mu} H) \right]^2/
(4 {\Lambda}^{\!2})$,
where:
${\vec \pi} {\partial}^{\mu} {\vec \pi}\!=\!Z_L 
{\partial}^{\mu} Z_L\!+\!(W_L^{+} {\partial}^{\mu} 
W_L^{-}\!+\!\mbox{h.c.})$.
Here, $g^2/{{\Lambda}^{\!2}}$ is the TSM
four-quark parameter. They calculated cross-sections
$d{\sigma}/d M$ for the processes 
$p p\!\to\!W_L^{+}W_L^{-}\!\to\! W_L^{+} W_L^{-}$ and
$p p\!\to\!Z_L Z_L\!\to\!W_L^{+} W_L^{-}$,
and concluded that appreciable deviations can be found
at LHC (SSC) energies, as long as $\Lambda $$<$$ 2$ TeV (5 TeV).
These limits are not high when compared with
the CMS energies at LHC (SSC).
The results may hence seem to be somewhat disappointing.

Such results are in accordance with the
nature of the compositeness in, e.g., ${\bar t} t$ condensation.
As stressed by Miransky (1990, 1991),
Higgs and NGB's in the TSM-type models are
tightly bound states and their composite
structure can hence be discerned only at energies of probes ${\mu}$
close to ${\Lambda}$ and well
above $E_{{\rm ew}} $$\sim$$ 10^2$ GeV.
This is in stark contrast with 
the DSB in QCD (or DSB in a confining TC) where
bound states are less tight and their structure starts to be
seen at $\mu$$\sim$${\Lambda}_{{\rm QCD}}$$\sim$$10^{-1}$ GeV.
This is well below the ``effective'' cutoff ${\Lambda}$$\sim$$1$ GeV
where the nonperturbative and perturbative QCD meet.
This difference is reflected also in the fact that QCD is confining
(gluons remain massless despite light quark condensation),
while TSM-type models are not (TSM terms are usually 
assumed to be mediated by massive gauge bosons).
Miransky pointed out that this difference can be understood
also by looking at the amputated BS
wave function 
$\chi({\bar q}^2)$$=$$-i {\Sigma}({\bar q}^2)/F_{\pi}$
of the composite pseudoscalar, where
$\Sigma({\bar q}^2)$ is
the dynamical mass of the constituent fermion.
$\Sigma({\bar q}^2)$ in the asymptotic region 
($|{\bar q}| $$\gg$$ v $ in TSM,
$|{\bar q}| $$\gg$$ {\Lambda}_{{\rm QCD}} $ in QCD) falls off
as $\Sigma({\bar q}) $$\sim$$ {\bar q}^{\gamma_m -2}$, where
$\gamma_m$ is called anomalous dimension.
In the TSM-type of binding, $\gamma_m $$\approx$$ 2$ and 
$\chi({\bar q}^2)$ falls off very slowly 
(at least in the leading-$N_{{\rm c}}$); 
in QCD, $\gamma_m $$=$$ 0$ and $\chi({\bar q}^2)$
falls off fast for ${\bar q}^2 $$>$$ {\Lambda}^{\!2}_{{\rm QCD}}$.

\subsubsection{Phenomenology of ${\bar t} t$ condensation models}
\label{SPTAQ11}
Phenomenological aspects of 
${\bar t} t$ condensation models with color-octet
isodoublet composite scalars (cf.~Sec.~\ref{EWESG3}) 
were investigated by
Kundu, De and Dutta-Roy (1994),
and by Kundu, Raychaudhuri {\em et al.\/} (1994).
First work investigated $K$-${\bar K}$ and $B_d$-${\bar B}_d$ mixing, 
CP-violating ${\epsilon}$ parameter, and 
${\Gamma}(b $$\to$$ s \gamma)$. Contribution of
colored scalars $\chi^{\pm}$ to $K$-${\bar K}$ mixing
was shown to be small. $B_d$-${\bar B}_d$
mixing and ${\epsilon}$
yielded important constraints on two parameters of the
framework -- on mass $m_{\chi\pm}$ and on Yukawa 
parameter $g^{\prime}_t$ (of $\chi$ with $t_R$).
For $m_{\chi\pm} $$\sim$$ 10^2$ GeV, these constraints
gave an upper bound 
$g_t^{\prime}(m_t) $$\stackrel{<}{\sim}$$ 0.8$. For lower
$m_{\chi\pm}$, upper bound 
$(g_t^{\prime})_{{\rm max}}$ decreases. 
Such restrictive bounds on $g_t^{\prime}$
are still marginally acceptable
as to the mass $m_t$, because 
measured $m_t$ is rather high
($\approx$$180$ GeV). For example, the fixed-point analysis
of one-loop RGE's for $g_t$ and $g_t^{\prime}$
(cf.~Sec.~\ref{EWESG3}) yielded $g_t^{\prime}(m_t) $$\approx$$ 1.06$
for $m_t $$=$$ 150$ GeV, and $g_t^{\prime}(m_t) $$\approx$$ 0.78$
for $m_t $$=$$ 190$ GeV. The authors emphasized
that the analysis is accompanied by large 
uncertainties connected mostly to 
experimental uncertainties of $B_d$-${\bar B}_d$
mixing parameter $x_d$, of CKM parameter ratio
$q$$=$$s_{13}/s_{23}$, CP-violating CKM phase $\delta$,
and especially to large
theoretical uncertainty of the hadronic bag parameter
$f_B \sqrt{B_B}$. They showed that presence of $\chi^{\pm}$
enhances ${\Gamma}(b $$\to$$ s \gamma)$
and that the resulting lower bound
$(m_{\chi\pm})_{{\rm min}}$ is several
hundred GeV, compatible with constraints on $m_{\chi\pm}$ obtained
from $B_d$-${\bar B}_d$ and from the value
of ${\epsilon}$. Analysis of 
$R_b $$\equiv$${\Gamma}(Z $$\to$$ b{\bar b})/
{\Gamma}(Z $$\to$$ \mbox{hadrons})$
by Kundu, Raychaudhuri {\em et al.\/}
gave negative contribution of colored
scalars. This was regarded as unfavorable
at that time when the measured $R_b$ was above
the MSM value by about $3 {\sigma}$. However, newer data 
\cite{Dawson96}
show that the world average is  $R_b $$=$$ 0.2178
\pm 0.0011$, which is less than $2 {\sigma}$
above the MSM value.
Kundu, Raychaudhuri {\em et al.\/} imposed the requirement
that the (negative) new contribution not exceed
typical values ${\sigma} $$\approx$$ 0.001$, 
and obtained rather restrictive
(high) values $(m_{\chi\pm})_{{\rm min}} 
$$\stackrel{>}{\approx}$$ 500$-$900$ GeV, for
$m_t $$=$$ 160$-$180$ GeV.

$B_d$-${\bar B}_d$ 
mixing was also studied within a modified
minimal framework, by
Kimura {\em et al.\/} (1992).
Calculation was performed in quark loop
(bubble) approximation. The authors took special care
to modify the minimal (truncated) TSM model (\ref{TSM}) so
that phenomenological introduction of CKM mixing
remained consistent with gauge invariance. They
imposed the requirement 
that $W$ propagator have a pole at
$M_W $$\approx$$ 80$ GeV and that quark loop relation
(\ref{mt}) be satisfied. This led them to nowadays 
unacceptable value $m_t $$\approx$$ 300$ GeV,
and $\Lambda $$\sim$$ 10^6$ GeV.
Quark-loop modified propagators of $W^{\pm}$ and
of NGB's ${\cal {G}}^{\pm}$ in the
relevant box diagrams for $B_d$-${\bar B}_d$ mixing  
resulted in sizable 
modifications of the MSM Inami-Lim functions --
by $1$-$10$\%. The modification was larger for smaller
$m_t$. Due to uncertain values of the bag parameter
$f_B \sqrt{B_B}$, they refrained from discussing
the subject further.

Peccei, Peris and Zhang (1991), motivated by the case
of low energy QCD, considered the possibility that appreciable 
residual (nonuniversal) interactions between $t$
and the composite NGB's arise in the course of the DEWSB, thus 
leading to extra interactions between $t$ and gauge
fields. Large additional axial $Z {\bar t} t$ couplings would
generate additional radiative corrections to low energy physical
quantities. Using data from LEP and deep inelastic
scattering experiments at that time, the authors showed that
there was enough room in the parameter space for deviations
of up to 10\% 
for the strength of the axial $Z {\bar t} t$ coupling.

Zhang (1995) proposed a scenario where the constituent $t$
of the broken phase, arising after the DEWSB, is a topological
color soliton. This was motivated by a work of Kaplan (1991)
who had shown, in the framework of a Skyrme model, that
constituent quarks could be considered as color solitons
(``qualitons''). In the picture, the constituent top quark
is a valence (current) top quark surrounded by a deformed 
$\langle {\bar t} t \rangle_0$ background, where $t$ and
${\bar t}$ interact through exchanges of NGB's. Berger
{\em et al.\/} (1996) carried out calculations in this scenario, 
based on the color chiral $SU(3)_L\!\times\!SU(3)_R$ symmetric 
four-quark interaction represented by the second term (sum) on the
right of Eq.~(\ref{Fierz2n}) 
(${\alpha} = 1, \ldots, 8$). Soliton solutions
were obtained by embedding an $SU(2)$ ``hedgehog'' ansatz for the
top-pion color octet and minimizing the classical energy functional.
The authors showed that the top quark soliton can acquire
values which would lead to predictions for the production
cross-section ${\sigma}_{\bar t t}$ in accordance with the
data at the Collider detector at Fermilab (CDF).

\subsubsection{Phenomenology of coloron (topcolor) and TC2 frameworks}
\label{SPTAQ12}
Hill and Parke (1994)
studied in DSB schemes
the production cross-section 
${\sigma}({\bar p}p $$\to$$ {\bar t}t) $$\equiv$
$ {\sigma}_{\bar t t}$
and top quark distributions in 
${\bar t} t$ production at the Tevatron Collider, at the
center-of-mass (c.m.s.) energy $\sqrt{s} $$=$$ 1.8$ TeV. 
They investigated
vector color-singlet (cf.~Sec.~\ref{RMUP2})
and vector color-octet channels (coloron, cf.~Sec.~\ref{RMUP3}).
For both types of channels, 
they concluded that ${\sigma}_{\bar t t}$
could be more than doubled in comparison with the pure QCD case, 
if the compositeness scale is $\Lambda $$\stackrel{<}{\sim}$$ 1$ TeV. 
Lane (1995)
subsequently investigated, among other
models, predictions of a variant of the coloron (topcolor) model
introduced by Hill and Parke 
(cf.~end of Sec. \ref{RMUP3}).
He calculated invariant-mass distribution
$d{\sigma}_{\bar t t}/d {\cal {M}}_{\bar t t}$
and the c.m.s.~angular 
distribution $d{\sigma}_{\bar t t}/d \cos \theta$
of the top quark at the Tevatron at $\sqrt{s}$$=$$1.8$ TeV.
His results also imply that the coloron resonances, via the process
$q \bar q \to V_8 \to t \bar t$, 
may easily double ${\sigma}_{\bar t t}$
at $\sqrt{s} $$=$$ 1.8$ TeV.
Such an enlarged ${\sigma}_{\bar t t}$, in comparison with the
pure QCD case, may be in accordance with the evidence at the
Collider Detector at Fermilab (CDF).
Hill and Parke, and Lane, apparently
employed the lowest order calculation at the parton level.

Hill and Zhang (1995)
later investigated the two types of frameworks
(cf.~Sec.~\ref{RMUP2}, Sec.~\ref{RMUP3})
as to the process $Z$$\to$${\bar b} b$. They found out that
the topcolor variant introduced and 
applied by Hill and Parke predicts
a substantially increased 
$R_b\!\equiv\!{\Gamma}(Z$$\to$$b{\bar b})/
{\Gamma}(Z$$\to$$\mbox{hadrons})$
(by up to $2 {\sigma}$ from $R_b^{{\rm SM}}$) if coloron mass is
low: $M_B $$\stackrel{<}{\approx}$$ 600$ GeV.
However, in such a case ${\sigma}_{\bar t t}$
appears to be too large -- at least four times larger than
the standard QCD prediction. On the other hand,
$M_B $$\geq$$ 800$ GeV seems to be compatible with
${\sigma}_{\bar t t}$. In such a case, $R_b$ is 
within $1 {\sigma}$ of the MSM prediction. 
In the context, we mention that the measured $R_b$ 
world average has recently decreased
to values $R_b\!=\!0.2178$$\pm$$0.0011$ which are now about
$1.8 {\sigma}$ above the Standard Model value
(cf.~Dawson, 1996).

While Hill and Zhang (1995) considered only
contributions of colorons (top-gluons) to ${\triangle R}_b$
[$({\triangle R}_b)_{{\rm coloron}} $$>$$ 0$],
Burdman and Kominis (1997)
included in their analysis also top-pions ${\tilde \pi}^a$
which arise in TC2 frameworks (cf.~Sec.~\ref{RMUP6}).
They found out that ${\tilde \pi}$ contributions to
${\triangle R}_b$ are negative and can be dangerously strong
for low $m_{\tilde \pi} $$<$$ 300$ GeV and low
decay constants $f_{\tilde \pi} $$\approx$$ 60$ GeV.
However, Hill (1997)
pointed out that their results are quite sensitive to
the precise values of $m_{\tilde \pi}$ and $f_{\tilde \pi}$.
For example, for $f_{\tilde \pi} $$\approx$$ 120$ GeV, all
$m_{\tilde \pi}$'s are compatible with measured values of
$R_b$; the range $f_{\tilde \pi} $$\stackrel{>}{\approx}$$ 100$ GeV
and $m_{\tilde \pi} $$\stackrel{>}{\approx}$$ 300 $ GeV is also
compatible. He also argued that ${\epsilon}$
factor (${\epsilon} m_t$ is the small ETC-generated component
of $m_t$) in general gets enhanced by radiative corrections of the
topcolor $SU(3)_1$ and $U(1)_{Y1}$ sectors by a factor
$\sim$$10^1$, and therefore the fermion-loop-induced mass
$m_{\tilde \pi}$ ($\propto$${\epsilon}$) gets enhanced, too.

Chivukula, Cohen and Simmons (1996)
constructed a flavor-universal variant of the
coloron model of Hill and Parke (cf.~end of Sec.~\ref{RMUP3}
on the latter). 
With such a variant they were able to explain
recent results for ${\bar p} p$ collisions
at $\sqrt{s} $$=$$ 1.8$ TeV
from CDF 
\cite{Abeetal96}
which indicate that the inclusive cross-section for jets with
$E_T $$>$$ 200$ GeV is substantially higher than the one predicted
by QCD. The model does not appear to lead to
any contradiction with other experimental evidence.
In contrast to Hill and Parke, they assigned
{\em all\/} quarks to triplet representations of the strong
$SU(3)_2$ group appearing in the full coloron group
$SU(3)_1\!\times\!SU(3)_2$.
Subsequently, Simmons (1997)
continued the analysis of the flavor-universal coloron
model and showed that
constraints from searches for new particles
decaying to dijets and from measurements of the
electroweak $\rho$ parameter imply that
colorons in the framework must have masses
$M_B $$\stackrel{>}{\approx}$$ 900$ GeV. 

Kominis (1995)
investigated questions of 
flavor-changing neutral currents (FCNC's) in
Hill's topcolor I model (cf.~Sec.~\ref{RMUP6},
topcolor assisted technicolor -- TC2).
He pointed out that the model, 
if without additional constraints,
may lead to dangerously light composite scalars
of the $b$-type: ${\Phi}_d $$\sim$$ {\bar Q}_L b_R$
[$Q_L $$\equiv$$ (t_L, b_L)^T$]. This may occur when the
strong $U(1)_{Y1}$ is not strong enough
to push the four-quark parameter
${\kappa}_b^{{\rm (eff.)}} $$\stackrel{<}{\sim}$$ 
{\kappa}_{{\rm crit.}}$ substantially below 
${\kappa}_{{\rm crit.}}$. Then
isodoublet ${\Phi}_d$ ($\langle {\Phi}_d \rangle_0 $$=$$ 0$)
may become dynamical 
(detectable) in the sense that its mass is substantially
below the cutoff ${\Lambda} $$\approx$$ 1$ TeV 
[-- cf.~also (\ref{kapbbelow})]. 
Such scalars can become relatively light ($\sim$$ 10^2$ GeV)
when ${\kappa}_b^{{\rm (eff.)}}$ is close to
${\kappa}_{{\rm crit.}}$ from below --
then they are dangerous because their exchanges result in 
a too strong $B$-${\bar B}$ mixing (an FCNC effect). 

Effects of these scalars were subsequently
taken into account by
Buchalla, Burdman, Hill and Kominis (1996)
who investigated two types of topcolor scenarios
(topcolor I, II), as already mentioned in
Sec.~\ref{RMUP6}. Substantial part of their work was
devoted to low-energy phenomenological implications
of these models. They showed that these frameworks offer
a natural possibility to suppress the dangerous
FCNC's, in particular the dangerous $B$-${\bar B}$
mixing contributions induced by exchanges of relatively light
composite scalars. They showed that the suppression 
emerges through a chiral-triangular
texture of the mass matrices. This texture is a consequence
of the fact that in these models gauge quantum numbers
distinguish generations. Topcolor I models contain
two $U(1)$ group factors and consequently a
massive isosinglet color-singlet $Z^{\prime}_{\mu}$.
They showed that
exchanges of $Z^{\prime}_{\mu}$ in topcolor I can
lead to substantial deviations in several semileptonic processes:
$\Upsilon(1S) \to \ell^{+}\ell^{-}$; $B_s \to {\ell}^{+} {\ell}^{-}$;
$B \to X_s {\ell}^{+} {\ell}^{-}$; $B \to X_s {\nu}{\bar \nu}$;
$K^{+} \to {\pi}^{+} {\nu} {\bar \nu}$. Future experiments
involving these processes could give some important clues
about the nature of new physics at the onset scale
${\Lambda} $$\sim$$ M_{Z^{\prime}} $$\sim$$ 1$ TeV, once they
show any significant deviation from the SM values.      
On the other hand, topcolor II gives no novel effects
in semileptonic processes, and only a few such effects in
low-energy nonleptonic processes, e.g., in $D^0$-${\bar D}^0$ mixing.
Since low-energy nonleptonic effects are hard to
disentangle from SM physics,
the authors suggested that high energy experiments
($\mu $$\stackrel{>}{\sim}$$ E_{{\rm ew}}$)
are the most promising ones to look for the new physics.

Chivukula and Terning (1996)
performed global fits to precision data for three
examples of natural TC2 scenarios proposed by
Lane and Eichten (Lane and Eichten, 1995; Lane, 1996
-- cf.~Sec.~\ref{RMUP6}), 
and came to the conclusion that
mass of new massive $Z^{\prime}$ bosons
[originating from the 
$U(1)_{Y1}\!\times\!U(1)_{Y2}\!\to\!U(1)_Y$ breaking] 
must be larger than roughly 2 TeV. 
Recently, Su, Bonini and Lane (1997) came to
the conclusion that TC2 models proposed by
Lane (1996) must have in general an even stronger
fine-tuning\footnote
{Such fine-tuning was regarded
by the authors as unacceptable.} 
$m_t $$\ll$$  {\Lambda}$ as a result of stringent bounds 
coming from data on Drell-Yan processes at the Tevatron.
Hill (1997)
stressed that the analysis of Chivukula and Terning, and of 
Chivukula, Dobrescu and Terning (1995) 
(see Sec.~\ref{RMUP6}), should be extended
to include, in addition to heavy
gauge bosons, also top-pions and possibly
other low mass bound states.

Eichten and Lane (1996a)
pointed out that conventional multiscale technicolor models (TC)
in general result in light technipions ${\tilde \pi}_T$ of mass
$\simeq$$ 100$ GeV and lead in such scenarios
to large decay rates $\Gamma(t $$\to$$ {\tilde \pi}_T^{+} b)$,
due to the large coupling of ${\tilde \pi}_T$ in these decays:
$m_t \sqrt{2}/F_{{\tilde \pi}_T}$, where the decay
width $F_{{\tilde \pi}_T} $$\approx$$ v $$\approx$$ 246$ GeV. 
The authors stressed that topcolor assisted technicolor scenarios
(TC2) can in general suppress this decay to phenomenologically
tolerable levels, since in such scenarios the relevant
coupling is substantially smaller: ${\epsilon} m_t \sqrt{2}/
F_{{\tilde \pi}_T}$, where ${\epsilon} m_t$
$\approx$ (0.01$-$0.1)$m_t$ is the ETC-induced part of $m_t$
(cf. Sec.~\ref{RMUP6}).

Balaji (1997)
investigated further this question, within a
natural TC2 introduced previously by
Lane and Eichten (1995) (cf. Sec.~\ref{RMUP6}). 
Top-pions ${\tilde \pi}_t$
(they originate from ${\bar t} t$ condensation
in TC2 models; cf.~Sec.~\ref{RMUP6})
in the framework are found to have 
$m_{{\tilde \pi}_t}$$>$$200$ GeV.
If there were no mixing between the heavy ${\tilde \pi}_t$ 
and the light technipion ${\tilde \pi}_T$,
there would be no decays $t$$\to$${\tilde \pi}_t^{+} b$.
However, there is mixing due to
the (walking) ETC, leading to
existence of pseudo-NGB's
which are mass eigenstates with
masses possibly lower than $m_t$. This opens a
potentially dangerous possibility that $t$
could decay into such a pseudo-NGB (plus $b$) 
at unacceptably high rate. Balaji
showed that the decay rate in such frameworks
can be within phenomenologically acceptable limits
due to the large mass of ${\tilde \pi}_t$'s and weak mixing
of them with ${\tilde \pi}_T$'s.

Eichten and Lane (1996b),
and Lane (1997),
discussed possible future direct TC2 signatures connected with the 
scenario's topcolor part at the Tevatron and at
the Large Hadron Collider (LHC) -- the detection of
top-pions ${\tilde \pi}_t^{+}$, of massive
colorons ($M $$\approx$$ 0.5$-$1$ TeV),
and color-singlets $Z^{\prime}_{\mu}$
($M_{Z^{\prime}} $$\approx$$ 1$-$3$ TeV). 
If $m_{{\tilde \pi}_t}$$\stackrel{>}{\sim}$$150$ GeV, 
${\Gamma}(t $$\to$$ {\tilde \pi}_t^{+} b)$ is
phenomenologically acceptable 
and can be sought with high (moderate) luminosity 
at the Tevatron (LHC).
Colorons may appear as resonances in $b {\bar b}$
and $t {\bar t}$ production. Heavy $Z^{\prime}_{\mu}$, due 
to its strong couplings to fermions, can lead to an
excess of jets at high $E_T$ and at high $\sqrt{s}$.

Wells (1997),
on the other hand, discussed prospects for detecting
signatures for the scalar degrees of freedom relevant to
the EWSB (e.g., originating from
technicolor), in scenarios where ${\bar t} t$ condensation
is mainly a spectator to the EWSB 
($F_t $$\equiv$$ f_t \sqrt{2} $$\ll$$
v $$\approx$$ 246$ GeV), for example in TC2 scenarios.
He parametrized such scalar d.o.f.'s by
an unspecified, possibly composite Higgs field
$h^0_{{\rm ew}}$: 
$\langle h^0_{{\rm ew}} \rangle_0 $$\approx$$ v$. 
Since in such scenarios
most of $m_t$ comes from ${\bar t} t$ condensation, 
Yukawa coupling of $h^0_{{\rm ew}}$ 
to $t$ is weak ($g_t $$\ll$$ 1$). 
He argued that a possible detection of 
$h^0_{{\rm ew}}$ at the LHC
comes largely from the production cross-section due to
top quark Yukawa processes, and may thus be very difficult 
because $g_t $$\ll$$ 1$. On the other hand, the search capabilities
for such a
$h^0_{{\rm ew}}$ 
at the Tevatron and LEPII should be similar to
those for the MSM Higgs, since the searches there
do not rely mainly on $g_t$.

Del\'epine, G\'erard, Gonz\'alez Felipe and Weyers (1997)
considered a model similar to a simplified version
of topcolor I of Hill (1995) and Buchalla {\em et al.\/} (1996).
They assumed that strong $SU(3)_1$ of this framework
involves also the strong topcolor CP phase $\theta$
arising from topcolor instantons (this possibility
was also mentioned by Buchalla {\em et al.\/}, 1996).
Their framework involves the usual scalar sector of topcolor I
-- an elementary Higgs doublet $H$, and two composite
doublets ${\tilde \phi}_1 $$\sim$$ {\bar t}_R {\Psi}_L$,
$\phi_2 $$\sim$$ {\bar b}_R {\Psi}_L$. In contrast to
topcolor I, the model doesn't contain a strong $U(1)_{Y1}$
which would discriminate between $b_R$ and $t_R$
and would thus prevent a nonzero $\langle {\bar b} b \rangle_0$
from appearing. Thus they allowed $\langle \phi_2 \rangle_0
$$\not=$$ 0$. They showed that in such scenario
there is a possibility that $\theta$-term triggers simultaneously
the mass hierarchy ($0$$ \not=$)$m_b $$\ll$$ m_t$ {\em and\/}
a large CP-violating phase in the CKM matrix required
phenomenologically by $K$-${\bar K}$ physics.
They did not investigate the dynamical role of
condensation scale ${\Lambda}$ and the question of
whether their framework can account for the full EWSB.

\subsection{Some theoretical issues and questions of
${\bar t}t$ condensation frameworks}
\label{SPTAQ2}
There are several outstanding theoretical questions about the
NJLVL-types of ${\bar t} t$ condensation. In the
early investigations of ${\bar t} t$
condensation (1989-1991), 
these questions were either left unanswered
or weren't addressed.
Some were later clarified, others
remain open. Some of these problems, e.g., 
effects of the next-to-leading terms
in $1/N_{{\rm c}}$ expansion 
(cf.~Sec.~\ref{NTLEE}), and
of higher than 6-dimensional quark contact operators
(cf.~Sec.~\ref{CCVAMF4}), were discussed in
previous parts of this review. Below, we briefly
discuss some other theoretical points.

Using a simple spherical covariant cutoff
in the quark loop (bubble) approximation of the TSM model,
a momentum routing ambiguity appears, as noted by
Willey (1993).
This is shown in Fig.~\ref{rmd12f} 
for the case of a four-point function.
\begin{figure}[htb]
\mbox{}
\vskip4.9cm\relax\noindent\hskip.0cm\relax
\includegraphics{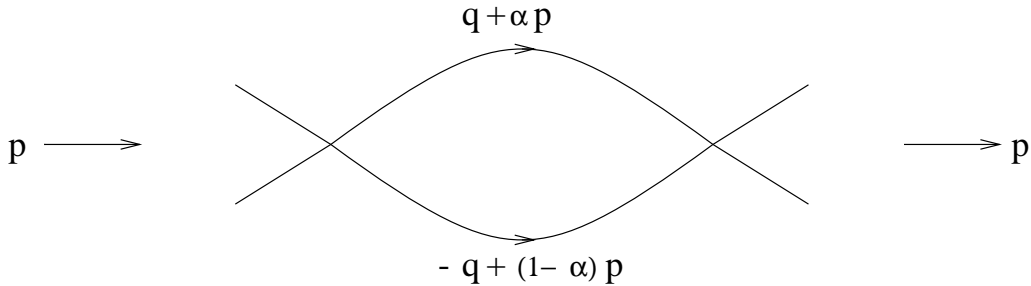} \vskip0.4cm
\caption{\footnotesize Momentum routing ambiguity in a
quark loop. Routing parameter $\alpha$
is in principle arbitrary, it influences predictions
for $M_H/m_t$ when spherical covariant cutoff is employed
for the quark momenta.}
\label{rmd12f}
\end{figure}
When the otherwise arbitrary
routing parameter ${\alpha}$ is set to zero,
the pole of the scalar channel in the bubble sum of such
four-point functions gives
the known TSM result ${\bar p}^2_{\mbox
{\scriptsize pole}} $$=$$ 4 m_t^2$, i.e., $M_H $$=$$ 2 m_t$.
He pointed out that $M_H/(2 m_t)$ depends on the 
routing parameter ${\alpha}$ and can thus acquire any value. 
Therefore, he argued, the TSM model (\ref{TSM}), 
or any similar type of NJLVL model, is not predictive.

Gherghetta (1994)
investigated whether there can be compatibility of the
$SU(2)_L\!\times\!U(1)_Y$ gauge invariance with
the necessary existence of quadratic (${\Lambda}^{\!2}$) terms
in the TSM gap equation in quark loop (bubble)
approximation. He showed that the two requirements are
incompatible when using the usual
spherical covariant cutoff for
the (quark) loop momenta. Gauge invariance is
destroyed by ${\Lambda}^{\!2}$-terms in propagators of EW gauge
bosons. He pointed out that BHL (1990),
in their calculation of various relations
in bubble approximation, used
somewhat inconsistently spherical covariant
cutoff in quark loops for the gap equation
and dimensional regularization in quark loops
for propagators of EW gauge bosons.
Dimensional regularization ignores
${\Lambda}^{\!2}$-terms.
The gap equation without ${\Lambda}^{\!2}$-terms does not
lead to DSB at quark loop level.
Gherghetta then showed, motivated partly by work 
of Nambu and Jona-Lasinio (NJL, 1961),
that use of dispersion relations to regulate
quark loops leads to a satisfactory solution of
this problem. Such a regularization maintains gauge invariance
and does not possess any routing ambiguity.
Therefore, it also solves
the routing ambiguity issue raised
by Willey.
As a matter of fact, such regularization
had already been used by NJL (1961) to investigate poles
of various channels in the four-quark functions.
Gherghetta also applied this same regularization
procedure to the $W$ and $Z$ propagators and 
showed that they then don't have any routing ambiguity and
don't contain any ${\Lambda}^{\!2}$-terms, 
i.e., they respect gauge invariance.
It appears, however, that the gap equation itself
cannot be regularized by this method. Gherghetta therefore
took the position of NJL (1961) who had declared that the requirement
of masslessness for the pole in the pseudoscalar channel
of the four-quark function is the gap equation.
Stated differently, imposition of the
Goldstone theorem at the quark loop level is then regarded
to be the gap equation. Interestingly, this led to a relation
very similar to the usual gap equation in the
spherical covariant cutoff approach (\ref{gaplead}). 
The two relations in fact become identical in the 
high-${\Lambda}$ limit $m_t/\Lambda \!\to\! 0$.

A related question is: when beyond the 
leading-$N_{{\rm c}}$
(quark loop) approximation, can we find a regularization
procedure which regularizes all quark loop
momenta at the leading-$N_{{\rm c}}$ 
and at the next-to-leading (NTL)
order in a mutually consistent manner, 
{\em and\/} possesses no routing ambiguities? It has been
pointed out (Cveti\v c, 1997) 
that the proper time regularization procedures,
when applied within the approach of 
effective potential ($V_{{\rm eff}}$),
satisfy these two requirements. Such procedures
are invariant under translation in the quark momentum space, 
and can probably be applied even when transverse
gauge bosonic d.o.f.'s are included. On the other hand,
it is unclear how to regularize NTL contributions to
$V_{{\rm eff}}$ and to the gap equation  
with the method of dispersion relations.

Chivukula, Golden and Simmons (CGS, 1993)
studied the question of whether and when
we can have in models of DSB (those containing more than one
quartic self-coupling of scalars, like
nonminimal ${\bar t}t$ condensation and strong ETC frameworks) 
a fine-tuning leading to hierarchy
$v/{\Lambda} $$\ll$$ 1$, where $v$ is VEV of the composite scalar
at low energies $\mu $($\sim$$ v $$\ll$$ {\Lambda}$): 
$v/\sqrt{2} $$\equiv$$ \langle {\Phi}_0^{(\mu)} \rangle_0$.
A point emphasized by them was that quantum fluctuations
may drive the transition between the symmetric and
the broken phase to become sudden and violent (``first order'').
Such a transition in DSB scenarios represents
the following behavior: VEV $v(\mu)/\sqrt{2} $$\equiv$$
\langle {\Phi}_0^{(\mu)} \rangle_0$  and the corresponding 
$m_f^{{\rm dyn.}}(\mu)$ of fermions at low energies 
$\mu $ ($\ll$$ {\Lambda}$) experience 
a sudden (discontinuous) jump from zero 
to large values $\sim$$ {\Lambda}$
when the corresponding four-fermion parameters
${\kappa}_f(\Lambda)$ of the 
(NJLVL-type) theory at high energy ${\Lambda}$
cross the critical values from below.
This phenomenon, although not present in the minimal
framework, may occur in
theories which have more than one quartic self-coupling
of composite scalars at low $\mu$.
This would include, for example, also NJLVL-type of effective models
leading to composite two-Higgs doublet scenarios discussed
in Sec.~\ref{EWESG1}. CGS considered specifically an effective
composite (Ginzburg-Landau) theory with
chiral symmetry 
$U(N_{{\rm f}})_L\!\times\!U(N_{{\rm f}})_R$, i.e., with
$N_{{\rm f}}$ left- and right-handed fermion flavors 
${\Psi}_L^{(j)},\!{\Psi}_R^{(j)}$,
($j$$=$$1,$$\ldots,$$N_{{\rm f}}$) 
in an $N_{{\rm c}}$-dimensional representation and with the 
(${\overline N}_{{\rm f}}, N_{{\rm f}}$)
order parameter ${\Phi}_{ij}$$\sim$${\overline {\Psi}}_R^{(j)} 
{\Psi}_L^{(i)}$
\begin{eqnarray}
{\cal {L}}^{(\mu)} &= &
{\overline {\Psi}} i {{\partial} \llap /} {\Psi}
+ \frac{\pi y(\mu)}{N_{{\rm f}}^{1/2} }
\left( {\overline \Psi}_L {\Phi} {\Psi}_R + \mbox{ h.c.} \right)
- M^2(\mu) \mbox{tr}({\Phi}^{\dagger} {\Phi}) 
\nonumber\\
&& + \mbox{tr}(\partial^{\nu} 
{\Phi}^{\dagger} {\partial}_{\nu}{\Phi} )
-\frac{\pi^2}{3} \frac{\lambda_1(\mu)}{N_{{\rm f}}^2}
\left( \mbox{tr} {\Phi}^{\dagger} {\Phi} \right)^2 
-\frac{\pi^2}{3} \frac{\lambda_2(\mu)}{N_{{\rm f}}}
 \mbox{tr} ({\Phi}^{\dagger} {\Phi} )^2 \ .
\label{chivukula1}
\end{eqnarray}
Superscripts $(\mu)$ at the fields, 
where $\mu$ is a finite effective UV cutoff, are omitted.
Such a theory can arise from a flavor-democratic NJLVL
term at $\mu $$=$$ {\Lambda}$
\begin{equation}
{\cal {L}}^{(\Lambda)}_{4{\rm f}} = 
{\overline {\Psi}} i {{\partial} \llap /} {\Psi}^ +
\frac{\alpha}{{\Lambda}^{\!2}}
({\overline \Psi}_L^{(i)} {\Psi}_R^{(j)})
({\overline \Psi}_R^{(j)} {\Psi}_L^{(i)}) \ .
\label{UU4f}
\end{equation}
Namely, application of equations of motion 
for ${\Phi}$ and ${\Phi}^{\dagger}$ to the
density consisting of the first line in (\ref{chivukula1}),
i.e., when ${\Phi}$ is an auxiliary field (at $\mu $$=$$ {\Lambda}$),
leads to ${\Phi}_{ij} $$\propto$$ 
{\overline {\Psi}}_R^{(j)} {\Psi}_L^{(i)}$
and to the NJLVL term (\ref{UU4f}).
The other terms on the right of 
(\ref{chivukula1}) appear after inclusion of
quantum effects in energy interval $(\mu,\Lambda)$ and after an
appropriate rescaling of composite
fields ${\Phi}$ so that the induced kinetic energy term of ${\Phi}$ 
has factor $1$ in front of it. 

For large hierarchy, we need $\langle {\Phi}^{(\mu)} \rangle_0
$$\sim$$ E_{{\rm ew}}$, i.e., 
$\langle {\Phi}^{(\mu)} \rangle_0 $$\ll$$ {\Lambda}$
at $\mu $$\sim$$ E_{{\rm ew}}$. 
CGS pointed out that such a
prediction can be spoiled by Coleman-Weinberg phenomenon
\cite{ColemanWeinberg73}.
Specifically, when CGS calculated trajectories
of ${\lambda}_1(\mu)$ and ${\lambda}_2(\mu)$ by one-loop RGE's 
[taking $N_{{\rm f}} $$=$$2$, $N_{{\rm c}} $$=$$ 3$,
and Yukawa parameter $y(\mu)$$=$$\mbox{const.}$], 
they found out that for initial
choices of small ${\lambda}_1(\Lambda)$ 
and large ${\lambda}_2(\Lambda)$
these parameters evolve very quickly and, after
a small change in $\mu$ [$\ln(\Lambda/\mu) $$\sim$$ 1$], 
acquire such values that the VEV $\langle {\Phi} \rangle_0$ 
changes from zero to a large value $\sim$$ \mu$.
Stated differently, quantum fluctuations don't allow a fine-tuning
of parameters at high energies $ {\Lambda}$ 
so as to obtain a small nonzero VEV 
$\langle {\Phi}^{(\mu)} \rangle_0
$$\sim$$ E_{{\rm ew}} $$\ll$$ {\Lambda}$ at low energies
$\mu $$\ll$$ {\Lambda}$.
Since the initial values for 
$\lambda_i(\Lambda)$ ($i$$=$$1,2$) were chosen
so as to correspond approximately to the compositeness condition
at $ {\Lambda}$, the authors argued that the 
discussed dynamical framework
does not allow large hierarchies $E_{{\rm ew}}/{\Lambda}
$$\ll$$ 1$. They also argued that the problem may
appear generally in models of the DEWSB with more than one
${\Phi}^4$-coupling.

Since coupling parameters at ${\Lambda}$
are large, several authors subsequently
performed calculations in the model (with $N_{{\rm f}}
$$=$$2$, $N_{{\rm c}}$$=$$3$) aimed at improving the
(one-loop) perturbative approach of CGS, primarily by employing
various nonperturbative approximations in the energy region
$\mu$$\sim$${\Lambda}$
(Shen, 1993; Bardeen, Hill and Jungnickel, 1994; 
Clark and Love, 1995; Khlebnikov and
Schnathorst, 1995). 

Bardeen {\em et al.\/} (1994) calculated
in the NJLVL $U(2)_L\!\times\!U(2)_R$ framework (\ref{UU4f})
the explicit leading-$N_{{\rm c}}$ solution.
They applied this solution [${\lambda}_1(\mu) $$=$$0$,
${\lambda}_2(\mu)$$=$$16/\ln({\Lambda}/{\mu})$,
$y^2(\mu)/{\lambda}_2(\mu) $$=$$ 1/3$] to a nonperturbative
region (logarithmically) close to ${\Lambda}$,
and matched it onto the perturbative two-loop RGE's
at a scale $\mu_i$ where the perturbative
approach is about to break down. They chose
$\mu_i/{\Lambda} $$\approx$$ 0.05$, although
the results were reasonably insensitive to the precise 
value of this ratio. Yukawa parameter was
evolving, too. They showed that Coleman-Weinberg
instabilities occur at substantially lower energies
${\mu}_{{\rm trans.}}$ already as a result
of the leading-$N_{{\rm c}}$  
compositeness condition alone. Further, 
inclusion of two-loop effects in RGE's (for
${\mu} $$<$$ {\mu}_i$) decreases 
${\mu}_{{\rm trans.}}$ by several orders
of magnitude, and inclusion of QCD in RGE's
tends to completely eliminate such
${\mu}_{{\rm trans.}}$. Therefore,
the authors argued that the framework in general
does allow fine-tuning hierarchies for the
NJLVL-motivated choices of the bare parameters
[i.e., for the ``strong Yukawa coupling'' choice:
$y^2({\Lambda})/{\lambda}_2({\Lambda}) $$\sim$$ 1$].

Shen (1993) investigated the case of negligible
Yukawa parameter ($y^2 $$=$$ 0$, thus ignoring fermions), 
by employing a lattice Monte-Carlo technique to deal
with the nonperturbative compositeness region.
He found out that the framework turns first order
and hence does not allow hierarchies $v/{\Lambda} $$\ll$$ 1$.
Clark and Love (1995), on the other hand, investigated
the case of weak but non-negligible bare Yukawa
parameters [$y^2({\Lambda})/{\lambda}_2({\Lambda})
$$\sim$$ 10^{-1}$] -- i.e., the case intermediate
between the lattice bosonic case of Shen and
the (relatively) strong Yukawa coupling case of 
Bardeen {\em et al.\/}. Clark and Love employed
a nonperturbative continuous Wilson renormalization
group equation approach (WRGE) near ${\Lambda}$,
including the chiral fermions. Since the WRGE formalism is
in general very complicated, they worked in a
so called local action approximation (which ignores
anomalous dimensions and derivative interactions),
neglected the operators higher than bilinear in 
fermion fields, and restricted
themselves to a fixed Yukawa parameter.
In general, they found out that the phase transition
turns first order near ${\Lambda}$
and does not allow hierarchies
$v/{\Lambda} $$\ll$$ 1$. Further, Khlebnikov and
Schnathorst (1995) considered the choice of very
(infinitely) large $y$, employed expansion
in inverse powers of $y$, and found out that
the model resembles closely the case of purely
bosonic theory (cf.~Shen, 1993) and that it
consequently exhibits also first order phase transitions,
i.e., the appearance of Coleman-Weinberg instability which does
not allow a hierarchy ${\Lambda} $$\gg$$ E_{{\rm ew}}$.

All in all, the mentioned investigations of the
nature of phase transitions in the chiral
$U(2)_L\!\times\!U(2)_R$ model suggest that in
models involving more than one ${\Phi}^4$-coupling
the appearance of Coleman-Weinberg (quantum) effects in general
tends to drive the transition first order and
thus to prohibit hierarchies ${\Lambda}/v $$\gg$$ 1$,
except possibly in models where new physics above
${\Lambda}$ results primarily in NJLVL-type 
of effective interactions [e.g., in (\ref{UU4f}); in
nonminimal NJLVL ${\bar t} t$ condensation;
in some strong ETC models].
These investigations rely on RGE 
analyses or on other RG-related
nonperturbative methods, and in one case partly on
a {\em leading\/}-$N_{{\rm c}}$
approximation. Such methods imply that
the minimal (TSM) ${\bar t}t$ condensation model,
now generally believed to be phenomenologically
untenable, does not exhibit Coleman-Weinberg instabilities
and its DSB phase transition is second order.
It is interesting that the latter conclusion
for the TSM is also suggested by
pure $1/N_{{\rm c}}$ expansion approach
at the next-to-leading level (cf.~Sec.~\ref{NTLEE5}),
although this approach is essentially different
from the RG-related ones.

Martin (1992a)
investigated the following problem which we are facing when
dealing with NJLVL four-fermion interactions: For any given
(set of) NJLVL term(s) with
given coupling strength(s), how can we find out which 
fermions actually participate in condensation, and
what kind of composite scalars with nonzero VEV (DSB)
such a condensation would lead to?
He applied two approaches to this problem:
\begin{itemize}
\item
A direct comparison of the strengths of relevant
attractive four-fermion channels (cf. Sec.~\ref{RMUP2}) in given
NJLVL Lagrangian density. For this, care must be given to normalizing
appropriately the four-fermion operators [for example
${\rho}^{(\beta)}$ in (\ref{Fierz2n})-(\ref{normrho})].
\item
Regarding a corresponding full renormalizable underlying gauge
theory. The two fermions (left-handed two-component Weyl
fermions) involved in condensation, and the resulting condensate,
transform in this gauge theory in irreducible 
representations $R_1$, $R_2$ and $R_3$, respectively.
Thus, a decomposition of direct product 
$R_1$$ \times$$ R_2$ into a direct sum holds: 
$R_1$$ \times$$ R_2 $$\equiv$$ R_3 $$+$$ \cdots$. 
In single-boson-exchange approximation, the condensate appears
in the most attractive channel $R_3$ for which the difference
$V$$=$$C_3$$-$$C_1$$-$$C_2$ is the most negative
($C_j$ is the quadratic Casimir invariant for $R_j$).
The approach is based on
work by~\citeasnoun{RabyDimopoulosSusskind80}.
\end{itemize} 

Dudas (1993)
focused on a somewhat related question.
He investigated an $SU(2)$-invariant NJLVL model
with two $SU(2)$ doublets $\psi_1$ and $\psi_2$
\begin{equation}
{\cal {L}} = {\bar \psi}_1 i {\partial \llap /} {\psi}_1
+ {\bar \psi}_2 i {\partial \llap /} {\psi}_2
+ \frac{g}{4 {\Lambda}^{\!2}} ( {\bar \psi}_1 {\tau}_{\alpha} {\psi}_2)
( {\bar \psi}_2 {\tau}_{\alpha} {\psi}_1) \ ,
\label{dudas1}
\end{equation}
where ${\tau}_{\alpha}$ ($\alpha$$=$$1,2,3$) are Pauli matrices.
He investigated in which way fermions would
condense and result in nonzero VEV's (DSB). 
The model possesses the symmetries: $SU(2)$; $U(1)_A$ 
[$\psi_1\!\to\!e^{i \beta \gamma_5} 
\psi_1,\!\psi_2\!\to\!e^{-i \beta \gamma_5}\psi_2$];
$U(1)_1$ [$\psi_1\!\to\!e^{i \beta}\psi_1,\!\psi_2\!\to\!\psi_2$];
$U(1)_2$ [$\psi_1\!\to\!\psi_1,\!\psi_2\!\to\!e^{i \beta}\psi_2$].
He applied the method of effective potential ($V_{{\rm eff}}$), 
at quark loop level, and made two choices for the auxiliary field:
$\varphi^{(\alpha)} $$\propto$$ {\bar \psi}_2 {\tau}_{\alpha} \psi_1$,
and $(\phi_1,\!\phi_2,\!\phi_3)\!\propto\!({\bar \psi}_2 
{\psi}_1,\!{\bar \psi}_2{\psi}_2,\!{\bar \psi}_1{\psi}_1)$. 
By minimizing the corresponding $V_{{\rm eff}}$, he obtained
in the first case a VEV 
$\langle {\bar \psi}_2 {\tau}_3 {\psi}_1 \rangle_0 $$\not=$$ 0$,
which breaks $SU(2)\!\times\!U(1)_{1-2}$, and in the second
case VEV's $\langle {\bar \psi}_1 {\psi}_1 \rangle_0 $$=$$
- \langle {\bar \psi}_2 {\psi}_2 \rangle_0 $$\not=$$ 0$, which
break $U(1)_A$. By comparing the energy densities at the
minimum,\footnote{
The VEV-independent undetermined additive
constants in both energy densities were fixed so
as to give zero energy density for zero VEV's.} 
he concluded
that the first case gives a more negative value and is thus
the actual vacuum. Since this vacuum doesn't break $U(1)_A$, 
fermions remain massless in this vacuum. 
Dudas applied also the Dyson--Schwinger
(DS) equation for the fermionic self-energy $\Sigma$
in quark loop approximation. He could reproduce in this
way the second case of DSB, but not the first one.
This point may appear to be disconcerting, 
in view of the fact that precisely the first case was
the actual energetically preferred DSB scenario. Concerning
this point, Dudas
solely mentions that the actual DSB cannot be obtained
by the DS approach because there is no $\psi_1$-$\psi_2$
mixing at the tree level in the Lagrangian density. However, it 
seems important to emphasize that the DS approach has no
inherent deficiencies when compared with the $V_{{\rm eff}}$
approach. Even more so, it is even more powerful than the 
latter one, once we go beyond quark loop approximation,
because it gives us information on the momentum dependence
of the dynamical quark mass $\Sigma({\bar p}^2)$ in an
inherently nonperturbative way (cf.~Sec.~\ref{MDSPS}). The reason
why the DS approach didn't give the vacuum in the model
lies in the fact that in the specific vacuum the fermions
remain massless, while the DS approach searches for
nonzero dynamical masses of fermions.

\subsection{Question of suppression of ${\Lambda}^{\!2}$-terms 
in the composite Higgs self-energy}
\label{SPTAQ3}
Blumhofer (1994, 1995)
investigated the question of whether radiative ${\Lambda}^{\!2}$
contributions that appear in the coefficient of the Higgs
${\Phi}^{\dagger} {\Phi}$-term (Higgs self-energy) can get
suppressed in frameworks of ${\bar t} t$ condensation.
The MSM is plagued by such terms already at one-loop level, where
$\Lambda$ is a large UV cutoff of the theory.
It is some-times postulated that these contributions cancel
(Decker and Pestieau, 1979, 1989; Veltman, 1981; Nambu, 1988).
This condition, however, requires an unnatural fine-tuning of
otherwise independent masses of particles in the loops
($t,\!b,\!\tau,\!\ldots$; 
${\cal {H}},\!{\cal {G}}^0,\!{\cal {G}}^{\pm}$; 
$Z,\!W^{\pm}$). It is not enforced by any known
symmetries; and it is scale dependent
(Al-sarhi, Jack and Jones, 1990; Chaichian, Gonzalez Felipe
and Huitu, 1995).

The same type of ${\Lambda}^{\!2}$ contributions appear also
in the TSM model (\ref{TSM}), where $\Lambda$ has the meaning
of compositeness scale.
Blumhofer raised the question of whether the {\em dynamics\/} of 
the TSM (\ref{TSM}), without invoking any
additional symmetry, leads to cancellation or at 
least a strong suppression of such terms in Higgs self-energy.
This question is motivated in a picture in which the underlying
physics is far away  (e.g., $\Lambda $$\sim$$ 
{\Lambda}_{{\rm GUT}}$), if we
want to avoid an excessive {\em ad hoc\/} fine-tuning of
parameters. Composite scalars in this framework 
have the following form
of propagators at quark loop level (leading-$N_{{\rm c}}$ with
no QCD):
\begin{equation}
D_{\cal {H}} = {\zeta}^{-1} \frac{i}{p^2 - 4 m_t^2} \ ,
\quad D_{{\cal {G}}^0} = D_{{\cal {G}}^{\pm}} = 
{\zeta}^{-1} \frac{i}{p^2} \ , \quad \mbox{where:   }
\ {\zeta} = \frac{ N_{{\rm c}} g_t^2}{(4 \pi)^2} \left[
\ln \frac{{\Lambda}^{\!2}}{p^2} + {\cal {O}}(1) \right] \ .
\label{lncprop}
\end{equation}
These propagators are obtained by
adding up the bubble diagrams of Fig.~\ref{rmd5f}(b) 
(in Sec.~\ref{NTLEE}),
$g_t $$=$$ M_0 \sqrt{G}$ of the TSM (\ref{TSM2}), and the mass
$m_t^{(0)}$ in the bubbles is given by the
quark loop gap equation (\ref{gaplead1})-(\ref{gaplead}) 
(with ${\Lambda}_{{\rm f}}$$=$$ {\Lambda}$). 
The ${\Lambda}^{\!2}$-term in the composite Higgs self-energy
in quark loop approximation
is in fact still there and is 
$\propto$$ 4 {N_{{\rm c}}} m_t^2 {\Lambda}^{\!2}/3$, just
as in the MSM. Blumhofer (1994) noted
the following interesting property: if we include in
the composite Higgs self-energy also corrections from
exchange of the composite scalars with leading-$N_{{\rm c}}$
propagators (\ref{lncprop}), as depicted in Fig.~\ref{rmd13f}, 
{\em and\/}
set $N_{{\rm c}}$$=$$3$, the ${\Lambda}^{\!2}$-terms cancel
\begin{equation}
\left[ \frac{4 N_{{\rm c}}}{3} m_t^2 - M_H^2 {\zeta}^{-1}
{\zeta} \left( \frac{2 m_t}{M_H} \right)^2 \right] {\Lambda}^{\!2}
\quad \left( \mapsto 0 \mbox{ for } N_{{\rm c}}=3 \right) \ .
\label{hierBlum}
\end{equation}
The first term is formally leading-$N_{{\rm c}}$
[${\cal {O}}({N_{{\rm c}}}^1)$] 
and the second next-to-leading [NTL,
${\cal {O}}({N_{{\rm c}}}^0)$] in $1/N_{{\rm c}}$ expansion, 
and for $N_{{\rm c}} $$\not=$$ 3$ the two terms don't cancel. 
\begin{figure}[htb]
\mbox{}
\vskip5.0cm\relax\noindent\hskip.0cm\relax
\includegraphics{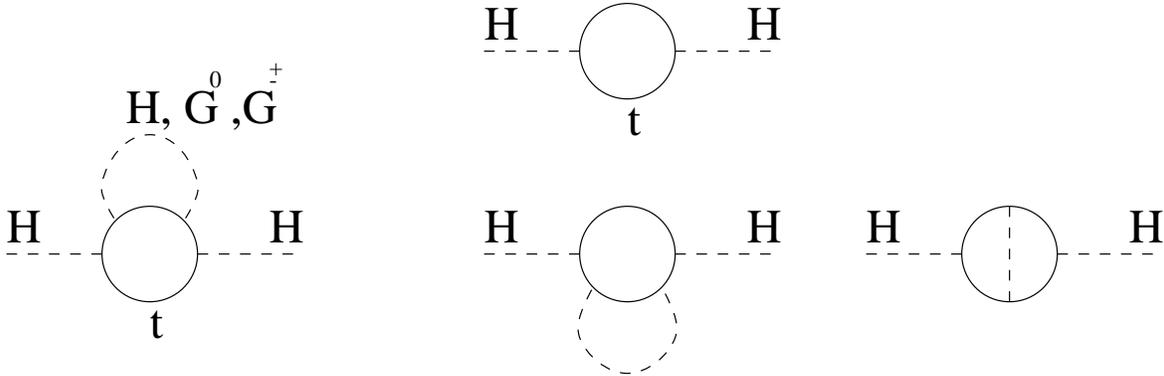} \vskip0.4cm
\caption{\footnotesize Diagrams for
self-energy of composite Higgs in the TSM: at the
leading-$N_{{\rm c}}$ (first graph), 
and at the NTL level (all four graphs).}
\label{rmd13f}
\end{figure}
In Sec.~\ref{NTLEE}, where the NTL contributions to the gap
equation were discussed, this effect was not investigated,
because only leading-$N_{{\rm c}}$ 
self-energies (propagators) of the composite scalars 
contributed to the NTL terms of the gap equation.
As seen from Fig.~\ref{rmd13f}, the diagrams corresponding to 
the NTL term in (\ref{hierBlum}) represent in fact an effective
one-loop-induced four-scalar vertex with two of the
four external legs connected into a (leading-$N_{{\rm c}}$)
propagator.
Blumhofer then showed in his subsequent work (1995) that the full
four-scalar vertex is induced just by the top quark-loop 
(cf.~Fig.~\ref{rmd1f}, where the top quark has $m_t $$\not=$$ 0$),
i.e., that all higher loop (higer in $1/N_{{\rm c}}$) corrections
cancel out. Consequently, also all higher than one-loop
(higher than NTL) corrections in the scalar self-energy cancel out. 
He showed this by means of a complicated system
of Dyson--Schwinger type of equations for the full induced
four-scalar vertices. This solution existed only for 
$N_{{\rm c}}$$=$$3$ and
$N_L $$=$$2$, where $N_L$ is isospin number
of the generalized weak isospin group $SU(N_L)_L$.
Therefore, the reason for cancellation of ${\Lambda}^{\!2}$-terms 
here has a fundamentally different character than 
in supersymmetry. It reminds one of the
conditions for the triangular anomaly cancellations.
The cancellation solution was obtained
by using a specific ansatz, but the hope is that it
represents the physical, i.e., the energetically most favored 
solution. Transverse degrees of freedom of the EW 
gauge bosons were not included in the analysis. 
Even if gauge bosons break cancellation
of ${\Lambda}^{\!2}$-terms, these terms can appear only as radiative
corrections and are therefore suppressed by factors
$g_1^2$ and/or $g_2^2$.
Hence, even in such a case the hierarchy problem
is substantially tamed and $\Lambda $$>$$ {\cal {O}}(1 \mbox{ TeV})$
could still appear without an excessive fine-tuning.

The described cancellation phenomenon of Blumhofer
suggests that the $1/N_{{\rm c}}$ expansion approach in
TSM should be regarded with caution.
Among other things, it appears to suggest that
the NTL effects are crucial
and cannot be ignored. Results
discussed in Sec.~\ref{NTLEE} offer a similar warning.
The analysis of Blumhofer suggests that, for at least
some predictions, it is wise to include the
leading-$N_{{\rm c}}$ and NTL contributions together,
and it holds out the prospect
that contributions beyond the NTL level could
be negligible. Implications of the 
discussed cancellation mechanism 
for $1/N_{{\rm c}}$ expansions in condensation frameworks
deserve further investigation. 
This type of cancellation and/or suppression may possibly occur also
in some other -- nonminimal -- frameworks, specifically those 
with an extended composite scalar sector. 
Unfortunately,
investigations in this direction have not been carried out.

\section{Summary and outlook}
\label{SO}
\setcounter{equation}{0}

The author hopes that this review has provided an outline of
research activities in the
physics of ${\bar t}t$ condensation.
It should be clear that there are
two main research directions:
\begin{enumerate}
\item
One direction focuses on effective four-fermion (NJLVL)
interactions at a compositeness scale ${\Lambda}$.
If some of these interactions
are strong enough, fermion-antifermion condensation
and dynamical generation of fermion (heavy quark) masses
as well as symmetry breaking (DSB) takes place.
\item
The other, more ambitious, approach focuses on constructing and 
investigating renormalizable models of underlying
physics above ${\Lambda}$, which effectively lead at 
``low energies'' $E $$\sim$$ {\Lambda}$ 
to four-fermion interactions and thus to condensation. 
\end{enumerate} 

The first group includes the
initially proposed minimal framework (e.g.,
modeled as truncated TSM)
in which the ${\bar t} t$ condensation is fully responsible
simultaneously for the mass $m_t$ and for the full
(dynamical) EWSB, thus leading to an effective
minimal SM (MSM) with one Higgs ${\cal {H}} $$\sim$$ {\bar t}t$.
This minimal framework appears to be ruled out
experimentally, by the measurement of
$m_t $$\approx$$ 175$ GeV. Namely, the minimal framework
predicts too high $m_t^{{\rm dyn.}} $$>$$ 200$ GeV
when the full DEWSB is implemented, as various
calculation methods indicate. 

To accommodate an acceptable value for
$m_t^{{\rm dyn.}}$ ($\approx$$ 175$ GeV)
while still saturating the EWSB ($v $$\approx$$ 246$ GeV),
various extensions of the minimal framework
involving additional scalars with nonzero VEV's
have been suggested (cf.~Secs. \ref{EWESG}--\ref{RMUP}).
The price one must pay is that the extended models
become more speculative and/or less predictive.
This is certainly the case for those
extensions of the minimal (TSM) framework which
don't involve an extension of the gauge symmetry
(cf.~Sec.~\ref{EWESG}). However, when the symmetry is enlarged
and the effective four-fermion NJLVL framework
is retained (cf.~Sec.~\ref{EGSG}), the physics
becomes very rich and can dynamically accommodate 
interesting scenarios, such as a fully dynamical
scheme for a L-R symmetric model.

The second major group of studies incorporating ${\bar t} t$
condensation represents more ambitious extensions of the minimal
framework -- those embedding ${\bar t}t$ condensation
in fully renormalizable models of underlying physics
at energies above the condensation scale ${\Lambda}$
(cf.~Sec.~\ref{RMUP}). Many of these extensions assume
that the effective four-quark coupling terms
responsible for $m_t^{{\rm dyn.}}$
emerge as a result of the exchange of very massive gauge
bosons ($M $$\sim$$ {\Lambda}$). While they
explain the large $m_t$ and possibly 
the EWSB dynamically, the question
of what mechanism is responsible for the large masses
of new gauge bosons is usually left open.
Hence, such models are not fully dynamical. Further,
most of these models do not address the issue
of too large $m_t^{{\rm dyn.}}$
when EWSB is fully saturated by the ${\bar t} t$
condensation. It appears that topcolor
assisted technicolor models (TC2, cf.~Secs. \ref{RMUP6}
and \ref{SPTAQ12}) are at this time the only renormalizable
frameworks which have been shown to satisfactorily cure
that problem. At the same time they are
more dynamical in the sense that DEWSB is induced
primarily by condensates (of technifermions) bound by
exchanges of {\em massless\/} new gauge bosons
(technigluons). 
TC2 models combine renormalizable models incorporating
${\bar t} t$ condensation [topcolor model(s), 
cf.~Sec.~\ref{RMUP3}] with the technicolor (TC)
and extended technicolor (ETC), curing simultaneously
some of the unsatisfactory features of the TC$+$ETC
models and of the minimal ${\bar t} t$ framework.
TC2 frameworks offer even scenarios explaining the
masses $M$$\sim$${\Lambda}$ of new heavy gauge
bosons\footnote{
Their exchanges lead to TSM four-quark interactions responsible
for ${\bar t}t$ condensation.} 
{\em dynamically\/} via DSB induced by condensation
of technifermion pairs.
Apparently the most intensive
investigations of resulting low energy phenomenology are
being carried out precisely for various TC2 models
(cf.~Subsec. \ref{SPTAQ12}), and some of these models
appear to be able to accommodate experimental phenomenology
(including the otherwise problematic FCNC
suppression constraints).
There are other renormalizable models which aim at
explaining dynamically mass hierarchies of 
SM fermions and the DEWSB {\em without\/} involving
TC$+$ETC, among them models possessing simultaneously
horizontal and vertical gauge symmetries (cf.~Sec.~\ref{RMUP5}).
However, it is not yet clear whether or not they
can be compatible with available (low energy) phenomenology.

Apart from the mentioned division of condensation models into
effective and renormalizable frameworks, there 
are also distinct classes 
of calculational approaches used in these models.
Applicability of perturbative 
RGE's plus compositeness conditions
(cf.~Sec.~\ref{MRGE}) in effective
frameworks is in general restricted
to the cases of high cutoff values ${\Lambda}$ because it
relies on an infrared fixed-point behavior.
This method, although being somewhat indirect, can be used 
for any particular realization of effective 
strong attraction at a large ${\Lambda}$, 
not just the TSM or other NJLVL four-quark interactions.
On the other hand, the approach with Dyson-Schwinger
(DS) and Bethe-Salpeter (BS) equations
($1/N_{{\rm c}}$
expansion) deals with the strong dynamics directly --
e.g., with NJLVL interactions in effective frameworks,
or with heavy particle exchanges in renormalizable frameworks.
This method has been generally applied to models 
with ${\bar t} t$ condensation only in the
leading-$N_{{\rm c}}$ approximation
(cf.~Sec.~\ref{MDSPS}),
employing instead of the leading-$N_{{\rm c}}$
BS equations their approximate sum rules
called Pagels-Stokar (PS) relations. 
The main drawback of the
DS$+$BS approach is the problem of
gauge noninvariance (or: viability of the
choice of the Landau gauge; cf.~Sec.~\ref{MDSPS5}),
and possibly a restricted application range
for the $1/N_{{\rm c}}$ expansion.
This expansion may either not be
predictive (if ${\Lambda} $$\gg$$ m_t$),
or the next-to-leading (NTL) effects can be reasonable 
but still very strong (for ${\Lambda} $$\sim$$ 1$ TeV)
-- cf.~Sec.~\ref{NTLEE}. A full NTL analysis
would have to include an NTL version of the Bethe-Salpeter
(BS) equation for the composite Nambu--Goldstone boson,
but this is still lacking.
For larger ${\Lambda}$'s 
($10^3$GeV $<$$ {\Lambda} $$<$$ 10^8$ GeV),
the analysis could either use at NTL order the DS$+$BS
equations in its functional form beyond the
``hard mass'' framework [${\Sigma}_t $$\equiv$$ m_t({\Lambda}) 
\mapsto {\Sigma}_t({\bar p}^2)$], or 
combine NTL gap equation with perturbative RGE evolution.
In principle, 
the viability of the minimal framework for low ${\Lambda}$'s,
although unlikely,
is still an open question, since
a consistent nonperturbative analysis applicable
to the range of low ${\Lambda}$'s (${\Lambda} $$<$$ 10^8$ GeV)
has not been carried out.
In this respect, a recently proposed
alternative method -- an approximation scheme
for the nonperturbative renormalization group 
(NPRG; cf.~Sec.~\ref{MRGE5}) approach --
is also promising, since it appears to avoid 
many of the
deficiencies of the $1/N_{{\rm c}}$
expansion and of the perturbative RGE approaches.

In conclusion, the author wishes to reemphasize 
the importance of studying
renormalizable and effective
models involving ${\bar t} t$ condensation.
It remains a challenge to
find a framework fully consistent with
experimental data while
offering a dynamical explanation of those
sectors of the SM which are at
the moment the most mysterious -- the
electroweak symmetry breaking (EWSB) and the mass
hierarchies of fermions, including possibly the
structure of CKM mixing. Of course, such investigations
will gain additional motivation
if a Higgs particle is detected {\em and\/}
direct or indirect experimental indications
for a composite nature of the Higgs are found.
The author also wishes to stress the need
for systematic nonperturbative calculational methods
which would be applicable to the cases
where the onset scale ${\Lambda}$ of new
physics responsible for the condensation
is not exceedingly far
above the electroweak scale, e.g., 
when ${\Lambda} $$<$$ 10^8$ GeV.

{\bf Note added.\/} 
Here we briefly mention some works that appeared
after the present article was accepted for publication.

Blumhofer, Dawid and Manus (1998) investigated, with RGE approach
arguments, the role of higher than six dimensional quark operators 
as well as the related role of heavy scalar and heavy vector exchanges
in ${\bar t} t$ condensation models. Dawid (1998) applied
a similar approach in studying a condensation mechanism in supersymmetry
such that the scalar composites are made entirely of sfermions.

In the approach with Dyson-Schwinger (DS) equation, and using
Ward-Takahashi identity, Hashimoto (1998a) derived a formula for 
the mass of the composite Higgs in terms of the fermionic 
$m_f^{\rm dyn.}$. Further, he (1998b) employed the DS$+$PS approach
(cf.~Sec.~III) to calculate $m_t^{\rm dyn.}$ in the minimal ${\bar t}t$
condensation framework, this time by including composite Higgs boson
loop effects in the DS equation (next-to-leading effects in $1/N_{\rm c}$)
and by employing simultaneously the Pagels-Stokar (PS) equation 
(i.e., leading-$N_{\rm c}$ effects only).

Holdom and Roux (1998) investigated the next-to-leading (in
$\alpha$) contributions to the gap equation when the exchanged
gauge boson is massive; they found out that these contributions
are large and thus throw doubt upon the expansion, as well as upon the
Most Attractive Channel hypothesis (cf.~Sec.~VIII.C).
When the gauge boson is massless, these contributions are small
(Appelquist {\em et al.\/}, 1988; cf.~Sec.~IV.E).

Dobrescu and Hill (1998), and Chivukula, Dobrescu, Georgi and
Hill (1999) introduced and studied a framework in which dynamical
EWSB is fully driven by a condensate involving the top quark
and an additional isosinglet quark $\chi$; the observed (i.e., low enough)
$m_t$ is then still allowed due to a seesaw mechanism between
the top quark and $\chi$. 

Chivukula and Georgi (1998a, 1998b) studied
topcolor and topcolor assisted technicolor (TC2) models by going
beyond the large-$N_{\rm c}$ and the NJLVL approximations; they
concluded that the strength of the ``tilting'' $U(1)$ gauge 
interaction has an upper bound, and that, as a consequence, the topcolor
coupling must be equal to the critical value to within a few per cent.

Lindner and Triantaphyllou (1998) studied a left-right symmetric
framework in which the ${\bar t}t$ condensate is assisted
in the dynamical EWSB by heavy mirror fermions which, in turn,
keep the electroweak parameters $S$ and $T$ under control.

Andrianov, Andrianov, Yudichev and Rodenberg (1997) developed further
the two-Higgs-doublet model in which the two composite Higgs particles
result from four-quark interactions which include derivatives
(cf.~also Sec.~VI.A.4).

Production rates and detection of
neutral scalars and charged (pseudo)scalars at colliders,
as predicted by various ${\bar t} t$ condensation frameworks (2HDM,
topcolor, TC2), were studied by 
Bal\'azs, He and Yuan (1998), 
Bal\'azs, Diaz-Cruz {\em et al.\/} (1999),
Choudhury, Datta and Raychaudhuri (1998),  
Diaz-Cruz, He, Tait and Yuan (1998), 
He and Yuan (1998), 
and Spira and Wells (1998).
Contributions of light scalars in such models to the ${\bar t}t$ 
production in photon colliders were investigated by 
Zhou {\em et al.\/} (1998).

\vspace{1.cm}

\section*{Acknowledgments}
\addcontentsline{toc}{section}{Acknowledgements}

\vspace{0.5cm}

The author acknowledges that E.~A.~Paschos had suggested 
to him to write this review article. The author acknowledges
useful discussions with U.~Sarkar and T.~Hambye
related to aspects of supersymmetry, and of
chiral perturbation theory
as low energy QCD, respectively. 

At middle stages of this work, the author was supported
in part by the Bundesministerium f\"ur Bildung,
Wissenschaft, Forschung und Technologie,
Project No.~057DO93P(7).

\end{document}